\documentclass[12pt,twoside,openright,a4paper]{book}

\usepackage{graphicx, hyperref, setspace}
\usepackage[portuguese,english]{babel}
\usepackage[acronym, toc]{glossaries}

\usepackage[utf8]{inputenc}
\usepackage{color,fancyhdr,framed,lettrine,setspace,listings,array}
\usepackage[OT1]{fontenc}   
\usepackage{mathtools,dsfont}
\usepackage{pdflscape}
\usepackage{authblk}
\usepackage{xcolor}
\usepackage{comment}
\usepackage{braket}
\usepackage{amsthm}
\usepackage{amsfonts}
\usepackage{amssymb}
\usepackage{amsmath,bm}
\usepackage{enumitem}
\usepackage{subeqnarray}
\usepackage{diagbox}
\usepackage{makecell}
\usepackage{physics}
\usepackage{float}
\usepackage[font=small,labelfont=bf]{caption}
\usepackage{geometry}

\usepackage{sectsty}
\usepackage{siunitx}
\DeclareSIUnit{\belmilliwatt}{Bm}
\DeclareSIUnit{\dBm}{\deci\belmilliwatt}
\newcommand*\diff{\mathop{}\!\mathrm{d}} 
\newcommand{\mathbbm}[1]{\text{\usefont{U}{bbm}{m}{n}#1}}
\newcommand{\Id}{\mathbbm{1}} 
\allsectionsfont{\normalfont\sffamily\bfseries}

\renewcommand\vec{\mathbf}
\newcommand{\trans}{\mathsf{T}} 
\newcommand{\reals}{\mathbb{R}}
\newcommand{\complex}{\mathbb{C}}
\newcommand{\naturals}{\mathbb{N}_0}
\newcommand{\hilbert}{\mathcal{H}} 

\newenvironment{dedication}
  {
   \thispagestyle{empty}
   \vspace*{\stretch{1}}
   \itshape             
   \raggedleft          
  }
  {\par 
   \vspace{\stretch{3}} 
   \clearpage           
  }

\makeglossaries
\providecommand{\keywords}[1]{\textbf{Keywords:} #1}
\newcommand*\NewPage{\newpage\null\thispagestyle{empty}\cleardoublepage}
\newcommand {\abstractEnglishPageNumber} {\thispagestyle{plain}\setcounter{page}{\abstractEnglishPage}}
\newcommand {\abstractPortuguesePageNumber} {\thispagestyle{plain}\setcounter{page}{\abstractPortuguesePage}}
\newglossaryentry{MSc}{name={MSc}, description={Masters degree in the area of Science.}}
\newglossaryentry{CRB}{name={CRB}, description={Cramér-Rao bound.}}
\newglossaryentry{qCRB}{name={CRB}, description={Quantum Cramér-Rao bound.}}
\newglossaryentry{QFI}{name={QFI}, description={Quantum Fisher information.}}
\newglossaryentry{MSE}{name={MSE}, description={Mean squared error.}}
\newglossaryentry{iid}{name={iid}, description={Independent and identically distributed (applies to measurements).}}
\newglossaryentry{ML}{name={ML}, description={Maximum likelihood.}}
\newglossaryentry{QFT}{name={QFT}, description={Quantum field theory.}}
\newglossaryentry{QM}{name={QM}, description={Quantum mechanics.}}

\newacronym{IST}{IST}{Instituto Superior T\'ecnico}

\newcommand\abstractname{Abstract}  
\makeatletter
\if@titlepage
  \newenvironment{abstract}{%
      \titlepage
      \null\vfil
      \@beginparpenalty\@lowpenalty
      \begin{center}%
        \bfseries \abstractname
        \@endparpenalty\@M
      \end{center}}%
     {\par\vfil\null\endtitlepage}
\else
  \newenvironment{abstract}{%
      \if@twocolumn
        \section*{\abstractname}%
      \else
        \small
        \begin{center}%
          {\bfseries \abstractname\vspace{-.5em}\vspace{\z@}}%
        \end{center}%
        \quotation
      \fi}
      {\if@twocolumn\else\endquotation\fi}
\fi
\makeatother

\begin{document}

\pagenumbering{gobble}
\clearpage
\thispagestyle{empty}

\newcommand {\Title} {\LARGE{\textbf{Sensing and Communication with
Quantum Microwaves}}}
\newcommand {\StudentName} {{Mateo Casariego Vias}}
\newcommand {\Supervisors} {{\large \textbf{Supervisor:} Doctor~Yasser~Rashid~Revez~Omar}}
\newcommand {\DegreeName} {Thesis approved in public session to obtain the PhD Degree in Physics}
\newcommand {\Classification} {{\large \textbf{Jury final classification:} Pass with Distinction}}

\def \includeAcknowledgments{1}

\def \includeGlossary{0}

\newcommand {\Chairperson} { \textbf{Chairperson:} Doctor José Luís Rodrigues Júlio Martins, Instituto Superior Técnico, Universidade de Lisboa}
\newcommand {\CommitteeMembers} {
\normalsize \textbf{Members of the Committee:}\\
{ \normalsize Doctor Stefano Pirandola, University of York, UK}\\
{ \normalsize Doctor Matteo Paris, Dipartamento di Fisica ``Aldro Pontremoli'', Università degli Studi di Milano, Itália}\\
{ \normalsize Doctor João Carlos Carvalho de Sá Seixas, Instituto Superior Técnico, Universidade de Lisboa}\\
{ \normalsize Doctor Yasser Rashid Revez Omar, Instituto Superior Técnico, Universidade de Lisboa}
}

\newcommand {\FundingInstitutions}{\large \textbf{Funding Institution:} Fundação para a Ciência e a Tecnologia (FCT)\\
}

\newcommand {\Month} {July}
\newcommand {\Year} {2023}

\def \acknowledgmentsPage{1}

\def \abstractEnglishPage{3}

\def \abstractPortuguesePage{5}


\newgeometry{margin=2cm}


\begin{titlepage}
\begin{figure}[t]
   \includegraphics[width=5cm]{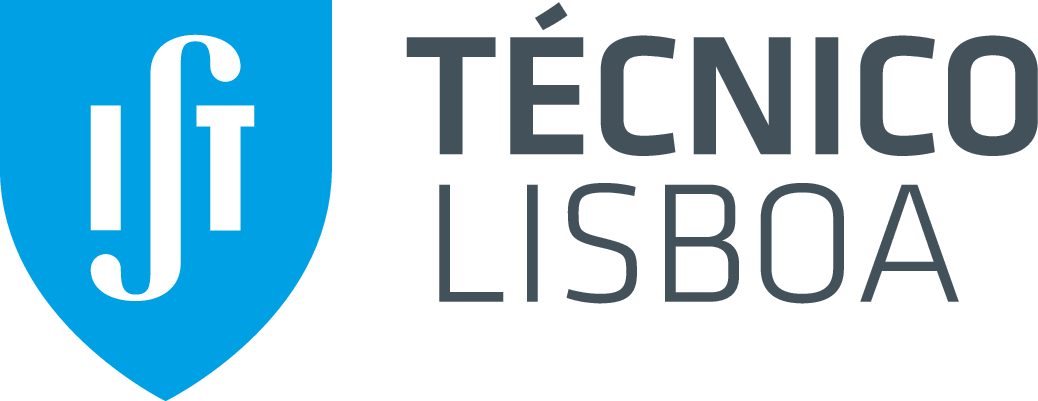}
\end{figure}

\centering
{\Large
\textbf{UNIVERSIDADE DE LISBOA}\\
\textbf{INSTITUTO SUPERIOR TÉCNICO}\\}

~\\[2.0cm]
\vspace{1cm}
{\textbf{\Title}}\\[1.0cm]

{\Large \textbf{\StudentName}}\\[2cm]


{\large \Supervisors}\\[4.0cm]

{\large \DegreeName }\\[1.0cm]

{\large \Classification }\\[3.0cm]

{\Large \textbf{\Year}}\\
\end{titlepage}
\NewPage


\begin{titlepage}
\begin{figure}[t]
   \includegraphics[width=5cm]{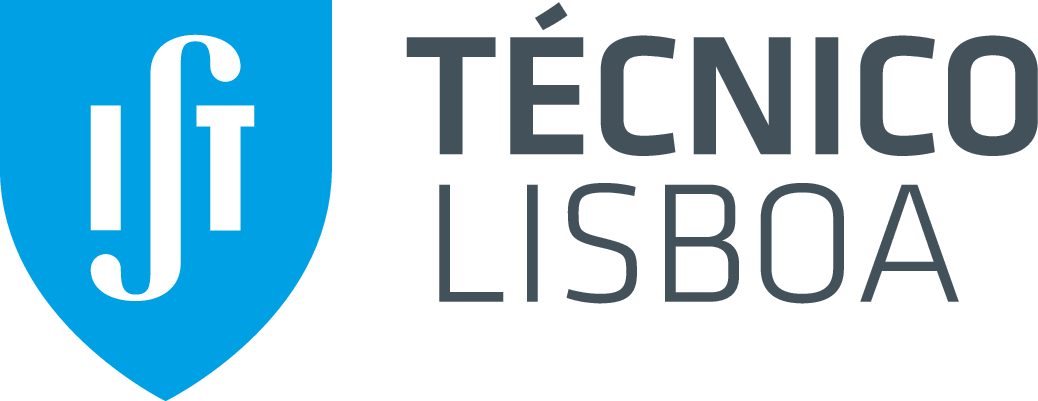}
\end{figure}

\centering
{\Large
\textbf{UNIVERSIDADE DE LISBOA}\\
\textbf{INSTITUTO SUPERIOR TÉCNICO}\\}

~\\[1.5cm]

{\textbf{\Title}}\\[1.0cm]

{\Large \textbf{\StudentName}}\\[1.2cm]


{\large \Supervisors}\\[1.4cm]

{\large \DegreeName }\\[1.0cm]
%
%
{\large \Classification }\\[1.0cm]
%
\begin{minipage}{0.9\textwidth}
{\begin{center}
    \large \textbf{Jury}
\end{center}} 
\flushleft
    {\Chairperson}
\end{minipage}\\[0.5cm]
%
%
\begin{minipage}{0.7\textwidth}
\flushleft
    {\CommitteeMembers }
\end{minipage}\\[1.0cm]
%
%
{\FundingInstitutions}\vspace{0.5cm}
{\Large {\Year}}\\
\end{titlepage}
\restoregeometry

\NewPage
   \begin{dedication}
    To Teo, who doesn't care either
    \par   
  \end{dedication}
\NewPage

\pagenumbering{roman}

\if\includeAcknowledgments 1

\chapter*{Acknowledgments}

As I write this, the thesis is still unfinished, it's a Wednesday, my back hurts, and my hands are slightly cold\footnote{As I write this footnote, the thesis is finished, it's a Tuesday, my back hurts a little, but my hands are warm.}. I recently opened a box that remained closed for years; one of those boxes where we lock memories, special objects. I have never been particularly good at retaining memories: I am one of those people for whom the help of a particular piece of paper, a photograph, or a song, is sometimes key to understanding themselves, their past, their history\footnote{Consequently, keep in mind the contents of this section form a strictly non-Abelian set (in a hierarchy-like sense), and that  I have not applied any time-ordering scheme to them. I also apologise for any omission. It is my subconscious that should be made responsible for these.}. If I was a Turing machine, we could say that the opening of this box triggered emotions and started a process that eventually halted in the materialisation of this very piece of text. Being a human, rather than a Turing machine, I leave some room as to whether I have really halted or not. I'm undecided. 

During the past few months, when the idea of writing the thesis was starting to leave the Platonic world to become real, I kept thinking that I would leave the Acknowledgments section blank. I was focusing on all the difficult moments I lived through, and the lonesome of my particular experience as a PhD student. I kept saying to myself that the human value I had found in these last years was so small, that I would simply skip saying thanks. There was nobody to really thank, because I was never really proud of my work. 

 My experience with Physics --and with pretty much anything, really, has been dominated by the impostor syndrome: never good enough, never hard-working enough, I was simply an outsider that could pretend very well. I kept acting. Eventually, I forgot that, indeed, there's always been something beating inside of me. Something that just wanted to understand things, and that found an unrivalled pleasure in anything that could be regarded as an approximation to some sort of truth.
 
 Then, years later, after an accident that left me with 4/5 senses, and having taken a break from the PhD, I decided to visit a friend that lived on the other side of the planet. One of the amazing things of this weird thing we call conscious existence is the fact that we can experience time in a different way to the strictly linear, strictly forward way Newtonian physics forced us to. I started tracing things back, somewhat unconsciously. I am an entirely different person from the one that started this journey, and at the same time, I'm not\footnote{\texttt{M.~(18:41): Dirías que lo que describo es un proceso dialéctico?  L.~(18:42): Pues, de alguna manera sí}}. I have decided I want to say thanks too, as everybody in their theses. I am no different from the rest, and so here we are. Writing this section is a healing process. Because I will forget some people, let me adapt a text from Pe Cas Cor, and say: gracias \textit{al que leído sea por esta dedicatoria}.

Let me start with the boring bits, the ones that have to do with money: I thank the support from the 
Doctoral Programme in the Physics and Mathematics of Information (DP-PMI),  the PQI -- Portuguese Quantum Institute, the Center of Physics and Engineering of Advanced Materials (CeFEMA), and Fundação para a Ciência e a Tecnologia (FCT) through scholarship PD/BD/135186/2017,  and IT -- Instituto de Telecomunicações Lisboa via the EU H2020 Quantum Flagship project QMiCS (820505). Ok, done. Now we continue with the story.
 
My supervisor, Professor Yasser Omar, should be the first person to say thanks to. During all these years, you have been very supportive, and I know for a fact that you have tried your best in order to enhance the quality of the academic environment. I'm particularly grateful for your quickness when it came to critical moments, and for your general openness and patience with me and my process. I think I've gained maturity and I thank you for that.
 
The day I knew I was accepted in the Doctoral Program that led to this thesis I remember there was a storm\footnote{I've been told that the day I was born there was a storm too, a similar one. Stormy beginnings!}. One of those storms we enjoy during the summer in Madrid. I was with friends, and they shared my happiness for getting the scholarship. I begin by thanking those friends: Juan, Paloma, Elena, Aga.  I thank my parents. I thank you together: thanks for the best childhood I could dream of. And I thank you apart. 
Mother: you taught me to be curious, to learn to experience by myself the joy of understanding, the joy of being alive. The hours-long conversations with you virtually everything, from philosophy to quantum phenomena, from history to artificial intelligence, have been the best training camp for enthusiastic, scientific discussions. Gracias, madre. Father: thanks for making me love insects, and for spending so much time with my sister and I, making us appreciate nature. But most importantly, thanks for showing me that love is stronger than anything, for showing me that a non-negative approach to reality is always there, for making me believe in individuals within families. Sometimes it felt that your lessons were taught via the technique of weak measurements. If you don't understand this analogy, ask me in person. Gracias, oso, eres un ejemplo en (casi) todo. I thank my sister, Clara, who came to the realm of existence bringing a dinosaur with her, from wherever she came. She was and continues to be the best example of an authentic enthusiast for life. You lack the pedantry of your older brother, and have an intuition for human emotions and relations that, beyond the great psychologist you are, makes you a person to look up to. You care about things, and you are brave to show the rest of the world all the things that matter. Clus, la psicóloga más cool que existe. Gracias también a Juan (tal), y a Tuku (guau).

I want to say thanks to the rest of my (very large) family in the way our grand mother taught us to cheer: \textit{a los ausentes, a los presentes, y a los futuros venideros}. Primero: Omi, Manolo, Pedro y Carmen. Some of you will never read these words, but my hope is that you will feel them somehow. I've been thinking about you, and I am proud to carry some (qu)bits of your personalities in me. Thanks to Cook, who left the most beautiful mark on the grass I've ever seen. I thank Antón and Berta \footnote{Esteeselcontestadordeantónybertayasabesloquetienesquehacer-ycuándo!-tantantataan.}: If I can do half of the good stuff you've done with me, with my (future) nephews, I'll be satisfied. I think you are close to being the perfect couple. I thank my uncle Juan for educating me in photography, through a notebook he gave me back in 2007. This created a passion that has served me to balance my analytic verbiage. Juan, an example of generosity and elegance within simplicity. I thank Tía María for, Juana and Manuela: three strong women that are a pillar of the family. Gracias a Sira, por regalarme libros divulgativos de ciencia, por tu interés; gracias a Tío Martín, por hacernos reír; gracias a Nico, por recordarme que no se puede estar sino atentos.
I thank Ana, Julieta, and Reshma, especially in these last times when their help was so important for the conclusion of the thesis. 
Thanks to Marcos, Miguel and Juan: cousins that had the bad luck of having me as a math teacher sometimes. Gracias a mis hermanos Darío y Elías: your company has made me realise many important things in life, maybe without you knowing; I admire you and I have always been pulled upwards in your company. I also thank Sara, their mother, for her  interesting thoughts on people, including myself. Gracias a Maruja también, siempre algo pilla, y a la familia Font por tratarme tan bien verano tras verano. Gracias también a Ignacio Opacio, amigo mío desde el comienzo de esto.

Teo, to whom this thesis is dedicated, has been an important witness of many of my struggles during the last years. You were a tiny puppy back then, but didn't think twice when the world appeared all covered in snow that morning of January 2021, when Filomena taught us that Madrid was not prepared for that amount of snow, but also that most people that inhabit the `accelerated city' decided it was better to enjoy than to complain about the unpredictable. Te dedico esta tesis porque te importa un pimiento, como debe ser.
María is also a very important part of this chapter of my life: you brought the music back to me. I owe you a helmet. Gracias también al fantasma de la Tía Cris, que creo que habitó durante un tiempo mi casa y fue divertido.

Thanks to Marco, who taught me to play backgammon when I arrived to Lisbon, among many other, less logical things. Gracias, Marco, por tu acogida, y ese colchón en tu salón, y también por ese cuarto de supuesto artista cuántico que me conseguiste en ZDB (thanks to Natxo Checa as well). Together with Borja, and Ignacio, we shared good parties and good music. From this time, I thank Claudinha too, that showed me around, and shared her dreams with us. Gracias también a Fati, por ser casa, por tu fuerza. 

I want to thank all the people I met at Técnico back in 2018: Alireza Tavanfar, Marco Pezzutto, João Moutinho, Gonçalo Frazão, João Sabino, Diogo Cruz, Duarte Magano, Bruno Mera: you all are really inspiring company. Thanks for sharing coffee, thoughts, dinners, and late-night office hours during the first years. Thanks to Fátima Barata, who was always there to help with a smile. Thanks to the people I met in the parking lot of Santos, especially to Manuel and Rafa. I hope our paths will cross again, who knows. I also thank Prof. José Leitão for inviting me from time to time to his office at the eleventh floor of Torre Norte, to have a coffee and talk about sculpture, old Lisbon stories. I will always remember your smiling eyes, and the calmness of those silences.
Thanks to Conrad:good memories, running along the Tejo at night, from our flat in Alcântara to Torre de Belém and back; having to take cold showers because it was the end of the month and we had run out of money.
Two great individuals need special mention here as well: Riccardo Gozzi and Francisco Lima. I hear your voices as I write your names. Your personalities, charisma, and intelligence were a really fun challenge to take. Ricardo Faleiro also falls in this category of smart guys that can out-think you in a matter of seconds. I enjoyed a lot every one of our conversations, having lunch, probably fish, and pretending to be two intellectuals concerned with the interpretation of relativistic quantum phenomena\footnote{Well, I should speak for myself only\ldots}. 
I also thank Sol, that suffered some of my spatio-temporal contradictions during those times. Gracias por tu vitalidad y tu risa, contagiosas las dos.
I thank the organisers of the Gravity in the Quantum Lab Workshop that took place in Benasque in the summer of 2018. And one of my heroes, Carlo Rovelli, for agreeing with me, a nobody, during one dinner when by chance I sat next to the big fish: I interrupted a quite renowned philosopher of science who was not really getting the point Carlo was trying to make about the nature of entanglement, and tried to give a practical example. Carlo looked at me and said: `he's right!'. That meant a tiny but important boost in my confidence, but more importantly, I took it as a sign that physics is still an area where the authority fallacy can be challenged, something which sadly not all of us can do when dealing with tricky situations of power within Academia. I thank Thiago Lucena de Macedo Guedes and Yuval Rosenberg for those nice hikes. 

My visits to the Walther Meissner Institut, in Garching, were also rewarding, for which I thank Frank Deppe and Kirill Fedorov; Michael Renger was very welcoming, and made the stays more fun. I want to thank Lea Vajda as well for taking me to the Schumann's, there in Munich, on that summer night. In the context of the QMiCS collaboration, I'd like to thank all of my co-authors, especially Emmanuel Zambrini and Tasio González-Raya, for being so actively involved while having a million other, more important things to do.

Special thanks to Mikel Sanz: every time we had an interaction, both in Bilbao and here in Lisbon, I would regain confidence, interest, and will to finish the damn PhD. Without your help I would not have done it, nor learned many of the techniques used in the derivation of the results. Thanks for inviting me so many times to Bilbao, and for believing in me. It has meant a lot, and I will always be grateful. 

Thanks as well to Vahid Salari for being in such a good mood no matter what. It was nice to share an office there in Bilbao. I thank Javier, Carlos, Ana, Tasio, Max, Arturo, Íñigo, Cristina, for the good times; and to Lucas Céleri for discussing GR-related stuff while having GR-sized beers, whatever that means. 

My beloved Serdamen, who has been by my side for so many years, deserves a special mention here: gracias por hacerme reír tanto siempre. Thanks to my BF Ángela Suárez, for making me grow. To Lati, for giving me energy in key moments: thanks to you I met Raymundo, a new friend (eeeeeh). Thanks to Lucas for your warmth, to Felipe for your openness and your sticker, to Neuza for taking such good care of me in Berlin the same way you took care of that bird in Colares. To Karol, for our long friendship. Thanks to Matteo Guarnaccia, Guillermo Trapiello, Fer Valenti, and again Ángela and Juan Feo for being my allies.
Before and after my accident, I want to thank Lucía Bayón (eran seis motoristas que eran motos).
I want to thank Toni Abril for believing in me and for being patient with a physicist that wanted to sell property using crocs instead of normal shoes. I really had a good time at Soane Praz, and both you and Ana were key in my recovery. 

Special thanks to Inês: I would not be here writing this without your help. You know. Obrigado sempre. 

Gracias también a Semba. The mere thought of the fact that you were there, somewhere, existing, gave me strength, again and again. You make me believe in myself.

I felt lucky during these last, key days, sharing an office with Nadir, Sabrina and Griffith. Though brief, our interactions have meant a lot to me, and I look forward to having more time-travel conversations in a year or so, in Brazil, as promised. Again, thanks to Diogo Cruz, Marco, Bruno, Matteo Turco, and Miguel Murça, for including me in all the cool telegram groups. At the moment I cannot add much content to the one dedicated to hortas, but hopefully I will, as soon as I find a place with a garden.

Obrigado, Paula. Obrigado, Zé. Obrigado, Portugal.

And\footnote{Last but not least.} I thank you, Laura. Gracias por devolverme la vida, por recordarme que soy alguien divertido, por empujarme a cuidarme, a prestar atención, a disfrutar de no tener razón. Por ir à frente. Amo-te.


\NewPage
\fi


\begin{otherlanguage}{english}
\begin{abstract}
\thispagestyle{plain}
\abstractEnglishPageNumber

Research efforts in quantum technologies are proven to establish interdisciplinary connections, where theory, experiment and industry meet. Teams across the globe are actively harnessing quantum theory to solve real-world problems, eventually leading to the emergence of other, new quantum technologies, applications of which span from faster and more efficient computation to higher communication security and enhanced sensing capabilities.  

The thesis establishes a link between theoretical foundations and practical applications in the emerging field of propagating quantum microwaves. Although the concrete focus of the main results lies in specific quantum communication and sensing protocols, the thesis also gives a self-contained  introduction to quantum parameter estimation and Gaussian quantum continuous variables, justifying the theoretical results used. Motivated, firstly, by the compatibility between superconducting circuits (a promising quantum computing platform), and the microwave frequency range; and, secondly, by the transparency window of the atmosphere to these frequencies, the thesis contains two concrete contributions to the fields of microwave quantum sensing and communication: a novel protocol that uses frequency entanglement to measure the first order dependence in frequency of the reflectivity of an object; and a large investigation on different aspects of entanglement degradation due to loss and atmospheric absorption in the context of continuous-variable entanglement distribution for the task of quantum teleportation in open-air.  

The thesis includes, as a last chapter, an article that reviews the state-of-the-art in quantum microwave technology, and proposes different research lines, including quantum communication between Earth-based stations and satellites, quantum radar, direct dark matter detection, and exploration of the quantum properties of the cosmic microwave background.
This more speculative, yet rigorous chapter closes with a roadmap for possible future research lines in the field of propagating quantum microwaves, that can serve as an outlook of the thesis.

\begin{flushleft}

\keywords{Quantum information, quantum sensing, quantum communication, quantum microwaves, quantum parameter estimation, quantum continuous variables.}

\end{flushleft}

\end{abstract}
\end{otherlanguage}
\NewPage

\begin{otherlanguage}{portuguese}
\begin{abstract}
\abstractPortuguesePageNumber

Os esforços de investigação na área de tecnologias quânticas têm  comprovadamente estabelecido ligações interdisciplinares, através das quais, teoria, experimentação e indústria se cruzam. Por todo o globo encontramos equipas que recorrem à teoria quântica para resolver problemas do mundo real, eventualmente conduzindo ao aparecimento de novas tecnologias quânticas, cujas aplicações compreendem uma computação mais rápida e eficiente, uma maior segurança comunicativa, melhoradas capacidades de sensoriamento, entre outras.

A presente tese procura estabelecer uma ligação entre a base teórica e as aplicações práticas na área emergente da propagação das microondas quânticas. Apesar dos resultados principais se focarem essencialmente em protocolos  específicos de comunicação e metrologia quânticas, a tese também contém em si mesma uma introdução à estimação de parâmetros quânticos e às variáveis contínuas Guassianas, justificando assim os resultados teóricos usados.

Motivada, em primeiro lugar, pela compatibilidade entre os circuitos supercondutores (uma das plataformas de computação quântica mais promissoras) e a gama de frequências de microondas; e, em segundo lugar, pela transparência natural da atmosfera a estas frequências, a tese contém duas contribuições específicas em metrologia e comunicação quântica por microondas: um novo protocolo teórico que utiliza o entrelaçamento em frequência para medir a dependência em frequência de primeira ordem da reflexividade de um objeto; e uma investigação sobre como evitar a perda de entrelaçamento associada à absorção atmosférica no contexto da distribuição continuamente variável no entrelaçamento livre. 

A tese inclui, como último capítulo, um artigo que revê o atual estado da arte da tecnologia quântica de microondas, e propõe diferentes linhas de investigação, incluindo a comunicação quântica entre estações terrestres e satélites, o radar quântico, a detecção direta de matéria escura, ou a exploração das propriedades quânticas do fundo cósmico de microondas.

Este capítulo, mais especulativo mas rigoroso, termina com um plano pormenorizado de possíveis linhas de investigação futuras no domínio da propagação quântica de microondas, que pode servir de epílogo à tese.

\begin{flushleft}

\textbf{Palavras-chave: }Informação quântica, detecção quântica, comunicação quântica, microondas quânticas, estimação quântica de parâmetros, variáveis contínuas quânticas

\end{flushleft}

\end{abstract}
\end{otherlanguage}
\NewPage

\tableofcontents
\NewPage


\addcontentsline{toc}{chapter}{\listfigurename}
\listoffigures
\NewPage

\printglossary[type=\acronymtype]
\NewPage

\pagenumbering{arabic}




\chapter{Introduction}
\label{chapter:introduction}
A common phenomenon in theoretical physics history is the fact that mathematical machinery needs to wait several years for a physicist to use it to try and solve their problems. A classic example is non-Euclidean differential geometry and general relativity: around 1912 Einstein was in desperate need of some \textit{tool} to help him represent his revolutionary ideas about spacetime. The mathematics needed came to him through his friend Grossmann, when he learned about the results by Riemann (who died in 1866) \cite{pais1982subtle}. It was only then that he could formulate the ideas in a precise way, accommodating more recent results by Levi-Civita, and Ricci, among other mathematicians. In a way, mathematics was `ahead' of physics. Another example is the work of Lie and Galois on group theory, and the study of symmetries, which was done mostly during the 19th century, and that had an enormous impact in helping build the `rigorous' formulation of quantum mechanics, pursued by Hilbert, Weyl, and others. A third, more recent example is information theory and statistics: the work by Shannon, Fisher, and others, in the 1940s and 1950s, has had a tremendous impact in quantum science, creating a new way of formulating quantum mechanics using information-theoretic concepts, together with the previous algebraic results: this led to quantum information (1980s) and quantum computation (1990s). 

In parallel, experimental demonstrations of quantum phenomena had been taken place: from the pioneering work on the photoelectric effect, and particle-wave duality by Millikan and others in the early 20th century, to the sequence of experimental tests to violate Bell's inequalities that culminated in the celebrated loophole-free 1.3 \si{\km} Bell test experiment in 2015 \cite{Hensen2015}, we can say today, with quite some confidence, that quantum theory is sufficiently well `proven' experimentally.
The 2022 Nobel prize in physics acknowledged the efforts of Clauser, Aspect and Zeillinger to bring quantum theory, and in particular quantum entanglement, out of the theory papers, and into the lab.

Currently, quantum theory continues to be subject of an immense collective   investigative effort. Theorists keep refining the theory, the concepts, and the methods, digging deeper into the meaning and reach of the ideas, and sometimes connecting different parts that seemed disjoint at the beginning; experimentalists, on the other hand, nowadays even lifted by industry, are starting to try and make \textit{use} of the theory, thinking of ways to solve real-life problems with the new tools. When this happens, new technologies are born. So the sequence: mathematics-physics-technology is, in a way, a materialisation of the idea of how scientific ontology can become useful for society. Technology, then, can be used to set the grounds for the next step in this strange, iterative, collective effort humanity has been doing over the last two centuries and that, hopefully, will be put in use with ethical considerations upfront. 

This is precisely the point we are now: quantum technologies are becoming a reality. They rest on three pillars: computation, communication, and sensing. Industrial and political interest is justified: quantum computation has an enormous potential, not only in the promise of faster-than-classical, efficient computation, but also for simulating quantum systems (drug design comes to mind); quantum communication could bring unprecedented security due to its inherent sensitivity towards eavesdropping, hence the interest from banks, and some of its protocols will have to be incorporated in the quantum internet; and quantum sensing, which goes beyond any classically-achievable resolution, promises sci-fi-like ideas like quantum radar, direct detection of dark matter, or exploring the quantum properties of the cosmic microwave background.

In this thesis we aim at giving our modest contribution to quantum communication and sensing, with focus on quantum microwaves. The field of propagating quantum microwaves is an emerging technological platform that finds its most direct motivation in the fact that superconducting circuits, which operate at cryogenic (millikelvin), emit radiation that fits in the GHz window. And superconducting circuits are one of the most promising quantum computing platforms. Therefore, if one wishes to set up a communication network between such chips, it makes sense to be able to have a technology toolbox that is able to exploit the quantumness of these propagating fields. Moreover, the atmospheric transparency window for microwave frequencies makes them suitable candidates for other tasks, such as quantum communication between Earth-based stations and satellites, or for sensing, including, but not limited to, quantum radar, or imaging. Recent advances in microwave-dedicated photocounters encourages us more to investigate this relatively new research field, as counting photons out of an electromagnetic field is one of the key aspects of any quantum-related protocol, be it in communication or in sensing.

The first part of the thesis is devoted to setting a sufficiently rigorous and self-contained introduction to quantum theory, continuous variables, and quantum parameter estimation.  Hence, in Chapter \ref{chapter:QuantumIntro}, we introduce the mathematical foundations of Hilbert spaces, and then proceed to assign a quantum interpretation to them, focusing on Bosonic continuous variables. Although the experienced reader can skip most of this chapter, it helps at providing a good theoretical background. 
Along these lines, in Chapter \ref{chapter:Estimation} we move on to present the theory of parameter estimation, starting from the classical version, based on many important information-theoretic results that were then used to build up the theory of quantum parameter estimation, an extremely convenient tool both from the pragmatic and conceptual point of view: indeed, the theory is very much connected with the fundamental problem of what a measurement is, what is the maximum achievable precision, and so on. Naturally, these areas pose very interesting questions regarding the very nature of statistics and the inherent probabilistic aspect of quantum mechanics. In particular, we review some important results regarding Gaussian quantum parameter estimation, a tool that can simplify enormously the computation of certain quantities, such as the quantum Fisher information.

As a link between this chapter and the following, we use the case example of quantum illumination to confirm that the methods are able to reproduce known results with somewhat less effort than previous works.

The first original results of the thesis are presented in Chapter \ref{chapter:biFreq}. These results have been published in Ref.~\cite{casariego2020}, and consist on a novel quantum-enhanced protocol for the measurement of a parameter related to the reflectivity of an object. It is somewhat inspired by quantum illumination, and it represents a link between the two approaches mentioned previously in this introduction: combining advanced topics in Gaussian quantum parameter estimation with a pragmatic approach, we find optimal observables for the quantities to be measured, and propose a simple realisation in terms of quantum optics transformations and operations, with microwaves in mind, but not excluding other implementations.

In Chapter \ref{chapter:MWentanglementDistribution} we present work done in collaboration with T. Gonzalez-Raya \cite{GonzalezRaya2022}, and published in Ref.~\cite{GonzalezRaya2022}, on the problem of distributing continuous-variable entanglement for the task of implementing quantum teleportation in open-air. This is our contribution to quantum communications. Specific considerations for microwaves, to find ways to overcome atmospheric entanglement degradation are given, and in particular, we discuss entanglement distillation, and swapping protocols, making use of photon-subtraction and taking into account the state of the art in microwave photon-counters for propagating modes.

Finally, the last chapter, Chapter \ref{chapter:Roadmap}, taken from the published work in Ref.~\cite{Casariego2022}, explores the field of propagating quantum microwaves and its applications as of today, both in communication and sensing. We argue the necessity for a controllable quantum microwave interface, and explain why  is still in early stages. The paper emphasizes the need for a fully functional toolbox and highlights potential research directions, including microwave quantum key distribution, quantum radar, bath-system learning, and direct dark matter detection. By examining these applications, the paper underscores the wide range of possibilities and future prospects in the field of quantum microwaves. For this, we construct a roadmap for future research lines in the field of propagating quantum microwaves.  This chapter may be seen as an epilogue for the thesis. Finally, all figures which are not exclusively mine have been credited using the initials of the  authors in their corresponding papers. I thank all of them again for letting me use them.

\chapter{Quantum mechanics and quantum continuous variables}
\label{chapter:QuantumIntro}

In this chapter we introduce the basic notions of quantum mechanics and quantum continuous variables. Regarding the latter, in this chapter we will simply give the most basic definitions, since more results can be found in section \ref{section:quantum_CV} of Chapter \ref{chapter:MWentanglementDistribution}. The purpose of the chapter is twofold: to fix the notation and to have as self-contained a thesis as possible. 

Most traditional texts on quantum mechanics build the mathematical foundation using Hilbert spaces and self-adjoint operators. We shall not depart from this formulation, since it is good enough for our purposes and, after all, the work presented in this thesis is not of any foundational nature. Yet, we would like to stress that other, more axiomatic approaches do exist, like the one based on $C^*$-algebras \cite{landsman1998, Gleason2009}. Algebraic approaches to quantum theory, as carefully explained in the book by R. Haag (see Ref. \cite{haagBook}), attain more generality, at the cost of higher mathematical sophistication. They,  are closer in spirit to quantum gravity, since they tend to make explicit the existence of a causal background on top of which one starts defining things. 

The goal of this chapter is, again, to give an introduction to quantum continuous variables that is easy, but by no means comprehensive. The book by Serafini \cite{serafini} covers the topic with a very nice balance between pragmatism and rigor, both conceptually and mathematically speaking. Quantum continuous variables are a mathematical tool, not a physical theory. As such, they can be used to describe a wide range of phenomena, quantum light or optics being its most notable example.
Loosely speaking, quantum optics is a laboratory-oriented realisation of a particular quantum field theory (QFT): quantum electrodynamics (QED). Quantum field theory is often taught in theoretical courses, and even in its most basic form (say a single, free Bosonic field), it usually fails in making sense of the theory from a pragmatic, experimental point of view: measurements are sometimes blurred by an overwhelming succession of elegant techniques and Gaussian integrals. Quantum optics, while at heart uses the same fundamental assumptions of QED (after all, it talks about photons, the quanta of the electromagnetic field), puts a lot of emphasis on operations, time evolution, and, ultimately \textit{detection}. Naturally, detection involves light-matter interaction, which can be studied using, say, Feynman diagrams, hence also `included' in QED; yet, the focus of quantum optics is not generality, but usefulness. Quantum optics can be seen as a pragmatic theory, because it `voluntarily' removes Poincaré invariance, brings attention to \textit{what} and \textit{how} things are measured, assuming, as traditional quantum mechanics, the existence of a stationary frame of reference.  

So, to summarize: let's not forget that quantum optics is \textit{not} a Lorentz-invariant QFT: its foundations can be derived from --the more fundamental-- theory of QED. If one wants to really understand QFT, not to mention to bring general-relativistic considerations into the picture, they \textit{must} read elsewhere. In my opinion, the following four books should form a good team, together with a curious reader, for the task of trying to make sense of relativistic quantum theory \cite{serafini, haagBook, peresBook, waldBook}. 

\section{Mathematical prerequisites}
Quantum mechanics is a physical theory whose mathematical arena is represented by some \textit{space}. Elements of this space are then given some physical meaning, and everything that happens in the theory can be seen as maps and relations between these elements. In the same way as Newtonian mechanics takes place --mathematically speaking-- in a six-dimensional Euclidean space, quantum mechanics takes place in Hilbert space\footnote{But do keep in mind what Asher Peres wrote: \textit{``Quantum phenomena do not occur in a Hilbert space. They occur in a laboratory.''} \cite{peresBook}}. 

\subsection{Hilbert space}
Hilbert spaces are special types of vector spaces.  We define Hilbert space $\mathcal{H}$ as a \textit{complete}\footnote{A complete vector space is a space such that every Cauchy sequence converges to a limit, and this limit also lies in the vector space. Whether this is naturally achieved, or done `by hand' depends on the specific case.} vector space on the complex field\footnote{The complex field is the set of complex numbers together with addition and multiplication operations; that is, a set satisfying the closure, associativity, and commutativity axioms, as well as containing both the identity operation (0 for addition, 1 for multiplication), and an inverse for every element $a$ (except $a=0$).} that is endowed with an inner product. We will use Dirac's notation, so a generic element (vector) of $\mathcal{H}$ will be denoted by a `ket' $\ket{\cdot}$. For example, if the dimension of $\mathcal{H}$ is $n$, then $\ket{a}$ may be identified with a (row) vector $\mathbf{a} = (a_1, \ldots, a_n)^\trans$ and components $a_j\in \mathbb{C}$ with $j=1, \ldots, n$. The inner product maps any two elements to a scalar and is denoted $\braket{\cdot}{\cdot}: \mathcal{H}\times  \mathcal{H} \rightarrow \mathbb{C}$, where the `bra' $\bra{\cdot}$ is actually the \textit{dual} of $\ket{\cdot}$. The inner product satisfies the following properties for all $\alpha, \beta\in \mathbb{C}$, and $\ket{a}, \ket{b} \in \mathcal{H}$:
\begin{enumerate}
    \item $\braket{a}{a} \geq 0$
    \item $\braket{a}{b+c} = \braket{a}{b} + \braket{a}{c}$
    \item $\braket{a}{\alpha b} = \alpha \braket{ a}{ b} $
    \item $\braket{a}{b} = \braket{b}{a}^*$
\end{enumerate}
Equality of property 1. holds only when $\ket{a} =0$; properties 4 and 5 imply $\braket{\alpha a}{\beta b} = \alpha^* \beta\braket{ a}{ b}$, with $\alpha, \beta\in \mathbb{C}$, a property that sometimes goes by the name of `sesquilinearity'. Importantly, the fact that there exists an inner product automatically makes a Hilbert space a \textit{normed} space. Indeed, the inner product induces a natural norm for vectors: $\norm{\ket{a}} := \sqrt{{\braket{a}}}$.  This means that Hilbert spaces are also complete metric spaces, which incidentally  allows to introduce the notion of `distance' between vectors in the Hilbert space. However, the immediate distance $d(\ket{a}, \ket{b}):= \norm{\ket{a}-\ket{b}}$ will not be the one of our choice, since we aim at a more general, \textit{operator} norm that allows for more flexibility and that, eventually, can be related to problems such as quantum parameter estimation, state discrimination, \textit{etc.}, relevant for the contents of this thesis and further explained in Chapter \ref{chapter:Estimation}.

\subsubsection{Completeness}
With the definition of Hilbert space given above, its dimension may be finite, countable infinite, or non-countable. 
It is quite common in introductory textbooks and courses on quantum information to take for granted that the Hilbert space is $\mathbb{C}^2$, because the usual focus there is on two-level systems (\textit{qubits}, the unit of binary quantum information), and so continuous variables  are often left out. When $\mathcal{H} = \mathbb{C}^{2n}$ (\textit{e.g.} the case of $n$ qbits) a possible inner product is 
\begin{equation}
    \braket{a}{b} = \sum_{j=1}^{2n} a^*_j b_j.
\end{equation}
In the case of a countable infinite dimension, like when the elements of $\mathcal{H}$ are functions: 

\begin{align*}
     f: \reals &\longrightarrow \complex \\
     x &\longmapsto f(x),
\end{align*}
a possible inner product is given by 
\begin{equation}
    \braket{f}{g} = \int_\reals \diff x f^*(x) g(x).
\end{equation}
Note that for this to be an inner product, integrals should converge for any two $f, g \in \mathcal{H}$ and therefore we must require (property 1. above):
\begin{equation}
    \braket{f}{f} = \int_\reals \diff x f^*(x) f(x) \equiv \int_\reals \diff x \abs{f(x)}^2
\end{equation}
to exist and be non-negative. 
Consequently, we will further require that in the case of countably many infinite dimensions every $f$ to be properly `completed' by `adding' to its corresponding Cauchy sequence a well-defined limit. Further details can be found in Chapter 4 of \cite{peresBook} and Chapter 3 of \cite{friedrichs1973spectral}.

\subsubsection{Separability}
In principle, both when  the dimension of the Hilbert space is finite or (countably) infinite, there should exist a numerable basis of vectors, which we denote  $\lbrace \ket{\mathsf{e}_j} \rbrace_{j\in \mathbb{N}}$, and whose elements are assumed to be orthonormal: $\braket{\mathsf{e}_j}{\mathsf{e}_k} = \delta_{jk}$. Bear in mind that this is an extra requirement: not all Hilbert spaces are separable! Technically, separable Hilbert spaces are the ones that contain a countable, dense\footnote{This is a limitation in `size' requirement, but not necessarily in the Haussdorf sense. An interesting result, out of the scope of this introduction but perhaps worth mentioning for further reference is the Baire category theorem, which in a way strengthens the ideas of Hilbert spaces as metric spaces.} subset.
This means that there exists a subset of $\hilbert$ in which every element is arbitrarily `close' to other elements in $\hilbert$. In other words, there exists a sequence of vectors in $\hilbert$ such that any other vector can be approximated as `well as needed' by some linear combination of the sequence.

For example, the Hilbert space of square-integrable functions $L^2(I)$ where $I\subset \reals $ is a compact interval is a separable Hilbert space, whereas  $L^2(\reals)$ is not.

\subsection{Linear operators}\label{subsec:LinearOps}
Linear operators on a Hilbert space will be denoted by a hat: $\hat{A}$. They are simply maps from $\hilbert$ to itself satisfying linearity: $\hat{A}(\beta\ket{b}+\alpha\ket{a}) = \beta\hat{A}\ket{b}+\alpha \hat{A}\ket{a}$. Their properties are central for the quantum-mechanical interpretation of Hilbert spaces. For a separable Hilbert space of dimension $d$ we can always represent a linear operator $\hat{A}$ as a matrix, via an orthonormal basis $\lbrace \ket{\mathsf{e}_j} \rbrace_{j=1}^d$:
\begin{equation}\label{eq:LinearOperator}
    \hat{A} \ket{\mathsf{e}_j} = \sum_{k=1}^d A_{jk} \ket{\mathsf{e}_k}.
\end{equation}

\subsubsection{Types of linear operators relevant for quantum mechanics}

An operator $\hat{N}$ is \textit{normal} if $\hat{N}^\dagger \hat{N} = \hat{N} \hat{N}^\dagger$. Normal operators commute with their adjoint (see below), which means that they both share their spectra (\textit{i.e.} they can be simultaneously diagonalised).  An normal operator  $\hat{U}$ is \textit{unitary} if $\hat{U}^\dagger \hat{U} = \hat{\Id} = \hat{U} \hat{U}^\dagger$.

A \textit{self-adjoint}\footnote{In infinite dimensions, self-adjoint and `Hermitian' are not the same. A self-adjoint operator is Hermitian only on a dense domain of $\hilbert$. The topology with respect to which we say `dense' is that induced by the inner product.} linear operator $\hat{A}$ is a normal operator such that $\hat{A} = \hat{A}^\dagger$, where $\hat{A}^\dagger$ denotes the adjoint of $\hat{A}$ and defined via $\braket{a}{\hat{A}b} = \braket{\hat{A}^\dagger a}{b}^*$. In terms of matrices, the adjoint of $\hat{A}$ is
\begin{equation}
    \hat{A}^\dagger \ket{\mathsf{e}_j} = \sum_{k=1}^d A^*_{kj} \ket{\mathsf{e}_k},
\end{equation}
\textit{i.e.} the matrix coefficients of $\hat{A}^\dagger$ are obtained by transposing  $\hat{A}$ and then taking the complex conjugate of its coefficients. Hence, $\hat{A}$ is self-adjoint when $A_{jk}= A^*_{kj}$. Note that every self-adjoint operator is normal.

An operator$\hat{A}$ is \textit{positive} if $\bra{a}\hat{A}\ket{a}\geq 0$ for any $\ket{a}$ in $\hilbert$.

Finally an operator $\hat{A}$ is said to be \textit{bounded} when there exists a maximum $m$ such that $\norm{\hat{A}\ket{a}}\leq m \norm{\ket{a}}$ for any $\ket{a}\in \mathcal{H}$. 

\subsubsection{Trace and trace-class operators}\label{subsubsec:trace}
The trace of a matrix $A$ is simply the sum of its diagonal elements:
\begin{equation}
    \Tr [A] = \sum_{j=1}^d A_{jj}.
\end{equation}
Due to Eq. \eqref{eq:LinearOperator}, the trace of a linear operator $\hat{A}$ in a separable $\hilbert$ is
\begin{equation}\label{eq:TraceLinearOperator}
    \Tr [\hat{A}]:=  \Tr [A] = \sum_{j=1}^d  \bra{\mathsf{e}_j}\hat{A}\ket{\mathsf{e}_j}.
\end{equation}
The trace is linear, cyclic, and, most importantly for us, basis-independent. This means the trace of a linear operator is an invariant, \textit{if it exists}! Again, the infinite-dimensional case $d\rightarrow \infty$ requires us to be more careful, since for example the identity operator $\hat{\Id}$ of an infinite-dimensional Hilbert space has an infinite trace, according to definition in Eq. \eqref{eq:TraceLinearOperator}. For certain quantum mechanical operators in infinite dimensions, the further requirement of being `trace-class' will come in handy. 
An operator $\hat{A}$ is trace class with trace $c$ if
\begin{equation}
    \Tr \left[ \hat{A} \right] = c,
\end{equation}
where $c \in \mathbb{R}$. When $c=1$ we say that the operator is `normalised', or trace-class with trace one. The linear subspace of trace-class operators in $\hilbert$ with trace $c$ is denoted $\mathcal{T}_c(\hilbert)$. Note that trace-class is a stronger requirement that implies boundedness.

Let us stress that in the case of unbounded linear operators, the trace may not be even defined, a relevant fact for the case of the position and momentum operators, as we shall see.
\subsubsection{Spectral decomposition theorem}
Although the full potential and consequences of this theorem in its most general form, that is, for bounded or unbounded operators on a finite or infinite-dimensional Hilbert space lies outside the scope of this thesis, it's worth stating the theorem. It says that any normal operator $\hat{N}$ of a separable Hilbert space $\hilbert$ can be expressed as:
\begin{eqnarray}
    \hat{N} = \int_{\sigma(\hat{N})}  \lambda \hat{\pi}(\diff\lambda),
\end{eqnarray}
where $\sigma(\hat{N})$ is the \textit{spectrum} of the operator, $\lambda$ are the eigenvalues and $\hat{\pi}(\diff\lambda)$ is a \textit{projection-valued measure}, which loosely speaking, is a function that, to each (measurable) subset $s$ of the spectrum $\sigma(\hat{N})$, assigns an orthogonal projection (that is, a self-adjoint, idempotent operator). Essentially, this measure decomposes the space of operators in a Hilbert space in `sufficiently large bits' that are mutually orthogonal, allowing to assign an eigenvalue of a given (normal) operator to each of those `projective bits'. In other words, normal operators can be diagonalised.

The key here is that this result applies even to unbounded operators, which are at the core of quantum continuous variables (as we will see, position and momentum operators, whose commutation relation sets the starting point of quantum continuous variables, are unbounded), and for which we should be able to talk about their spectrum even if it is the continuum $\reals^{+}$. Moreover, this theorem allows us to compute expectation values of trace-class operators because it gives us a resolution of the identity, essential for computing expectation values of quantum mechanical objects in a trace fashion, as we shall see later.

Finally, the theorem can be extended to (suffiently well-behaved) functions of operators as well:
\begin{eqnarray}\label{eq:OperatorFunction}
    f(\hat{A}) = \int_{\sigma(\hat{A})}  f(\lambda) \hat{\pi}(\diff\lambda),
\end{eqnarray}

\subsubsection{Operator inner product, norms, and distances}
Because we shall not adopt the ray version of quantum states (which corresponds to identifying quantum states with vectors $\ket{\psi}$), but rather use the density operator formalism, a richer generalisation, we will now give some definitions in terms of operators that will be useful for the next sections.

The Hilbert-Schmidt inner product between two operators $\hat{A}$ and $\hat{B}$ is given by 
\begin{equation}
    \langle\hat{A}, \hat{B}\rangle := \Tr(\hat{A}^\dagger \hat{B}) = \sum_j \bra{\mathsf{e}_j}\hat{A}^\dagger \hat{B} \ket{\mathsf{e}_j},
\end{equation}
where we have assumed the Hilbert space to be separable (hence admitting a numerable ONB). This inner product, which is linear, conjugate-symmetric, and positive definite, automatically induces a norm, the Hilbert-Schmidt, or  Frobenius norm, which is simply:
\begin{equation}
\norm{\hat{A}}_{\text{HS}} := \sqrt{\Tr(\hat{A}^\dagger \hat{A})}
\end{equation}
Other relevant norm is the trace norm, which we will strictly reserve to \textit{bounded} operators, and that is the special case $p=1$ of the Schatten $p$-norms:
\begin{equation}\label{eq:TraceNorm}
    \norm{\hat{A}}_1 := \Tr\left( \sqrt{\hat{A}^\dagger \hat{A}}\right),
\end{equation}
where $\sqrt{\hat{A}^\dagger \hat{A}}$ denotes an operator $\hat{T}$ such that $\hat{T}^2 = \hat{A}^\dagger \hat{A}$. In the case of a diagonalisable $\hat{A}$ (such as a normal operator), $\norm{\hat{A}}_1$ equals the sum of the absolute values of its eigenvalues.

\section{Assigning quantum meaning to a Hilbert space}
We now can proceed to give a quantum-mechanical interpretation of the different structures described above. We shall not get into philosophic interpretations, and take a pragmatic, axiomatic approach that suffices for the purposes of this thesis.
\subsection{States: density operator}
Quantum states $\rho$ are associated with elements of $\mathcal{B}(\mathcal{H})$, which denotes the set of linear, self-adjoint, positive, trace-class with trace 1 operators on $\mathcal{H}$. Note that this is the only operator that will not wear a hat, but this is just an exception to the rule. The operator $\rho$ is called the density operator.

Note that sometimes it is assumed that positivity ($\bra{a}\rho\ket{b}\geq 0$) implies self-adjointness ($\rho = \rho^\dagger$), which striclty speaking is not always the case. As far as this thesis is concerned, we don't need to pay much attention to this technical nuance, but for further reference it seems worth mentioning that the context is that of `extensions' of operators, a topic that should be available in most functional analysis books. 

We can use the spectral decomposition theorem, being $\rho$ normal (as it is self-adjoint):
\begin{equation}\label{eq:SpectralDecompositionRho}
    \rho = \int_{\sigma(\rho)}  \lambda \hat{P}_\lambda\diff\lambda
\end{equation}
Positivity means that $\sigma(\rho)\ni \lambda \geq 0$ and the normalisation condition (trace-class with trace one) implies $\int_{\sigma(\rho)} \lambda  \diff\lambda = 1$. The projectors in Eq. \eqref{eq:SpectralDecompositionRho} can be rewritten as $\hat{P}_\lambda = \ket{\psi_\lambda}\bra{\psi_\lambda}$ using Dirac's notation. Each $\ket{\psi_\lambda}\bra{\psi_\lambda}$ represents what is called a `pure state', that is, a quantum state that has just one non-zero eigenvalue, equal to one. This means that, even when the representation in Eq. \eqref{eq:SpectralDecompositionRho} is not unique, the most general quantum state is a convex combination of pure states, interpreted as an ensemble, or statistical mixture of pure states.

 Note that the density operator formalism is not incompatible with the `usual' way of assigning a just a ket $\ket{\psi}$ to a quantum system. In fact, this is just a particular case which corresponds to 
a \textit{closed} systems. Closed quantum systems are therefore represented by a \textit{ray} in Hilbert space, which is an equivalence class of vectors: Two vectors $\ket{\psi}$ and $\ket{\psi^\prime}$ represent the same state if $\ket{\psi} = c\ket{\psi^\prime}$, where $c\in \mathbb{C}_{\ne 0}$. Since we require states to be normalised, \textit{i.e.} $\braket{\psi}{\psi}=1$, then the c-number $c$  that relates two vectors belonging to the same equivalence class takes the form $c =e^{i\phi}$. Because $\ket{\psi}$ and $e^{i\theta}\ket{\psi}$ represent the same state, we say that global phases like $\theta$ \textit{are physically meaningless} (they cannot be measured as such, which will become clearer after we introduce the notion of measurement).\\
So, a pure state is the quantum state of a closed system and it's represented by a ray in a Hilbert space.

A simple example to make the case for the --apparently redundant-- density operator formalism over the ray version, and to start seeing its full potential: Let's say that we prepare a collection of atoms in some energy level. The system is initially closed and characterized by some $\ket{\psi}$.  At some point, we open a small hole in its boundaries, and new atoms in a different state $\ket{\phi}$ leak into our system, mixing with  $\ket{\psi}$ (and not interacting with it). 
Let's say we are quite smart, and we are able to obtain an estimate of the relative abundances of  $\ket{\psi}$'s and $\ket{\phi}$'s. Our estimation can be arbitrarily good. \\
However, we are asked to pick an atom, randomly, so in principle we don't know whether a specific one will be in state $\ket{\psi}$ or $\ket{\phi}$. All we know is that it is in $\ket{\psi}$ with probability $p$ and in $\ket{\phi}$ with $1-p$. It seems tempting to say that our system \textit{is} in the state: 
\begin{equation}\label{eq:coh_sup}
\sqrt{p}\ket{\psi} + \sqrt{1-p}\ket{\phi}.
\end{equation}
But this is wrong. It is wrong because what happened has nothing to do with the preparation of a coherent superposition, and \eqref{eq:coh_sup} \textit{is} a coherent superposition, \textit{i.e.} something `very' quantum. What happened is that we were forced to use \textit{classical} probabilities to account for our ignorance. We could have done better. Let's not assume that due to our ignorance the physical state has suddenly become a coherent superposition!\\
The correct description of the situation is to consider a density operator like
\begin{equation}
{\rho} = p \ket{\psi}\bra{\psi} + (1-p) \ket{\phi}\bra{\phi},
\end{equation}
which is precisely a (classical) statistical mixture of two pure states, $\ket{\psi}\bra{\psi}$ and $\ket{\phi}\bra{\phi}$, each occuring with probability $p$ and $1-p$, respectively.

\subsection{Composite systems}
For simplicity, for what follows, we assume that out density operator takes the following form
\begin{equation}
    \rho = \sum_j p_j \ket{\mathsf{e}_j}\bra{\mathsf{e}_j}
\end{equation}
where $\lbrace \mathsf{e}_j \rbrace_j$ is an orthonormal basis and $\sum_j p_j =1$ with $p_j \geq 0$ for all $j$.

Now, composite systems are associated with a tensor-product-structured Hilbert space $\bigotimes_{j=1}^n \mathcal{H}_j$. We will only consider bipartite systems in this thesis (\textit{i.e.} $n=2$).
The physical motivation for studying composite systems comes from the idea of correlations: In most situations, when we look at a bipartite system is because we think the two parties share {something}. They either interacted in the past, or we think they are going to interact at some point. This may sound trivial, but it is just to illustrate the fact that this is very similar as what happens in classical physics: there is little interest in setting up a theoretical description for \textit{simultaneously} solving the equations of motion of a snooker ball in Bristol and a free-falling bifana \cite{bifana} at Lisbon's Estrela garden. Such a study only makes sense if we want to check whether or not these two events actually share something. So \textit{correlations}.
Thus, when we have two subsystems in quantum mechanics, the user's manual tells us that the Hilbert space of whole is described by the tensor product of the individual Hilbert spaces. \textit{Labelling} each Hilbert space $A$ and $B$, we have that the density operator describing both should live in the space $\mathcal{B}({H}_{AB})$ with
\begin{equation}
\mathcal{H}_{AB} = \mathcal{H}_{A} \otimes \mathcal{H}_{B}.
\end{equation}
It is important to note that the labels $A$ and $B$ need to correspond to physically distinguishable things, like a location in space, or any other degree of freedom that can be assigned a constant meaning during the time we study the system. The relation of this to the Kochen-Specken theorem \cite{KochenSpecker1968}, contextuality, \textit{etc} is beyond the scope of this thesis.

\textit{Partial states} of a bipartite system ${\rho}_{AB}$ are the parts that correspond to $A$ and $B$. These are obtained by \textit{tracing out}\footnote{Tracing out is \textit{not} a physical operation. The partial trace is mathematical tool that we interpret as (voluntary or involuntary) ignorance about the part that is traced out.} the other part.  Mathematically, the way to obtain the partial states is by partial tracing: ${\rho}_{A} = \Tr_B ({\rho}_{AB})$ and ${\rho}_{B} = \Tr_A ({\rho}_{AB})$. Partial trace over, say, subsystem $B$, is obtained by sandwiching and summing over an orthonormal basis for $B$, $\lbrace \ket{\mathsf{f}_i}_B \rbrace_{i=1}^{d_B}$
\begin{equation}
\rho_A = \sum_i ^{d_B}  {}_B\bra{\mathsf{f}_i}  {\rho}_{AB}  \ket{\mathsf{f}_i}_B,
\end{equation}
where $d_B := \text{dim}\hilbert_B$
Partial tracing and `ignoring' can be taken as synonyms: it is therefore a subjective operation that depends on the state of knowledge of the entity in charge of describing, not on the system per se.

\subsection{Separability, entropy and purity}
We say that a state ${\rho}_{AB}$ is \textit{separable}\footnote{This notion of separability of quantum states has nothing to do with the separability of Hilbert spaces sketched in the previous sections.} when ${\rho}_{AB} = {\rho}_{A}\otimes{\rho}_{B}$. When this is not possible, we say that the state is \textit{entangled}.

Imagine that we are given a bipartite state $\ket{\psi}_{AB}$ and we want to see if it's separable or not. Unless we immediately see a way to separate, the best thing to do is the following: i) Write down the density operator ${\rho}_{AB}=\ket{\psi}_{AB}\bra{\psi}_{AB}$; ii) Compute the partial traces  to get  ${\rho}_{A}$ and  $\rho_{B}$; iii) Compute $\Tr({\rho}^2_{A})$ and check whether it is one or less. Do the same for subsystem $B$; iv) If you get one it means the subsystems are pure states, meaning that the bipartite state is not entangled. Intutively: if tracing out the degrees of freedom of one subsystem does not affect the amount of information that we can have about the other part, then there are no `internal' correlations between both subsystems. On the other hand, if tracing out the other subsystem introduces some (classical) randomness into our description of the other subsystem (making it a mixed state), then we have lost something in the way: there was information that was shared meaning that the total information content is not the sum of the individual parts.

If $\lbrace \ket{\mathsf{e}_i}_A \rbrace_{i}$ and $\lbrace \ket{\mathsf{f}_j}_B \rbrace_{j}$ are two orthonormal bases, then an orthonormal basis for $\mathcal{H}_{AB}$ is $\lbrace \ket{\mathsf{e}_i,\mathsf{f}_j}_{AB} \rbrace_{(i,j)}$, where
\begin{equation}
  \ket{\mathsf{e}_i,\mathsf{f}_j}_{AB} \equiv \ket{\mathsf{e}_i}_A\otimes\ket{\mathsf{f}_j}_{B},
 \end{equation} 
 and $(i,j)$ stands for the Cartesian product of the sets of variables $i,j$.  Here, $i$ and $j$ are assumed to be elements of two finite sets of integers $I$ and $\Gamma$ with cardinalities $\abs{I} = \text{Dim}(\mathcal{H}_A)$ and $\abs{\Gamma} = \text{Dim}(\mathcal{H}_B)$. This is just a fancy way of saying that we are dealing with quantum states that can be put in a finite number of states. Note that there are exactly $\text{Dim}(\mathcal{H}_A) \times \text{Dim}(\mathcal{H}_B)$ different pairs $(i,j)$.\\
We can write an arbitrary state $\ket{\psi}_{AB}$ as
\begin{equation}
\ket{\psi}_{AB} = \sum_{i,j} \alpha_{i j} \ket{\mathsf{e}_i,\mathsf{f}_j}_{AB},
\end{equation}
with the usual normalisation condition:
\begin{equation}
\sum_{i,j} \abs{\alpha_{i j}}^2 = 1.
\end{equation}

There is a nice theorem in linear algebra called the Schmidt decomposition theorem. It will prove very useful here. It says that every bipartite pure state $\ket{\psi}_{AB}$ can be expressed in its Schmidt decomposed form: 
\begin{equation}
\ket{\psi}_{AB} = \sum_{i=1}^r\sqrt{p_i} \ket{\mathsf{e}_i,\mathsf{f}_j}_{AB}.
\end{equation}
where the $\lbrace\ket{\mathsf{e}_i}_A\rbrace$ and $\lbrace\ket{\mathsf{f}_j}_B\rbrace$ are bases of their respective Hilbert spaces,  $p_i>0$, and $r\leq \min (d_A, d_B)\geq 1$, with $d_A$ and $d_B$ the dimensions of the Hilbert spaces. The number of nonzero $p_i$'s (that is, $r$) is usually called the \textit{Schmidt number} (or \textit{Schmidt rank}). The proof of the Schmidt decomposition theorem can be found in any textbook. It is usually obtained using the singular value decomposition of invertible matrices. 

For the specific case of a pure bipartite system $\ket{\psi}_{AB}$  written in Schmidt form we have:
\begin{equation}
{\rho}_A = \sum_{i=1}^r  p_i  \ket{\mathsf{e}_i} \bra{\mathsf{e}_i}
\end{equation}
and 
\begin{equation}
{\rho}_B = \sum_{i=1}^r  p_i  \ket{\mathsf{f}_i} \bra{\mathsf{f}_i},
\end{equation}
that is, they have the \textit{same} non-zero eigenvalues in the Schmidt bases, which are the ones that \textit{diagonalise} the partial states. \\
This gives a way to explicitly find the Schmidt decomposition for a given pure bipartite system: find the reduced states, and then diagonalise their corresponding matrices, finding the eigenvalues and the eigenvectors.

The quantity that tells us how pure a state ${\rho}$ is, is the trace of its square, and it is called \textit{purity}:
\begin{equation}
\mu(\rho):=\Tr ({\rho}^2).
\end{equation}
For pure states, this must equal one trivially, as we have seen. For mixed states, it has to be less than one:
\begin{equation}
\Tr ({\rho}^2)=\Tr \left( \sum_{a,b} p_a p_b \ket{a}\braket{a}{b}\bra{b}\right)=\Tr \left( \sum_a p^2_a \ket{a}\bra{a} \right)=\sum_a p^2<\sum_a p =1.
\end{equation}

\subsection{Measurements, statistics, and quantum channels}\label{subsec:MeasurementsStatisticsQuantumChannels}
Quantum-mechanical measurements are associated with positive operator-valued measures (POVMs), \textit{i.e.} a set $\{ \hat{K}_m \}_m$ of operators (called Kraus operators), that is \textit{complete}, that is, such that:
\begin{equation}
    \sum_m \hat{K}_m^\dagger \hat{K}_m =\hat{\Id}.
\end{equation}

If a state is prepared in $\rho$, the probability of obtaining measurement outcome $m\in \reals$ is
\begin{equation}\label{eq:BornRule}
p(m) = \Tr(\hat{K}_m^\dagger\hat{K}_m{\rho}),
\end{equation}
commonly known as the Born rule. The conditional state after an ideal, non-destructive measurement with outcome $m$ is
\begin{equation}
{\rho}^\prime (m) = \frac{\hat{K}_m{\rho}\hat{K}^\dagger_m}{p(m)}.
\end{equation}

Therefore, a quantum measurement can be seen as a (positive, convex) map from quantum states $\rho$ to classical probabilities $p(m)$, that is, as a quantum-to-classical randomisation process. A more general, mathematically rigorous definition of POVM does exist, which is more infinite-dimensional-friendly. As usual, there are conceptual benefits at the cost of a higher abstraction level. Again, we point the reader to the previously given references on (quantum) functional analysis for these. 

Additionally, for evolution, as well as for quantifying entanglement, a more general notion of \textit{quantum channel} may be introduced, as a linear, trace-preserving (normalised), completely positive\footnote{Which implies positivity. It turns out that the stronger requirement of complete positivity is needed in order not to run into non-physical transformations, as we will see below when explaining transposition.},  map that sends density operators to density operators, \textit{i.e.}, a superoperator $\Phi: \mathcal{B}(\mathcal{H}) \longrightarrow \mathcal{B}(\mathcal{H})$. $\Phi$ is completely positive if $\mathcal{I}_n\otimes \Phi$ is positive for all $n\geq 0$, where $\mathcal{I}_n$ is the identity superoperator of dimension $n$. $\Phi$ is trace-preserving when $\Tr[\Phi[\rho]] = \Tr[\rho] = 1$. Quantum channels are often referred to as CP-maps (completely positive maps), or CPTP-maps (completely positive, trace-preserving maps). They can account for every `deterministic'\footnote{This means excluding things like postselection, which leads to heralded phenomena, which are by nature not linear.} quantum process that sends physical states to physical states, and they hence include the special case of unitary evolution of closed quantum systems. A more infinite-dimensional-friendly version is given in terms of the Choi-Jamio{\l}koswki isomorphism \cite{CHOI1975285, JAMIOLKOWSKI1972275}, in section 5.5.2 of \cite{serafini}.

\subsection{Entanglement measures: partial transposition, and logarithmic negativity}

Although the definition of entanglement is clear and applies to both pure and mixed states, \textit{quantifying} bipartite entanglement is quite delicate when it comes to mixed states.

The von Neumann entropy of a \textit{pure}, bipartite state $\rho$, that can be seen as a quantum-mechanical version of Shannon's entropy for classical random variables, can be expressed as
\begin{equation}
    S (\rho) := - \Tr[\rho \log_2 \rho],
\end{equation}
is an entanglement measure and it's quite easy to compute.
But what happens if $\rho$ is a general, bipartite density operator? In the context of continuous variables and Gaussian phenomena, transposition, or partial transposition --which does \textit{not} represent a physical operation, plays an important role, and, for the purposes of the thesis, defining the (logarithmic) negativity will suffice. If we are given a bipartite densitiy operator $\rho_{AB}$, its partial transposed version is denoted $\tilde{\rho}_{AB}$. As in most cases involving entanglement inspection, we work by contradiction. Imagine that our bipartite state is separable and in some statistical mixture

\begin{equation}\label{eq:separableBipartite}
    \rho_{AB} = \sum_{j=1}^d p_j (\rho^{(j)}_A\otimes \rho^{(j)}_B),
\end{equation}
with $\sum_{j=1}^d p_j=1$. Now, define transposition $\mathcal{T}:\mathcal{H}_{k} \longrightarrow \mathcal{H}_{x}$ as a positive superoperator acting on either one of the subsystems ($k = A, B$). It is clear that transposition sends states to states, since if we first diagonalise the density operator, transposition leaves it unchanged. Hence transposition is a \textit{positive} map $\mathcal{T}_x[\rho_x] =  \rho^\trans_x \geq 0$. However, it is not completely positive. Partial transposition can be then defined as a positive but not completely positive superoperator acting on a bipartite system via $\mathcal{I}_A\otimes\mathcal{T}_B$ \footnote{Or $\mathcal{T}_A\otimes\mathcal{I}_B$: it is actually irrelevant which of the subsystem gets transposed, so we will assume it is sybsystem $B$}. The point here is that if we act with this bipartite superoperator on a \textit{separable} state of the form in Eq. \eqref{eq:separableBipartite} we will clearly obtain a positive result, as $\mathcal{T}_B$ does not `mess' with subsystem $A$'s eigenvalues. But if we find that partial transposition $\mathcal{I}_A\otimes\mathcal{T}_B$ produces a negative operator, we can say with confidence that the initial state was not separable, and therefore entangled. This is known as the positivity of the partial transposition (PPT) criterion, and there is an entanglement measure, called `negativity' that quantifies this criterion making use of the trace norm (defined in Eq. \eqref{eq:TraceNorm} :
\begin{eqnarray}
    \mathcal{N}(\rho) := \frac{\norm{\tilde{\rho}}_1 -1}{2}.
\end{eqnarray}
The negativity has been shown to be an entanglement monotone\footnote{That is, a non-negative quantity that does not increase under local operations and classical communication (LOCC).}, and it takes values between 0 and 1; however, the \textit{logarithmic negativity}, $E_\mathcal{N}(\rho)$, defined as
\begin{equation}
    E_\mathcal{N}(\rho) := \log_2\norm{\tilde{\rho}}_1
\end{equation}
and also a monotone (in this case unbounded) is favoured in many quantum information scenarios, and will be used later in this thesis.
Bear in mind that there are limitations to the logarithmic negativity: in higher dimensions, there exist entangled states that have vanishing logarithmic negativity. These are called `bounded entangled states', and their study and applications is beyond this thesis. However, a good reference for their use and relation to quantum metrology can be found in Ref. \cite{PhysRevResearch.3.023101}.

\section{Quantum continuous variables}
Quantum continuous variables (qCVs) are not only a theoretical necessity in quantum optics: they also provide a very rich conceptual bridge between classical and quantum mechanics through  the  common language of functional analysis in \textit{phase space}. Let us briefly describe classical and quantum phase space:
\subsection{Phase space}
In classical mechanics, phase space of a single dynamical degree of freedom (`particle') is the space generated by the position and momentum variables, which are represented by two vectors $(\mathbf{q};\mathbf{p})$, each in $\reals^3$. So classical phase space is isomorphic to $\reals^6$. A point in this space represents a classical state, and its evolution in time ($t\in \reals^+$, and extra continuous parameter) is governed, essentially, by the definition of the canonical Poisson bracket $\lbrace \cdot, \cdot \rbrace$ (there is no risk of confusing it with the anticommutator in this thesis) and its use in Hamilton's equations of motion. More generally, the properties of a dynamical degree of symbol are captured by some $f(\mathbf{q};\mathbf{p}; t)$ (called a Liouville density), and the equations of motion are dictated by the Liouville equation:
\begin{equation}
    \frac{\diff f}{\diff t} \equiv \frac{\partial f}{\partial t} + \lbrace f, H \rbrace,
\end{equation}
where $H = H(\mathbf{q};\mathbf{p}; t)$ is the Hamiltonian. Alternative versions of this equation, for instance the one that makes use of the Liouville \textit{operator}, or the one that utilises symplectic geometry methods, can be used to gain more intuition and to both bring closer and separate classical mechanics from quantum mechanics\footnote{If a formulation is able to separate and bring closer two concepts  at the same time, I would argue that it succeeds at bringing clarity and rigor. This is in fact the case here: some of the weirdness of quantum (phenomena, correlations) are conceptually relaxed by `going back' to classical mechanics, exhausting its possibilities, seeing what has been left unexplained, and then trying to explain \textit{that} with quantum theory. This may sound philosophical, but I believe that an iterative sequence of visits to theoretical classical mechanics is advisable if one wishes to `understand' quantum mechanics, especially if each iteration increases slightly the mathematical abstraction level.}. For more details, see Chapter 10, section 4 of Ref. \cite{peresBook} and references therein. For two classic textbooks on classical theoretical mechanics, see Refs. \cite{saletan1971theoretical, MarsdenBook1999}.

When $\mathbf{q}$ and $\mathbf{p}$ are promoted to quantum operators, phase space becomes quantum phase space. Quantum phase space, due to the fundamental non-commutative nature of the dynamical variables, is to be approached with care if one wishes to visualize it as a `place' where a system evolves. It is \textit{not}\footnote{If it was, then where would be the fun of quantum mechanics?} isomorphic to $\reals^6$. 
Moreover, and loosely speaking, because the quantum mechanical operators corresponding to position and momentum are, in principle, unbounded, their spectra is a continuum, hence the name quantum continuous variables. They arise when the Hilbert space under consideration is $L^2(\mathbb{R}^n)$ with finite $n$, \textit{i.e.}, the space of square-integrable complex-valued functions \textit{over} the reals\footnote{If $n$ was infinite, we would be in QFT territory.}. While the completeness requirement for a space to meet the definition of Hilbert space is naturally met in the qubit case, things become more nuanced when moving to infinite dimensions, and the level of mathematical rigor one needs to correctly address the quantum mechanical interpretation of the theory slightly increases. 

We will restrain to Bosonic qCVs. The (canonical) recipe for quantizing a  (non-relativistic) dynamical, Bosonic, degree of freedom starts from considering two self-adjoint operators $\hat{x}$ and $\hat{p}$, called `position' and `momentum', and making them satisfy the canonical commutation relation (ccr):
\begin{equation}\label{eq:ccr}
    [\hat{x}, \hat{p}] = i \hbar \hat{\Id},
\end{equation}
where the reduced Planck constant takes a numerical value $\hbar := 1.054571817\ldots \times 10^{-34} \si{\joule \second}$. It is an exact constant corresponding to one quantum of action. From now on, we adopt natural units, and set $\hbar \equiv 1$.
The $\hat{x}$ and $\hat{p}$ operators are unbounded operators exhibiting a continuous spectrum. Let's take the position operator $\hat{x}$ for a one-dimensional case. Even when its `eigenvectors' $\ket{x}$ with $x \in \reals$ are not proper, normalised vectors in $\hilbert$, we can still write
\begin{equation}
    \hat{x} = \int_\reals x \ket{x}\bra{x}\diff x,
\end{equation}
where $\braket{x}{y} = \delta (x-y)$. Therefore, we can write, by virtue of Eq. \eqref{eq:OperatorFunction}:
\begin{equation}
    f(\hat{x}) = \int_\reals f(x) \ket{x}\bra{x}\diff x,
\end{equation}
which in turn allows one to compute the expectation value of a self-adjoint operator $\hat{A}=\sum_a a\ket{a}\bra{a}$:
\begin{equation}
    \langle\hat{A}\rangle_{\hat{x}} \equiv \Tr(\hat{A}f(\hat{x})) = \int_\reals \sum_a f(x) \bra{x}\hat{A}\ket{x}\diff x,
\end{equation}
with $$\bra{x}\hat{A}\ket{x} = \sum_a a \abs{\braket{a}{x}}^2.$$

Going back to Eq. \eqref{eq:ccr}, it turns out that there is no finite matrix representation of the algebra generated by the ccr. The proof works by contradiction: suppose there exists a finite matrix representation. Then, the trace is well-defined, and so we can take the trace on both sides. Because of the cyclic property of the trace, the trace of a commutator is zero, which, would necessarily imply either $\hbar = 0$ or $\Tr(\Id)=0$, both false \cite{weyl1950theory}. Luckily, infinite-dimensional representations of the ccr algebra do exist\footnote{There are some nuances here, in the sense that in order to guarantee that there are indeed irreducible representations, one shoud use the exponentiated version of the ccr,  known as Weyl relation: $\exp(i\lambda \hat{x}) \exp(i\mu \hat{x}) = \exp(i\lambda\mu/2) \exp(i\mu \hat{x}) \exp(i\lambda \hat{x})$ with $\lambda, \mu \in \reals$.
}. For a system in $L^2(\mathbb{R}^n)$, we can define `eigen' equations for the position and momentum operators via
\begin{align}
    \hat{x}_j\ket{f} &= x f(\mathbf{x})\\
    \hat{p}_j\ket{f} &= -i \hbar \frac{\diff}{\diff x_j} f(\mathbf{x}),
\end{align}
where $\ket{f} \equiv f(\mathbf{x})$ are equivalently in  $L^2(\mathbb{R}^n)$, and $\mathbf{x} = (x_1,\ldots, x_n)^\trans$. Other representations may be in place, but the Stone--von Neumann theorem states that they are all unitarily equivalent for finite $n$. (For an infinite number of dynamical variables we get into quantum field theory, which requires additional ingredients such as Poincaré invariance. For more details, see for example the book of Ref. \cite{dubin1990quantum}, or Haag's classic book \cite{haagBook}.) Finally, for a nice exposition of the (dis)analogies between classical and quantum mechanics, via the paralellism between the classical Liouville density and the Wigner function, and including interesting passages about the notion of a trajectory in quantum phase space\footnote{The one notable exception --that I know of-- for a quantum system that does exhibit a trajectory-like behaviour in phase space is the harmonic oscillator, a fact that can be both misleading or illuminating depending on the background one has.}, as well as classical vs quantum irreversibility, can be found in section 10-5 of Peres' book \cite{peresBook}.
\subsubsection{Displacement and squeezing operators}
It is convenient at this point to introduce the  operators $\hat{a}^\dagger$ and $\hat{a}$. Though sometimes they are referred to as `field' operators, they are also known as raising and lowering, or creation and annihilation operators, respectively. In terms of the canonical operators for a single dynamical, \textit{theoretically massless}, Bosonic degree of freedom\footnote{For a proper exposition of the quantisation of the electromagnetic theory, that sheds light into the connections between quantum optics and QFT, see for example Chapter 9 of Ref. \cite{serafini}, or any other book on quantum optics.}, they can be defined in natural units via
\begin{align}
    \hat{a} &:= \frac{1}{\sqrt{2}}(\hat{x}+i\hat{p})\\
    \hat{a}^\dagger &:= \frac{1}{\sqrt{2}}(\hat{x}-i\hat{p}).
\end{align}
Their commutation relation is immediate to check: $[\hat{a}, \hat{a}^\dagger] = \hat{\Id} $. Leaving the (fascinating) structural properties of Fock spaces aside, if we assume there exists an ONB $\lbrace \ket{n} \rbrace_n$ with $n\in \naturals$, where $\naturals$ is the natural numbers set including 0, the (raising/lowering)/(creation/annihilation) names become obvious:
\begin{align}
    \hat{a}^\dagger\ket{n} &:= \sqrt{n+1}\ket{n+1}\\
    \hat{a} \ket{n} &:= \sqrt{n}\ket{n-1},
\end{align}
with $\hat{a}\ket{0} := 0\ket{0}$. We call $\ket{n}$ a `number state' and $\hat{N}:= \hat{a}^\dagger \hat{a}$ the `number operator', because $\hat{a}^\dagger \hat{a}\ket{n} = n\ket{n}$. The element $\ket{0}$ is called the vacuum in this context, and loosely speaking is interpreted as a quantum mechanical state that produces `no click' upon entering a well-functioning, appropriate-for-the-field, particle\footnote{For an interesting discussion about the nature of particles, see Ref. \cite{colosi2004particle}.} detector.

A very relevant operator that is constructed out of the field operators is the \textit{displacement}\footnote{The name `displacement' becomes immediate when visualizing the action of the operator in (quantum) phase space.} operator:
\begin{equation}
    \hat{D}(\alpha):= e^{\alpha \hat{a}^\dagger - \alpha^*\hat{a}}\equiv e^{-\abs{\alpha}^2/2}e^{\alpha \hat{a}^\dagger}e^{-\alpha^* \hat{a}},
\end{equation}
with $\alpha \in \complex$.
The action of the displacement operator on the vacuum produces a pure state $\ket{\alpha}$ that we call \textit{coherent state}\footnote{The coherent state can be equivalently defined as the eigenstate of the annihilation operator: $\hat{a}\ket{\alpha} = \alpha\ket{\alpha} $.}: $\hat{D}(\alpha)\ket{\alpha} = \ket{0}$. The set of vectors$\lbrace \ket{\alpha} \rbrace_{\alpha\in \complex}$ forms an overcomplete ONB that nevertheless allows for a resolution of the identity:
\begin{equation}
\hat{\Id} = \frac{1}{\pi}\int_\complex \diff^{2}\alpha |\alpha\rangle\langle\alpha |, 
\end{equation}
where $\pi^{-1}\diff^{2}\alpha \equiv \diff \left(\Re{\alpha}\right)\diff \left(\Im{\alpha}\right)$. 
Physically, in optics, coherent states are `semiclassical' states in the sense that, while they exhibit quantum properties (one can express them in terms of states containing $n$ `photons': $\ket{n}$), they do not exploit the `other' special feature of quantum fields (which essentially is non-classical correlations). The displacement operator, hence, can be seen as a device that is able to create photons out of a vacuum but that puts them in a quantum-correlation-wise `harmless' state, sometimes known as laser (which can definitely be \textit{non}-harmless tissues-wise).
Some useful formulas follow, that will be relevant for computing fidelities of the teleportation protocol presented in Chapter \ref{chapter:MWentanglementDistribution}:

Overlap of a coherent state and a Fock number state:
\begin{equation}
\langle n|\alpha\rangle = e^{-\frac{|\alpha|^{2}}{2}}\frac{\alpha^{n}}{\sqrt{n!}}
\end{equation}

Overlap of a coherent state and a position `eigenstate':
\begin{equation}
\langle x|\alpha\rangle = \pi^{-1/4} \exp\left[ -\frac{x^{2}}{2} + \sqrt{2}x\alpha - \frac{\alpha^{2}}{2} - \frac{|\alpha|^{2}}{2} \right].
\end{equation}
Action of the displacement on a Fock basis element:
\begin{equation}
\hat{D}(\alpha)|n\rangle = \frac{1}{\sqrt{n!}}(\hat{a}^{\dagger}-\overline{\alpha}\hat{\Id})^{n}|\alpha\rangle.
\end{equation}
Elements of the displacement operator in the Fock basis:
\begin{eqnarray}
\langle m|\hat{D}(\alpha)|n\rangle &=& e^{-|\alpha|^{2}/2}\sqrt{\frac{m!}{n!}} \times \\
\nonumber && \sum_{k=0}^{n}\begin{pmatrix}n\\k\end{pmatrix}(-1)^{n-k}\theta(m-k)\frac{\alpha^{m-k}\bar{\alpha}^{n-k}}{(m-k)!}.
\end{eqnarray}

Composition of two displacement operations, with $\alpha, \beta \in \complex$:
\begin{equation}
\hat{D}(\alpha)\hat{D}(\beta) = e^{\frac{1}{2}(\alpha\overline{\beta}-\overline{\alpha}\beta)}\hat{D}(\alpha+\beta)
\end{equation}
A formula that may interest somebody, sometime:
\begin{equation}
\langle n |\hat{D}(\alpha)|\beta\rangle = e^{-\frac{|\alpha|^{2}}{2}}e^{-\frac{|\beta|^{2}}{2}}e^{-\overline{\alpha}\beta} \frac{(\alpha+\beta)^{n}}{\sqrt{n!}}.
\end{equation}
Some integrals: 
\begin{equation}
\int_{-\infty}^{\infty} \diff x e^{-x^{2}\pm\beta x} = \sqrt{\pi} e^{\frac{\beta^{2}}{4}}
\end{equation}
\begin{equation}
\int_{-\infty}^{\infty} \diff x e^{-x^{2}\pm i\beta x} = \sqrt{\pi} e^{-\frac{\beta^{2}}{4}}.
\end{equation}

Another (complete) representation of states is given by the Wigner function, also known as the characteristic function (CF). For an $n$-mode state $\rho$: 
\begin{equation}
\chi(\bm{\mathrm{d}}) = \tr\left[ \rho\hat{D}_{-\bm{\mathrm{d}}}\right],
\end{equation}
with normalisation condition given by $\chi(\bm{0}) =1$.

Another relevant operation is squeezing. The single-mode squeezing operation $\hat{S}_1(z)$ acts on the vacuum:
\begin{equation}
    \ket{z}\equiv \hat{S}(z)\ket{0} := \exp(z\hat{a}^{\dagger\;2}-z^*{\hat{a}}^2)\ket{0},
\end{equation}
where $z = r e^{i\theta}$ is the squeezing parameter. However, it is more relevant for this thesis the \textit{two-mode} squeezing operation, as it entangles the two (vacuum) modes. Two-mode squeezing is described by a very similar operator, but this time  $\hat{S}_2(z): \mathcal{H}_A\otimes \mathcal{H}_B \longrightarrow \mathcal{H}_A\otimes \mathcal{H}_B$, where labels $A$ and $B$ identify the two modes. In the case of two vacuum modes, the resulting state is called two-mode squeezed vacuum (TMSV) state:
\begin{equation}
    \ket{\psi}_{\text{TMSV}}\equiv \hat{S}_2(z)\ket{0,0} := \exp(z{\hat{a}^\dagger}{\hat{b}^\dagger}-z^*{\hat{a}}\hat{b})\ket{0,0}.
\end{equation}
Taking $z=r\in\reals$, we can write
\begin{equation}
    \ket{\psi}_{\text{TMSV}} = \frac{1}{\cosh(r)}\sum_{j=0}^\infty (\tanh(r))^j  \ket{j,j}.
\end{equation}

More details and interesting formulae can be found in Chapter 5 of Ref. \cite{serafini}, as well as in the book in Ref. \cite{barnett2002methods}.

\subsubsection{Gaussian states}
Within the bosonic CV quantum systems, quantum Gaussian states are defined as the ones arising from Hamiltonians that are at most quadratic in the field operators, which we list in the vector $\hat{\bm{A}}:=(\hat{a}_1, \hat{a}_2, \ldots, \hat{a}_N, \hat{a}^\dagger_1, \hat{a}^\dagger_2, \ldots \hat{a}^\dagger_N)^\trans$, where $N$ is the number of modes. The $a$-th element of this vector is denoted $\hat{A}_a$. This ordering of the creation and annihilation operators is commonly referred to as the `complex basis' or `complex form' \cite{adesso}, and allows for a compact way of writing down the commutation relations: $\left[ \hat{{A}}_{{a}}, \hat{{A}}^\dagger_{{b}}\right]= K_{{a}{b}} \hat{\bm{\text{I}}}$, where ${a, b}=1, \ldots, N$, $\hat{{\Id}}$ is the identity operator, and $K_{ab}$ are elements of a symplectic form $\hat{K}$ satisfying: $\hat{K}^{-1} = \hat{K}^\dagger = \hat{K}$ and that can have a diagonal matrix representation $K=\text{diag}(\Id_N, -\Id_N) $, where $\Id_N$ is the $N\times N$ identity matrix (note it has no hat to distinguish it from an identity operator).

 Instead of having to resort to the infinite-dimensional density operator in order to describe a state, Gaussian systems are fully characterized by an $N$-vector called the \textit{displacement vector} and a $N \times N$ matrix, the \textit{covariance matrix}. These correspond to the first and second moments of the state.  We can construct the displacement vector $\mathbf{d}$ with entries given by:
\begin{equation}
{d}_a:=\Tr \left[ \rho \hat{A}_a\right],
 \end{equation} 
 and the covariance matrix $\Sigma$ with entries:
\begin{equation}
\Sigma_{ab} := \Tr \left[ \rho \lbrace \Delta \hat{{A}}_a, \Delta \hat{{A}}_b^\dagger \rbrace \right],
 \end{equation} 
where $\rho$ is the density operator, $\lbrace \cdot, \cdot \rbrace$ denotes the anticommutator, and $\Delta \hat{{A}}:= \hat{{A}}-\mathbf{d}\hat{\Id}$. It is important to bear in mind that other choices of basis lead to different, but equivalent definitions. In fact, in the following sections we will start by writing down covariance matrices in the so-called `quadrature basis' $\left( \hat{x}_1, \ldots, \hat{x}_N, \hat{p}_1, \ldots, \hat{p}_N\right)$ with the \textit{canonical} position and momentum operators  defined by the choice $\kappa_1=2^{-1/2}$ in $\hat{a}_k = \kappa_1(\hat{x}_k + i \hat{p}_k)$\footnote{The `quantum optics' convention takes $\kappa_1=1$. }. A key result with important consequences in the context of Gaussian states is the normal mode decomposition \cite{ arvind, simon}, which follows the more general theorem due to Williamson \cite{williamson} and that, from a physical point of view, establishes that any Gaussian Hamiltonian (i.e., quadratic) is equivalent --up to a unitary-- to a set of free, non-coupled harmonic oscillators.  This apparent simplicity of Gaussian states, however, has a rich structure when it comes to analyzing their Hilbert space properties, as well as information-theoretic quantities such as the quantum Fisher information, entropies, and so on.  We can state the result in the following way: any positive-definite Hermitian matrix $\Sigma$ of size $2N\times 2N$ can be diagonalized with a symplectic matrix $S$: $\Sigma = S D S^\dagger$, where $D = \text{diag}\left( \nu_1,  \ldots, \nu_N, \nu_1, \ldots, \nu_N\right)$ with $\nu_{a}$ the symplectic eigenvalues of $\Sigma$, that are the positive eigenvalues of matrix $K\Sigma$. An important result for what follows is that a state is pure if and only if all the symplectic eigenvalues are one: $\nu_{a}=1$ $\forall {a}$, and $\nu_{a} \geq 1$ for any Gaussian state.

Let's now consider a bipartite system with covariance matrix
\begin{equation}\label{eq:twoModeCovariance}
\Sigma = \begin{pmatrix}
\Sigma_A & \varepsilon_{AB}\\
\varepsilon_{AB}^\intercal & \Sigma_B
\end{pmatrix},
\end{equation}
where $\Sigma_A$, $\Sigma_B$ and $\varepsilon_{AB}$ are all $2\times 2$ matrices.
One of the advantages of the symplectic formalism of Gaussian states is that computations become quite straightforward. For example, the negativity of a bipartite Gaussian state is 
\begin{equation}\label{eq:GausssianNegativity}
\mathcal{N}(\rho) =\max \left\{0, \frac{1-\tilde{\nu}_{-}}{2\tilde{\nu}_{-}}\right\},
\end{equation} 
where $\tilde{\nu}_{-}$ is the smaller of the two partially transposed symplectic eigenvalues:
\begin{equation}\label{eq:ptseigen}
    \tilde{\nu}_{\mp}:=\sqrt{\frac{\tilde{\Delta} \mp \sqrt{\tilde{\Delta}^2- 4 \det\Sigma}}{2}},
\end{equation}
and $\tilde{\Delta} = \det \Sigma_A + \det \Sigma_B - 2\det \varepsilon_{AB}$ is a symplectic invariant (because determinants are basis-independent). Therefore, a bipartite Gaussian state is entangled when $\tilde{\nu}_{-} < 1$.

A Gaussian state $(\bm{\mathrm{d}}, \Sigma)$ has a characteristic function given by
\begin{equation}
\chi_G(\bm{\mathrm{r}}) = e^{-\frac{1}{4}\bm{\mathrm{r}} \Omega \Sigma \Omega^\intercal \bm{\mathrm{r}}^\intercal}e^{-i\bm{\mathrm{r}} \Omega \bm{\mathrm{d}}^\intercal},
\end{equation}
where $\bm{\mathrm{r}} = (x_1, p_1, \ldots, x_N, p_N) \in \mathbb{R}^{2N}$.
In this case we used the ``real basis'': $\hat{\bm{r}}:= ( \hat{x}_1, \hat{p}_1, \hat{x}_2, \hat{p}_2, \ldots, \hat{x}_N, \hat{p}_N)$, for which the ccr are $\left[\hat{\bm{r}}, \hat{\bm{r}}^\intercal\right] = i\Omega$, where $\Omega = \bigoplus_{j=1}^N \Omega_1$ is the \textit{real} quadratic (or symplectic) form, and 
\begin{equation}
 \Omega_1 =
    \begin{pmatrix}
       0 & 1\\
       -1 & 0
    \end{pmatrix}.
   \end{equation}

\chapter{Parameter estimation}
\label{chapter:Estimation}
In this chapter we motivate some of the machinery that is used later in the thesis, especially for the protocol described in Chap. \ref{chapter:biFreq}. The field of (quantum) parameter estimation is a fascinating one. Given the simplicity of the questions it asks --how, and with what accuracy can I measure things that may not correspond to observables, it comes as no surprise that it is deeply linked to foundational problems. We will not dwell too much into those here, but there are excellent texts to seek for such connections. Classic references for further reading include work by Fano, Holevo and Helstrom, and can be found in Refs. \cite{Fano1957, holevo, helstrom, HelstromBook}. Additionally, at the end of the chapter we include a case example that, using the formulae obtained for the estimation of a single parameter encoded quantum-mechanically in our system, reproduces known results in quantum illumination.

Because quantum parameter estimation can be seen as a more nuanced version of classical parameter estimation, in the sense that it cannot avoid the fundamental probabilistic nature of the outcomes of the quantum mechanical measurement, we start with the classical theory, and then move on to the quantum version. We will deal with the asymptotic theory, which assumes availability of large data sets, among other things. Moreover, maximum likelihood methods to obtain (unbiased, local) quantum estimators will be the preferred ones, because in this case the computation of an optimal observable can be made explicit. The reader should bear in mind that these methods, which essentially involve Cramér-Rao inequalities, quantum Fisher information (matrices), and symmetric logarithmic derivatives, are by no means the one and only story in quantum estimation. Bayesian methods for constructing estimators out of limited data, or when prior knowledge of the region where the true value of the parameter may lie is missing, have been intensively investigated, and represent an active area of research \cite{parisBook, RubioPhDThesis}. Bayesian methods sometimes are referred to as `global', as opposed to the frequentist, local estimation. This is the case of the estimation of a completely unknown parameter \cite{Martinez2017}. Likewise, in the presence of imperfect data, or estimation of unknown parameters, the information-theoretic approach of maximum entropy methods can prove useful \cite{Jaynes1957, parisBook}. We shall also omit from our discussion the results associated to \textit{adaptive} methods, which have interesting connections to machine learning, variance-bias trade-off analysis, and adaptive interferometry \cite{Berry2000}, but go beyond the scope of this thesis \cite{parisBook, Lumino2018}. Good reviews of results in these areas can be found in Refs. \cite{parisBook, HayashiBook2005}. For the classical theory of statistical inference, we highly recommend the book by Casella and Berger \cite{CasellaBerger}. Finally, when restricting to quantum Gaussian states and Gaussian-preserving channels, the results obtained by {\v S}afr{\'a}nek (\textit{e.g.} in Refs. \cite{SafranekThesis, Safranek_2018}) will be used extensively in Chapter \ref{chapter:biFreq}. 

\section{Asymptotic classical parameter estimation}
Let's consider a  set of continuous parameters put into a vector $${\bm{\lambda}} = (\lambda_1, \ldots, \lambda_d).$$ Assume that these parameters correspond to pieces of information about some physical system that we wish to extract: its size, its color, its temperature, etc.
The idea behind any measurement is to make the system interact with another system, often called the `probe', which gets ``imprinted'' with some qualities from the system. The type of probe will depend on the nature of the parameter we wish to extract. In the case of attributes that correspond to different physical quantities, the probe may  actually be a set of probes: a ruler, a spectrometer, a thermometer, etc. Upon exiting the system, we aim at extracting the value of ${\bm{\lambda}}$ from our probe. We may repeat this process any number of times, to refine the information extraction, and to minimize the statistical deviations coming from possible imperfections, or noise. Let's call the number of experimental runs $M$.
Let us remark that, because in the scenarios presented later in this thesis the classical strategy will be compared with a quantum one, it is convenient to:
i) fix the energy per run of the experiment and ii) fix the (expected) number of probes per experimental run to $N$.
We then construct a measurement of ${\bm{\lambda}}$ with stochastic outcomes $\vec{x}_i$ for each run (this being an $d$-dimensional vector), and associated probability density function $p_{\bm{\lambda}}(\vec{x}_i) \equiv p(\vec{x}_i |{\bm{\lambda}})$, where $i=1, \ldots M$. Additionally, it is natural to consider independent and identically distributed (iid) measurements, which means
\begin{equation}
    p(\vec{x}_i |{\bm{\lambda}}) = p(\vec{x}_j |{\bm{\lambda}})
\end{equation}
for any $i,j \in 1, \ldots, M $.

Crucially, classical measurements are ideal, that is, they have no impact on the system. In other words, classicality allows to extract $p(\vec{x}_i |{\bm{\lambda}})\Rightarrow \vec{x}_i$ perfectly, in principle. This means no headaches: classically, the parameter \textit{is} measurable in itself, which is in sharp contrast with respect to the quantum case, as we shall see.\\
After all the $M$ runs we compute an \textit{estimator}, $\tilde{{\bm{\lambda}}}$, which is a function of all the stochastic outcomes $\vec{x}_i$:
\begin{equation}
    \tilde{{\bm{\lambda}}} = \tilde{{\bm{\lambda}}}(\vec{x}_1, \ldots, \vec{x}_M), 
\end{equation}
 and we assume that its units are the same as the parameter ${\bm{\lambda}}$'s itself.
Bear in mind that, if we treat each probe as a `particle', which will be a natural choice in the context of quantum optics, the total number of particles is then $N\cdot M$.

A figure of merit in parameter estimation is the mean squared error (MSE). For an estimator $\tilde{{\bm{\lambda}}}$, it is defined as:
\begin{equation}
    \text{MSE}(\tilde{{\bm{\lambda}}}) := \mathrm{E}_{\bm{\lambda}}\left[\left(   {\bm{\lambda}} - \tilde{{\bm{\lambda}}} \right)^2\right],
\end{equation}
 where $\mathrm{E}_{\bm{\lambda}}[\cdot]$ denotes the statistical expected value over the $M$ rounds.
 The expected value of a random variable, be it discrete or continuous, is simply its ``average'', weighted accordingly to its associated probability distribution. There are certain nuances regarding the convergence of the expected value, both in the discrete and the continuous case. For the curious reader, we just give here a general definition of MSE that should satisfy even a mathematician. Take the probability space $(\mathcal{X}, \mathcal{F}, \mathrm{P})$ with $\mathcal{X}$ being the sample space for the experiment, that is, the set of \textit{all} possible outcomes that the random variable(s) can take (for example, for the experiment of rolling a die three times, this would be the set given by the Cartesian product of the sample space for rolling the die once, that is, $\lbrace 1,2,3,4,5,6 \rbrace$, three times); $\mathcal{F}$ the event space, \textit{i.e.}, a set of actual outcomes after a given number of repetitions of the experiment (in the die example, after repeating the three rolls a couple of times, the event space could be the set $\lbrace \lbrace 3,2,6 \rbrace, \lbrace 4, 5,1 \rbrace \rbrace$, and more generally, it could contain all sets which are subsets of the sample space, being a $\sigma$-algebra\footnote{Which is, in simple terms, a collection of sets that is closed under countable unions, intersections, and complements.} on the set $\mathcal{X}$\footnote{There are certain subtleties with respect how one may \textit{choose} the event space, because sometimes certain subsets may not be available to be ``measured'', but let's not get into those. }); and $\mathrm{P}$ the probability function, assigning a definite probability to each of the events (in our example, since each die roll is independent from the previous one, and, assuming a fair die, every result is equiprobable, and so, both the events $\lbrace 3,2,6 \rbrace$ and $\lbrace 4, 5,1 \rbrace$ would be assigned a probability $1/6^3$). With all this, the expected value for the random variable $X$ would be defined as a Lebesgue integral
 \begin{equation}
     \mathrm{E}[X] = \int_\mathcal{X} \diff \mathrm{P} X.
 \end{equation}
 In any case, even when in general one may need to resort to measure theory, Lebesgue integration, and funny-named things like the law of the unconscious statistician (LOTUS) \cite{Billingsley1995}, here we will take the pragmatic, physicist approach and assume all integrals converge, and go on to unveil further interesting results. 
 
 The MSE is a quantity that measures how good our estimator is, since it computes the (squared) difference between the true value of the parameter, and the estimated one. Naturally, it is not the only measure for the quality of an estimator (the mean absolute error, for example, would also do the job), but it is a particularly good candidate because its quite tractable analytically, and it captures both the variance and the bias of an estimator:
 \begin{equation}
     \text{MSE}(\tilde{{\bm{\lambda}}}) = 
     \mathrm{Var}_{\bm{\lambda}}(\tilde{{\bm{\lambda}}}) + \mathrm{Bias}^2_{\bm{\lambda}}(\tilde{{\bm{\lambda}}}),
\end{equation}
where the variance\footnote{Note that in some literature the variance is written $(\Delta \tilde{{\bm{\lambda}}})^2$.} takes the usual definition (as the second (central) moment of the probability distribution:
\begin{align}
    \mathrm{Var}_{\bm{\lambda}}(\tilde{{\bm{\lambda}}})&:= \mathrm{E}_{\bm{\lambda}}\left[\left(\tilde{{\bm{\lambda}}} - \mathrm{E}_{\bm{\lambda}}[\tilde{{\bm{\lambda}}}]\right)^2\right]\\
    & = \mathrm{E}_{\bm{\lambda}} [ \tilde{{\bm{\lambda}}}^2] - \mathrm{E}_{\bm{\lambda}} [ \tilde{{\bm{\lambda}}}]^2
\end{align}
and the bias is defined as
\begin{align}
    \mathrm{Bias}^2_{\bm{\lambda}}(\tilde{{\bm{\lambda}}})&:= \left({\bm{\lambda}} - \mathrm{E}_{\bm{\lambda}}[\tilde{{\bm{\lambda}}}]\right)^2.
\end{align}
However interesting the problem of simultaneous optimization of the bias and the variance may be, in the results presented in this thesis, we will deal with unbiased estimators, that is, those that are expected to take the true value of the parameters: 
\begin{equation}
    \mathrm{E}_{\bm{\lambda}}[\tilde{{\bm{\lambda}}}]={\bm{\lambda}}.
\end{equation}
Then the MSE becomes the variance:
\begin{equation}
    \left. \text{MSE}(\tilde{{\bm{\lambda}}})\right|_\text{Unbiased} = \left. \text{Var}_{\bm{\lambda}}(\tilde{{\bm{\lambda}}})\right|_\text{Unbiased} = 
\mathrm{E}_{\bm{\lambda}}\left[\left(\tilde{{\bm{\lambda}}} - \bm{\lambda}
\right)^2\right].
\end{equation}
(From now on, we will drop the `unbiased' specification.)

Let us focus on the single-parameter case from now on, so that we have a continuous parameter $\lambda$ and a scalar estimator $\tilde{\lambda}$. We assume we have a measurement with probability density function $p(\vec{x}|\lambda)$, where $\vec{x} = (x_1, \ldots, x_M)$, and each data point $x_i\in\mathcal{X}$ for all $i=1, \ldots, M$. The goal is to minimize the (unbiased) variance of the estimator $\tilde{{\lambda}}$, which, making explicit use of the integral definition of the MSE, reads

\begin{equation}\label{eq:UnbiasedMSE}
      \text{Var}_\lambda(\tilde{\lambda}) =  \mathrm{E}_{{\lambda}}\left[\left(\tilde{\lambda} - {\lambda}
\right)^2\right] = \int_{\mathcal{X}^M} \diff  \vec{x} \;p(\vec{x}|\lambda) \left(\tilde{\lambda} - {\lambda}
\right)^2,
\end{equation}
where the dependence on the data is implicit in the estimator $\tilde{\lambda} = \tilde{\lambda} (\vec{x})$, and 

\begin{equation}
    \int_{\mathcal{X}^M} \diff  \vec{x} \equiv\prod_{i=1}^M \int_{\mathcal{X}}\diff x_i.
\end{equation}
Minimising Eq. \eqref{eq:UnbiasedMSE} is a priori hard, because we would need to consider \textit{all} the estimation strategies. The good news is that there exists, in principle, a lower bound to $\text{Var}(\tilde{\lambda})$. The celebrated inequality, the Cramér-Rao bound (CRB) is
\begin{equation}\label{eq:CRBound}
    \text{Var}_\lambda(\tilde{\lambda}) \geq \frac{1}{F({\lambda})},
\end{equation}
where 
 $F({\lambda})$ is the Fisher information (also known as the `information number'), defined as:
\begin{equation}\label{eq:FisherInfo}
    F({\lambda}) : = \mathrm{E}_\lambda \left[  \left( 
 \frac{\partial \ln p(\vec{x}|\lambda)}{\partial \lambda}
 \right)^2\right]
    \equiv
    \int_{\mathcal{X}^M} \diff  \vec{x} \;p(\vec{x}|\lambda) \left( 
 \frac{\partial \ln p(\vec{x}|\lambda)}{\partial \lambda}
 \right)^2.
\end{equation}
It is very important to note that the Fisher information explicitly depends on the measurement used, via the probability density function $p(\vec{x}|\lambda)$. This will become more relevant in the discussion of its quantum version, as we will see. 

In the case of independent random variables we have
\begin{equation}
    p(\vec{x}|\lambda) = \prod_{i = 1}^M p(x_i|\lambda).
\end{equation}
Additionally, if they are identically distributed, $p(x_i|\lambda) = p(x_j|\lambda)\equiv p(x|\lambda)$ for any $i,j$, and, so, $p(\vec{x}|\lambda) = (p(x|\lambda))^M$.
Therefore, in the case of iid measurements, the Fisher information is
\begin{align}
    \label{eq:FisherInfo}
    F({\lambda}) &= \mathrm{E}_\lambda \left[  \left( 
 \frac{\partial \ln (p(x|\lambda))^M}{\partial \lambda}
 \right)^2\right]\\
 &= M \mathrm{E}_\lambda \left[  \left( 
 \frac{\partial \ln  p(x|\lambda)}{\partial \lambda}
 \right)^2\right] \\
 &\equiv M F_\text{iid}({\lambda}),
\end{align}
due to the log property, and the linearity of the expected value. The Cramér-Rao bound, for this case, then reads\footnote{In some books and papers, the Fisher information is directly defined as the iid version below. In general, we will follow this convention, hence dropping the iid subscript.}
\begin{equation}\label{eq:iidCRBound}
    \text{Var}_\lambda(\tilde{\lambda}) \geq \frac{1}{M F_\text{iid}({\lambda})}.
\end{equation}
This inequality means that there exists at least one estimator that saturates the CRB. We call this estimator \textit{best}.
The question that remains is whether the bound can be \textit{achieved}, which forces us to look into asymptotic theory. How can we construct $\tilde{\lambda}$ so that it saturates Eq. \eqref{eq:iidCRBound}? In the next section we address a very popular method.
\subsection{Maximum likelihood estimation}
A common method relies on maximum likelihood (ML) estimators. Simply put, an ML estimator is the one that best fits the data. The likelihood function is defined as 
\begin{equation}\label{eq:LikelihoodFunction}
    \mathcal{L}(\lambda|\vec{x}) := \prod_{i = 1}^M p(x_i|\lambda).
\end{equation}
Note that this is a function of $\lambda$ alone, with the data $\vec{x}$ held constant and `already observed'. It measures how likely different values of $\lambda$ are with respect to the observed data. Although we won't use Bayesian methods, some intuition can be gained using Bayes' rule for updating one's knowledge about a given event. Consider two events $A$ and $B$. 
\begin{equation}\label{eq:Bayes}
    p(A|B) = \frac{p(B|A)p(A)}{p(B)}.
\end{equation}
Let us call $p(A|B)$ the `posterior', $p(B|A)$ the `likelihood', $p(A)$ the `prior', and $p(B)$ the `evidence'. In this setting, let's say that we \textit{know} $B$ has happened, or at least we know its probability, and we want to gain information about event $A$. $p(A|B)$ is the posterior, because, having observed $B$, we are, through Eq. \eqref{eq:Bayes}, one step closer to obtaining information about $A$. In a way, the posterior $p(A|B)$ is a more down to Earth version of the true value we are seeking, that is $p(A)$.

Let's use this rule to the case of the likelihood function. Using the definition above and Bayes' rule we find:
\begin{equation}
    \mathcal{L}(\lambda|\vec{x}) = p(\vec{x}|\lambda) = \frac{ p(\lambda |\vec{x}) p(\vec{x})}{p(\lambda)}.
\end{equation}
We rewrite this:
\begin{equation}
     p(\lambda | \vec{x}) = \frac{ \mathcal{L}(\lambda|\vec{x})p(\lambda) }{p(\vec{x})}.
\end{equation}
This equation can be interpreted as follows: $p(\lambda |\vec{x})$ is the posterior, $\mathcal{L}(\lambda|\vec{x})$ is the likelihood, $p(\lambda)$ is the prior, and $p(\vec{x})$ the observed data, or evidence. In practice, $p(\vec{x})$ is an extremely elusive quantity, which should be replaced by
\begin{equation}
    p(\vec{x}) \rightarrow \int \diff \lambda p(\vec{x}|\lambda) p(\lambda).
\end{equation}
The above-described method is recommended when the available data is limited, or small. However, if the dataset is large, an ML estimator $\tilde{\lambda}_{ML}$ can be defined as the argument of the ML function that maximizes it:
\begin{equation}\label{eq:MaxLikelihood}
    \tilde{\lambda}_{ML} \in \{\text{arg } \max_{\lambda \in \Lambda}\mathcal{L}(\lambda|\vec{x}) \},
\end{equation}
where $\Lambda$ is the range of parameter $\lambda$.
The maximization problem implicit in  Eq. \eqref{eq:MaxLikelihood} can be a bit problematic sometimes, especially if differentiation is to be used. In these cases, maximizing the so-called \textit{log likelihood}, $\log \mathcal{L}(\lambda|\vec{x})$ can be easier, since the natural logarithm is a striclty increasing function in the positive reals. In any case, a maximum likelihood estimator guarantees asymptotic CRB achievability, in the sense that these estimators satisfy the following:
\begin{equation}
    \lim_{M\rightarrow \infty} M \;\text{Var}_\lambda\left( \tilde{\lambda}_{ML}\right) = F^{-1}_\text{iid}({\lambda}).
\end{equation}
Accordingly, ML estimators are called `almost optimal'.
In a more precise way, ML estimators  $\tilde{\lambda}_{ML}$ are: i) Consistent: As $M$ grows, estimator sequences  $\lbrace 
\tilde{\lambda}_M\rbrace_M$ converge (in probability) to $\lambda$; and ii) Efficient: Fisher's theorem \cite{fisher1925}
states that as $M\rightarrow \infty$, an ML estimator approaches a Gaussian distribution in $\lambda$. This is what we mean by the statement that the CR bound is asymptotically achievable.

Let us remark that even when the multiparametric case will not be used in the thesis, it is, as far as classical theory is concerned, a trivial generalisation of the single parameter case. This is \textit{not} the case for the quantum situation, for non-Abelian considerations need to be accounted for. However, in what follows, we focus on the single parameter estimation. 
\section{Quantum Parameter Estimation}

Perhaps the main difference between classical and quantum parameter estimation lies in the fact that extracting information out of a system seems to be `free' of cost in classical mechanics, but rather tricky in the quantum realm. What to measure, and how to measure it, so that we minimize the impact on the system under observation, are key aspects of quantum parameter estimation, as the very notion of the prior existence of `measurable quantities' can even become meaningless in the quantum mechanical sense.

\begin{figure}
\centering
\includegraphics[width = 0.8\textwidth]{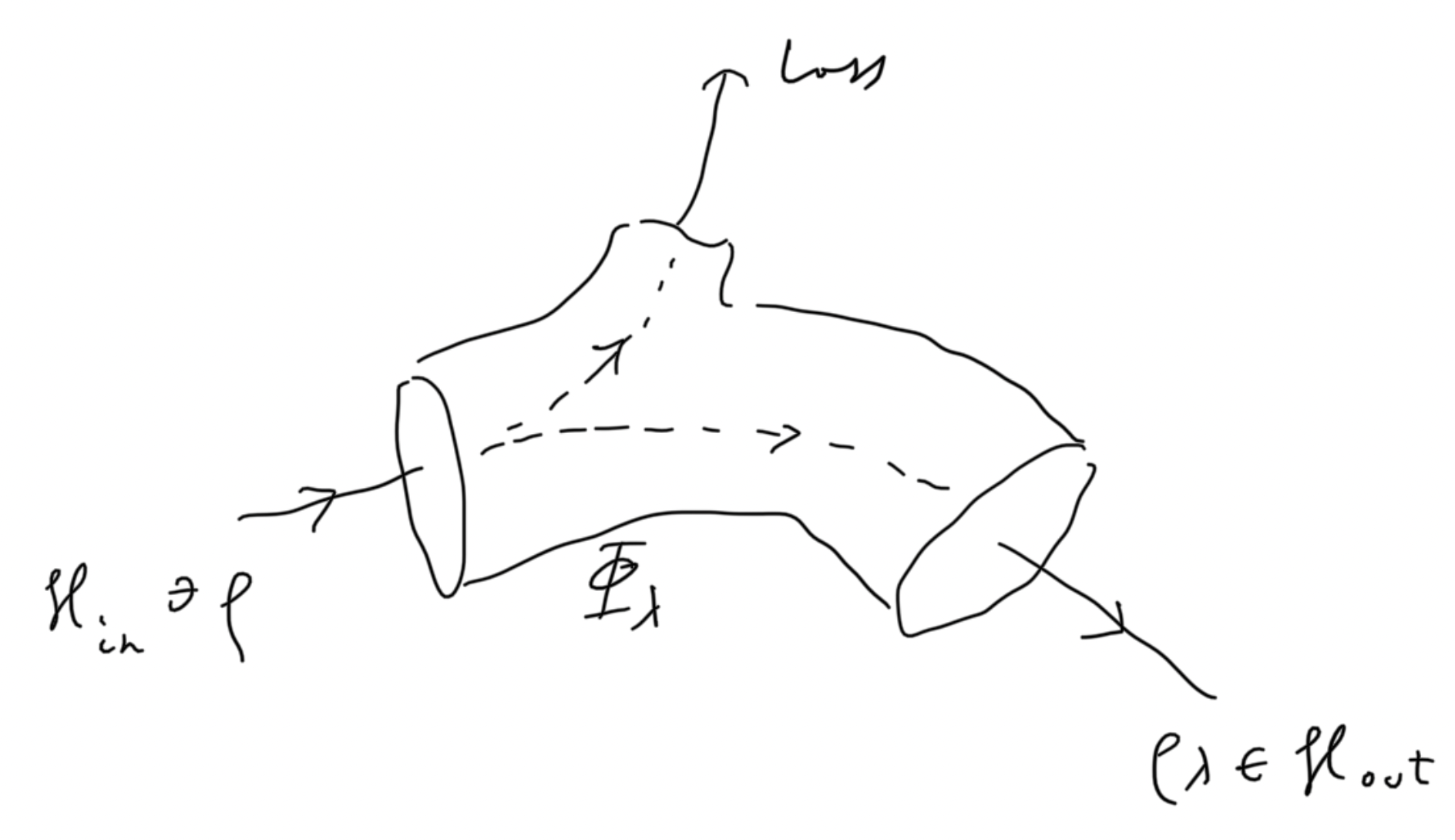}
\caption[Pictorial representation of a quantum parameter encoding through a quantum channel with loss.]{A quantum channel $\Phi_\lambda$ takes an input state ${\rho} \in\mathcal{B}(\mathcal{H}_{in})$ and maps it to $\Phi_\lambda[{\rho}] \equiv {\rho}_\lambda \in \mathcal{B}(\mathcal{H}_{out})$, where $\mathcal{B}(\mathcal{H})$ stands for the space of density matrices associated to Hilbert space $\mathcal{H}$. Since the dimensions of input and output states need not coincide, we have included some possible loss. }
\label{fig:estimation}
\end{figure}

Making use of the notion of quantum channels and POVMs introduced in Section \ref{subsec:MeasurementsStatisticsQuantumChannels}, the idea is quite straightforward, and outlined in Fig. \ref{fig:estimation}: we take a \textit{probe} ${\rho}$, generally containing $n$ particles or degrees of freedom\footnote{In the qCVs case, ${\rho}\in \mathcal{B}(L^2(\reals^n))$}, we make it go through a quantum channel:
\begin{equation}
   \mathcal{D}(\mathcal{H}_{\text{in}})
\ni
{\rho} 
\mapsto 
\Phi_\lambda[\bm{\rho}] \equiv {\rho}_\lambda \in \mathcal{D}(\mathcal{H}_{\text{out}}) 
\end{equation}
where the dimensions of the input and output Hilbert spaces need not be the same (to account, mostly, for a possible loss of information or particles). The channel depends on some parameter $\lambda\in \reals$ that we wish to estimate.

Upon exiting the channel, the probe should get some dependence on $\lambda$: the parameter is now encoded in the quantum state ${\rho}_\lambda$. The first task is then to find a POVM that can extract some information about it. Bear in mind that $\lambda$ itself may not correspond to an `observable' in the usual, von Neumann way. For example, $\lambda$ could simply be a \textit{phase}. But what POVM to choose? Intuitively, we would like to find a measurement scheme specific for \textit{that particular} state ${\rho}_\lambda$, whose statistics --after $M$ iid repetitions and (maybe) classical data processing-- approach the value of $\lambda$. Then, a classical estimator ${\tilde{\lambda}}_{\text{POVM}}$ for the parameter $\lambda$ is constructed, and we are finished.

Assume we have found a suitable POVM and constructed its corresponding estimator. This means we can use the classical CR-bound:
\begin{equation}
    (\Delta{\tilde{\lambda}})^2_{\text{POVM}} \geq \frac{1}{M F_{\text{POVM}}({\lambda})},
\end{equation}
where we stress that statistics for $\Delta{\tilde{\lambda}}$ are obtained with iid processes with the same POVM, and $F_{\text{POVM}}$ corresponds to the Fisher information with $p(x|\lambda)$ provided by Born's rule (see Eq. \eqref{eq:BornRule}) associated with \textit{that} POVM, that is, we take the POVM as a collection o $\{ \Pi_x \}_{x\in\mathcal{X}}$ of  operators satisfying
\begin{equation}
    \int_\mathcal{X} \diff x \Pi_x = \Id 
\end{equation}
where $\mathcal{X}$ describes the set of possible outcomes, and $x$ is one such value, and identify
\begin{equation}
    p(x|\lambda) = \Tr (\Pi_x {\rho}_\lambda)
\end{equation}
Now the trick is to ask for the \textit{best} POVM\footnote{Best POVM may require ancillary d.o.f.} as the one maximizing $F_{\text{POVM}}$.

This gives us the {quantum Cramér-Rao} bound:
\begin{equation}\label{eq:qCRBound}
  \text{Var}_\lambda\left(\tilde{\lambda}_{\hat{O}}\right) \geq \frac{1}{M H({\lambda})}
\end{equation}
where 
\begin{equation}\label{eq:QFI}
    H({\lambda}) := \max_{\text{POVM}} F_{\text{POVM}}({\lambda})
\end{equation}
is the \textit{quantum Fisher information} (QFI); $\tilde{\lambda}_{\hat{O}}$ is the optimal estimator; and ${\hat{O}}$ is the optimal \textit{quantum} estimator \cite{paris2009}. The variance in the qCR bound is computed against the probability distribution provided by the $M$ repetitions of the POVM associated with the optimal quantum estimator.

Starting from a pure state probe, we can get\footnote{For basis-independent formula, see the last reference by M. Paris \cite{paris2009}.}
\begin{equation}\label{eq:QFI-basis}
H(\lambda) =   2 \sum_{m,n}\frac{|\bra{\psi_m}\partial_\lambda {\rho}_\lambda \ket{\psi_n}|^2}{\psi_m + \psi_n},
\end{equation}
 $\lbrace \psi_{m}, \ket{\psi_{m}} \rbrace$ are the eigensolutions to ${\rho}_\lambda \ket{\psi_{m}} = \psi_{m} \ket{\psi_{m}}$, and ${\rho}_\lambda$ is the measured, or received state.
Moreover, the theory also provides an explicit way of finding one such optimal observable, whose outcomes allow us to explicitly find the estimator \cite{paris2009}:
\begin{equation}\label{eq:opt-obs}
\hat{O}_{\tilde{\lambda}} = \lambda \hat{\Id} + \frac{\hat{L}_\lambda}{H(\lambda)},
\end{equation}
 where $\hat{L}_\lambda$ is a symmetric logarithmic derivative (SLD) that solves the equation $\lbrace\hat{L}_\lambda, {\rho}_\lambda\rbrace = 2\partial_\lambda {\rho}_\lambda$, where $\lbrace \cdot, \cdot \rbrace$ is the anticommutator. When the estimator $\tilde{\lambda}$ is constructed using a maximum likelihood method, the  quantum Cramér-Rao bound (qCRB) \cite{cramer, rao} is  asymptotically achieved meaning that the observable in \eqref{eq:opt-obs} has the smallest possible variance.

Note that in Eq. \eqref{eq:QFI-basis} the QFI implicitly depends on \textit{the way} parameter $\lambda$ is imprinted onto ${\rho}$ by means of the quantum channel $\Phi_\lambda$.
 This suggests we can further optimize over the probes ${\rho}$. Intuitively, this means that some probes should be more \textit{sensitive} to the channel than others, as can be clearly seen in the direct relation between a growing QFI and the sensitivity of the probe to small changes in the parameter via the derivative $\partial_\lambda {\rho}_\lambda$. This fact can be related to geometric results, which motivates the following digression:
 
\subsubsection{Geometry of quantum parameter estimation}\label{subsubsec:GeometryOfQuantumParameterEstimation}

The QFI can be interpreted geometrically by means of a notion of distance in the Hilbert space spanned by density operators\footnote{As mentioned in the introductory chapter, the inner product required to exist in a Hilbert space naturally gives rise to a notion of norm, which can then be connected to a metric. A Hilbert space can then be seen as a complete metric space, where completeness is understood in terms of convergence of any Cauchy sequence.}. Among the many candidates, the Bures distance
\begin{equation}
D^2_B(\rho_1, \rho_2) := 2\left( 1-\sqrt{F(\rho_1, \rho_2)}\right),
\end{equation}
{where $F(\rho_1, \rho_2) := \left(\Tr \left[ \sqrt{\sqrt{\rho_1}\rho_2\sqrt{\rho_1}}\right]\right)^2$ is the Uhlmann fidelity} between states $\rho_1$ and $\rho_2$, the one correctly linking estimation to geometry \cite{Sommers-Zyczkowski-2003}. This makes the interpretation of quantum estimation straightforward: it depends on the distinguishability between states. For the sake of generality, let's take the multiparametric case: if $\bm{\lambda}$ is a vector of parameters that defines a (possibly continuous) family of states $\lbrace \rho_{\bm{\lambda}}\rbrace$, then the Bures distance between two states $\rho_{\bm{\lambda}}$ and $\rho_{\bm{\lambda} + \diff \bm{\lambda} }$ that are infinitesimally close in terms of the parameter $\bm{\lambda}$ can be related to a metric tensor $\mathsf{g}_{ab}$:
\begin{equation}
D^2_B(\rho_{\bm{\lambda}}, \rho_{\bm{\lambda} + \diff \bm{\lambda} }) \equiv
\mathsf{g}_{ab} \diff \lambda_a \diff \lambda_b.
\end{equation}
Now, it can be shown  that 
\begin{equation}
 \mathsf{g}_{ab}   = \frac{H_{ab}(\bm{\lambda})}{4},
\end{equation}
 where $H_{ab}(\bm{\lambda}) := \Tr(\partial_a \rho_{\bm{\lambda}}\hat{L}_{b})$ is the quantum Fisher information \textit{matrix}, and the subindices $a$ and $b$ refer to parameters $\lambda_a$ and $\lambda_b$, so that $\partial_a \equiv \partial/\partial\lambda_a$, and  $\hat{L}_{b}$ is the symmetric logarithmic derivative for parameter $\lambda_b$ (see \textit{e.g.} Ref. \cite{paris2009} for more details).

A large QFI translates in a large distinguishability between states. The problem of computing the QFI, however, can become mathematically challenging due to the diagonalization implicit in Eq. \eqref{eq:QFI-basis}. In the case of Gaussian states and Gaussian preserving operations, the diagonalization of the infinite-dimensional density operator simplifies considerably.

\subsection{Gaussian quantum parameter estimation}\label{subsec:GaussianQuantumEstimation}
As shown in \cite{Safranek_2018}, when we are in the presence of Gaussian states and Gaussian-preserving channels, there is no need to diagonalize the density matrix  in order to find the QFI. For a single parameter, and a bipartite state, the QFI can be computed using
\begin{align}\label{eq:QFI}
\begin{split}
H(\lambda) &= \frac{1}{2(\det[A]-1)}\left[
\det[A] \Tr\left[ (A^{-1}\partial_{\lambda}A)^2\right]\right. \\
&+  \sqrt{\det[\Id_2 + A^2]}\Tr\left[\left((\Id_2 + A^2)^{-1}\partial_{\lambda}A\right)^2 \right]  \\
&-\left. 4\left( \nu_+^2 - \nu_-^2 \right) \left(\frac{(\partial_\lambda \nu_+)^2}{\nu_+^4-1} - \frac{(\partial_\lambda \nu_-)^2}{\nu_-^4-1}\right)\,\right]+ 2\partial_\lambda \vec{d\,}^\dagger\Sigma_\lambda^{-1}\partial_\lambda \vec{d},
\end{split}
\end{align}
where $\nu_{\pm}$ are the \textit{symplectic eigenvalues} of ${\Sigma}_\lambda$, defined following Ref.  \cite{_afr_nek_2015}
\begin{equation}\label{eq:symp-eigenv}
2\nu_{\pm}^2 := \Tr[A^2] \pm \sqrt{\left(\Tr[A^2]\right)^2-16 \det[A]},
\end{equation}
with the matrix $A$  given by 
$A := i {\Omega}{T}{\Sigma}_\lambda{T}^\intercal$,  $\Omega:=\text{antidiag}(\Id_2, -\Id_2)$, and ${T}_{ij}:=\delta _{j+4,2 i}+\delta _{j,2 i-1}$ is the matrix that changes the basis to the \textit{quadrature basis}
$$(\hat{x}^{\text{th}}_1,  \hat{x}^{\text{S}}_1, \hat{x}^{\text{th}}_2, \hat{x}^{\text{S}}_2, \hat{p}^{\text{th}}_1, \hat{p}^{\text{S}}_1, \hat{p}^{\text{th}}_2, \hat{p}^{\text{S}}_2)^{\intercal}.$$
For a Gaussian state $\left( {\Sigma}_\lambda, \vec{d}_\lambda\right)$ written in the complex basis,  the symmetric logarithmic derivative in Eq. \eqref{eq:opt-obs} can be obtained as in  Ref. \cite{Safranek_2018}: 
\begin{equation}\label{eq:sld}
\hat{L}_\lambda = \Delta \vec{\hat{A}}^\dagger {\mathcal{A}}_\lambda  \Delta \vec{\hat{A}} -\frac{1}{2}\Tr [{\Sigma_\lambda}{\mathcal{A}}_\lambda] + 2\Delta \vec{\hat{A}}^\dagger {\Sigma}_\lambda^{-1}\partial_\lambda\vec{d}_\lambda,
\end{equation}
where $\Delta \vec{\hat{A}} := \vec{\hat{A}} - \vec{d}_\lambda \hat{\Id}$,
 $\vec{\hat{A}}$  the \textit{complex basis} vector of bosonic operators, ${\mathcal{A}}_\lambda:= {\mathcal{M}}^{-1}\partial_\lambda \vec{d}_\lambda$, where   ${\mathcal{M}} =\bar{ {\Sigma}}_\lambda \otimes {\Sigma}_\lambda - {K}\otimes {K}$, where the bar denotes complex conjugate, and ${K} := \text{diag}\left(\Id_2, -\Id_2\right)$.

\section{Case example: Quantum illumination with loss}
\label{section:QuantumIllumination}
Here we reproduce some well-known results with a slightly different approach than the usual. This section, hence, shall be used as a link to Chapter \ref{chapter:biFreq}, where  novel results are presented.

Quantum illumination (QI) is a protocol that makes use of entanglement in order to enhance the results in a detection problem. We refer the reader to Ref. \cite{shapiroStory} and references therein for more details. Here we will use the methods described in the previous section to confirm the quantum illumination results, with the explicit inclusion of absorption loss.
Absorption losses in a homogeneous medium can be modeled as an array of $N$ `infinitesimal beams splitters'  spanned over length $L$, each of them with the same transmittivity per unit length $\mu_i$, where $i = 1, \ldots, N$.

Let's take two such beam splitters. Then, if they are embedded in the same environment, the combined transmittivity is $\tau^{(2)} =\tau_1 + \tau_2 - \tau_1\tau_2 $. Now, assuming that all transmittivities are equal, $\tau_i\equiv \tilde{\tau}=\mu L/N$ for all $i$, we get
\begin{equation}
\tau^{(2)}  = 2 \frac{\mu L}{N} \left( 1- \frac{\mu L}{2N} \right) =  2 \tilde{\tau} \left( 1- \frac{\tilde{\tau}}{2} \right).
\end{equation}
Treating this relation as an iterative equation, and using the logistic map, 
\begin{equation}
    x_{n+1} = r x_{n}(1-x_{n}),
\end{equation}
whose solution in the case $r=2$ is 
\begin{equation}
    x_{n} = \frac{1}{2} \left[ 1 - (1-2x_{0})^{2^{n}}\right],
\end{equation}
 we find that after $k$ iterations
\begin{equation}
\frac{\eta^{(k)}}{2}  =  \frac{1}{2} \left[ 1 - \left(1-\frac{\tilde{\eta} L}{N}\right)^{2^{k}}\right].
\end{equation}
Given that $N=2^{k}$, we take the limit on $k$, obtaining the following
\begin{equation}
\lim_{k\rightarrow\infty} \left(1-\frac{\tilde{\tau} L}{2^{k}}\right)^{2^{k}} = e^{-\tilde{\tau} L} \equiv e^{-\gamma},
\end{equation}
Therefore, the total reflectivity $\eta = 1-\tau$ of an array of infinitesimal beam-splitters, each having the same reflectivity per unit length is
\begin{equation}
\eta_{\text{loss}} = e^{-\gamma},
\end{equation}
and this shall represent a lossy medium in what follows.

\begin{figure}
\centering
\includegraphics[width = 0.8\textwidth]{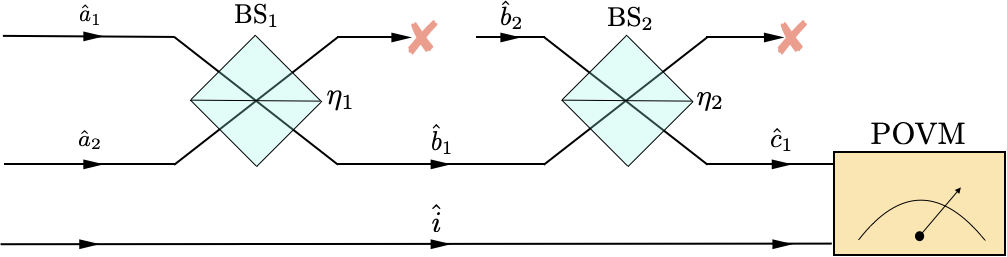}
\caption[Circuit representation of the quantum illumination protocol with absorption loss.]{Quantum illumination protocol with two beam splitter interactions made explicit: the first one corresponds to the object to be detected, while the second summarizes absorption loss (the order is irrelevant). The modes $\hat{a}_2$ and $\hat{i}$ correspond, respectively to the signal and the idler, which in QI are entangled. If the two beam splitters are embedded in the same environment, a requirement made explicit in the text, then $\hat{a}_1 \equiv \hat{b}_2$. The joint measurement is performed over the reflected signal $\hat{c}_1$ and the idler. Quantum illumination shows that even when the evolution may correspond to an entanglement-breaking channel, the initial stronger-than-classical correlations between signal and idler pairs are good enough to produce a quantum-enhancement in the estimation of the object's reflectivity, which is equivalent to say that the detection problem gets solved with a quantum strategy that beats the classical approach.}
\label{fig:2bs}
\end{figure}

Now, consider the situation depicted in  FIG. \ref{fig:2bs}. It is straightforward to compute the output $\hat{c}_2 = \sqrt{\eta_2 \left(1-\eta_1\right)}\hat{a}_1 + \sqrt{\eta_1 \eta_2}\hat{a}_2 + \sqrt{1-\eta_2}\hat{b}_2$. If we were to equate this to a single, effective beam splitter of reflectivity $\eta_\text{eff}$ mixing the same initial mode $\hat{a}_1$ with an effective bath $\hat{b}_\text{eff}$ it is clear that we must have: $\eta_\text{eff} = \eta_ 1 \eta_2 $ and $\hat{b}_\text{eff} = \hat{h}_\text{th} = \hat{a}_1 = \hat{b}_2$, such that $\langle \hat{h}^\dagger_\text{th} \hat{h}_\text{th} \rangle = N_\text{th}$. Therefore,  we set $\eta_\text{eff} = \eta_1 e^{-\gamma}\equiv \eta e^{-\gamma}$ as the reflectivity of the combined beam splitter.

As we have seen, QI can be presented as a quantum estimation problem, where the target's reflectivity $\eta$ is the parameter of interest.
Here, the model is as follows: the target object is represented by a low reflectivity beam splitter, which for simplicity introduces no phase. Additionally, absorption losses are modeled by a second beam splitter of reflectivity $\exp(-\gamma)$, where $\gamma := \alpha L$ is a dimensionless factor, and where $\alpha$ represents the loss per unit length, and $L$ is the linear total distance the signal travels. Note that this is a one dimensional-problem. As we have seen, the combination of two beam splitters of reflectivities $\eta_1$ and $\eta_2$ embedded in the \textit{same} environment (here a thermal state with $N_\text{th}$ photons) can be reduced to a single beam splitter with an effective reflectivity given by $\eta_\text{eff} = \eta_1\eta_2 $. In our case
\begin{equation}
\eta_\text{eff} = \eta e^{-\gamma}.
\label{eq:etaeff}
\end{equation}
 In the symplectic formalism, beam splitters  are represented by the  transformation
\begin{equation}
{S}_{\text{BS}}(x) = 
\left(
\begin{array}{cc}
x\Id_2 & \sqrt{1-x^2}\Id_2 \\
 -\sqrt{1-x^2}\Id_2 & x\Id_2
\end{array}
\right),
\label{eq:BS}
\end{equation}
where $x$ is the reflectivity.

QI can be seen as a procedure where a tripartite state of: thermal bath, signal, and idler, exhibiting bipartite entanglement between signal and idler, undergoes a transformation where the received state is bipartite and carries some dependence on $\eta$. An estimation of $\eta$ follows, performing some measurement and --perhaps-- post-processing the outcomes to construct an optimal estimator. Symbolically, an initial state characterized by a covariance matrix $\Sigma$ and a displacement vector $\vec{d}$ is mapped to an output state:
\begin{equation}
\left( \Sigma^\text{ABC}, \vec{d}^\text{ABC}\right) \mapsto \left( \Sigma^\text{DC}_\eta, \vec{d}^\text{DC}_\eta\right),
\end{equation}
where $A$, $B$, $C$ are the initial Hilbert spaces (bath, signal, and idler), and $D$ represents the Hilbert space where the reflected signal, mixed with the thermal noise, lives.

\begin{figure}[h!]
\centering
\includegraphics[width=0.8\textwidth]{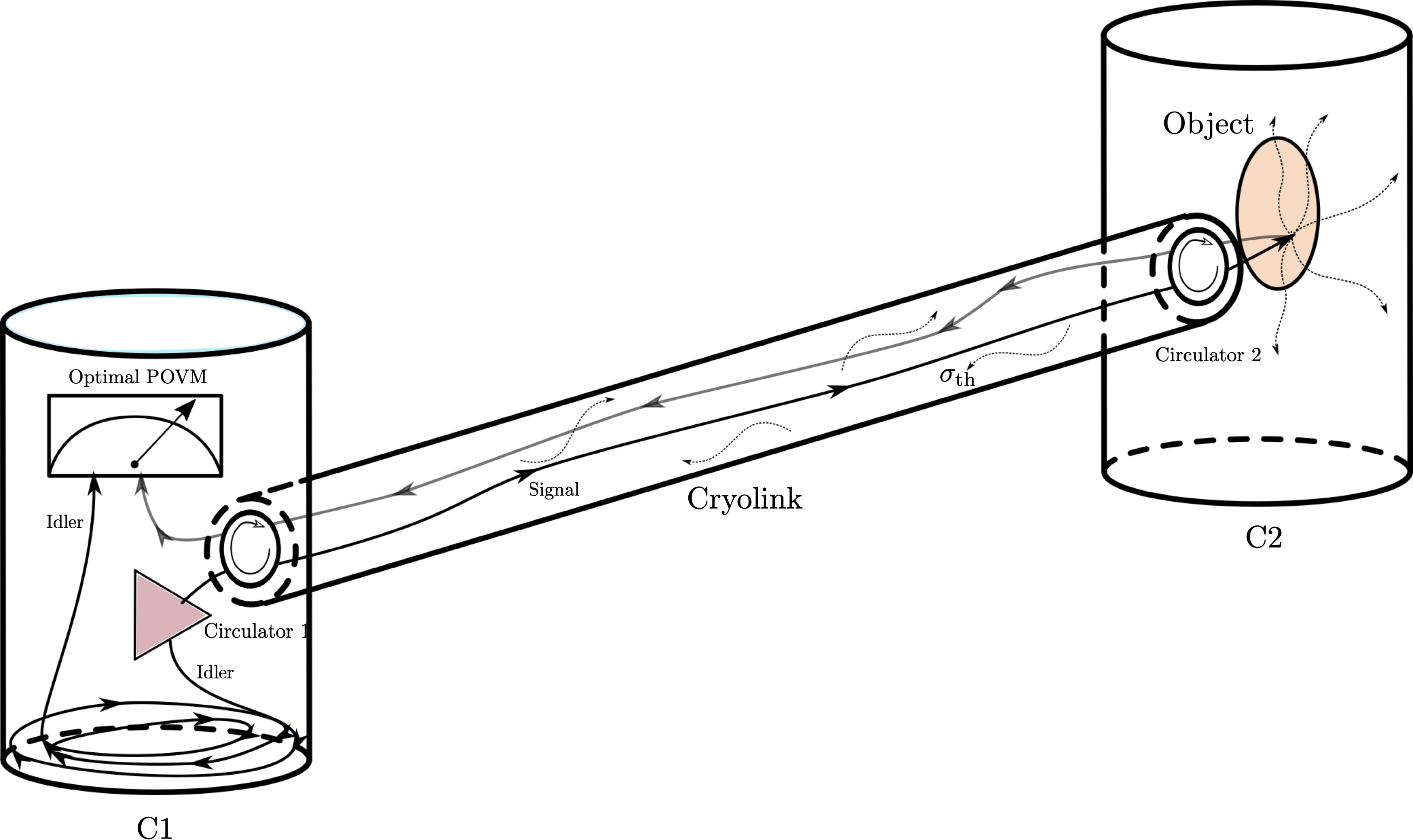}
\caption[Possible scheme for the implementation of quantum illumination with losses between two cryogenic stations linked via a cryogenic cable]{Possible scheme for the implementation of quantum illumination with losses between two cryogenic stations linked via a cryogenic cable. C1 contains the state generator for both the signal and the idler. The signal is sent through the lossy medium of the cryolink, over to the other fridge, C2, where it interacts with a low-reflectivity object embedded in a thermal background of tunable temperature. The reflected, noisy signal makes its way back to the first fridge, where the idler has been kept in a delay line for the joint measurement. This setup, highly idealised, would allow to test QI in a full-microwaves fashion.}\label{fig:QI2fridges}
\end{figure}

The classical strategy could be seen in a similar fashion, but not allowing for any non-classical correlations between any subsystems. In this case, the idler mode is clearly irrelevant, since it can't enhance the detection of $\eta$.  
In Fig. \ref{fig:QI2fridges} we attempt at an idealised implementation of the QI protocol between two cryogenic refrigerators.

\subsection{Quantum probe}
\begin{figure}
\centering
x\includegraphics[width=0.8\textwidth]{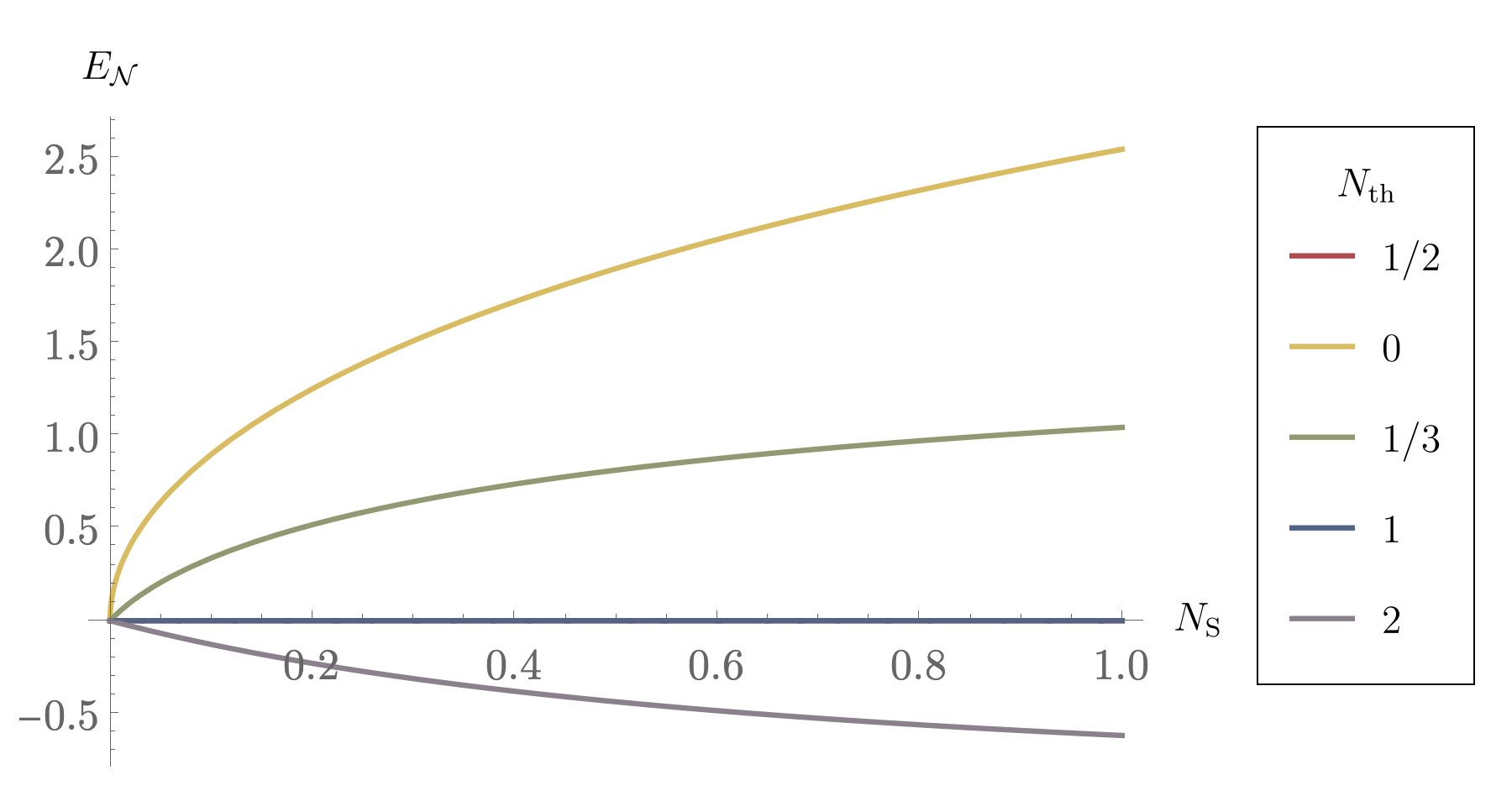}
\caption[Logarithmic negativity $E_\mathcal{N}$ of the state in Eq. \eqref{eq:MikelState} as a function of the signal photon number $N_\text{S}$ for different thermal photon number $N_\text{th}$ present in the state.]{Logarithmic negativity $E_\mathcal{N}$ of the state in Eq. \eqref{eq:MikelState} as a function of the signal photon number $N_\text{S}$ for different thermal photon number $N_\text{th}$ present in the state. Entanglement is shown for every $N_\text{th} < 1$. }
\label{fig:LogNeg}
\end{figure}

We use the following tripartite Gaussian state, which preserves the number of thermal photons and that is defined by a null displacement and a real covariance matrix given by
\begin{equation}
\Sigma=
\left(
\begin{array}{ccc}
 \Sigma_\text{A}  & 0 & 0  \\
 0 &  \Sigma_B &  \varepsilon_{BC}  \\
 0 &   \varepsilon_{BC}  &  \Sigma_C  
\end{array}
\right),
\label{eq:MikelState}
\end{equation}
where $\Sigma_\text{A} : = (1+ 2 N_\text{th})\Id$ is a thermal state with $N_\text{th}$ photons, $\Sigma_B = \left(1+2 N_\text{S}+2 N_\text{th}\right)\Id$, $\varepsilon_{BC} = 2 \sqrt{N_\text{S} (N_\text{S}+1)}\Sigma_{Z}$ are the correlations between subsystems $B$ and $C$, and $\Sigma_C = (1+2 N_\text{S})\Id$, where $\Id$ is the $2 \times 2$ identity matrix, and $\sigma_{Z}$ is the third Pauli matrix. In our setting, subsystem $C$ represents the idler mode, and $A$ and $B$ are the quantum entangled subsystems comprising the signal with photon number $N_\text{S}$ and the idler.
This covariance matrix satisfies the uncertainty principle, and as such represents a valid quantum  state. The motivation for using such a state is as follows: we want to have a state that has a \textit{fixed} thermal photon number $N_\text{th}$ while not exhibiting a shadow effect. Moreover, this state can also be obtained by mixing the signal part of a TMSV state with a modified thermal state with a $2\times 2$ covariance matrix $\left(1+2N_\text{th}(1-x)^{-1}\right)\Id$ via a beam splitter of reflectivity $x$, as already proposed in \cite{shapiroStory}.
The smaller symplectic eigenvalue  of the \textit{partially transposed} version  of the state in Eq. \eqref{eq:MikelState} is
\begin{landscape}
\begin{equation}
\tilde{\nu}_{-} = \sqrt{8 N_\text{S}^2-2 (2 N_\text{S}+N_\text{th}+1) \sqrt{4 N_\text{S} (N_\text{S}+1)+N_\text{th}^2}+4 N_\text{S} N_\text{th}+8 N_\text{S}+2 N_\text{th}^2+2 N_\text{th}+1}.
\end{equation}
\end{landscape}
The entanglement condition $\tilde{\nu}_{-} < 1$ reduces to $N_\text{S} > 0 \wedge N_\text{th} <1$. In Fig. \ref{fig:LogNeg} we plot the logarithmic negativity $E_\mathcal{N} := -\log_2 \tilde{\nu}_{-}$ as a function of the signal photon number $N_\text{S}$ for different $N_\text{th}$. In particular, $N_\text{th} = 0$ reproduces the negativity of the two-mode squeezed vacuum state, as expected. 

The  thermal bath and the signal interact via Eq. \eqref{eq:BS} setting $x = \eta_\text{eff}$, while the idler mode is kept in a lossless delay line, represented by the identity operation. The total symplectic transformation for the full state is thus 
\begin{equation}
{S}_T = {S}_{\text{BS}}(\eta_\text{eff}) \oplus \Id.
\end{equation}
The transformed state is then represented by covariance matrix $\Sigma^\prime_\eta = {S}_T \Sigma_T {S}_T^\intercal$, which already carries the dependence on $\eta$, the parameter that we want to estimate using the Gaussian estimation techniques described above. Essentially, we want to discriminate between two channels: one where there is no object ($\eta = 0$) and another where there is a slightly reflective object ($\eta \ll 1$) \cite{Sanz_2017}. The goal is to show that the suggested state of Eq. \eqref{eq:MikelState} performs better than the classical optimal probe. The transmitted part of the signal is lost, which is implemented by removing the corresponding rows and columns of the covariance matrix. We call the received covariance matrix $\Sigma_\eta$, given by
\begin{equation}
\Sigma_\eta = 
\left(
\begin{array}{cc}
f\Id   & g\sigma_Z  \\
g\sigma_Z &  \Sigma_C
\end{array}
\right),
\end{equation}
where $f =1 +2N_\text{th} + 2N_\text{S} e^{-2\gamma}\eta^2$, $g = 2\sqrt{N_\text{S}\left(1+N_\text{S} \right)}e^{-\gamma}\eta$, and $\Sigma_C = 1 + 2N_\text{S}$. It can be seen that in the limit of infinite losses $\gamma \rightarrow \infty$ this state becomes an uncorrelated bath and an coherent idler state.

We compute the quantum Fisher information relative to $\eta$ using the Gaussian formulas found in \cite{Safranek_2018, SafranekThesis}, working in the assumed neighborhood of $\eta \sim 0$. 
In order to avoid regularization schemes, the two symplectic eigenvalues of $\Sigma_\eta$ need to be larger than one. These conditions are equivalent to having $N_\text{S} > 0 \wedge N_\text{th} > 0$
The QFI is then found to be
\begin{equation}
 H_Q =  \frac{4 N_\text{S} e^{-2\gamma}(1+N_\text{S})}{1+2 N_\text{S} N_\text{th}+N_\text{S}+N_\text{th}}.
 \label{eq:HQ}
 \end{equation} 
As expected, when absorption losses tend to infinity, the QFI vanishes. It also vanishes for $N_\text{S}=0$, indicating that the protocol is free from shadow effects.
 \subsection{Classical probe}
 Here we use a coherent state as probe: $\ket{\psi} =\ket{\alpha}$, where we set $\alpha \in \mathbb{R}$ for simplicity. The expected photon number in this state is $\left| \alpha^2\right|$, and we set it equal to the signal photon number $N_\text{S}$ in the quantum state, for a matching-resources comparison.
The initial displacement vector of the total state (thermal bath, and signal) in the real basis is $\vec{d}_{0}^\intercal = (0,\sqrt{2}\alpha,0,0,0)$ which leads --after the interaction and the trace of the losses-- to  $\vec{d}_\eta^\intercal = (\sqrt{2}e^{-\gamma}\alpha\eta,0, 0,0)$.
We follow the same procedure as before, and compute the QFI $H_C$ for this state, inserting the displacement vector, and the covariance matrix in Eq. \eqref{eq:QFI}. We find 
\begin{equation}
H_C = \frac{4 e^{-2 \gamma } N_\text{S}}{2 N_\text{th}+1}
\label{eq:HC}
\end{equation}
In the next section, we compare the ratio $H_Q/ H_C$ to the one obtained in non-lossy QI.
\subsection{Comparison with quantum illumination}
In this section we compare our results to the ones obtained in \cite{Sanz_2017} for quantum illumination.  
The gain $R \equiv H_Q/H_C$ of the lossy protocol turns out to be exactly that of the original, non lossy QI:

\begin{equation}
R = \frac{(1+N_\text{S}) (1+2 N_\text{th})}{1+2 N_\text{S} N_\text{th}+N_\text{S}+N_\text{th}}.
\end{equation}

\begin{figure}
\centering
\includegraphics[width=0.7\textwidth]{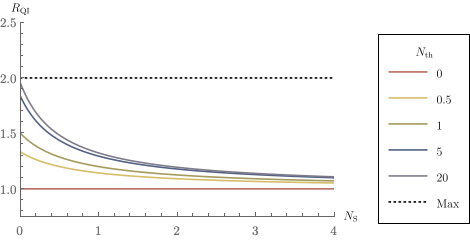}
\caption[Quantum Fisher information (QFI) ratio $R_{\text{QI}}:=H_Q/H_C$ for quantum illumination (QI) in the presence of losses.]{Quantum Fisher information (QFI) ratio $H_Q/H_C$ for quantum illumination (QI) in the presence of losses as a function of the signal photon number $N_\text{S}$ for different bath photon number $N_\text{th}$. In the low signal photon number and high noise photon number, the  protocol converges to the 3dB limit, that is, an overall factor of 2, represented by the plane $H_Q = 2H_C$. }
\label{fig:2dQI}
\end{figure}

This comes as no surprise, and it is actually a trivial result, coming from the additivity of the QFI. However, it proves that the methods are quite robust, and reproduce well-known results with a very straightforward calculation fashion. 

\chapter{Bi-frequency illumination: a quantum-enhanced protocol}
\label{chapter:biFreq}
This chapter is based on the following article: Casariego, M., Omar, Y., Sanz, M., \textit{Bi-Frequency Illumination: A Quantum-Enhanced Protocol}, Adv Quantum Technol.  \textbf{5}, 2100051 (2022). 

\section{Introduction}
Quantum information technologies are opening very promising prospects for faster computation, securer communications, and more precise detection and measuring systems, surpassing the capabilities and limits of classical information technologies \cite{nielsen_chuang_2010, google-quantum-supremacy, PhysRevLett.69.3598, Pirandola_2018, katori}. Namely, in the domain of quantum sensing and metrology \cite{ Giovannetti_2011}, we are currently witnessing a boost of applications to a wide spectrum of physical problems: from gravimetry and geodesy \cite{bongs, menoret, flury, grotti, zeuthen}, gravitational waves \cite{ligo}, clock synchronisation \cite{katori}, thermometry \cite{thermometry} and bio-sensors \cite{bowen2013, omar, plenio, biosensors, jensen}, to experimental proposals to seek quantum behavior in macroscopic gravity \cite{milburn}, to name just a few. 

While many of the quantum metrology studies focus on unlossy and noiseless (unitary) scenarios, the more realistic, lossy case has also been investigated \cite{lee, changhun, zhang, zhang2, guta2012, davidovich2012, acin2013, sekatski, sekatski2}. Equivalently, one can talk about quantum metrology with open quantum systems. Understanding what are the precision limits of measurements in the presence of loss is a fundamental endeavour in quantum metrology \cite{huelga, friis}. Certain noise properties have been found to be beneficial in some scenarios \cite{huelga2014, sekatski2017}, and quantum error correction schemes have been proposed to overcome decoherence and restore the quantum-enhancement \cite{muschik2017}. 
Quantum illumination (QI) \cite{Lloyd1463, PhysRevLett.101.253601, guha2009, shapiroStory, alsing2019, pirandolaMicrowave, shabir, borre, genovese, gregory, cai2021} is a particularly interesting example of a lossy and noisy protocol where the use of entanglement proves useful even in an entanglement-breaking scenario. QI shows that the detection of a low-reflectivity object in a noisy thermal environment with a low-intensity signal is enhanced when the signal is entangled to an idler that is kept for a future joint measurement with the reflected state. This makes QI a candidate for a quantum radar \cite{lanzagorta}, although a more involved protocol is needed \cite{macconeRadar, Zhuang2021}. The decision problem of whether there is an object or not can be rephrased as a quantum estimation of the object's reflectivity $\eta$,  in order to discriminate an absence ($\eta =0$) from a presence ($\eta \ll 1$) of a low-reflectivity object \cite{Sanz_2017}.

The goal of quantum estimation \cite{holevo, helstrom, caves, parisBook} is to construct an estimator $\tilde{\lambda}$ for certain parameter $\lambda$ characterizing the system. It is noteworthy that not every parameter in a system corresponds to an observable, and this may imply the need for data post-processing. Either way, the theory provides techniques to obtain an optimal observable --not necessarily unique, \textit{i.e.} whose mean square error is minimal. The estimator $\tilde{\lambda}$ is nothing but a map from the results of measuring the optimal observable to the set of possible values of the parameter $\lambda$. One of the main results of this theory is the quantum Cramér-Rao (qCR) bound, which sets the ultimate precision of any estimator. Whether this bound is achievable or not depends on the data-analysis method used, and on the statistical distribution of the outcomes of different runs of the experiment. In most practical situations, maximum likelihood methods for unbiased estimators, together with a Gaussian distribution of the outcomes of the (independent) runs of the experiment, make the bound achievable. 
In order to find the qCR --and the explicit form of the optimal observable-- one needs to compute the quantum Fisher information\footnote{In a multiparameter scenario the quantum Fisher information is actually a matrix}(QFI), which roughly speaking quantifies how much information about $\lambda$ can be extracted from the system, provided that an optimal measurement is performed. In general, computing the QFI involves diagonalisation of the density matrix, which makes the obtention of analytical results challenging. However, if one restricts to Gaussian states and Gaussian-preserving operations \cite{parisOlivares, adesso, olivares, serafini}, the so-called symplectic approach simplifies the task considerably \cite{Safranek_2018, _afr_nek_2015, pinel, friis2015, pinel2013, monras, jiang2013, marian2016, gao, nichols, banchi}.
As the QFI is by definition optimized over all POVMs, it only depends on the initial state, often called \textit{probe}. This means that a second optimization of the QFI can be pursued, this time over all possible probes. Moreover, this approach allows us to quantitatively compare different protocols, \textit{e.g.} with and without entanglement in the probe, since an increase in the QFI when the same resources are used --which typically translates into fixing the particle number, or the energy-- directly means an improvement in precision.

{In this article, we propose an idler-free quantum-enhanced, lossy protocol to estimate the reflectivity $\eta(\omega)$ of an object as a function of the frequency when the object is embedded in a noisy environment. In particular, we propose a  method where a bi-frequency state is sent to probe a target --modeled as a beam splitter with a frequency-dependent reflectivity $\eta(\omega)$ and embedded in a thermal environment.  The goal is to obtain an estimator for the parameter $\lambda = \eta(\omega_2)-\eta(\omega_1)$, that captures information about the linear frequency dependence of the object. For simplicity, it is assumed that the frequencies are sufficiently close  so that we can work in a neighborhood of $\lambda\sim 0$.} 

By imposing that the expected photon number is the same in quantum and classical scenarios, we find the QFI ratio between them, and analyze when it is greater than one. We find that the maximum enhancement is obtained for highly reflective targets, and derive explicit limits in the highly noisy case.
We also provide expressions for the optimal observables, proposing a general experimental scheme described in Figure 2 , and motivating applications in microwave technology \cite{Sanz_2018}.  

The article is structured as follows. First, we introduce the model, along with the main concepts and formulas from quantum estimation theory, motivating the use of Gaussian states. Then, we compute the QFI and show the quantum enhancement. Finally, we compute the optimal observables for both the quantum and the classical probes, and briefly discuss applications.

\section{Model}

\begin{figure}[ht]
\centering
\includegraphics[width=0.6\linewidth]{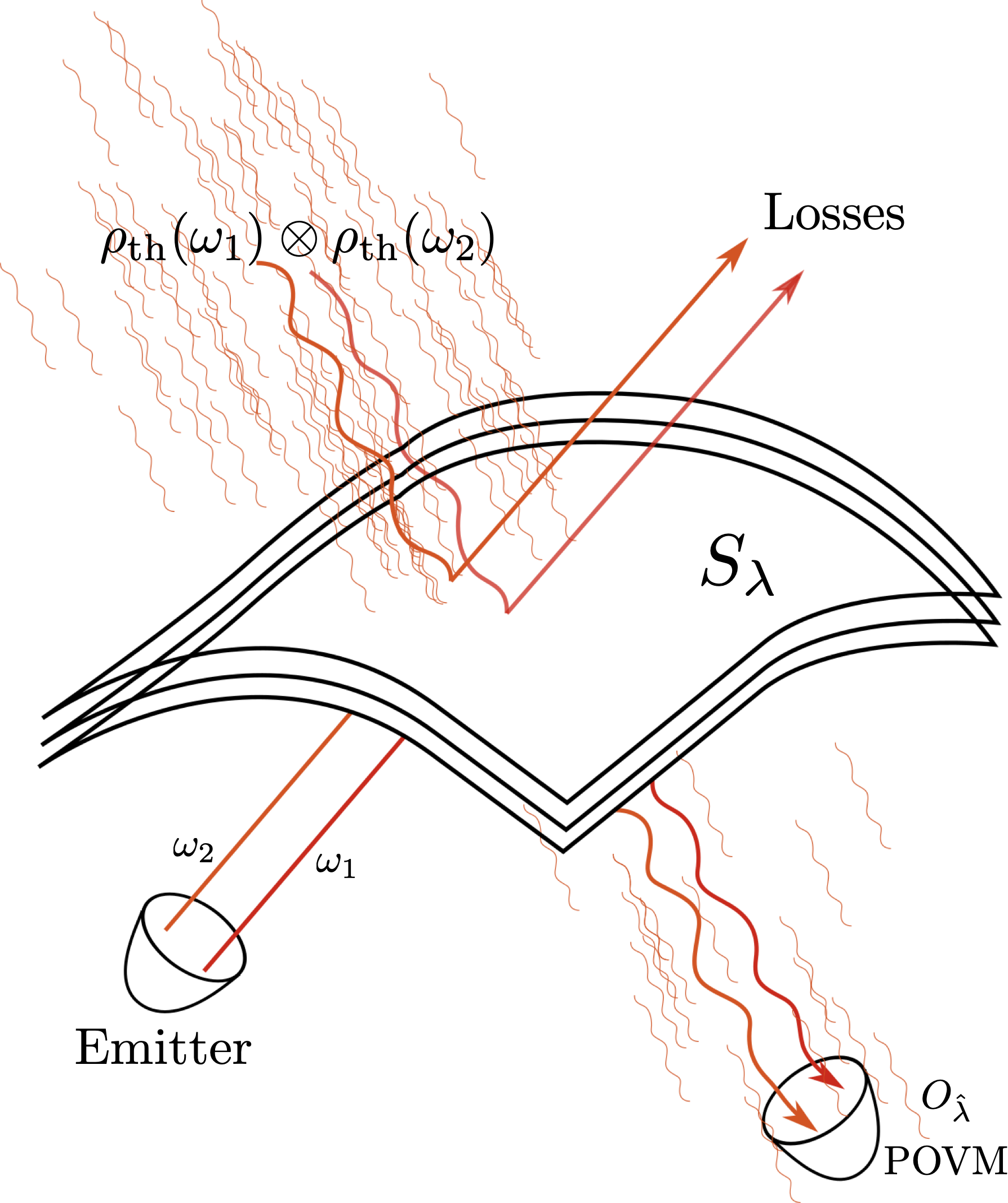}
\caption[Bi-frequency illumination on a reflective, layered object]{An object reflects a bi-frequency beam (notice the similar but different colors of the two beams coming out of the emitter), and mixes it with a thermal bath for each frequency with the same expected photon number, coming from the upper left. The transmitted signal and reflected thermal state are lost, and a measurement $O_{\hat{\lambda}}$ is performed onto the available part (lower right corner), whose expectation values converge, after classical data processing, to an estimator of the parameter $\lambda$ encoded in the object. In our case, $S_\lambda$ represents the transformation associated to a multi-layered object, modeled as a beam splitter, where $\lambda := \eta_2 - \eta_1$ is the parameter to be estimated, where $\eta_i =\eta(\omega_i)$ are the reflectivities for the different frequencies. The emitter can produce either a pair of coherent states (classical strategy) or an entangled, two-mode squeezed state. The latter proves advantageous in the parameter estimation, giving a strictly larger quantum Fisher information.}
\label{fig:diag}
\end{figure}

The model is synthesized in Figure \ref{fig:diag}: the target object, modeled as a beam splitter with a frequency-dependent reflectivity is subject to an illumination with a bi-frequency probe. The transmitted signal is lost, and a only the reflected part is collected for measurement. For a single frequency, a beam splitter is characterised by a unitary operator
\begin{equation}\label{eq:BS}
{U}({\omega})
\equiv   
\exp
\left[
\arcsin\left(\sqrt{\eta(\omega)}\right)(
\hat{s}_\omega^\dagger \hat{b}_\omega e^{i\varphi} - \hat{s}_\omega \hat{b}_\omega^\dagger e^{-i\varphi}
)
\right],
\end{equation}
where $\eta(\omega)$ is a frequency-dependent reflectivity, related to transmittivity $\tau$ via $\eta(\omega) + \tau(\omega) = 1$.

We assume for simplicity that $\varphi = 0$, \textit{i.e.} there is no phase difference between transmitted and reflected signals.
This unitary maps states (density matrices) that live in the density matrix space associated with Hilbert space $\mathcal{H}$, $\mathcal{D}(\mathcal{H})$ to itself.
Formulating the problem from a density operator perspective, we have that the received state is
\begin{equation}\label{eq:rholambda}
{\rho}_\lambda = 
\Tr_{S_1} \Tr_{S_2}
\left[{U}_\lambda{\rho}{U}^\dagger_\lambda
\right],
\end{equation}
where the parameter is defined as
$\lambda = \eta(\omega_2)-\eta(\omega_1)$, and ${\rho}\in \mathcal{H}_{S_1, S_2, B_1, B_2}$ is a four-mode state that includes the two signals (the two-mode state that we can control) and two thermal environments of the form ${\rho}^{th}_1\otimes {\rho}^{th}_2$, where the subscript indicates the frequency, \textit{i.e.}
${\rho}^{th}_a = (1+N_{\text{th}})^{-1}\sum_{n=0}^\infty (N_{\text{th}}/(1+N_{\text{th}}))^n
\ket{n}_a\bra{n}$
	where $N_{\text{th}} = \Tr ({\rho}^{th}_a \hat{b}^\dagger_{\omega_a}\hat{b}_{\omega_a})$ is the average number of thermal photons, which we assume to be the same for the two modes. 
Note that in order to obtain the explicit form of the interaction ${U}_\lambda$ in Equation \eqref{eq:rholambda} one just needs to reparametrize the four-mode unitary ${U}(\omega_1)\otimes{U}(\omega_2)$ using the difference of reflectivities $\lambda \equiv \eta_2-\eta_1$ and Equation \eqref{eq:BS}.	
The equal thermal photon number is an accurate approximation as long as the frequency difference $\Delta\omega \equiv\omega_2 - \omega_1$ is sufficiently small. To make this statement more quantitative, let us assume two different thermal photon densities, $N_1$ and $N_2$. The Bose-Einstein distribution for photons is
$N_i  \propto 1/(e^{\beta \omega_i }-1)$ where $\beta \equiv \hbar /k_B T$ is a function of the temperature $T$. Then,
\begin{equation}
\frac{N_1}{N_2} = \frac{e^{\beta \omega_1 }-1}{e^{\beta \omega_2 }-1} = \frac{1}{1 + \frac{\beta \Delta\omega e^{\beta \omega_1}}{e^{\beta \omega_1}-1}},
\end{equation}
we see that up to first order in $\beta \Delta\omega$, the last expression reduces to $1 -\Delta\omega / \omega_1$. This means that $N_1 \approx N_2$ if $\Delta\omega / \omega_1 \ll 1$. In particular, for $T= 300 \text{ K}$ and $\omega_1/2\pi = 5 \text{ GHz}$ the expected thermal photon number is roughly 1250. It is straightforward to check that for these frequencies and temperatures, the above approximations are good (\textit{i.e.} $\sim 4$\% of relative error) for frequency differences up to 20\%.

{
Because we are working within the local estimation approach and our goal is to find observables that saturate the qCRB, we shall take the true value of $\lambda$ to be  exactly zero. This means that the goal of the protocol is to increase one's confidence about this initial ansatz of the parameter being zero, and be able to tell when it is close but not exactly zero. Hence, we work in a neighborhood of $\lambda \sim 0$ --~which can be implemented by taking the limit $\lambda \rightarrow 0$ in the derived expressions. Moreover, this relies on a physical assumption, since we are interested in probing regions of  $\eta(\omega)$ that do not change drastically, \textit{i.e.} that are well approximated by a linear function with either no slope or a small one.} In this sense, the protocol is a quantum sensing one, since we are interested in answering the question of whether the parameter either vanishes or is small.

 {It is also worth discussing briefly the effect of absorption loss due to the medium through which the signal travels. These  can be accommodated in the model by means of an additional beam splitter. The medium through which the signal travels can be seen as an array of infinitesimal beam splitters, each of which having the same reflectivity, and mixing some incoming signal with the same thermal state. For a travel distance $L$, the flying mode will see a reflectivity
\begin{equation}
\eta_{\text{abs}} = 1-e^{-\mu L},
\end{equation}
where $\mu$ is a parameter characterizing the photon-loss of the medium. A concatenation of beam splitters can be easily put into a single one, as long as they are embedded in the same environment, which is our case. For beam splitters of transmittivities $\tau_1$ and $\tau_2$ the resulting transmittivity of them combined is simply the product: $\tau = \tau_1 \tau_2$. Thus, acommodating absorption losses into our model is trivially obtained by the transformation $\tau \mapsto e^{-\mu L} \tau $. Since the QFI deals with derivatives with respect to the parameter to be estimated, and ultimately we are interested in QFI  \textit{ratios} between a quantum protocol and its classical counterpart, the above transformation will not affect the overall results, since multiplicative factors will cancel out.}
\section{Results: Quantum Fisher information}
In this section we compute the QFI for two different probes: an entangled two-mode squeezed (TMS) state, and a pair of coherent beams. {The choice of the TMS state over other possible entangled states is motivated by the fact that these are customarily produced in labs, both in optical --e.g. with non-linear crystals, and in microwave frequencies --using Josephson parametric amplifiers (JPAs).}
\subsection{Two-mode squeezed vacuum state}
The TMS vacuum (TMSV) state is the continuous-variable equivalent of the Bell state, being the Gaussian state that optimally transforms classical resources (light, or photons) into quantum correlations. The TMSV state is a cornerstone in  experiments with quantum microwaves \cite{PhysRevLett.107.113601, PhysRevLett.109.183901, RevModPhys.77.513, Casariego2022}.  In our case, we are interested in states produced via nondegenerate parametric amplification, in order to have two distinguishable frequencies.
The state can be formally written as:
$\ket{\psi}_{\text{TMSV}}:=(\cosh r)^{-1}\sum_{n=0}^\infty \left(
-e^{i\phi}\tanh r \right)^n \ket{n,n}$,
where $r \in \mathbb{R}_{\geq 0}$ is the \textit{squeezing parameter}.  For simplicity we take $\phi=0$. {In any realistic application, the TMSV state should be replaced by a TMS thermal state, which can be defined as the one obtained by applying the two-mode squeezing operation to a pair of uncorrelated thermal states $\rho_\text{th, 1}$, and $\rho_\text{th, 2}$ with mean thermal photon numbers $n_1$ and $n_2$, respectively, and hence resulting in a mixed state \cite{TMST}. The expected total photon number in these states is given by $N_\text{TMST}=\braket{\hat{N}_1+\hat{N}_2} = (n_1+n_2)\cosh 2r + 2 \sinh ^2 r$, where $ \hat{N}_i \equiv \hat{a}^\dagger_{S_i} \hat{a}_{S_i}$ for $i=1,2$.
Typically, one has $n_1 = n_2 \equiv n$, which gives us a symmetric TMST state. In this case we define the signal photon number $N_\text{S}$ as the photon number in each of the modes,  $ N_{\text{S}} \equiv N_\text{TMST} /2 = n(1+2{N_r})+{N_r}$, where $N_r \equiv \sinh^2 r$\footnote{Note that $N_\text{S}$ is defined as the expected photon number in a \textit{single} mode, meaning that the actual photon number in the signal is $2N_\text{S}$.}. In microwaves, a squeezing level $S = -10 \log_{10}\left[(1+2 n)\exp\left(-2r\right)\right]$ of 9.1 dB has been reported \cite{finite-time} for $n=0.34$ and $r\sim 1.3$, using JPAs operating at roughly 5 GHz with a filter bandwidth of 430 kHz. This corresponds to $N_\text{S} \sim 8$.}

The total initial (real) covariance matrix, written in the \textit{real basis} $$(\hat{x}^{\text{th}}_1, \hat{p}^{\text{th}}_1, \hat{x}^{\text{S}}_1, \hat{p}^{\text{S}}_1,\hat{x}^{\text{th}}_2, \hat{p}^{\text{th}}_2,\hat{x}^{\text{S}}_2, \hat{p}^{\text{S}}_2)^{\intercal},$$ is given by

{
\begin{equation}
{\Sigma} = 
N\left(
\begin{array}{cccc}
 N^{-1}{\Sigma}_{\text{th}} & 0 & 0 & 0\\
 0 & {\Sigma}_{r} & 0 & {\varepsilon}_r \\
 0 & 0 & N^{-1}{\Sigma}_{\text{th}} & 0\\
 0 & {\varepsilon}^{\intercal}_r & 0 &{\Sigma}_{r}
\end{array}
\right),
\end{equation}
where $N\equiv 1+2n$, ${\Sigma}_{\text{th}} = (1+2N_{\text{th}})\Id_2$ is the real covariance matrix of a thermal state, ${\Sigma}_{r} = \cosh(2r)\Id_2$ corresponds to the diagonal part of one of the modes in a TMSV state, and ${\varepsilon}_r = \sinh(2r)\sigma_Z$ is the correlation between the two modes, where $\sigma_Z$ is the Z Pauli matrix. Note that the covariance matrix of the thermal TMS state is simply $N$ times the one of the TMSV state.

The displacement vector of a TMST state is identically zero $\bm{d}_{\text{TMST}}=\bm{0}$, so the last term of Eq. \eqref{eq:QFI} vanishes.
Under the assumption that the object does not entangle the two modes,  we have that the symplectic transformation is  ${S}(\eta_1, \eta_2)  = {S}_{\text{BS}}(\eta_1) \oplus {S}_{\text{BS}}(\eta_2)$\footnote{In a state language this translates to the total unitary being the tensor product of two beam splitters: ${U}_T(\eta_1,\eta_2)= {U}(\eta_1) \otimes {U}(\eta_2)$, with ${U}(\eta_i)\equiv \exp \left[ \arcsin\sqrt{\eta(\omega_i)}(\hat{s}^\dagger_i \hat{b}_i-\hat{s}_i \hat{b}^\dagger_i)\right]$.}, where 
\begin{equation}
{S}_{\text{BS}}(x) = 
\left(
\begin{array}{cc}
 \sqrt{x}\Id_2 & \sqrt{1-x}\Id_2 \\
 -\sqrt{1-x}\Id_2 & \sqrt{x}\Id_2
\end{array}
\right)
\end{equation}
is the real symplectic transformation associated with a beam splitter of reflectivity $x$. We define the parameter of interest as $\lambda \equiv \eta_2 - \eta_1$.
With this, ${S}(\eta_1, \eta_2)$ becomes a function of $\lambda$. For simplicity, we define ${S}_\lambda := {S}(\eta_1,  \eta_1 + \lambda)$.
The full state after the signals get mixed with the thermal noise is given by 
$\tilde{{\Sigma}}_\lambda \equiv {S}_\lambda {\Sigma} {S}_\lambda^\intercal$.}
In covariance matrix formalism, partial traces are implemented by removing the corresponding rows and columns \cite{adesso}; in our case the rows and columns 1, 2, 5, and 6. The resulting \textit{received} covariance matrix reads as follows
\begin{equation}
{\Sigma}_\lambda
=
\left(
\begin{array}{cc}
a  \Id  & b \sigma_Z \\
b \sigma_Z & c \Id
\end{array}
\right),
\end{equation}
{with $a \equiv 1+ 2 N_\text{th}+2\eta _1 (2 N_{r} +4 nN_{r}  - N_\text{th})$, $b \equiv 2(1+2n) \sqrt{2\eta _1N_\text{S}(\eta _1+\lambda ) (2 N_\text{r}+1)} $, and $c\equiv  (1+2 n)\left(1+4 \lambda  N_{r}+\eta _1 (4 N_{r}-2 N_\text{th})+2 (1-\lambda) N_\text{th}\right)$.

For this state, the symplectic eigenvalues $\nu_{\pm}$ defined in Eq. \eqref{eq:symp-eigenv} are strictly larger than one for any value of the parameters $n$, $N_r$, $N_\text{th}$, and $\eta_1$, other than $\eta_1=1 \land N_\text{th}=0$ so there is no need of any regularization scheme  \cite{_afr_nek_2015}. Indeed, this is due to the mixedness of the received state: regularization is only needed for pure states.

We obtain the function $H_Q (\lambda)$ from Eq. \eqref{eq:QFI}, and compute the two-sided limit $H_Q\equiv \lim_{\lambda \rightarrow 0}H_Q (\lambda)$ when the parameter $\lambda$ goes to zero, finding
\begin{equation}
\begin{split}
H_Q =\kappa &\left[\eta _1 \bar{n} \left(\bar{N}_\text{th} (4 n N_r+n+4 N_r-2 N_\text{th})-2 \eta _1 (2 (n+1) N_r \bar{N}_\text{th}+N_\text{th} (n-N_\text{th}))\right) \right.\\
&\left. +\beta\right) \left(\eta _1 \bar{n} \left(\bar{N}_\text{th} (4 n N_r+n+4 N_r-2 N_\text{th})-2 \eta _1 (2 (n+1) N_r \bar{N}_\text{th}+N_\text{th} (n-N_\text{th}))\right) \right.\\
&\left.+\beta+1\right]
\end{split}
\end{equation}
where
\begin{equation}
\begin{split}
\kappa^{-1}\equiv \bar{n}^2 &\left[2 \eta _1 (\bar{N}_\text{th} (N_r (\bar{n} (8 (n+1) N_r^2+6 n N_r+n)-4 N_r)+2 N_\text{th}^2 (4 n N_r+n+6 N_r)\right.\\
&-\left. 2 N_r N_\text{th} (2 (6 n+5) N_r+2 n-1)-2 N_\text{th}^3)-2 \eta _1 (-N_\text{th} (4 (n+1) N_r+n)\right.\\
&+\left.N_r ((4 n+2) N_r-1)+N_\text{th}^2) (2 (n+1) N_r \bar{N}_\text{th}+N_\text{th} (n-N_\text{th})))\right.\\
&\left.+2 n (2 N_r+1) N_r \bar{N}_\text{th}^2+4 N_r^2 (6 N_\text{th} (N_\text{th}+1)+1)-4 N_r N_\text{th}^2 (4 N_\text{th}+3)+N_\text{th}^2 \bar{N}_\text{th}^2\right]
\end{split},
\end{equation}

and  $\bar{N}_\text{th} \equiv 1+2 N_\text{th}$,   $\bar{n}\equiv 1+2n$, and $\beta \equiv n \bar{N}_\text{th}^2+2 N_\text{th} (N_\text{th}+1)$.

}

\subsection{Coherent states}
{
Here we use a pair of coherent states as probe: $\ket{\psi} =\ket{\alpha}\otimes\ket{\alpha}$. The total expected photon number in this state is $2N_\text{C}:= 2\left| \alpha^2\right|$. For simplicity we take $\alpha \in \mathbb{R}$. Moreover, since we will compare with the TMST state, we set $\alpha^2 = n(1+2N_r)+N_r$.
The initial covariance matrix is simply given by the direct sum of two identity matrices (corresponding to each of the coherent states), and two thermal states. After the interaction and the losses, the measured covariance matrix is 

\begin{equation}
{\Sigma}_\lambda
=
\left(
\begin{array}{cc}
d \Id  &0 \\
0 & f \Id
\end{array}
\right),
\end{equation}
where $d =1+ 2 N_\text{th}\tau _1 $, $f = 1+ 2 N_\text{th}(\tau _1-\lambda)$.

The initial displacement vector in the real basis is $\bm{d}_{0}^\intercal = (0,0,\sqrt{2}\alpha,0,0,0,\sqrt{2}\alpha,0)$ which leads --after the interaction and the trace of the losses-- to  $\bm{d}^\intercal = \alpha(\sqrt{2\eta_1},0, \sqrt{2(\eta_1 + \lambda)})$.
 The symplectic eigenvalues are also larger than one here. 
Inserting these in Eq. \eqref{eq:QFI}, and taking the limit $\lambda\rightarrow 0 $, we find that the QFI for the coherent state is
\begin{equation}
H_C = \frac{4 N_\text{th}^2 \left((1+2 N_\text{th} \tau_1)^2+1\right)}{(1+2 N_\text{th} \tau_1)^4-1}+\frac{n(1+2N_r)+N_r	}{\eta_1(1+2 N_\text{th} \tau_1)},
\end{equation}
where $\tau_1 = 1-\eta_1$ is the transmittivity. Having computed both the quantum and the classical QFIs, in the next section we analyze their ratio $H_Q / H_C$, a quantifier for the quantum enhancement.}

\subsection{Comparison: Quantum enhancement}
\begin{figure*}
\centering
\includegraphics[width=\linewidth]{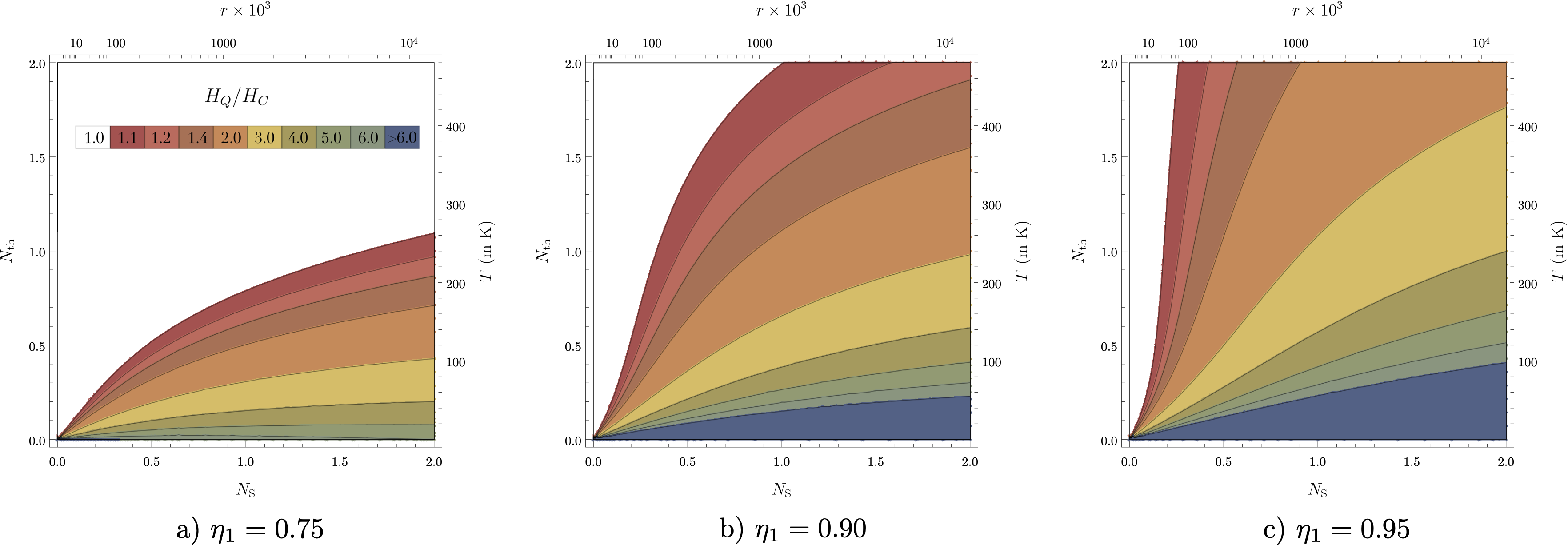}
\caption[Ratio of the quantum Fisher information associated to the classical and quantum approaches, for the bi-frequency illumination protocol]{Values, represented by a non-linear color grading, of the quantum enhancement given by the ratio $H_Q/H_C$ of the quantum Fisher information of the two-mode squeezed vacuum state probe $H_Q$ by the quantum Fisher information of the coherent states probe $H_Q$ as a function of the photon numbers of the signal ($N_\text{S}$) and of the thermal bath ($N_\text{th}$), for a reflectivity $\eta_1$ of a) 0.75, b) 0.90, and c) 0.95.
 Equivalently, scales of squeezing, $r$, given by $\sqrt{N_{\text{S}}}=\sinh r$, and temperature $T$ in Kelvin, are provided. The relation between temperature and mean thermal photon number is obtained via the usual Bose-Einstein distribution $N_\text{th} = 1/(\exp(E/k_B T) - 1)$ when the energy is set to $E=\hbar \omega = h \nu$, which requires a choice of the frequency $\nu$. We have taken $\nu = 5$ GHz, a typical frequency of microwaves.
White represents no quantum enhancement, \textit{i.e.} $H_Q / H_C =1$. We clearly see that as $\eta_1$ grows, the quantum enhancement becomes not only more significant, but also easier to achieve with less signal photons. Importantly, as the reference reflectivity $\eta_1$ grows,  the protocol becomes more resilient to thermal noise.}
\label{fig:QFIratios}
\end{figure*}

We analyze the ratio between the TMST state's QFI ($H_Q$) and the coherent pair's QFI ($H_C$) for different situations. As a first approximation and to simplify the discussion, we take the limit where $n\rightarrow 0$, which corresponds to a TMSV state input.  Finding values of $(\eta_1, N_\text{th}, N_\text{S})$  such that the ratio $H_Q / H_C$ is larger than one means that one can extract more information about parameter $\lambda$ using a TMST state than using a coherent pair, provided an optimal measurement is performed in both cases. In Figure \ref{fig:QFIratios} we plot the results for various values of $\eta_1$.
We can immediately see that the ratio gets larger  for large values of $\eta_1$, \textit{i.e.} for highly reflective materials. In particular, we find the high-reflectivity limit the ratio converges even when the individual QFIs do not (since they correspond to a pure state being transmitted):
\begin{equation}
\lim_{\eta_1 \rightarrow 1}\frac{H_\text{Q}}{H_\text{C}} = \frac{N_\text{S}^2 \left(8 N_\text{th} (N_\text{th}+1)+4\right)+4 N_\text{S} N_\text{th}^2+N_\text{th}^2}{N_\text{th} \left(N_\text{S} (4 N_\text{th}+2)+N_\text{th}\right)}
\end{equation}
which converges to 
$1 + 8 N_\text{S}^2/(4N_\text{S}+1)$
in the highly noisy scenario $N_\text{th}\gg 1$. Using a squeezing of $r\sim 1.3$ which is experimentally realistic for microwave quantum states, and that corresponds to an expected photon number of $N_\text{S}\sim2.9$, we expect  to find a quantum-enhancement of roughly a factor of 6, \textit{i.e.}, $ H_\text{Q}/H_\text{C} \sim 6.4$ in the highly reflective limit.

In the next section we explicitly compute the observables that lead to an optimal extraction of $\lambda$'s value for both the classical and the quantum probes.

\section{Optimal observables}
Here we address the question of how to extract the maximum information about parameter $\lambda$ for each of the probes. The theory provides us with explicit ways to compute an optimal POVM, which despite being not unique, provides us with an optimal measurement strategy: upon measuring the outcomes and possibly after some classical data-processing, the results asymptotically tend towards the true value of the parameter to be estimated.
\subsection{Optimal observable for the TMSV state probe}
 Computing the SLD in Eq. \eqref{eq:QFI} and inserting it in Eq. \eqref{eq:opt-obs} we find 
\begin{equation}\label{eq:QObserv}
\hat{O}_Q = L_{11}\hat{a}_1^\dagger \hat{a}_1 + L_{22}\hat{a}_2^\dagger \hat{a}_2 + L_{12}\left( \hat{a}_1^\dagger \hat{a}_2^\dagger + \hat{a}_1 \hat{a}_2\right) + L_0 \Id_{12},
\end{equation}
where the general expressions  for the coefficients can be found in Appendix \ref{appendix:OptimalObs}.
{ The variance of this operator is found to be $\text{var}\left(\hat{O}_Q\right) = 2N_\text{S}^2L_{12}\left(1+N_\text{S}\right)$. We can numerically test the validity of the qCR bound for this observable by examining the bound itself for the extreme choice of $M=1$. The saturation of the bound produces the following relation:
\begin{equation}
\text{var}\left(\hat{O}_Q\right) H_Q = 1.
\end{equation}
Now, as the left hand side a function of $(N_\text{S}, N_\text{th}, \eta_1)$, we can give different values to the reflectivity and find the limiting condition between $N_\text{S}$ and $N_\text{th}$, which is depicted in FIG. \ref{fig:qCR}. Naturally, the larger $M$, the better results we can achieve, but $M=1$ proves the existence of a choice of parameters for which the bound is saturated. 
\begin{figure}[ht]
\centering
\includegraphics[width=0.6\textwidth]{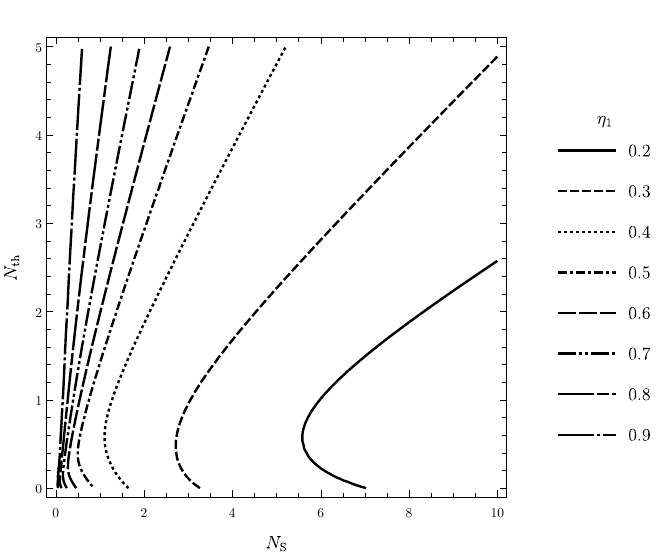}
\caption[Saturation of the quantum Cramér-Rao bound for the optimal observable in the bi-frequency illumination protocol]{ Proof of the saturation of the quantum Cramér-Rao bound for the optimal observable $\hat{O}_Q$ given in Eq.~\eqref{eq:QObserv} for different values of the reflectivity $\eta_1$, expressed as the existence of a real function $N_\text{th} =  N_\text{th}(N_\text{S})$, for the extreme case of just one experimental run ($M=1$). As the reflectivity grows, we observe an interesting behavior: the best choice of $N_\text{th}$ --defined as the one that saturates the bound while keeping $N_\text{S}$ as low as possible-- is actually non-vanishing.}
\label{fig:qCR}
\end{figure}
}

\begin{figure}[ht]
\centering
\includegraphics[scale=.15]{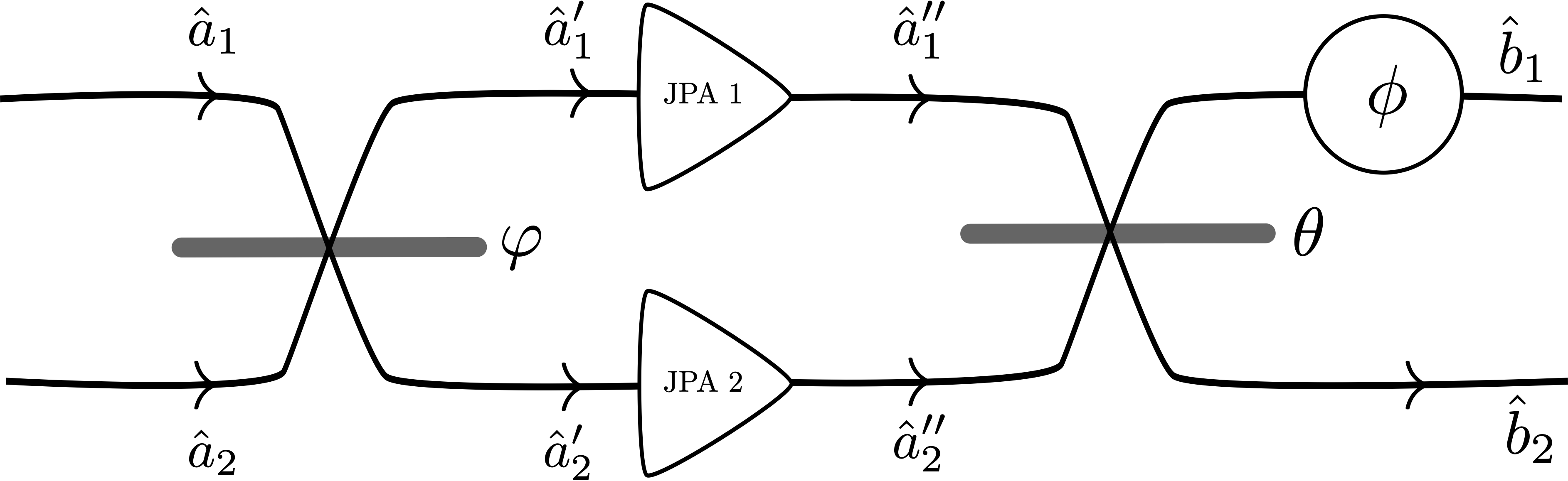}
\caption[Implementation of the optimal observable for the quantum bi-frequency illumination protocol]{Schematic circuit for the generation of mode $\hat{b}_1$, needed to correctly implement the optimal observable in the two-mode squeezed vacuum (TMSV) state. The original $\hat{a}_i$ modes mix at a $\varphi$ beam splitter, and the outputs go through a single-mode squeezing operator --Josephson parametric amplifier (JPA) in microwaves, with parameters $r_i$ and $\theta_i$ corresponding to squeezing and phase. Then they mix at a second beam splitter $\theta$. A phase shift $\phi$ is applied at the end, to cancel undesired terms. This scheme is  technology-independent, and could be applied to optics by replacing the JPAs with the corresponding squeezing device, \textit{e.g.} a non-linear crystal that performs spontaneous parametric down-conversion.}
\label{fig:jpas-supp}
\end{figure}

 Moreover, it is illustrative to study a possible implementation of the noiseless case, since this captures the essence of what is being measured. When $N_\text{th}\rightarrow 0 $ we have that  $\hat{O}_Q ^{\text{Lim}} =
-\mu^2\hat{a}_1^\dagger \hat{a}_1 - \hat{a}_2^\dagger \hat{a}_2 
 + \mu( \hat{a}_1^\dagger \hat{a}_2^\dagger + \hat{a}_1 \hat{a}_2)  -\nu \Id_{12}$
where $\mu^2 \equiv \left(1+1/2 N_\text{S}\right)$ and $\nu \equiv \left(1+1/4 N_\text{S}\right)$, and we have taken the limit of vanishing $N_\text{th}$. Notice that we can rewrite this observable as $\hat{b}_1^\dagger \hat{b}_1 - 1$, \textit{i.e.} implementing photon-counting on the operator $\hat{b}_1\equiv -i (\hat{a}_2^\dagger - \mu\hat{a}_1)$. This is achieved by means of the transformations captured in FIG. \ref{fig:jpas-supp}.

{ Following that scheme, we have that after the first beam splitter
\begin{align}
\begin{split}
\hat{a}_1^\prime &=  \hat{a}_1\cos\varphi + \hat{a}_2\sin\varphi \\
\hat{a}_2^\prime &=  	-\hat{a}_1\sin\varphi   + \hat{a}_2\cos\varphi ,
\end{split}\label{eq:aprimes-supp}
\end{align}
then the Josephson parametric amplifiers (JPA) --ideally squeezing operators-- produce $\hat{a}_i^{\prime \prime} = S^\dagger(r_i, \theta_i) \hat{a}_i^\prime S(r_i, \theta_i)$
where $S(r_i, \theta_i)$ is the squeezing operator, acting as
\begin{align*}
\begin{split}
\hat{a}_i^{\prime \prime}=S^\dagger(r_i, \theta_i) \hat{a}_i^\prime S(r_i, \theta_i) &= \hat{a}_i^\prime \cosh r_i  - e^{i\theta_i} \hat{a}^{\prime \dagger}_i \sinh r_i \\
{\hat{a}^{\prime \prime \dagger}}_i=S^\dagger(r_i, \theta_i) \hat{a}_i^{\prime\dagger} S(r_i, \theta_i) &=  \hat{a}_i^{\prime\dagger} \cosh r_i  - e^{-i\theta_i} \hat{a}^\prime_i \sinh r_i .
\end{split}
\end{align*}
Assuming that the phase shifter $\phi$ acts as $\hat{c} \mapsto e^{-i\phi}\hat{c}$ we find the following output modes
\begin{align*}
\begin{split}
e^{i\phi}\hat{b}_1 &= \cos \theta \left(\hat{a}_1^\prime \cosh r_1  - e^{i\theta_1} \hat{a}^{\prime \dagger}_1 \sinh r_1 \right) 
 + \sin\theta  \left(\hat{a}_2^\prime \cosh r_2  - e^{i\theta_2} \hat{a}^{\prime \dagger}_2 \sinh r_2 \right) \\
 \hat{b}_2 &= -\sin \theta \left(\hat{a}_1^\prime \cosh r_1  - e^{i\theta_1} \hat{a}^{\prime \dagger}_1 \sinh r_1 \right) 
 + \cos\theta  \left(\hat{a}_2^\prime \cosh r_2  - e^{i\theta_2} \hat{a}^{\prime \dagger}_2 \sinh r_2 \right).
\end{split}
\end{align*}
We insert \eqref{eq:aprimes-supp} in the last expression and regroup, finding
\begin{align*}
\begin{split}
e^{i\phi}\hat{b}_1 
&= \hat{a}_1 \left(   \cos\theta  \cos \varphi \cosh r_1- \sin \theta \sin \varphi  \cosh r_2 \right) \\ 
&+ \hat{a}_2 \left(   \cos\theta  \sin \varphi \cosh r_1+ \sin \theta \cos \varphi  \cosh r_2 \right) \\ 
&+ \hat{a}_1^\dagger \left(   -e^{i\theta_1}\cos\theta \cos\varphi  \sinh r_1+ e^{i\theta_2}\sin \theta  \sin \varphi \sinh r_2\right) \\ 
&+ \hat{a}_2^\dagger \left(   -e^{i\theta_1}\cos\theta \sin\varphi  \sinh r_1- e^{i\theta_2}\sin \theta  \cos \varphi \sinh r_2\right)
\end{split}
\end{align*} 
Because we want to perform photon-counting over the operator $\hat{b}_1\equiv -i (\hat{a}_2^\dagger - \mu\hat{a}_1)$, we identify:
\begin{align}
i \mu&= \cos\theta  \cos \varphi \cosh r_1- \sin \theta \sin \varphi  \cosh r_2\\
i &= e^{i\theta_1}\cos\theta \sin\varphi  \sinh r_1+ e^{i\theta_2}\sin \theta  \cos \varphi \sinh r_2.
\end{align}
}
\subsection{Optimal observable for the coherent state probe}
The optimal observable in this case is given by
$\hat{O}_C =  A \Id_{(1)}\otimes[( \hat{a}_2^\dagger - \eta_1 \sqrt{\alpha} ) (\hat{a}_2 - \eta_1 \sqrt{\alpha} ) + \frac{1}{2}]$,
where $A=1/(\eta_1-1)(1-N_\text{th}(\eta_1-1))$, and $\Id_{(1)}$ is the absence of active measurement of mode 1. 
This expression can then be put as 
$\hat{O}_C =  A\Id_{(1)}\otimes\left[\left( \hat{a}_2^\dagger - \eta_1 \sqrt{\alpha} \right) \Big(\hat{a}_2 - \eta_1 \sqrt{\alpha} \Big) + \frac{1}{2}\right]$. This operator can be experimentally performed with a displacement $D(-\eta_1 \sqrt{\alpha})$ \cite{Paris1996} and photon-counting in the resulting mode.
The interpretation is simple: because $\eta_1$ is known (it serves as a reference), there is nothing to be gained by measuring the first mode in the absence of entanglement. Moreover, the observable is separable, as one should expect, and the experimental implementation is straightforward: photon-counting in the --locally displaced-- second mode.

We have seen that both quantum and classical observables are non-Gaussian measurements, since they can be related to photon-counting, as expected in order to obtain quantum enhancement \cite{serafini}. Current photoncounters in microwave technologies can resolve up to 3 photons with an efficiency of 96\% \cite{dassonneville2020numberresolved}. Inefficiencies in the photon-counters can be accounted for with a simple model of an additional beam splitter that mixes the signal with either a vacuum or a low-tempereature thermal state. Additionally, the fact that real digital filters are not perfectly sharp  should also be accounted for in a full experimental proposal, which we leave for future work \cite{PhysRevLett.117.020502}.

\section{Conclusions}
We have proposed a novel protocol for achieving a quantum enhancement in the decision problem of whether a target's reflectivity depends or not on the frequency, using a bi-frequency, entangled probe, in the presence of  noise and losses. Crucially, our protocol needs no idler mode, avoiding the necessity of coherently storing a quantum state in a memory. 
The scaling of the quantum Fisher information (QFI) associated to the estimation problem for the entangled probe is faster than in the case of a coherent signal.
This quantum enhancement is more significant in the high reflectivity regime. Moreover, we have derived analytic expressions for the optimal observables,  which allow extraction of the maximum available information about the parameter of interest, sketching an implementation with quantum microwaves.

This information can be related to the electromagnetic response of a reflective object to changes in frequency, and, consequently, the protocol can be applied to a wide spectrum of situations. Although the results are general, we suggest two applications withing quantum microwave technology: radar physics, motivated by the atmospheric transparency window in the microwaves regime, together with the naturally noisy character of open-air  \cite{Sanz_2018, dassonneville2020numberresolved, GonzalezRaya2020, Zhang2020, munuera2020, dambach2107}; and quantum-enhanced microwave medical contrast-imaging of low penetration depth tissues, motivated not only by the non-inoizing nature of these frequencies, but also because resorting to methods that increase the precision and/or resolution without increasing the intensity of radiation is crucial in order not to heat the sample.

Our work paves the way for extensions of the protocol to accommodate both thermal effects in the input modes, and continuous-variable frequency entanglement \cite{PhysRevLett.84.5304}, where a more realistic model for a beam containing a given distribution of frequencies could be used instead of sharp, ideal bi-frequency states. It also serves as reminder that quantum enhancement provided by entanglement can survive noisy, lossy channels.


\chapter{Open-Air Microwave Entanglement Distribution for Quantum Teleportation}
\label{chapter:MWentanglementDistribution}
In this chapter, we present the results of Ref. \cite{GonzalezRaya2022}: Tasio Gonzalez-Raya, Mateo Casariego, Florian Fesquet, Michael Renger, Vahid Salari, Mikko Möttönen, Yasser Omar, Frank Deppe, Kirill G. Fedorov, and Mikel Sanz,
Phys. Rev. Applied \textbf{18}, 044002 (2022). This was a collaborative paper, and not all the results are mine. The main efforts were done between the two first authors, and overall I was in charge of much of the writing (especially in the first sections). It seemed more coherent to us to present the article as a whole, and the authors, and in particular the first author, Tasio Gonzalez-Raya, kindly agreed to let all the results be included in the thesis. After all, it represents an interesting application of many of the techniques described in the introductory chapters of this thesis to open-air quantum microwaves. Credits for the figures that are not mine are given in their corresponding caption.  The abstract reads as follows:
\begin{quote}
    Microwave technology plays a central role in current wireless communications, including mobile communication and local area networks. The microwave range shows relevant advantages with respect to other frequencies in open-air transmission, such as low absorption losses and low-energy consumption, and in addition, it is the natural working frequency in superconducting quantum technologies. Entanglement distribution between separate parties is at the core of secure quantum communications. Therefore, understanding its limitations in realistic open-air settings, especially in the rather unexplored microwave regime, is crucial for transforming microwave quantum communications into a mainstream technology. Here, we investigate the feasibility of an open-air entanglement distribution scheme with microwave two-mode squeezed states. First, we study the reach of direct entanglement transmission in open air, obtaining a maximum distance of approximately 500 m with parameters feasible for state-of-the-art experiments. Subsequently, we adapt entanglement distillation and entanglement swapping protocols to microwave technology in order to reduce the environment-induced entanglement degradation. The employed entanglement distillation helps to increase quantum correlations in the short-distance low-squeezing regime by up to 46\%, and the reach of entanglement increases by 14\% with entanglement swapping. Importantly, we compute the fidelity of a continuous-variable quantum teleportation protocol using open-air-distributed entanglement as a resource. Finally, we adapt this machinery to explore the limitations of quantum communication between satellites, where the impact of thermal noise is substantially reduced and diffraction losses are dominant.
\end{quote}

\section{Introduction}\label{section:introduction}
Quantum communication~\cite{Gisin2007,Yuan2010,Krenn2016} represents an application of quantum information theory that tackles the subject of information transfer by taking advantage of purely quantum resources, such as superposition and entanglement. By establishing quantum channels to share quantum resources, namely quantum states, as well as secure classical channels, it aims at outperforming classical communication protocols in both efficiency and security. 

Among the best-known quantum communication protocols, quantum teleportation~\cite{Bennett1993,Pirandola2015} aims at transferring information of an unknown quantum state held by one party, to a second party at a remote location, by means of an entangled resource and classical communication. Initially proposed for discrete-variable quantum states~\cite{Baur2012,Steffen2013}, this protocol has also been studied in continuous-variable settings~\cite{Pirandola2006,Braunstein1998,Fedorov2021}.

Another notorious advantage of quantum communication is quantum key distribution~\cite{Diamanti2016,Wang2021}, whose foundation has been set by two distinct protocols: BB84~\cite{Bennett1984} and E91~\cite{Ekert1991}. These protocols allow two distant parties to develop a shared random key, which is unconditionally secure against eavesdroppers by virtue of quantum laws. 

At the heart of many quantum communication protocols lies entanglement distribution~\cite{Cirac1997,Chou2007,Mista2009,Dias2020}, also a key point for the famous quantum internet idea~\cite{Kimble2008,Wehner2018,Gyongyosi2019}; quantum entanglement distribution represents the act of sharing entangled states between communication parties. This has been experimentally attained~\cite{Herbst2015,Wengerowsky2019,Yin2017}, as well as quantum key distribution~\cite{Liao2017,Liao2018,Minder2019} and quantum teleportation~\cite{Bouwmeester1997,Furusawa1998,Boschi1998,Jin2010,Ma2012,Ren2017}, through optical fibers and through open air. 

The subject of quantum communication has been developed parallel to other areas such as quantum computing, quantum sensing~\cite{Degen2017,Pirandola2018}, or quantum metrology~\cite{Giovannetti2011,Polino2020}, to which it is usually tangential. Quantum computing, for example, is expected to benefit from efficient transfer of quantum information between processing units, a proposal that is categorized as distributed quantum computing~\cite{Rodrigo2020,DiAdamo2021}. Quantum sensing and quantum metrology profit from the use of entanglement in their attempt to perform high-resolution measurements on various systems. By using quantum resources, they can reach ultimate measurement limits set by quantum mechanics, thus outperforming classical strategies. Respective examples can be found in the detection of gravitational waves at LIGO~\cite{LIGO2011,Oelker2016}, in the measurement of biological systems~\cite{Taylor2016,Mauranyapin2017}, quantum imaging~\cite{Moreau2019,Berchera2019}, navigation~\cite{Fink2019}, and synchronization~\cite{Quan2016}, among others.

The flourishing of these quantum-information-based fields has gone hand in hand with the development of quantum technologies, among which superconducting circuits stand out in terms of controllability, scalability, and coherence. Partially, this is due to the development of Josephson junctions (JJs)~\cite{Josephson1969}, nonlinear elements with essential applications in quantum computation~\cite{Buttiker1987,Bouchiat1998,Koch2007}, and quantum information processing~\cite{Yamamoto2008,Castellano-Beltran2008}. This has led to different experiments in quantum state transfer and remote entanglement preparation between various JJ-based superconducting devices~\cite{Axline2018, Campagne2018, Kurpiers2018, Leung2019, Roch2015, Narla2016,Dickel2018,Magnard2020}, as well as to sensitive noise analysis~\cite{Goetz2017, Goetz2017_2}. These devices naturally work at microwave frequencies ($1$--$100$ GHz), for which the number of thermal photons per mode at room temperature ($T=300$ K) is around $1250$ at $5$ GHz, thus creating the need for cryogenic cooling in order to shield superconducting circuits from thermal noise. This problem is somewhat nonexisting in the optical regime, where most quantum communication experiments have been performed so far. However, in this regime there are many other sources of error and inefficiency~\cite{Sanz2018}: large absorption losses in open air and significant power consumption requirements. At the same time, in order to establish a quantum communication channel between superconducting quantum circuits, one requires either converting microwave photons to the optical domain~\cite{Forsch2020, Rueda2019} or using microwave quantum signals directly. The former approach still suffers from huge conversion quantum inefficiencies of the order of $10^{-5}$. Therefore, it is natural to consider the purely microwave quantum communication approach, its advantages, and limitations. 

Another interesting application of the Josephson junction is the Josephson parametric amplifier (JPA), a device that can generate squeezed states~\cite{Fedorov2016}, from which entangled resources~\cite{Pogorzalek2019,Fedorov2018} can be produced for quantum communication with microwaves. Among the various applications of open-air entanglement distribution, quantum teleportation represents one of the fundamental quantum communication protocols, since its associated network has just two nodes: Alice and Bob. This protocol has been explored in the microwave cryogenic environment, both theoretically~\cite{DiCandia2015} and experimentally~\cite{Fedorov2021}. However, a realistic model for open-air quantum teleportation with microwaves should take into account additional challenges associated with impedance mismatches and absorption losses in order to find ways to mitigate these imperfections.

In this article, we address two pragmatic questions: which is the maximum distance for open-air microwave Gaussian entanglement distribution in a realistic scenario, and which technological and engineering challenges remain to be faced? In particular, we adapt the Braunstein-Kimble quantum teleportation protocol employing entangled resources previously distributed through open air, adapted to microwave technology. Performing this protocol is possible due to the recent breakthrough in the development of microwave homodyning~\cite{Fedorov2021} and photocounting~\cite{Dassonneville2020} schemes. When formulated in continuous variables, teleportation assumes a previously shared entangled state, ideally a two-mode squeezed vacuum (TMSV) state with infinite squeezing. In real life, however, only a finite squeezing level can be produced, making the state sensitive to entanglement degradation whenever either one or both modes are exposed to decoherence processes like thermal noise and/or photon losses. As the thermal microwave background in space is smaller, we investigate the distances for entanglement preservation in microwave quantum communication between satellites. Based on previous works~\cite{Pirandola2021,Pirandola2021_2}, we neglect environmental attenuation and focus on diffraction losses, a powerful loss mechanism in microwaves. The article is organized as follows. In section~\ref{section:quantum_CV}, we introduce the main concepts in quantum continuous variables, which are needed for describing quantum microwaves, and briefly discuss the quantum teleportation protocol of Braunstein and Kimble~\cite{Braunstein1998}. In section~\ref{section:open_air_entanglement_distribution}, we study the generation of two-mode squeezed states and the challenges of their subsequent distribution through open air, and compute maximum distances of entanglement preservation for various physical situations. In section~\ref{subsection:distillation_and_swapping}, we review entanglement distillation and entanglement swapping techniques for reducing the effects of noise and losses in open air. In section~\ref{section:homodyning_and_photocounting} we consider recent advances in microwave photodetection and homodyning, and address their current limitations. In section~\ref{section:teleportation}, we investigate open-air microwave quantum teleportation fidelities using the various quantum states derived in the manuscript, and we conclude by addressing the same concerns in quantum communication between satellites in section~\ref{section:satellites}. 

\section{Quantum Continuous-Variable Formalism}\label{section:quantum_CV}
\subsection{Review of Gaussian states}
When some degree of freedom of a quantum system is described by a continuous-spectrum operator, we say that it is a `continuous variable' (CV). Bosonic CV states are those whose quadratures (or, equivalently, their creation and annihilation operators) have a continuous spectrum or, equivalently, the complete description of the Hilbert space requires an infinite-dimensional basis (typically, the Fock basis). Gaussian states are CV states associated with Hamiltonians that are, at most, quadratic in the field operators. As such, their full description does not require the infinite-dimensional density matrix, and can be compressed into a vector and a matrix, called the displacement vector and the covariance matrix, respectively. These are related to the first and second moments of a Gaussian distribution; hence their name ``Gaussian states". For a system with density matrix $\rho$ describing $N$ distinguishable modes, or particles, the displacement vector $\bm{d}$ is a $2N$ vector and the covariance matrix $\Sigma$ is a $2N\times 2N$ square matrix: 
\begin{align}
\bm{d} &:=\tr\left[ \rho \hat{\bm{r}}\right]\\
\Sigma &:= \tr\left[ \rho \lbrace (\hat{\bm{r}} - \bm{d}), (\hat{\bm{r}} - \bm{d})^\intercal\rbrace\right].
\end{align}
Here $\hat{\bm{r}}:= ( \hat{x}_1, \hat{p}_1, \hat{x}_2, \hat{p}_2, \ldots, \hat{x}_N, \hat{p}_N)$ defines the so-called ``real basis'', for which canonical commutation relations read $\left[\hat{\bm{r}}, \hat{\bm{r}}^\intercal\right] = i\Omega$, where $\Omega = \bigoplus_{j=1}^N \Omega_1$ is the quadratic (or symplectic) form, and 
\begin{equation}
 \Omega_1 =
    \begin{pmatrix}
       0 & 1\\
       -1 & 0
    \end{pmatrix},
   \end{equation}
where we have chosen natural units, $\hbar = 1$. Note that the \textit{canonical} position and momentum operators  are defined by the choice $\kappa=2^{-1/2}$ in $\hat{a}_j = \kappa(\hat{x}_j + i \hat{p}_j)$.

The normal mode decomposition theorem~\cite{Arvind1995}, which follows from Williamson's seminal work \cite{Williamson1936,Williamson1937,Williamson1939}, can be stated as every positive-definite Hermitian matrix $\Sigma$ of dimension $2N\times2N$ can be diagonalized with a symplectic matrix $S$: $D = S \Sigma S^\intercal$, with $D = \text{diag}\left( \nu_1, \nu_1,  \ldots, \nu_N, \nu_N \right)$, where the $\nu_{a}$, for $a \in \{1,...,N\}$, are the symplectic eigenvalues of $\Sigma$, defined as the positive eigenvalues of matrix $i\Omega\Sigma$. A Gaussian state satisfies $\nu_{a} \geq 1$, with equality for all $a$ strictly for the pure state case (which meets $\det\Sigma=1$). As a measure of bipartite, mixed state entanglement, the negativity is the most commonly used entanglement monotone, and is defined as $2\mathcal{N}(\rho):=\norm{\tilde{\rho}}_1-1$, where $\norm{\tilde{\rho}}_{1}:=\tr\sqrt{\tilde{\rho}^\dagger \tilde{\rho}}$ is the trace norm of the partially transposed density operator. In general, $\mathcal{N}(\rho) =\abs{ \sum_j \lambda_j}$ with the $\lambda_j$ the negative eigenvalues of $\tilde{\rho}$. For a bipartite Gaussian state with covariance matrix 
\begin{equation}\label{eq:twoMode}
\Sigma = \begin{pmatrix}
\Sigma_A & \varepsilon_{AB}\\
\varepsilon_{AB}^\intercal & \Sigma_B
\end{pmatrix},
\end{equation}
one defines the two partially transposed symplectic eigenvalues as
\begin{equation}\label{eq:ptseigen}
    \tilde{\nu}_{\mp}:=\sqrt{\frac{\tilde{\Delta} \mp \sqrt{\tilde{\Delta}^2- 4 \det\Sigma}}{2}},
\end{equation}
where the partially transposed symplectic invariant is $\tilde{\Delta} = \det \Sigma_A + \det \Sigma_B - 2\det \varepsilon_{AB}$. The negativity can then be obtained as
\begin{equation}\label{eq:negativity}
\mathcal{N}(\rho) =\max \left\{0, \frac{1-\tilde{\nu}_{-}}{2\tilde{\nu}_{-}}\right\}.
\end{equation} 
Hence, a bipartite Gaussian state is separable when the smaller partially transposed symplectic eigenvalue meets the condition $\tilde{\nu}_{-} \geq 1$. Alternatively, it is entangled when $\tilde{\nu}_{-} < 1$ is met.

Coherent states $\lbrace\ket{\alpha}\rbrace_{\alpha\in \mathbb{C}}$ are defined as the eigenstates of the annihilation operator $\hat{a}$ with eigenvalue $\sqrt{2}\alpha =x + ip$, where $x, p\in \mathbb{R}$ are the eigenvalues of the canonical position and momentum  operators, respectively. They play an important role in quantum CVs, as they allow for a straightforward phase space description of Gaussian states. The displacement operator $\hat{D}_{-\bm{d}}:=\exp[-i\bm{d}\Omega_1\hat{\bm{d}}^\intercal]=\hat{D}_\alpha = \exp \left[ \alpha \hat{a}^\dagger - \bar{\alpha} \hat{a}\right]$ acts on the vacuum as $\hat{D}_\alpha\ket{0} = \ket{\alpha}$, and satisfies $\hat{D}_\alpha^\dagger = \hat{D}_{-\alpha}$. Coherent states are not orthogonal, and their overlap can be computed as $\bra{\beta}\ket{\alpha} = \exp\left[ -\frac{1}{2}(\alpha\bar{\beta} -\bar{\alpha}\beta -\abs{\alpha-\beta}^2)\right]$. This does not prevent the set of all coherent states from forming a basis, which, though overcomplete, allows one to find the coherent states resolution of the identity $\Id = \pi^{-1}\int \diff^2 \alpha \ket{\alpha}\bra{\alpha}$, where $\diff^2 \alpha \equiv \diff \Re{\alpha} \diff \Im{\alpha}$, enabling the computation of  traces of operators in an integral fashion: $\tr [ \hat{O}] =\pi^{-1} \int \diff^2 \alpha \bra{\alpha}\hat{O}\ket{\alpha}$. In this context, it is common to use the fact that, for a coherent state $\bm{d} = (x, p) = \sqrt{2}(\Re{\alpha}, \Im{\alpha})$, and so $2 \diff^2 \alpha  = \diff x \diff p$. More generally, an $n$-mode displacement operator may be defined via $\hat{D}_{-\bm{\mathrm{d}}} =\bigotimes_{j=1}^n  \hat{D}_{-\bm{d}_j} =\hat{D}_{-{\bigoplus_{j=0}^n}\bm{d}_j} $, where $\bm{d}_j:=(x_j, p_j)$, and $\Omega :=\bigoplus_{j=1}^n \Omega_1$.
A complete representation of states that is closely related to coherent states is given by the (Wigner) characteristic function, normally referred to simply as the characteristic function (CF), and for an $n$-mode state $\rho$ (not necessarily Gaussian), is given by 
\begin{equation}
\chi(\bm{\mathrm{d}}) = \tr\left[ \rho\hat{D}_{-\bm{\mathrm{d}}}\right],
\end{equation}
with normalization condition given by $\chi(\bm{0}) =1$. A Gaussian state of first and second moments $(\bm{\mathrm{d}}, \Sigma)$ has a CF given by
\begin{equation}
\chi_G(\bm{\mathrm{r}}) = e^{-\frac{1}{4}\bm{\mathrm{r}} \Omega \Sigma \Omega^\intercal \bm{\mathrm{r}}^\intercal}e^{-i\bm{\mathrm{r}} \Omega \bm{\mathrm{d}}^\intercal},
\end{equation}
where $\bm{\mathrm{r}} = (x_1, p_1, \ldots, x_N, p_N) \in \mathbb{R}^{2N}$.

\subsection{Quantum teleportation with CVs}
Quantum teleportation is a quantum communication protocol that, in principle, allows one to achieve perfect transfer of quantum information between two parties by means of previously shared entanglement, combined with local operations and classical communication. The protocol was first proposed in 1993 by Bennett and collaborators~\cite{Bennett1993}, as a way to take advantage of an entangled resource for the task of sending an unknown quantum state from one place to another, using discrete-variable quantum states. The original idea was simple, yet powerful: assuming that a maximally entangled, bipartite Bell state was shared between two parties (Alice and Bob) prior to the start of the protocol, Alice, in possession of some \textit{unknown} state $\ket{\psi}=\alpha\ket{0} + e^{i\beta}\sqrt{1-|\alpha|^2}\ket{1}$ couples her part of the Bell state to $\ket{\psi}$ by means of a Bell measurement, whose 2-bit output she communicates classically to Bob. Upon receiving the message, Bob performs a conditional unitary on his part of the shared Bell state, recovering $\ket{\psi}$ modulo a global phase in his location.

A year later, Vaidman extended the idea to the transmission of a CV state by means of a perfectly correlated (singular) position-momentum EPR state shared by Alice and Bob~\cite{Vaidman1994}. In 1998, Braunstein and Kimble~\cite{Braunstein1998} made this idea more realistic by relaxing the correlation condition to more experimentally accessible states, such as finitely squeezed states. Their protocol, known as the  Braunstein-Kimble protocol, was first realised in 1998 by A. Furusawa \textit{et al.} in the optical domain~\cite{Furusawa1998}. We review the protocol here for convenience.

Kimble and Braunstein derived an expression for fidelity between an unknown state of a single-mode Bosonic field and a teleported copy, when imperfect quantum entanglement is shared between the two parties. A generalization to a broadband version, where the modes have finite bandwidths, followed quite directly~\cite{Braunstein2005}. In the Braunstein-Kimble protocol, Alice and Bob share a TMSV state, which enables them to teleport the complete state of a single mode of the electromagnetic field, where two orthogonal field quadratures play the role of position and momentum. Shortly after, quantum teleportation of an unknown coherent state was demonstrated, showing an average fidelity (see Eq. \eqref{eq:Pure-PS-fidelity} below)  $\overline{F}=0.58 \pm 0.02$~\cite{Furusawa1998}, which beat the maximum classical fidelity of $\overline{F} = 0.5$ for Gaussian states~\cite{Braunstein2005, Serafini2017, Weedbrook2012}. Other works followed, where the Bell measurement of two orthogonal quadratures was replaced by the photon-number difference and phase sum, and the question of an optimal quantum teleportation protocol depending on the entangled resource was raised~\cite{Milburn1999}. Subtraction of single photons from TMS states has been shown to enhance the fidelity of teleportation \cite{Opatrny2000, Cochrane2002}. We review here the Braunstein-Kimble protocol, replacing the Wigner function approach with its Fourier transform, the characteristic function. 
\begin{figure}[H][t]
\centering
\includegraphics[width=0.5 \textwidth]{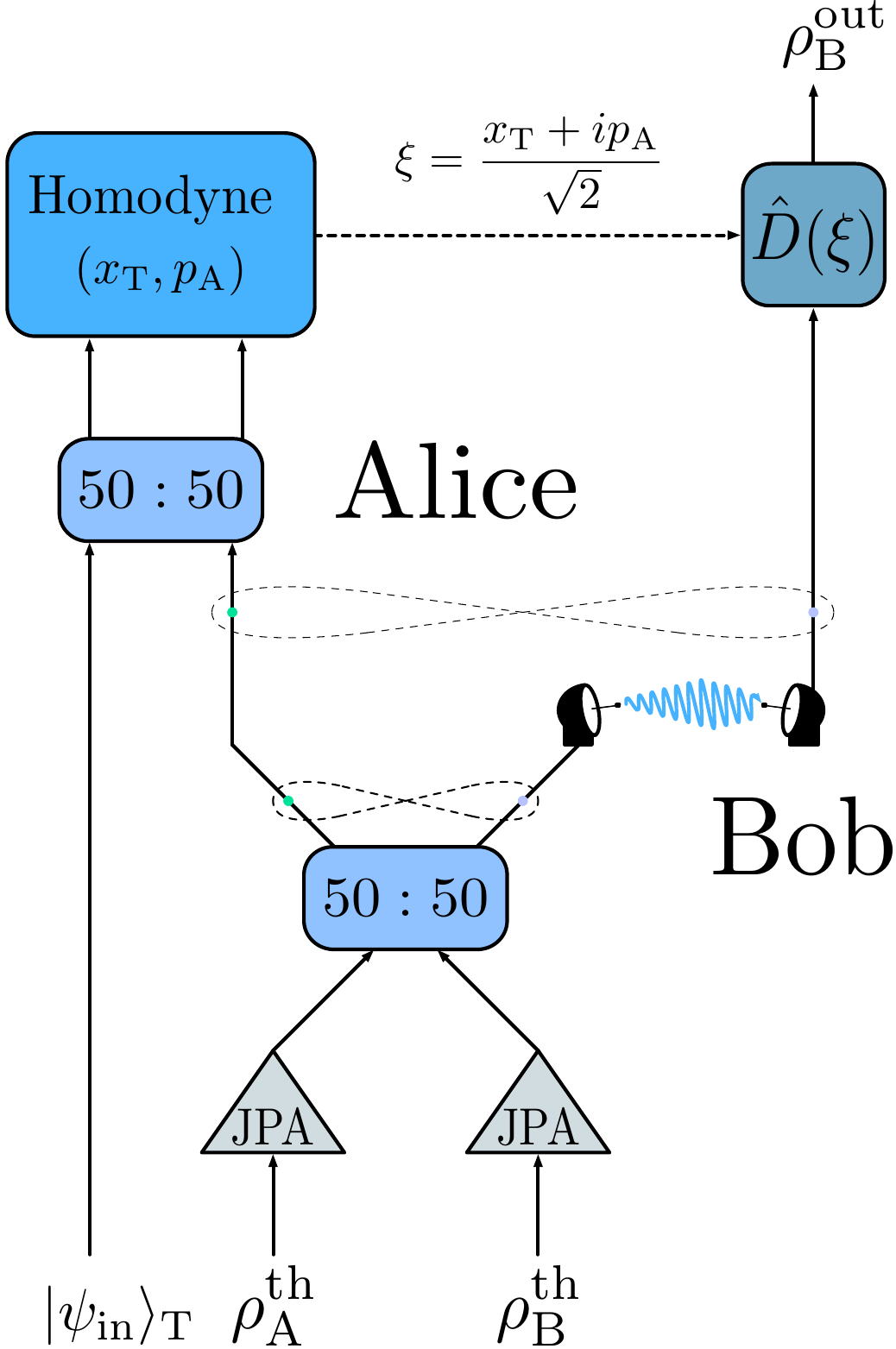}
\caption[Circuit representation of a CV microwave quantum teleportation protocol with Gaussian states]{Circuit representation of a CV microwave quantum teleportation protocol with Gaussian states. The entangled resource is harvested from two single-mode squeezed thermal states, generated from identical JPAs, which are then combined on a balanced beam splitter. Assuming this state is generated by Alice, one of the modes has to be sent to Bob, represented here by the presence of antennae, in order for the two parties to share the entangled resource. Following this, Alice combines the target state to be teleported $|\psi_{\text{in}}\rangle_{\text{T}}$ with the mode of the entangled state she holds in a balanced beam splitter, which is then subject to two homodyne detections, $x_\text{T}$ and $p_\text{A}$. The measurement results $\xi$ are communicated to Bob, who applies a displacement $\hat{D}(\xi)$ on his part of the entangled resource, resulting in the state $\rho^{\text{out}}_{\text{B}}$. \textit{Figure credit: T.G.-R. \& \textbf{M. C.}}}
\label{fig1}
\end{figure}
The protocol goes as follows: 
\begin{enumerate}
\item Alice uses a 50:50 beam splitter to couple her part of the resource state $\rho_{AB}$ with an incoming unknown state $\rho^{\text{in}}_{T}$. The output Hilbert spaces of this beam splitter are labeled $A$ and $T$.
\item Alice performs two homodyne detections, where each of the local oscillator phases are set in order to measure photocurrents, whose differences are integrated over some time, and proportional to quadratures $\hat{x}_T := (\hat{x}_1  + \hat{x}_\text{in})/\sqrt{2}$ and $\hat{p}_A := (\hat{p}_1  - \hat{p}_\text{in})/\sqrt{2}$. She sends the outcomes $(x_T, p_A)$ to Bob via a classical communication channel.
\item Bob, upon reception of the signal $(x_T, p_A)$, performs a displacement $\hat{D}(\xi)$ to his part of $\rho_{AB}$, with $\xi := (x_T + ip_A)/\sqrt{2}$. The state at Bob's location is now, in average, closer to $\rho_{\text{in}}$ than what it would be if no entanglement was present in $\rho_{AB}$.
\end{enumerate}
This protocol is depicted in Fig.~\ref{fig1}, where we also sketch the sequence that leads to an entangled resource shared through open air by Alice and Bob, which is then consumed in the teleportation process. For simplicity, we define $(x_T, p_A) = (x,p)$. The conditional state that Bob has after knowing the outcomes of Alice's homodyne measurements is
\begin{equation}
\rho_{B}(x,p) = \frac{1}{P_{B}(x,p)} \langle \Pi(x,p)| \rho_{T}^{\text{in}}\otimes \rho_{AB}|\Pi(x,p)\rangle_{TA},
\end{equation}
with $P_{B}(x,p) = \Tr_{B}\langle \Pi(x,p)| \rho_{\text{in}}\otimes \rho_{AB}|\Pi(x,p)\rangle_{TA}$ and 
\begin{equation}
\ket{\Pi(x,p)}_{TA} = \frac{1}{\sqrt{2\pi}}\int_\mathbb{R}\diff y e^{ipy}\ket{x+y}_T\ket{y}_A,
\end{equation}
which is an element of the maximally entangled basis corresponding to Alice's Bell-like measurement. Now, this expectation value over the teleported ($T$) and the senders ($A$) modes is computed as 

\begin{equation}
\langle \Pi(x,p)| \rho_{\text{in}}\otimes \rho_{AB}|\Pi(x,p)\rangle_{TA} = \frac{1}{2\pi}\int_{-\infty}^{\infty}\int_{-\infty}^{\infty} \diff y\diff y' e^{ip(y-y')}\langle x+y'|\rho_{\text{in}}|x+y\rangle_{T} \langle y'|\rho_{AB}| y\rangle_{A}.
\end{equation}

Once we have computed $\rho_{B}(x,p)$, we need to compute the outcoming state after the receiver applies the displacements, and average over all possible measurement outcomes
\begin{equation}
\rho_{B}^{\text{out}} = \int_{-\infty}^{\infty} \diff x \int_{-\infty}^{\infty} \diff p P_{B}(x,p)\hat{D}_{B}\left(\xi\right)\rho_{B}(x,p)\hat{D}^{\dagger}_{B}\left(\xi\right).
\end{equation}
As a measure of the quality of the protocol, one typically uses an overlap fidelity $F(\rho_\text{in}, \rho_\text{out}) = \tr[\rho_{\text{in}}\rho_{\text{out}}]$, which represents a simplified version of the Uhlmann fidelity $\left(\tr[\sqrt{\sqrt{\rho_{\text{in}}}\rho_{\text{out}}\sqrt{\rho_{\text{in}}}}]\right)^{2}$ in the case when $\rho_{\text{in}}$ is pure. 

The figure of merit in quantum teleportation is the \textit{average} fidelity, which refers to the fact that we have averaged over all possible measurement outcomes,
\begin{equation}\label{eq:Pure-PS-fidelity}
\overline{F} = \tr[\rho_{T}^{\text{in}}\rho_{B}^{\text{out}}].
\end{equation}
Sometimes it can be useful to have it written in terms of the CFs~\cite{Marian2006}:
\begin{equation}
\overline{F} =  \frac{1}{\pi}\int \diff ^{2}\beta \chi_{T}^{\text{in}}(-\beta)\chi_{B}^{\text{out}}(\beta),
\end{equation}
with the average over $(x,p)$ having already been performed in $\rho_{B}^{\text{out}}$.

If the resource state $\rho_{AB}$ is a Gaussian state with the covariance matrix given in Eq.~\eqref{eq:twoMode}, and the teleported state is a coherent state $|\alpha_{0}\rangle\langle\alpha_{0}|$, the average fidelity can be written as
\begin{equation}\label{eq:Gaussian-fidelity}
\overline{F} = \frac{1}{\sqrt{\det[\Id_{2} + \frac{1}{2}\Gamma]}},
\end{equation}
with $\Gamma \equiv (\sigma_Z \Sigma_A\sigma_Z + \Sigma_B -\sigma_Z \varepsilon_{AB}-\varepsilon_{AB}^\intercal\sigma_Z)$. Coherent states are typically those chosen to be teleported due to the ease of their experimental generation. In theory, the result of the average fidelity does not depend on the displacement $\alpha_{0}$; therefore, it will suffice to use an unknown coherent state for a demonstration of quantum teleportation. In experiments, however, the teleportation fidelity may depend on $\alpha_{0}$. 

It is also interesting to see the average fidelity of a process in which $k$ teleportation protocols are concatenated, i.e.,
\begin{equation}
\overline{F}^{(k)} = \frac{1}{\sqrt{\det[\Id_{2} + \left(k-\frac{1}{2}\right)\Gamma)]}},
\end{equation}
assuming that, in each step, an entangled Gaussian resource with the covariance matrix that characterizes $\Gamma$ is used. 

Consider a symmetric covariance matrix with $\Sigma_{A}=\Sigma_{B}=\alpha\Id_{2}$ and $\varepsilon_{AB}=\gamma\sigma_{z}$. Then, we have $\Gamma = 2(\alpha-\gamma)\Id_{2}$ and $\tilde{\nu}_{-}=\alpha-\gamma$, which leads to
\begin{equation}\label{eq:Gaussian-fidelity}
\overline{F} = \frac{1}{1+\tilde{\nu}_{-}}.
\end{equation}
It is easy to check the two following limits for the average teleportation fidelity of an arbitrary coherent state: $\lim_{\tilde{\nu}_{-}\rightarrow 1} \overline{F} = 1/2$ and $\lim_{\tilde{\nu}_{-}\rightarrow 0} \overline{F} = 1$. The first limit corresponds to using no entanglement ($\tilde{\nu}_{-}\geq1$), and is interpreted as the `classical teleportation' threshold, meaning that any approach giving an average fidelity of 0.5 or less does not demonstrate quantum teleportation. The second limit corresponds to an idealized case of an infinite two-mode squeezing level ($\tilde{\nu}_{-}=0$), i.e., an EPR state, which realizes perfect quantum teleportation.

\begin{figure}[H]
\centering
\includegraphics[width=0.6\textwidth]{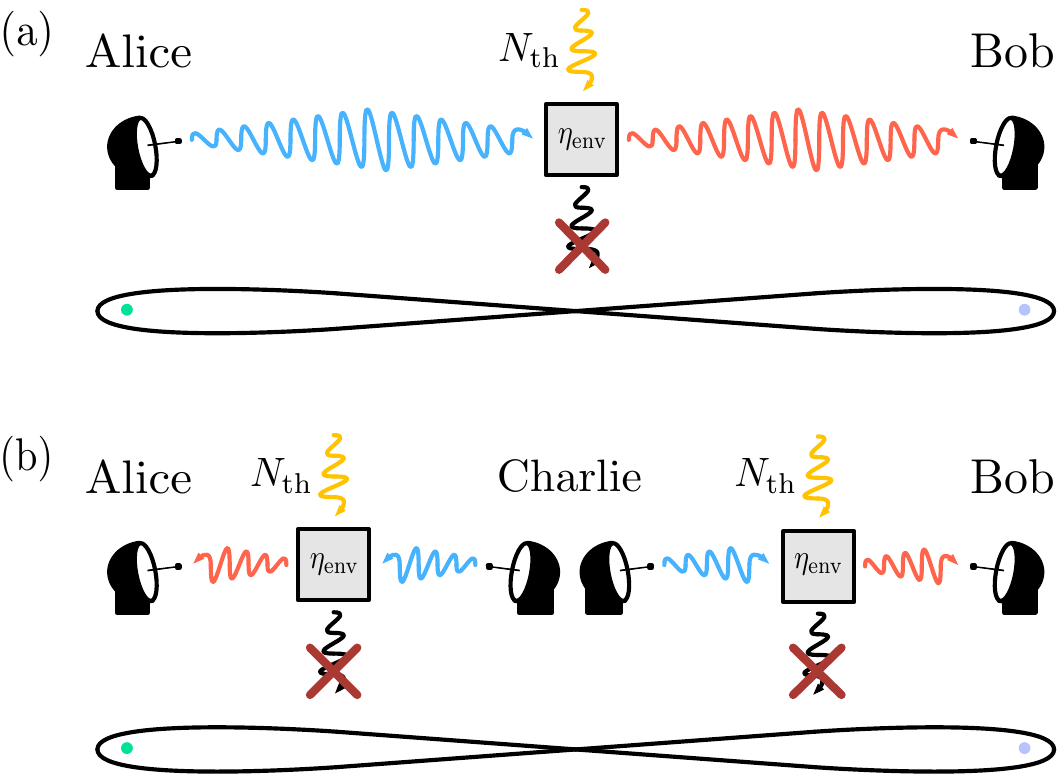}
\caption[Representation of an entanglement distribution protocol that uses antennae to efficiently transmit the quantum states into open air, where photon losses and thermal noise effects are described with a beam splitter with reflectivity $\eta_{\text{env}}$.]{Representation of an entanglement distribution protocol that uses antennae to efficiently transmit the quantum states into open air, where photon losses and thermal noise effects are described with a beam splitter with reflectivity $\eta_{\text{env}}$. We analyze two different scenarios: (a) Alice generates the entangled state, and attempts to share one of its modes with Bob by sending it through a noisy and lossy open-air channel that degrades the entanglement strength; (b) Charlie generates a two-mode entangled state, and sends one entangled mode to Alice and another to Bob. In this case, although both modes go through the same noisy and lossy channel, they travel half the distance compared to the previous case. \textit{Figure credit: T. G.-R.}}
\label{fig2}
\end{figure}

\section{Open-air microwave entanglement distribution}\label{section:open_air_entanglement_distribution}
The scheme we have envisioned (see Fig.~\ref{fig2}) for open-air entanglement distribution relies on a variety of things: first, generation of two-mode squeezed thermal (TMST) states with thermal photons $n < e^{-r}\sinh(r)$, given squeezing $r$, which will take place in a cryostat at $ T \sim 50$ mK temperatures; second, efficient transmission of states out of the cryostat and into open air, targeting optimal entanglement preservation; for this task, we rely on an antenna based on the design proposed in Ref.~\cite{GonzalezRaya2020}, with special attention given to the shape of the impedance function, which greatly affects entanglement preservation; third, estimation of losses in open air that describe the attenuation of the signal caused by the presence of thermal noise in the environment, with the objective of setting bounds on effective transmission distances. This protocol will prepare the foundation for an open-air microwave quantum teleportation protocol, whose fidelity will depend on the entanglement of the resource shared by the parties involved.

\subsection{State Generation}\label{subsection:state_generation}                    
As the first step, we discuss the generation of entangled states inside a cryostat, more precisely, two-mode squeezed states. Squeezing is an operation in which one of the variances of the electromagnetic field quadratures of a quantum state is reduced below the level of vacuum fluctuations, while the conjugate quadrature is amplified, satisfying the uncertainty principle. This can be achieved by sending the vacuum state to a JPA, a coplanar waveguide resonator line terminated by a direct-current superconducting interference device (dc SQUID). The dc SQUID provides magnetic flux tunability to the resonator and enables parametric phase-sensitive amplification, which is the key for generating squeezed microwave states~\cite{Zhong2013,Pogorzalek2017}.

The relation between the frequency of the external magnetic flux, $\Omega$, and the fundamental frequency of the JPA, $\omega_{c}$, determines whether the JPA operates in the phase-insensitive or phase-sensitive regime. The latter is achieved in the so-called degenerate regime, $\Omega = 2\omega_{c}$. A corresponding three-wave mixing process, when one pump photon splits into two signal photons, is described by the Hamiltonian
\begin{equation}
H = g\left( \beta^{*}a^{2} - \beta a^{\dagger 2} \right).
\end{equation}
It can be shown that the aforementioned Hamiltonian corresponds to  a single-mode squeezing operator
\begin{equation}
S(\xi) = \exp\left[ \frac{1}{2}(\xi^{*}a^{2}-\xi a^{\dagger 2})\right],
\end{equation}
with the squeezing parameter given by $|\xi| \propto 2g|\beta|t$. 

A symmetric two-mode squeezed state can be generated by combining two, orthogonally squeezed, states with equal squeezing levels at a hybrid ring. The latter element represents a symmetric 50:50 microwave beam splitter. Microwave squeezed states produced by JPAs are subject to various sources of imperfections and noise. Therefore, the output states can be effectively modeled as two-mode squeezed thermal states, whose second moments differ from those of ideal two-mode squeezed vacuum states by a factor of $1+2n$, where $n$ is the number of thermal photons.

Thermal photons in squeezed states may have various physical origins. One of the most trivial reasons for noise in the two-mode squeezed states is finite temperatures of the input JPA modes, which lead to the fact that one applies a squeezing operator to a thermal state rather than to a vacuum. Another important source of noise in squeezed states produced by flux-driven JPAs arises from Poisson photon-number fluctuations in the pump mode, which lead to extra quasithermal photons in the output squeezed states~\cite{Renger2020}. Last but not least, higher-order nonlinear effects also contribute to additional effective noise under the Gaussian approximation~\cite{Boutin2017}.  More experimental details on the microwave squeezing and related imperfections can be found elsewhere~\cite{Fedorov2018}.


\subsection{Antenna Model \& Open-air losses}\label{subsection:antenna_model}

\begin{figure}[H]
\centering
\includegraphics[width=0.8\textwidth]{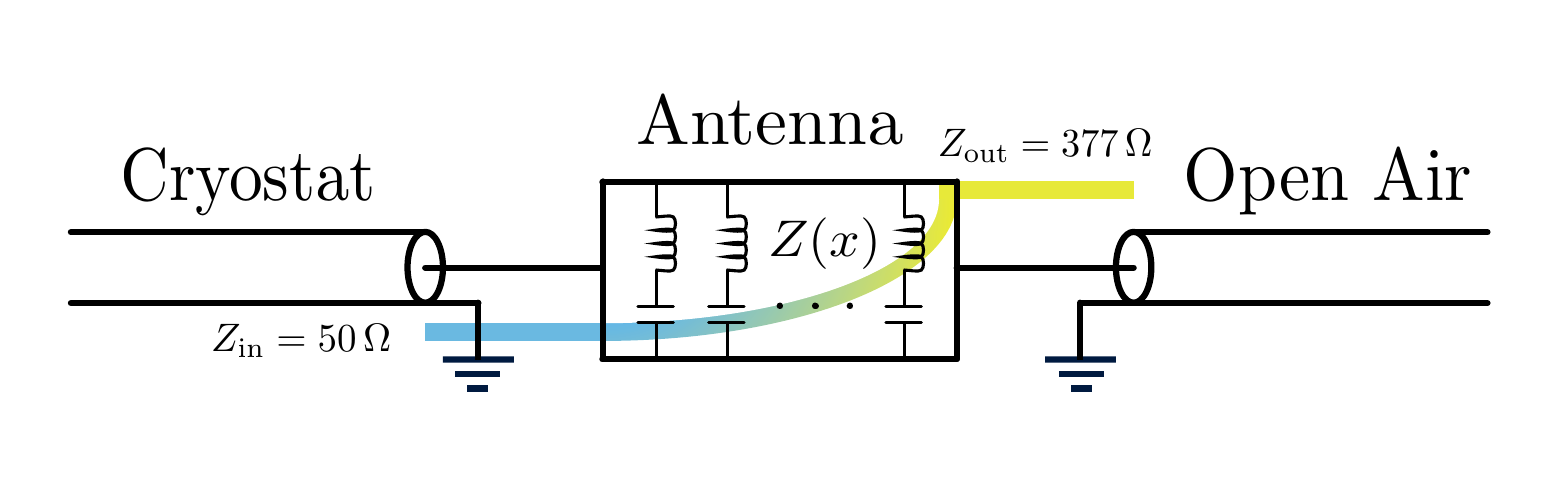}
\caption[Schematic representation of the circuit needed for a quantum communication protocol in which a quantum state is generated in a cryostat, and is then sent through an antenna into open air, where transmission is assumed to be focused.]{Schematic representation of the circuit needed for a quantum communication protocol in which a quantum state is generated in a cryostat, and is then sent through an antenna into open air, where transmission is assumed to be focused. As is shown, a waveguide at impedance $Z_{\text{in}}=50 \, \Omega$ connects the cryostat with the antenna. The latter is a finite transmission line with variable impedance, $Z(x)$, designed to match the impedance of the cryostat with that of the open air, $Z_{\text{out}}=377 \, \Omega$, maximizing the transmissivity. The final waveguide represents the directed propagation of the signal in open air. \textit{Figure credit: T. G.-R.}}
\label{fig3}
\end{figure}

In transmission of quantum states from a cryostat into the open air, an interface antenna comes into play as an inhomogeneous medium (as depicted in Fig.~\ref{fig3}) that connects those two very different environments. The main purpose of this antenna is to maximize transmission of the incident signal to the open-air medium. Here, we consider a transmission line coming out of the cryostat with an impedance of 50 $\Omega$, and assume focused transmission in open air described by a transmission line with impedance 377 $\Omega$. Such an antenna can be modeled by a finite transmission line with variable characteristic impedance and designed to match the impedances of the cryostat and of open air at its ends. This approach was discussed in a previous study~\cite{GonzalezRaya2020}, where the transmissivity of the antenna was optimized for the task of entanglement distribution with TMST states of 5 GHz in an antenna of 5 cm. With an exponential profile of the impedance inside the antenna, reflectivity can be reduced down to $\sqrt{\eta}< 10^{-9}$, qualitatively matching the classical result of a horn antenna. 

Using this description, we can obtain a reflectivity coefficient that depends on experimental parameters, such as the length of the antenna, the carrier frequency of quantum states, and the internal and external impedances, among others. This result is compatible with the description of the antenna as a distributed beam splitter, with its inputs being one of the modes of the TMST state together with a thermal state with $N_{\text{th}}$ thermal photons coming from the environment. From the two output modes, the reflected one is discarded, whereas the transmitted one is sent to Bob. The main antenna aim is to minimize the reflections of Alice's input signal in order to preserve entanglement between the transmitted mode and the retained one. 

Backscattered thermal photons entering the emitting antenna might represent a certain problem for Alice and must be filtered. A straightforward, albeit somewhat challenging, solution for this problem is to use nonreciprocal microwave devices such as isolators (circulators) to protect entanglement-generating circuits from the unwanted thermal radiation.

\begin{figure}[H]
\centering
\vspace{0.5cm}
\includegraphics[width=0.6\textwidth]{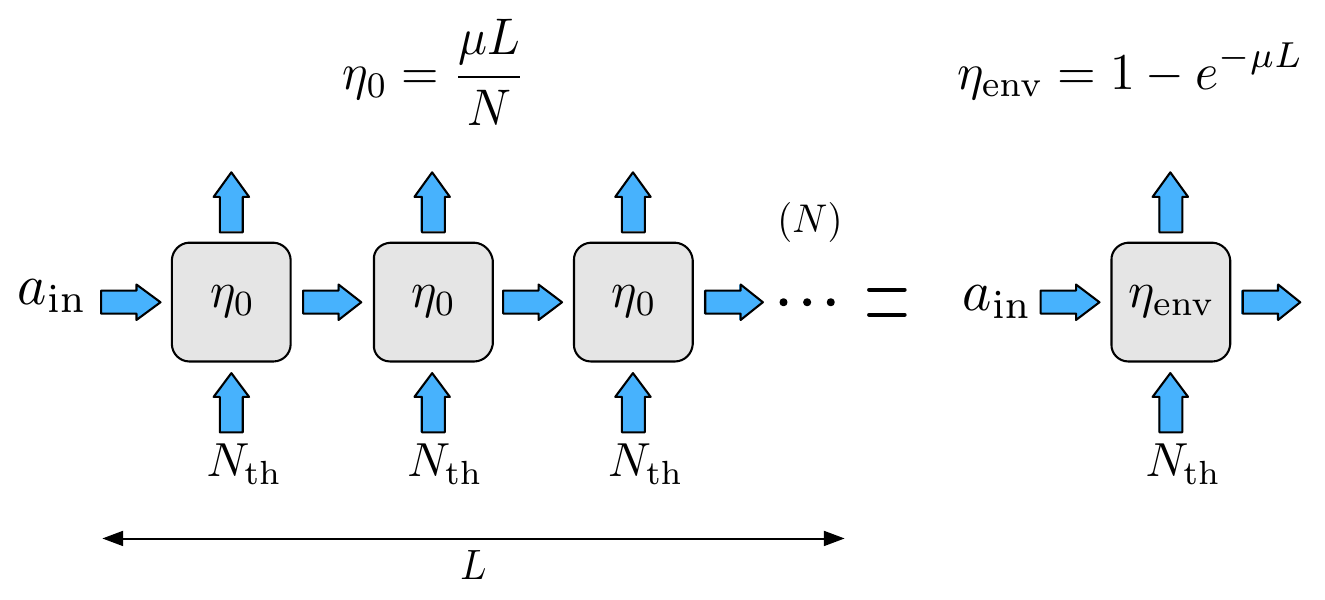}
\caption[Sketch of a beam splitter loss model of an open air quantum channel.]{Sketch of a beam splitter loss model of an open air quantum channel. Entanglement degradation of a state propagating in open air at a constant temperature is modeled by an array of $N$ beam splitters, each one introducing thermal noise characterized by $N_{\text{th}}$ thermal photons, assuming constant temperature throughout the path. An infinite array of beam splitters ($N\rightarrow\infty$) can be approximated by a single beam splitter with reflectivity $\eta_{\text{env}} = 1 - e^{-\mu L}$, where $L$ is the total channel length and $\mu$ is the reflectivity per unit length. \textit{Figure credit: M. S., T. G.-R. \& \textbf{M. C.}}}
\label{fig4}
\end{figure}

Once the state has been successfully sent out of the cryostat, we have to address the effects of entanglement degradation in open air. Considering directed transmission in open air, we envision an infinite array of beam splitters to describe losses in open air, as represented in Fig.~\ref{fig4}. Each one of these beam splitters will allow for the mixing of thermal noise with the state. Assuming constant temperature throughout the sequence of possible absorption events, meaning that the thermal noise in each of the beam splitters is characterized by $N_{\text{th}}$, we can obtain the reflectivity of an effective beam splitter based on an attenuation channel~\cite{Serafini2017}, which represents the decay of quantum correlations and amplitudes,
\begin{equation}
\eta_{\text{env}} = 1-e^{-\mu L}.
\end{equation}
Here, $\mu$ represents a density of reflectivity, which in turn models photon losses per unit length, and $L$ is the traveled distance. This density of reflectivity can be interpreted as an attenuation coefficient that quantifies the specific attenuation of signals in a given environment. In this work we consider $\mu = 1.44\times 10^{-6}\text{ m}^{-1}$ for the specific attenuation of 5 GHz signals caused by the presence of oxygen molecules in the environment (see Refs.~\cite{Ho2004,ITU-R}).

We could go further and assume that, attached to the antenna (at constant temperature), there is another transmission line where the temperature is not constant throughout the trajectory, which leads to an inhomogeneous absorption probability. This is represented by the density of reflectivity $\mu(x)$, and the number of thermal photons $n(x)$. The latter still follows the Bose-Einstein distribution. An infinite array of beam splitters that reproduce these features (see Ref.~\cite{GonzalezRaya2020}) can be replaced by a single beam splitter with an effective reflectivity and number of thermal photons given by
\begin{eqnarray}
\eta_{\text{env}} &=& 1-e^{-\int_{0}^{L} \diff x \mu(x)}, \\
\nonumber n_{\text{th}} &=& \frac{\int_{0}^{L} \diff x \mu(x)n(x) e^{-\int_{x}^{L}\diff x' \mu(x')}}{1-e^{-\int_{0}^{L} \diff x \mu(x)}},
\end{eqnarray}
where $L$ represents the total length of the array. Given that we are extending the length in which the transmission line remains at cryogenic temperatures, we see that $n_{\text{th}} \leq N_{\text{th}}$.

Now that we have discussed how the signal is processed into the environment, let us characterize the resulting states. Assume that Alice generates a TMST state with $n$ thermal photons, and sends one mode to Bob over a distance $L$ through open air, with a thermal background characterized by $N_{\text{th}}$ thermal photons. Then, the resulting state is what we call the ``asymmetric'' state
 
\begin{equation}\label{CM_asym}
\Sigma_{\text{Asym}}=(1+2n)\begin{pmatrix} \left[\left(\frac{1+2N_{\text{th}}}{1+2n}\right) \eta_{\text{eff}} + (1-\eta_{\text{eff}})\cosh 2r\right]\Id_2 &  \sqrt{1-\eta_{\text{eff}}} \sinh 2r \sigma_Z  \\ 
\sqrt{1-\eta_{\text{eff}}} \sinh 2r \sigma_Z &  \cosh 2r \Id_2 \end{pmatrix},
\end{equation}

where $\eta_{\text{eff}}=1- e^{-\mu L}(1-\eta_\text{ant})$ represents the combined reflectivities of the antenna $\eta_{\text{ant}}$ and of the environment $\eta_{\text{env}}$. A sketch of the layout that leads to this kind of state can be seen in Fig.~\ref{fig2}~(a). With this, using Eq. \eqref{eq:ptseigen}, we compute the partially transposed symplectic eigenvalue,
\begin{equation}
\tilde{\nu}_{-}^{\text{out}} = \tilde{\nu}_{-}^{\text{in}} + \left( \frac{1}{2} + N_{\text{th}}\right)\eta_{\text{ant}}
\end{equation}
for very low reflectivities, $\eta_{\text{ant}} N_{\text{th}} \ll 1$, with $\tilde{\nu}_{\text{in}} = (1+2n)e^{-2r}$. Note that, by reducing the reflectivity of the antenna, the impact of thermal noise is reduced, and the partially transposed symplectic eigenvalue approaches that of the input state. In this extreme case, entanglement is fully preserved. 

Let us use the partially transposed symplectic eigenvalue to compute the limit of entanglement. We use the negativity as a measure of Gaussian entanglement, $\mathcal{N} = \max\{ 0, \frac{1-\tilde{\nu}_{-}}{2\tilde{\nu}_{-}}\}$, such that this limit occurs for $\tilde{\nu}_{-}=1$. This constitutes a bound on the reflectivity; all smaller values of $\eta_{\text{eff}}$ will result in entanglement preservation. This result is
\begin{equation}\label{ref_bound}
\eta_\text{max} = \frac{1}{1+\frac{N_{\text{th}}}{1+\frac{2n(1+n)}{1-(1+2n)\cosh(2r)}}},
\vspace{0.15cm}
\end{equation}
together with the conditions $n < e^{-r}\sinh(r)$ and $r > 0$. With this bound, the maximum distance entanglement can survive is
\begin{equation}
L_\text{max} = -\frac{1}{\mu}\log(1-\eta_\text{max}).
\end{equation} 
Imagine that TMST states are generated in the cryostat at $50$ mK temperature, with thermal photons $n \sim 10^{-2}$,  and squeezing $r = 1$. In open air, at $300$ K, the number of thermal photons is $N_{\text{th}} \sim 1250$. Assuming a perfect antenna ($\eta_{\text{ant}}=0$), the maximum distance the state can travel before entanglement completely degrades is $L_\text{max} \sim 550$ m.


As a different approach to the entangled resource, we assume that a TMST state is generated at an intermediate spot between both parties, and that each mode is sent through an antenna and travels some distance $L_i$, with $i=\{1,2\}$, before reaching Alice and Bob. Then, each mode will see an effective reflectivity of $\eta^{(i)}_{\text{eff}}= 1 - e^{-\mu L_i}(1-\eta_\text{ant})$, combining the effects of the antenna and the environment. We assume for simplicity that $L_1+L_2 = L$, where $L$ is the linear distance between Alice and Bob. The covariance matrix of such a state, which we refer to as ``symmetric'':
 \begin{landscape}
     \begin{equation}\label{CM_sym}
\Sigma_{\text{Sym}} =(1+2n)\begin{pmatrix} \left[\left(\frac{1+2N_{\text{th}}}{1+2n}\right)\eta^{(1)}_{\text{eff}}  + (1-\eta^{(1)}_{\text{eff}})\cosh 2r\right]\Id_2 & \sqrt{\left(1-\eta^{(1)}_{\text{eff}}\right)\left(1-\eta^{(2)}_{\text{eff}}\right)} \sinh 2r \sigma_Z  \\ 
\sqrt{\left(1-\eta^{(1)}_{\text{eff}}\right)\left(1-\eta^{(2)}_{\text{eff}}\right)} \sinh 2r \sigma_Z & \left[ \left(\frac{1+2N_{\text{th}}}{1+2n}\right)\eta^{(2)}_{\text{eff}}  + (1-\eta^{(2)}_{\text{eff}})\cosh 2r\right]\Id_2 \end{pmatrix},
\end{equation}

and corresponds to the layout represented in Fig.~\ref{fig2}~(b). With this state, the maximum distance entanglement can survive is $L_\text{max} \sim 480$ m.
 \end{landscape}

Throughout this manuscript, we refer to these states as the (asymmetric and/or symmetric) lossy TMST states, the bare states, or the TMST states distributed through open air. 

Furthermore, we could also consider the specific attenuation caused by the presence of water vapor in the environment~\cite{Ho2004}. This would lead to higher attenuation coefficients, thus reducing the distances that entanglement can survive. For an average water vapor density, these distances are 450 and 390 m for asymmetric and symmetric states, respectively. They become 400 m for asymmetric states and 350 m for symmetric states in a maximum water vapor density scenario.


\subsection{Overcoming entanglement degradation: distillation and swapping with microwaves}\label{subsection:distillation_and_swapping}
We have seen that entanglement distribution between two parties is limited by environmental noise, as well as by photon losses. Considering we have a perfect antenna, these factors can limit the maximum distance entanglement can survive to a few hundred meters. Since an amplification protocol only contributes to the degradation of quantum correlations (see Appendix~\ref{app_A}), we investigate entanglement distillation and entanglement swapping, two techniques that could improve both the reach and the quality of entanglement, at the expense of efficiency~\cite{DiCandia2015}. 

\subsubsection{Entanglement distillation}
This technique aims at increasing entanglement in quantum states by means of local operations. Let us briefly review different ways to distill entanglement. One of them is noiseless linear amplification, a nondeterministic operation~\cite{Ralph2009,Xiang2010} that requires nonincreasing distinguishability of amplified states, and which has recently been achieved in the microwave regime~\cite{Dassonneville2020}. It also requires efficient photocounting, which is where the nondeterministic part comes into play. At the core of this protocol lies a process based on the quantum scissors~\cite{Pegg1998}. The gain of this procedure is inversely proportional to the success probability, which also decreases as the number of resources increases, making it very costly. 

Another widely known protocol is Gaussian distillation, which is also nondeterministic, but it requires only two initial copies of a state, as well as efficient photodetection. If the incoming entangled state is Gaussian then it is initially de-Gaussified by combining two copies of said state with balanced beam splitters and keeping the transmitted state when any number of photons has been detected at the reflected modes~\cite{Browne2003}. Another possible de-Gaussification protocol applies an operation $\hat{Z}=(1-\omega)a^{\dagger}a+\omega a a^{\dagger}$~\cite{Fiurasek2010} on a quantum state without requiring a copy. Gaussian distillation begins when two copies of the resulting state are mixed by $50:50$ beam splitters and, if no photons are reflected, the operation is applied again~\cite{Browne2003,Eisert2004}. Provided that the initial states were entangled, this process leads to a non-Gaussian state with higher entanglement. However, it is also costly in terms of the number of resources, and it only produces a state that is Gaussian (and with higher entanglement) in an infinite-application limit of the Gaussification channel.  

\begin{figure}[H]
\centering
\includegraphics[width=0.6\textwidth]{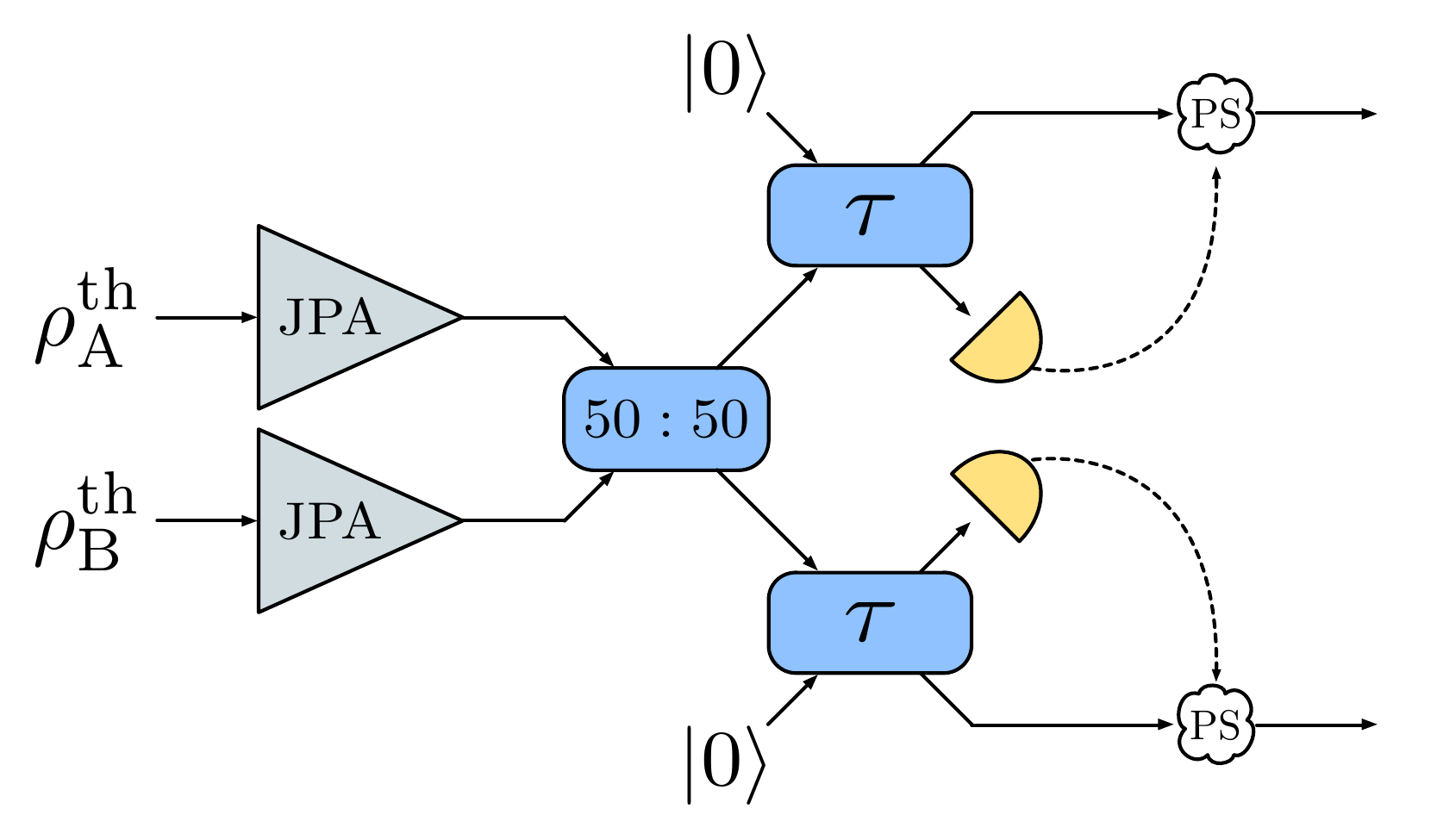}
\caption[Sketch of a photon subtraction scheme applied to two-mode squeezed thermal states, which are generated from single-mode squeezed thermal states by JPAs, subsequently combined in a balanced beam splitter.]{Sketch of a photon subtraction scheme applied to two-mode squeezed thermal states, which are generated from single-mode squeezed thermal states by JPAs, subsequently combined in a balanced beam splitter. Each mode of the resulting state is then combined with an ancillary vacuum state in high-transmissivity (we consider $\tau = 0.95$) beam splitters, with photocounters placed at each reflected path. The resulting state shows higher entanglement for low values of the squeezing parameter, where the limit for enhancement will vary depending on the number of photons detected. \textit{Figure credit: T. G.-R. \& \textbf{M. C.}}}
\label{fig5}
\end{figure}

Finally, we review another nondeterministic protocol, which does not require the storage or production of simultaneous copies of a quantum state, and whose gain is also inversely proportional to the success probability. This protocol is called probabilistic photon subtraction~\cite{Takahashi2010}, and it utilizes non-Gaussian operations in order to distill entanglement, as we have seen in the previous protocols. However, in this situation, we do not look to re-Gaussify the state afterwards. A discussion about the heuristic photon-subtraction protocol with TMSV states, the more theoretical approach that does not consider the effects of beam splitters and measurement, can be found in Appendix~\ref{app_B}.

Probabilistic photon subtraction starts with an entangled state, for example a TMSV state. This can be produced by two single-mode squeezed states with squeezing parameter $r$, which are combined by a $50:50$ beam splitter, as shown in Fig.~\ref{fig5}, resulting in a TMSV state,
\begin{equation}
\sqrt{1-\lambda^{2}}\sum_{n=0}^{\infty} \lambda^{n}|n,n\rangle_{AB}
\end{equation}
with $\lambda = \tanh(r)$. The next step of the protocol is to mix each mode with an ancillary vacuum state at two highly transmitting, identical beam splitters. The output photon-subtracted state is postselected depending on the outcome of the photocounts performed at each beam splitter. Here, we focus on photon-subtracted TMSV states where the same number of photons is subtracted from each mode. The resulting $2k$ photon-subtracted TMSV state is then
\begin{equation}\label{eq:kPSstate}
|\psi^{(2k)}\rangle_{AB} = P_{2k}^{-1/2}\sum_{n=0}^{\infty} a^{(k)}_n\ket{n,n}_{AB},
\end{equation}
with $a^{(k)}_n \equiv  \sqrt{1-\lambda^{2}}\lambda^{n+k} \begin{pmatrix}n+k\\k\end{pmatrix} (1-\tau)^{k} \tau^{n}$, and $P_{2k} \equiv \sum_{n=0}^\infty \abs{a^{(k)}_n}^2$, which can be interpreted as the probability of successfully subtracting $k$ photons from each mode of a TMSV state. The sum converges to $P_{2k} = \left(1-\lambda ^2\right) (\lambda -\lambda_\tau)^{2 k} \, _2F_1\left(k+1,k+1;1;\lambda_\tau ^2\right)$ where $\, _2F_1\left(a,b;c;z\right)$ is the Gaussian hypergeometric function and $\lambda_\tau\equiv \tau \lambda$.

In what follows, we focus on the cases $k=1,2$, which correspond to two-photon subtraction (2PS) and four-photon subtraction (4PS), respectively, and whose corresponding success probabilities are
\begin{eqnarray}\label{eq:success-prob-TMSV}
\nonumber P_{2} &=& (1-\lambda^{2})\lambda^{2}(1-\tau)^{2}\frac{(1+\lambda_{\tau}^{2})}{(1-\lambda_{\tau}^{2})^{3}}, \\
P_{4} &=& 4\left(1-\lambda ^2\right) \lambda ^4  (1-\tau)^4\frac{\left(1+\lambda_{\tau}^4+4 \lambda_{\tau} ^2\right)}{\left(1-\lambda_{\tau}^2\right)^5}.
\end{eqnarray}

\begin{figure}[H]
\centering
\includegraphics[width=0.6\textwidth]{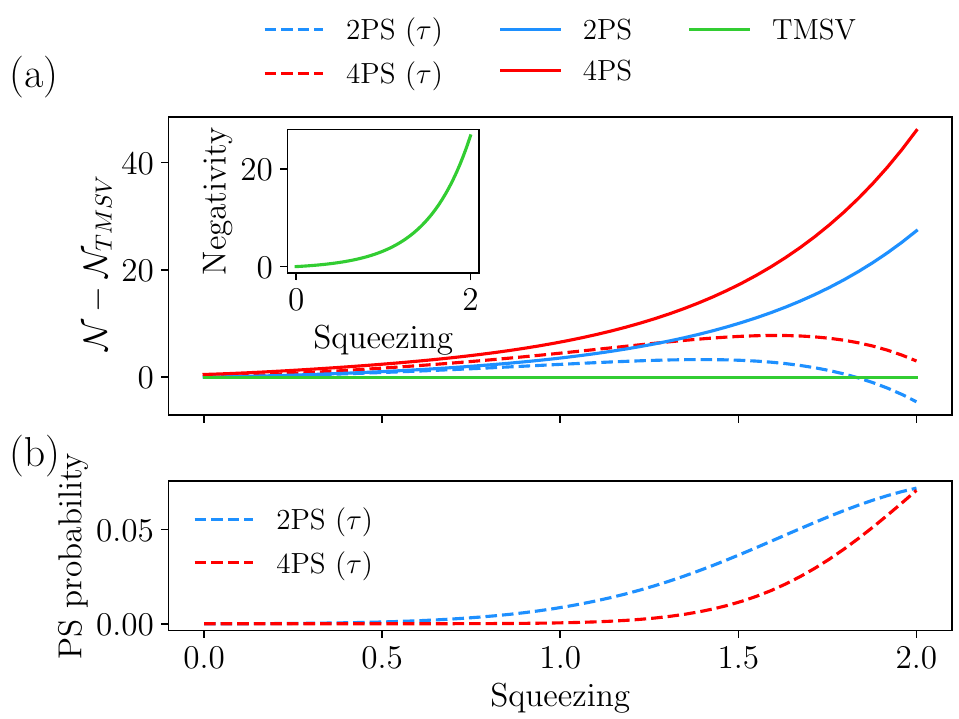}
\caption[Negativity difference between photon-subtracted and bare TMSV states, represented against the initial squeezing parameter, and success probability of symmetric photon-subtraction schemes]{(a) Negativity difference between photon-subtracted and bare TMSV states, represented against the initial squeezing parameter. The blue and red curves represent two- and four- photon subtraction, respectively, whereas the green curve represents the no-gain line, above which any point represents an improvement in negativity. Curves associated with probabilistic photon subtraction appear dashed, whereas the solid ones are associated with heuristic photon subtraction. We have considered the transmissivity of the beam splitters involved in probabilistic photon subtraction to be $\tau=0.95$. In the inset, we represent the negativity curve for the TMSV state versus the initial squeezing parameter. (b) Success probability of symmetric photon-subtraction schemes: two-photon subtraction (2PS) is represented with a blue dashed line, and four-photon subtraction (4PS) is represented with a red dashed line. \textit{Figure credit: T. G.-R.}}
\label{fig6}
\end{figure}

If photon subtraction is successful for any (nonzero) number of photons, the resulting state shows increased entanglement with respect to the TMSV state in a certain interval. This can be seen by computing the negativity $\mathcal{N}(\rho^{(2k)})$ of the family of states \eqref{eq:kPSstate}. We find that, for $\rho^{(2k)} \equiv \ket{\psi^{(2k)}}\bra{\psi^{(2k)}}$, the negativity is
\begin{equation}
\mathcal{N}(\rho^{(2k)}) =  \frac{A_k-1}{2},
\end{equation}
where \begin{equation}
A_k \equiv \frac{\left(\sum_{n=0}^\infty a^{(k)}_n\right)^2}{P_{2k}}.
\end{equation}
Performing the sum we obtain
\begin{equation}
\mathcal{N}(\rho^{(2k)})=\frac{1}{2} \left(\frac{(1- \lambda_\tau)^{-2 (k+1)}}{\, _2F_1\left(k+1,k+1;1; \lambda_\tau^2\right)}-1\right),
\end{equation}
which describes the negativity of the heuristic photon-subtraction protocol (see Appendix~\ref{app_B}) in the limit $\tau\rightarrow 1$, while reproducing the negativity of the TMSV state, $\mathcal{N}_{\text{TMSV}}=\lambda/(1-\lambda)$, in the case $\tau\rightarrow1$ and $k=0$.

In Fig.~\ref{fig6}~(a) we represent negativity differences as a function of the initial squeezing $r$. We subtract the negativity of the TMSV state from those of the two-photon (blue) and four-photon (red), heuristic (solid) and probabilistic (dashed), subtraction with a beam splitter transmissivity $\tau = 0.95$. Note that probabilistic photon subtraction works for lower squeezing, while heuristic photon subtraction is always advantageous. In Fig.~\ref{fig6}~(b), we display the success probability of two-photon (blue, dashed line) and four-photon (red, dashed line) subtraction. Observe that 2PS shows higher probability than 4PS, whereas the latter shows higher improvement than the former. As the squeezing parameter increases, both probabilities grow closer, as probabilistic PS loses its advantage.

The rate of two-mode squeezed state generation is defined by the effective bandwidth of JPAs. In the case of conventional resonator-based JPAs, these bandwidths are typically of the order of about $10$\, MHz~\cite{Pogorzalek2017}. By exploiting more advanced designs based on traveling-wave Josephson parametric amplifiers, one can hope to increase these bandwidth to about$1$\,GHz. However, the price for this increase is typically lower squeezing levels and higher noise photon numbers.

When dealing with our lossy TMST states, we consider a two-photon-subtraction protocol that is applied right before the teleportation experiment, in order to prepare the entangled resource for enhanced performance. This means that we apply photon subtraction on states that have already traveled through open air. In the case of TMSV states, photon subtraction is beneficial for low squeezing, which translates into low entanglement. In the case of lossy TMST states, entanglement is affected by the initial squeezing, but also by the distance, since reducing it means reducing photon loses and the presence of thermal noise in the state. Consequently, better performance of the teleportation protocol using photon-subtracted entangled states as the resource occurs for small distances. The submatrices of the covariance matrix that characterizes the entangled resource, written as $\Sigma_{A} = \alpha\Id_{2}$, $\Sigma_{B} = \beta\Id_{2}$, and $\varepsilon_{AB} = \gamma\sigma_{z}$, are modified by a symmetric two-photon-subtraction process as follows

\begin{eqnarray}\label{2PS_submatrices}
\nonumber \tilde{\Sigma}_{A} &=& \left[1-2\tau\frac{(1-\alpha)(1+\beta)+\gamma^{2}+((1-\alpha)(1-\beta)-\gamma^{2})\tau}{(1+\alpha)(1+\beta)-\gamma^{2}+2(1-\alpha\beta+\gamma^{2})\tau+((1-\alpha)(1-\beta)-\gamma^{2})\tau^{2}}\right]\Id_{2}, \\
\nonumber \tilde{\Sigma}_{B} &=& \left[1-2\tau\frac{(1+\alpha)(1-\beta)+\gamma^{2}+((1-\alpha)(1-\beta)-\gamma^{2})\tau}{(1+\alpha)(1+\beta)-\gamma^{2}+2(1-\alpha\beta+\gamma^{2})\tau+((1-\alpha)(1-\beta)-\gamma^{2})\tau^{2}}\right]\Id_{2}, \\
\tilde{\varepsilon}_{AB} &=& \frac{4\tau\gamma}{(1+\alpha)(1+\beta)-\gamma^{2}+2(1-\alpha\beta+\gamma^{2})\tau+((1-\alpha)(1-\beta)-\gamma^{2})\tau^{2}}\sigma_{z},
\end{eqnarray}
whose success probability is given by
\begin{equation}\label{eq:success-prob}
P = 4(1-\tau)^{2}\frac{\left[1-\alpha\beta+\gamma^{2}+((1-\alpha)(1-\beta)-\gamma^{2})\tau\right]^{2}-(\alpha-\beta)^{2}+4\gamma^{2}}{\left[(1+\alpha)(1+\beta)-\gamma^{2}+2(1-\alpha\beta+\gamma^{2})\tau+((1-\alpha)(1-\beta)-\gamma^{2})\tau^{2}\right]^{3}}. 
\end{equation}

See Appendix~\ref{app_C} for the general expressions. In order for these submatrices to characterize a covariance matrix, they need to satisfy a positivity condition, as well as the uncertainty principle. Both these requirements can be summarized as
\begin{equation}
\left|\sqrt{\det\tilde{\Sigma}} - 1 \right| \geq \left| \tilde{\alpha}-\tilde{\beta}\right|,
\end{equation}
given that we have used $\tilde{\Sigma}_{A} = \tilde{\alpha}\Id_{2}$, $\tilde{\Sigma}_{B} = \tilde{\beta}\Id_{2}$. Unfortunately, the state resulting from photon subtraction is not Gaussian, and thus it is not completely characterized by the covariance matrix. In this case, we use the characteristic function to describe the photon-subtracted TMST states. The general expression for the characteristic function of a probabilistically two-photon-subtracted Gaussian state (where one photon has been subtracted in each mode) can be found in Appendix~\ref{app_C}. Furthermore, see Appendix~\ref{app_B} for an equivalent discussion regarding heuristic photon subtraction.

\subsubsection{Entanglement swapping}
\begin{figure}[H]
\centering
\includegraphics[width=0.6\textwidth]{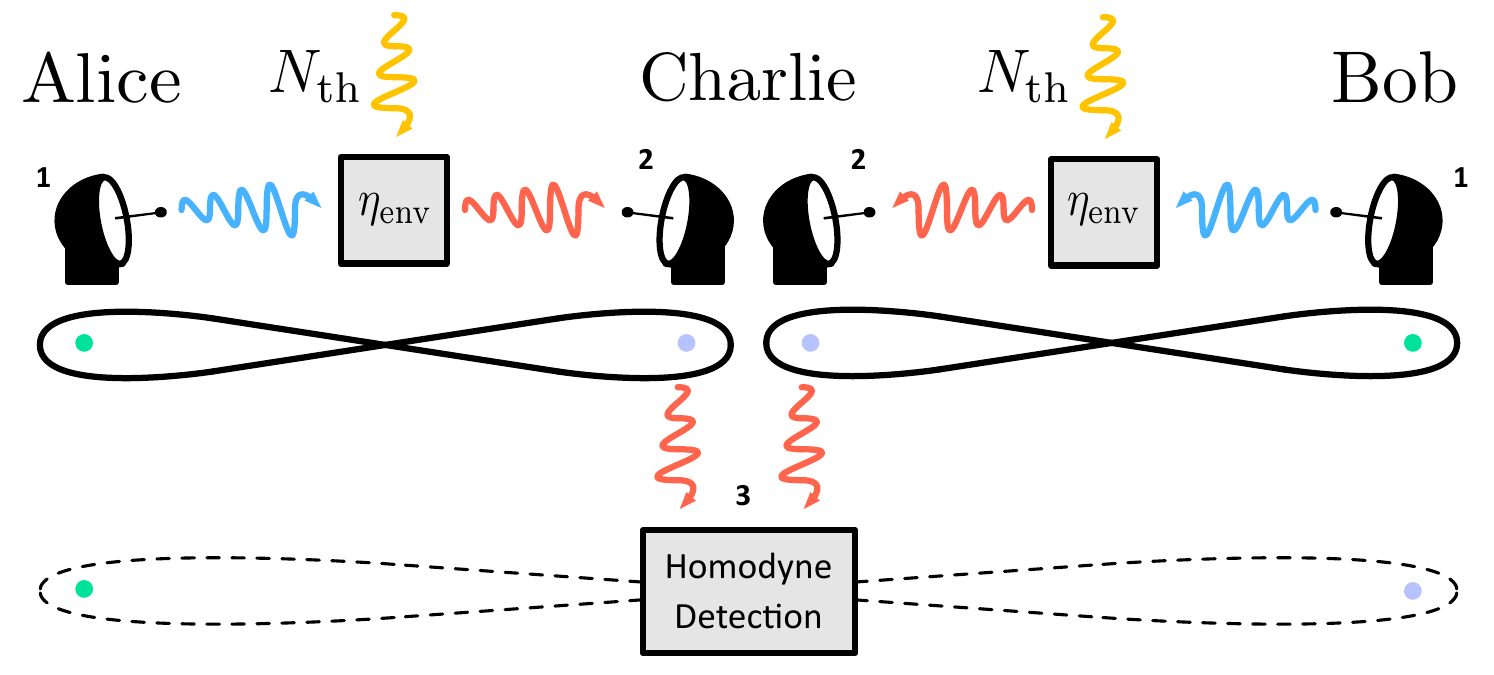}
\caption[Sketch of the optimal entanglement swapping scheme involving three parties, and three key steps.]{Sketch of the optimal entanglement swapping scheme involving three parties, and three key steps. First, Alice and Bob generate two-mode squeezed thermal states and, while keeping one of the modes each, send the others through open air, where they are subject to photon loss and get mixed with thermal noise. Second, Charlie receives and processes both modes, and third, he uses them to perform Homodyne detection. In the end, Charlie is able to transform the pairwisely entangled states he shares with Alice and with Bob independently into an entangled state held solely between Alice and Bob. \textit{Figure credit: T. G.-R.}}
\label{fig7}
\end{figure}

In this section we contemplate the CV version of entanglement swapping~\cite{Hoelscher-Obermaier2011}, a procedure that can be used to reduce the distance that states have to travel through the environment, and hence attenuate the effects of entanglement degradation. We consider the case in which we have two entangled states, shared by three parties pairwisely. That is, between Alice and Charlie, and between Charlie and Bob. Entanglement swapping is a technique that allows for the conversion of two bipartite entangled states into a single one shared by initially unconnected parties. By performing measurements in a maximally entangled basis, Charlie is able to transform the entangled resources he shares with Alice and with Bob into a single entangled state shared only by Alice and Bob. In the CV formalism, these measurements are described by Homodyne detection, and their effect on the state is computed as we have seen in the CV teleportation protocol. Consider that these states are Gaussian, with covariance matrices
\begin{eqnarray}
\nonumber \Sigma_{1} &=& \begin{pmatrix} \Sigma_{A} & \varepsilon_{AB} \\ \varepsilon_{AB}^{\intercal} & \Sigma_{B} \end{pmatrix}, \\
\Sigma_{2} &=& \begin{pmatrix} \Sigma_{C} & \varepsilon_{CD} \\ \varepsilon_{CD}^{\intercal} & \Sigma_{D} \end{pmatrix},
\end{eqnarray}
and null displacement vectors. Then, the covariance matrix of the remaining state,
\begin{equation}
\Sigma^{\text{ES}} = \begin{pmatrix} \tilde{\Sigma}_{A} & \tilde{\varepsilon}_{AD} \\ \tilde{\varepsilon}_{AD}^{\intercal} & \tilde{\Sigma}_{D} \end{pmatrix},
\end{equation}
conditioned by the measurement results is characterized by
\begin{eqnarray}
\nonumber \tilde{\Sigma}_{A} &=& \Sigma_{A} - \frac{\varepsilon_{AB}\Omega^{\intercal}\left(\Sigma_{B}+\sigma_{z}\Sigma_{C}\sigma_{z}\right)\Omega\varepsilon_{AB}^{\intercal}}{\det\left(\Sigma_{B}+\sigma_{z}\Sigma_{C}\sigma_{z}\right)}, \\
\nonumber \tilde{\Sigma}_{D} &=& \Sigma_{D} - \frac{\varepsilon_{CD}\Omega^{\intercal}\left(\Sigma_{C}+\sigma_{z}\Sigma_{B}\sigma_{z}\right)\Omega\varepsilon_{CD}^{\intercal}}{\det\left(\Sigma_{B}+\sigma_{z}\Sigma_{C}\sigma_{z}\right)}, \\
\tilde{\varepsilon}_{AD} &=& -\frac{\varepsilon_{AB}\Omega^{\intercal}\left(\Sigma_{B}\sigma_{z}+\sigma_{z}\Sigma_{C}\right)\Omega\varepsilon_{CD}^{\intercal}}{\det\left(\Sigma_{B}+\sigma_{z}\Sigma_{C}\sigma_{z}\right)}.
\end{eqnarray}
We observe that, in the setup we are considering, the only protocol that presents an improvement in negativity with respect to the bare states is that in which Alice and Bob generate the two-mode entangled states, and each send one of the modes to Charlie. This setup is represented in Fig.~\ref{fig7}. Then, the two modes used for entanglement swapping are those that have become mixed with environmental noise. Nevertheless, this enhancement occurs for large distances, which implies low negativities, and works significantly better in low-temperature environments, where $N_{\text{th}}$ is reduced. Considering $\Sigma_{A} = \Sigma_{D} = \alpha\Id_{2}$, $\Sigma_{B} = \Sigma_{C} = \beta\Id_{2}$, and $\varepsilon_{AB} = \varepsilon_{CD} = \gamma\sigma_{z}$, then we can characterize the covariance matrix of the resulting resource by
\begin{eqnarray}\label{ES_submatrices}
\nonumber \tilde{\Sigma}_{A} &=& \left( \alpha - \frac{\gamma^{2}}{2\beta}\right)\Id_{2}, \\
\nonumber \tilde{\Sigma}_{D} &=& \left( \alpha - \frac{\gamma^{2}}{2\beta}\right)\Id_{2}, \\
\tilde{\varepsilon}_{AD} &=& \frac{\gamma^{2}}{2\beta}\sigma_{z}.
\end{eqnarray}
The condition for this characterization to be appropriate is given by
\begin{equation}
\left|\sqrt{\det\Sigma_{1}}-\frac{\beta}{\alpha}\right| \geq 0.
\end{equation}
See Appendix~\ref{app_D} for further discussion. 

\section{Experimental limitations to photocounting and homodyning with microwaves}\label{section:homodyning_and_photocounting}
In this section, we review current advances on photocounting and homodyne detection techniques with microwave quantum technologies. These techniques are vital for photon subtraction, as well as for entanglement swapping and quantum teleportation, which are the processes described in this manuscript. 

\subsection{Photodetection}
Traditionally, the problem of detecting microwaves has been the low energy of the signals when compared to the optical regime. Any of the entanglement distillation protocols we have discussed will require some kind of photodetection scheme. In particular, for photon subtraction, a photocounter for microwave photons is required. In the current landscape of microwave quantum technologies, there have been recent proposals for nondemolition detection of itinerant single microwave photons~\cite{Kono2018,Besse2018,Lescanne2020} in circuit-QED setups, with detection efficiencies ranging from $58\%$ to $84\%$. Based on similar setups, a photocounter has been proposed~\cite{Dassonneville2020} that can detect up to three microwave photons. 

This device is able to catch an incoming wavepacket in a buffer resonator, which is then transferred into the memory by means of pumping a Josephson ring modulator. Then, the information about the number of photons in the memory is transferred to a transmon qubit, which is coupled to the memory modes, and from there it is read bit by bit. Consequently, this photocounter requires previous knowledge on the waveform and the arrival time of the incoming mode to be detected. Furthermore, this device is not characterized by a single quantum efficiency; rather, the detection efficiency varies depending on the number of photons. That is, $99\%$ for zero photons, $76\%\pm3\%$ for a single photon, $71\%\pm3\%$ for two photons and $54\%\pm2\%$ for three, assuming a dark count probability of $3\%\pm0.2\%$ and a dead time of $4.5\,\mu$s.


\subsection{Homodyne detection}
Homodyne detection allows one to extract information about a single quadrature. It can be used to perform CV-Bell measurements, i.e., a projective measurement in a maximally-quadrature-entangled basis for CV states. One way to perform Bell measurements with propagating CV states is to use the analog feedforward technique, as demonstrated in Ref.~\cite{Fedorov2021}. This approach requires operating two additional phase-sensitive amplifiers in combination with two hybrid rings and a directional coupler, which effectively implements a projection operation for conjugate quadratures of propagating electromagnetic fields. An alternative, more conventional approach can be implemented by adapting microwave single-photon detectors to the well-known optics homodyning techniques.

As we have seen, entanglement swapping provides an advantage if this measurement scheme is used without averaging over the results (single-shot homodyning), whereas the Braunstein-Kimble quantum teleportation protocol assumes that this average is performed, given an unknown coherent state. In theory, single-shot homodyning can be implemented by using quantum-limited superconducting amplifiers and standard demodulation techniques~\cite{Eichler2012}. However, some fundamental aspects of the ``projectiveness'' of this operation and its importance for the Bell detection measurements or for photon subtraction are still unclear and must be verified.

See Appendix~\ref{app_E} for more involved descriptions of errors associated with these measurement techniques.

\section{Open-Air Microwave Quantum Teleportation Fidelities}\label{section:teleportation}

In this section, we compute the average teleportation fidelity for different resource states. In all cases, the teleported state is a coherent state $\ket{\alpha_0}\bra{\alpha_0}$. 

\subsection{Two-mode squeezed vacuum resource}
The case of a TMSV state is particularly simple, as we can simply plug its covariance matrix into Eq.~\eqref{eq:Gaussian-fidelity},
\begin{equation}
\overline{F}_{{\text{TMSV}}} = \frac{1+\lambda}{2}.
\end{equation}
When symmetric $2k$-photon subtraction is performed, the formula for Gaussian average fidelity can no longer be invoked. The results for $k=1,2$ (2PS and 4PS, respectively) are:
\begin{eqnarray}
\overline{F}_{\text{2PS}} &=& \left(1-\lambda_{\tau} + \frac{\lambda_{\tau}^{2}}{2}\right)\frac{(1+\lambda_{\tau})^{3}}{2(1+\lambda_{\tau}^{2})}, \\
\nonumber \overline{F}_{\text{4PS}} &=& \frac{(1+\lambda_{\tau} )^{5}\left[ 8-\lambda_{\tau}(2-\lambda_{\tau})(8-3\lambda_{\tau}(2-\lambda_{\tau})) \right]}{16(1+4\lambda_{\tau}^2+\lambda_{\tau}^4)},
\end{eqnarray}
with $\lambda_{\tau}=\lambda\tau$.
\begin{figure}[H]
\centering
\includegraphics[width=0.6 \textwidth]{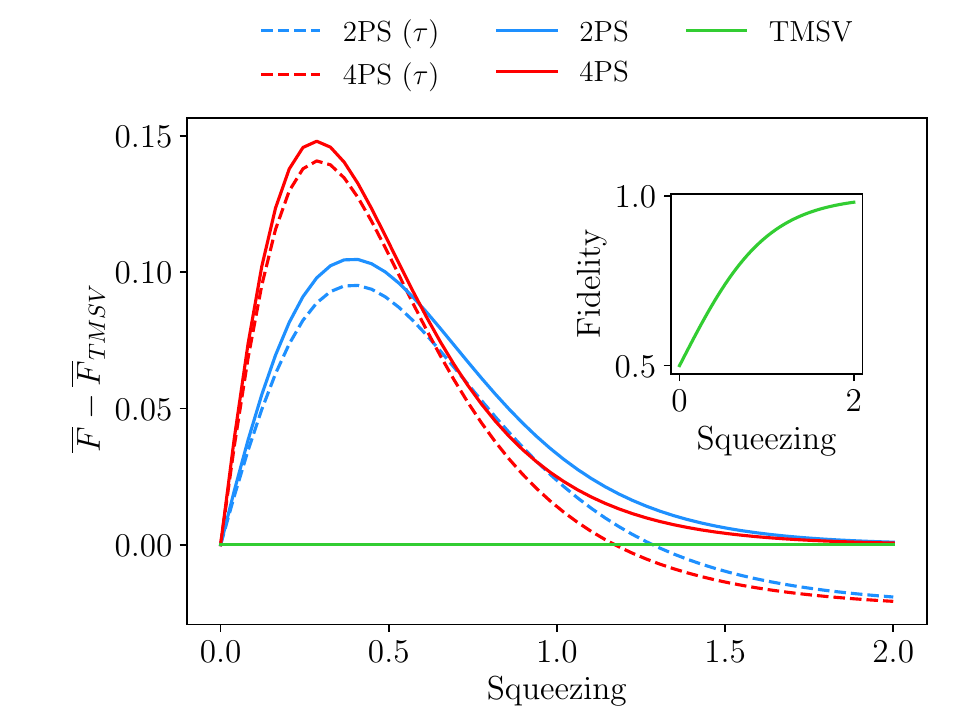}
\caption[Average fidelity of CV quantum teleportation of an unknown coherent state with respect to the initial squeezing parameter.]{Average fidelity of CV quantum teleportation of an unknown coherent state with respect to the initial squeezing parameter. We subtract the average fidelity associated with a TMSV state resource from the average fidelities of two-photon-subtracted (2PS, blue) and four-photon-subtracted (4PS, red) TMSV states. Curves associated with probabilistic photon subtraction appear dashed, whereas the solid curves are associated with heuristic photon subtraction. The green curve describes the TMSV case, which delimits the no-gain line, above which any point represents an improvement in fidelity due to photon subtraction. In the inset, we plot the average fidelity associated with a TMSV state against the initial squeezing parameter. We have considered the transmissivity of the beam splitters involved in probabilistic photon subtraction to be $\tau=0.95$. \textit{Figure credit: T. G.-R.}}
\label{fig8}
\end{figure}

In Fig.~\ref{fig8}, we represent the result of subtracting the fidelity associated with the bare TMSV state to those associated with two-photon-subtracted (2PS, blue) and four-photon-subtracted (4PS, red) TMSV states. Fidelity differences associated with heuristic photon subtraction appear as solid lines, whereas those associated with probabilistic photon subtraction appear dashed. The green solid line represents the no-gain line, above which any photon-subtracted state presents an advantage in fidelity. Note that photon subtraction works better for low squeezing, and as we increase it, we see that using the TMSV state as a resource for teleportation renders a higher fidelity than probabilistic photon subtraction, while heuristic photon subtraction tends to the TMSV result. 

\subsection{Two-mode squeezed thermal resource}
We now study the teleportation fidelity associated with a two-mode squeezed thermal state, sent through a lossy and noisy channel defined by the combination of the antenna and an environment with $N_\text{th}$ photons. By defining $\Sigma_{A}=\alpha\Id_{2}$, $\Sigma_{B}=\beta\Id_{2}$, and $\varepsilon_{AB}=\gamma\sigma_{z}$, we can write the average fidelity as
\begin{equation}
\overline{F}_{\text{TMST}} = \frac{1}{1+\frac{1}{2}\left(\alpha+\beta-2\gamma\right)}.
\end{equation}
If we consider the composition of $k$ teleportation protocols where each of the parties involved is separated by $L/k$, with $L$ the total distance aimed to cover. The final average fidelity is then given by
\begin{equation}
\overline{F}^{(k)}_{\text{TMST}} = \frac{1}{1+\left(k-\frac{1}{2}\right)\left(\alpha+\beta-2\gamma\right)},
\end{equation}
such that $\overline{F}_{\text{TMST}}>\overline{F}^{(k)}_{\text{TMST}}$ for $k>1$. Since the composition of teleportation protocols does not improve the overall fidelity, we study entanglement distillation and entanglement swapping in search for such gain. However, this fidelity composition may improve the overall fidelity when diffraction effects at the termination of the antenna come into play, which will reduce the reach of entanglement from the hundreds to the tens of meters.  

In Table~\ref{table1} we present the parameters we use to represent the different fidelity curves in this section.
\begin{table}
\centering
\begin{tabular}{|l|c|c|}
\hline
Parameter &Symbol &Value\\
\hline
Losses per unit of length  &$\mu$ & $1.44\times 10^{-6} \text{m}^{-1}$\\
Atmospheric temperature & $T$ &  300 K\\
Mean photon number & $N_\text{th}$ & 1250\\
Squeezing parameter & $r$ &  1\\
Thermal photon number (signal) & $n$ & $10^{-2}$ \\
Transmission coefficient & $\tau$ &  0.95\\
Antenna reflectivity & $\eta_{\text{ant}}$ & 0\\
\hline
\end{tabular}
\caption{Parameters for a terrestrial (1 atm of pressure, temperature of 300 K)  two-mode squeezed thermal state generated at a 50 mK cryostat, for a frequency of 5 GHz. These parameter values correspond to an Earth-based quantum teleportation scenario.}
\label{table1}
\end{table}

\subsubsection{Asymmetric case}
Assume that Alice generates a TMST state and sends one of the modes to Bob. Then, the covariance matrix of the state, given in Eq.~\eqref{CM_asym}, is characterized by
\begin{eqnarray}
\nonumber \alpha &=& (1+2N_{\text{th}})\eta_{\text{eff}}  + (1+2n)(1-\eta_{\text{eff}})\cosh 2r, \\
\nonumber \beta &=& (1+2n)\cosh 2r, \\
\gamma &=& (1+2n)\sqrt{1-\eta_{\text{eff}}}\sinh 2r,
\end{eqnarray}
which results in an average fidelity
\begin{eqnarray}
\nonumber \overline{F}_{\text{TMST}} &=& \bigg[ 1 + \left(\frac{1}{2}+N_{\text{th}}\right)\eta_{\text{eff}}\\
&+&\left(\frac{1}{2}+n\right)(2-\eta_{\text{eff}})\cosh2r \\
&-& (1+2n)\sqrt{1-\eta_{\text{eff}}}\sinh2r \bigg]^{-1},
\end{eqnarray}
with $\eta_{\text{eff}} = 1-e^{-\mu L}(1-\eta_{\text{ant}})$. 

\subsubsection{Symmetric case}
In this case, we consider that the resource state is generated at an intermediate point between Alice and Bob, and is sent to both of them, such that now both modes are affected by the lossy and noisy channel described above. The covariance matrix of this state, presented in Eq.~\eqref{CM_sym}, is characterized by
\begin{eqnarray}
\nonumber \alpha &=& (1+2N_{\text{th}})\eta_{\text{eff}}  + (1+2n)(1-\eta_{\text{eff}})\cosh 2r, \\
\nonumber \beta &=& (1+2N_{\text{th}})\eta_{\text{eff}}  + (1+2n)(1-\eta_{\text{eff}})\cosh 2r, \\
\gamma &=& (1+2n)(1-\eta_{\text{eff}})\sinh 2r
\end{eqnarray}
where we have assumed $L_{1}=L_{2}=L/2$, and thus $\eta^{(1)}_{\text{eff}} = \eta^{(2)}_{\text{eff}} = \eta_{\text{eff}} = 1-e^{-\mu\frac{L}{2}}(1-\eta_{\text{ant}})$ . Then, the average fidelity can be written as
\begin{eqnarray}
\nonumber \overline{F}_{\text{TMST}} &=& \bigg[ 1 + (1+2N_{\text{th}})\eta_{\text{eff}}+(1+2n)(1-\eta_{\text{eff}})\cosh2r \\
&-& (1+2n)(1-\eta_{\text{eff}})\sinh2r \bigg]^{-1}.
\end{eqnarray}
Note that, for short distances, the fidelities associated with the asymmetric and symmetric states coincide. That is, at first order in $\mu L\ll 1$, and with $\eta_{\text{ant}}=0$, 
\begin{eqnarray}
&&\overline{F}_{\text{TMST}} \approx \\
\nonumber && \left[ 1 + (1+2N_{\text{th}})\frac{\mu L}{2}+(1+2n)\left(1-\frac{\mu L}{2}\right)e^{-2r}\right]^{-1}.
\end{eqnarray}
When considering a lossy antenna, we observe higher entanglement degradation in the symmetric state due to the fact that both modes of the state are output by an antenna, whereas only one mode of the asymmetric state goes through it. Although $\sqrt{\eta_{\text{ant}}}$ can theoretically be reduced below $10^{-9}$~\cite{GonzalezRaya2020}, this leads to a slightly lower fidelity in the case of the symmetric state. In the figures appearing in this section, however, we consider $\eta_{\text{ant}}=0$ for simplicity. 

\subsection{Fidelity with photon subtraction}
If we consider a symmetric two-photon-subtraction process, in which the desired resource has lost a single photon in each mode, the average fidelity becomes
\begin{eqnarray}
\nonumber \overline{F}_{\text{2PS}} &=& \frac{1}{4}\left[ 1+\tau\frac{-\alpha\beta + (1+\gamma)^{2} +((1-\alpha)(1-\beta)-\gamma^{2})\tau}{(1+\alpha)(1+\beta)-\gamma^{2}-(\alpha\beta-(1-\gamma)^{2})\tau}\right]^{3} \times \\
&& \left[1 + \frac{(1-\alpha\beta+\gamma^{2})^{2}-(\alpha-\beta)^{2}+4\gamma^{2}+4\gamma((1-\alpha)(1-\beta)-\gamma^{2})\tau}{(1-\alpha\beta+\gamma^{2}+((1-\alpha)(1-\beta)-\gamma^{2})\tau)^{2}-(\alpha-\beta)^{2}+4\gamma^{2}}\right].
\end{eqnarray}

\begin{figure}[H]
\centering
\includegraphics[width=0.7\textwidth]{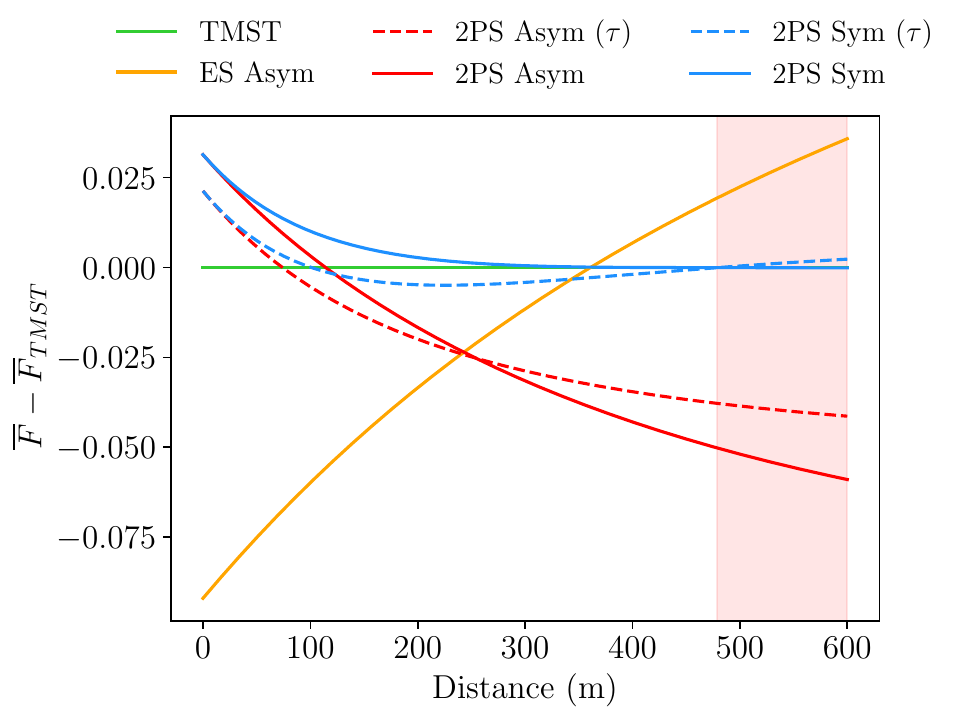}
\caption[Average fidelity of CV quantum teleportation of an unknown coherent state using an entangled resource distributed through open air, represented versus the traveled distance.]{Average fidelity of CV quantum teleportation of an unknown coherent state using an entangled resource distributed through open air, represented versus the traveled distance. We subtract the average fidelity associated with the TMST state distributed through open air (green), from the average fidelities of the two-photon-subtracted asymmetric (2PS asym) and symmetric (2PS sym) states, represented in red and blue, respectively, as well as from the average fidelity of the entanglement-swapped asymmetric (ES asym) state, in orange. We represent the states resulting from  probabilistic photon subtraction (dashed), as well as heuristic photon subtraction (solid). The pale red background represents the region where the fidelity is below the maximum classical fidelity of $1/2$, and the quantum advantage is lost. The green line then shows no gain, and any point above it corresponds to an improvement in fidelity. Parameters are $n=10^{-2}$, $N_{\text{th}}=1250$, $r=1$, $\mu=1.44\times^{-6}$ m$^{-1}$, $\eta_{\text{ant}}=0$, $\tau=0.95$. \textit{Figure credit: T. G.-R.}}
\label{fig9}
\end{figure}
\begin{figure}[H]
 \centering
\includegraphics[width=0.7 \textwidth]{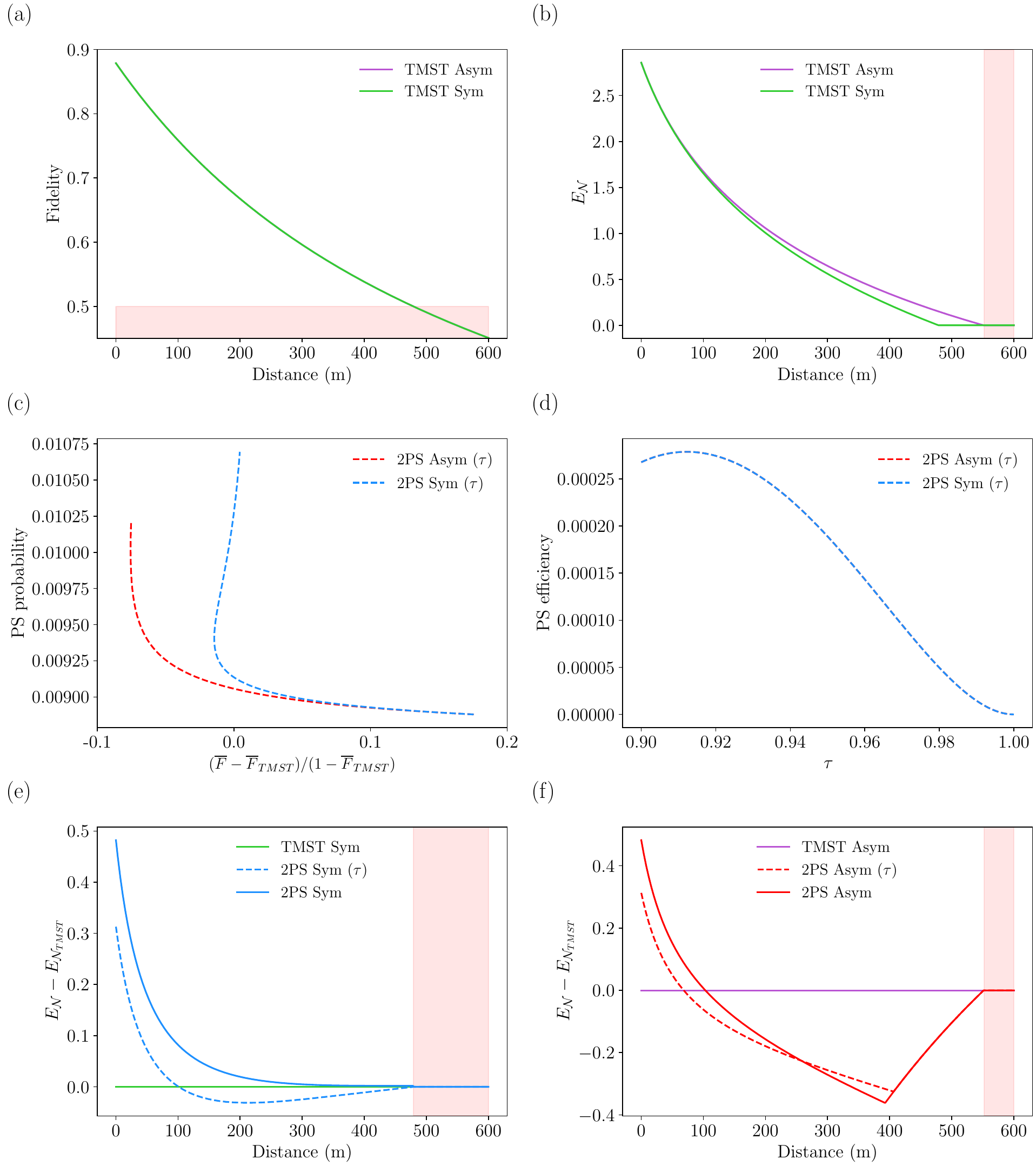}
\caption[Various features of the symmetric and asymmetric TMST states distributed through open air.]{Various features of the symmetric and asymmetric TMST states distributed through open air. (a) Average fidelity of the CV quantum teleportation protocol of an unknown coherent state using either symmetric or asymmetric lossy TMST states, represented against the traveled distance. (b) Logarithmic negativities $E_{\mathcal{N}}=\log_{2}(2\mathcal{N}+1)$ of the lossy TMST symmetric (green) and asymmetric (purple) states. (c) Probability of successful two-photon subtraction applied to lossy TMST symmetric (blue dashed) and asymmetric (red dashed) states, represented against the fidelity gain compared to the lossy TMST state, which is weighted to be larger for larger fidelities. (d) Efficiency of two-photon subtraction against the transmissivity, with $\tau\in[0.9,1]$, applied to TMST symmetric (blue dashed) and asymmetric (red dashed) states, at $x=0$. In panels (e) and (f), we represent the logarithmic negativity $E_{\mathcal{N}}=\log_{2}(2\mathcal{N}+1)$ of entangled resources distributed through open air, represented against the traveled distance. (e) We subtract the logarithmic negativity of the lossy TMST symmetric (TMST sym, in green) state from those of the probabilistic (dashed) and heuristic (solid) re-Gaussified two-photon-subtracted symmetric states (2PS sym, in blue). (f) We subtract the logarithmic negativity of the lossy TMST asymmetric (TMST asym, in purple) state from those of the probabilistic (dashed) and heuristic (solid) re-Gaussified two-photon-subtracted asymmetric states (2PS asym, in red). Any point above the green (purple) line represents an improvement in negativity for the re-Gaussified photon-subtracted symmetric (asymmetric) states. Parameters are $n=10^{-2}$, $N_{\text{th}}=1250$, $r=1$, $\mu=1.44\times^{-6}$ m$^{-1}$, $\eta_{\text{ant}}=0$, $\tau=0.95$. \textit{Figure credit: T. G.-R.}}
\label{fig10}
\end{figure}
In Fig.~\ref{fig9} we represent the difference in fidelities associated with a CV open-air quantum teleportation protocols for an unknown coherent state, using two-mode squeezed thermal states distributed through open air as a resource, against the traveled distance. We subtract the fidelity associated with the bare resource (TMST) to those related to two-photon-subtracted symmetric (blue) and asymmetric (red) states, as well as entanglement-swapped (orange) states. We consider both heuristic (solid lines) and probabilistic (dashed lines, labeled $\tau$) photon subtraction. In Fig.~\ref{fig10}~(a), we can see the fidelity associated with the bare resource, knowing that it coincides for the symmetric and asymmetric states in the region $\mu L\ll1$. The solid green line represents the no-gain line, above which any point represents an improvement in fidelity over the bare state. The former gives an enhancement for short distances, whereas the latter helps extend the point where the classical limit is reached. One of the reasons the gain related to photon subtraction is lost might be the increase of thermal photons in the state, which occurs for increasing $L$. This happens because, as photon losses are more relevant, the cost of doing photon subtraction is higher: if we subtract thermal photons, the entanglement hardly increases, whereas if we subtract photons from the signal, entanglement decreases. Using a pale red background, we represent the region in which the fidelity associated with the bare resource reaches the maximum classical value of $1/2$.

In Fig.~\ref{fig10}, we represent various features of the two-mode squeezed thermal states distributed through open air: (a) average fidelity, which coincides for the symmetric and asymmetric states for $\mu L\ll 1$; (b) logarithmic negativity $E_{\mathcal{N}}=\log_{2}(2\mathcal{N}+1)$ of the symmetric (green) and asymmetric (purple) states; (c) success probability of photon subtraction (see Eq.~\eqref{eq:success-prob}) for symmetric (blue, dashed) and asymmetric (red, dashed) states, against $(\overline{F}_{\text{2PS}}-\overline{F}_{\text{TMST}})/(1-\overline{F}_{\text{TMST}})$, which represents the gain in fidelity of the photon-subtraction schemes, weighted to show larger values when the gain occurs at larger fidelities; (d) efficiency of photon subtraction at $x=0$, computed as $P(\overline{F}_{\text{2PS}}-\overline{F}_{\text{TMST}})$, against different values of the transmissivity, with $\tau\in[0.9,1]$. Note that greater fidelity gains come at lower success probabilities for photon subtraction, which can be reflected in the efficiency (of the order of $10^{-4}$). The latter achieves maximum values for a transmissivity of $\tau\approx 0.92$, and goes to zero with the probability, as $\tau$ goes to 1.

In an attempt to explain the crossing that occurs between the probabilistically-photon-subtracted and bare fidelities, which delimits the region in which photon subtraction results in an enhanced teleportation fidelity, we consider the following approach: we attempt to find the Gaussian state that is related to our non-Gaussian photon-subtracted state by the same teleportation fidelity. Essentially, we are looking to identify the photon-subtracted states with Gaussian resources in order to compute the negativities from their covariance matrices, and investigate what happens to entanglement at the points where fidelity with photon-subtracted states loses its advantage. A similar approach can be found in Appendix~\ref{app_B} for heuristic photon subtraction. First, know that the fidelity with two-photon subtraction can be written as
\begin{equation}\label{eq:fidelity_2PS}
\overline{F}_{\text{2PS}} = \frac{1+ g}{\sqrt{\det\left[\Id_{2}+\frac{1}{2}\tilde{\Gamma}\right]}},
\end{equation}
where $\tilde{\Gamma} \equiv \sigma_z \tilde{\Sigma}_{A}\sigma_z + \tilde{\Sigma}_{B} -\sigma_z \tilde{\varepsilon}_{AB}-\tilde{\varepsilon}_{AB}^\intercal\sigma_z$, and $\tilde{\Sigma}_{A}$, $\tilde{\Sigma}_{B}$, and $\tilde{\varepsilon}_{AB}$ are defined in Eq.~\eqref{2PS_submatrices}. Here, $g$ is the result of integrating all the non-Gaussian corrections to the characteristic function, which enforces the non-Gaussianity of the state resulting from photon subtraction (see Appendix~\ref{app_C} for the general expression). We split the terms in the previous equation and write

\begin{eqnarray*}
\nonumber \frac{1}{\sqrt{\det\left[\Id_{2}+\frac{1}{2}\tilde{\Gamma}\right]}} &=& \frac{1}{2}\left[ 1+\tau\frac{-\alpha\beta + (1+\gamma)^{2} +((1-\alpha)(1-\beta)-\gamma^{2})\tau}{(1+\alpha)(1+\beta)-\gamma^{2}-(\alpha\beta-(1-\gamma)^{2})\tau}\right], \\
1+g &=& \frac{1}{2}\left[ 1+\tau\frac{-\alpha\beta + (1+\gamma)^{2} +((1-\alpha)(1-\beta)-\gamma^{2})\tau}{(1+\alpha)(1+\beta)-\gamma^{2}-(\alpha\beta-(1-\gamma)^{2})\tau}\right]^{2} \times \\
&& \left[1 + \frac{(1-\alpha\beta+\gamma^{2})^{2}-(\alpha-\beta)^{2}+4\gamma^{2}+4\gamma((1-\alpha)(1-\beta)-\gamma^{2})\tau}{(1-\alpha\beta+\gamma^{2}+((1-\alpha)(1-\beta)-\gamma^{2})\tau)^{2}-(\alpha-\beta)^{2}+4\gamma^{2}}\right].
\end{eqnarray*}

If we define a matrix $G = (1+g)\Id_{2}$ with $G^{-1} = \frac{1}{1+g}\Id_{2}$, then we can write $1+g = \sqrt{\det G}$, which leads to 
\begin{equation}
\frac{1+ g}{\sqrt{\det\left[\Id_{2}+\frac{1}{2}\tilde{\Gamma}\right]}} = \frac{1}{\sqrt{\det\left[\left(\Id_{2}+\frac{1}{2}\tilde{\Gamma}\right)G^{-1}\right]}}.
\end{equation}
By rearranging the terms resulting from the matrix product, we can obtain
\begin{equation}
\left(\Id_{2}+\frac{1}{2}\tilde{\Gamma}\right)G^{-1} = \Id_{2} + \frac{1}{2}\left(\frac{\tilde{\Gamma} - 2g\Id_{2}}{1+g}\right) \equiv \Id_{2} + \frac{1}{2}\tilde{\tilde{\Gamma}},
\end{equation}
where we have defined $\tilde{\tilde{\Gamma}} = \frac{\tilde{\Gamma} - 2g\Id_{2}}{1+g}$. Now, we want to incorporate the non-Gaussian corrections into the covariance matrix of the effective Gaussian state by using the formula
\begin{equation}
\tilde{\tilde{\Gamma}} = \sigma_{z}\tilde{\tilde{\Sigma}}_{A}\sigma_{z} + \tilde{\tilde{\Sigma}}_{B} - \sigma_{z}\tilde{\tilde{\varepsilon}}_{AB} - \tilde{\tilde{\varepsilon}}_{AB}^{\intercal}\sigma_{z}.
\end{equation}
We refer to the resulting state as the ``re-Gaussified'' state. Then, we define
\begin{eqnarray}\label{2PS_sym_eff_submatrices}
\nonumber \tilde{\tilde{\Sigma}}_{A} &=& \frac{1}{1+g}\left(\tilde{\Sigma}_{A} - g\Id_{2}\right), \\
\nonumber \tilde{\tilde{\Sigma}}_{B} &=& \frac{1}{1+g}\left(\tilde{\Sigma}_{B} - g\Id_{2}\right), \\
\tilde{\tilde{\varepsilon}}_{AB} &=& \frac{1}{1+g}\tilde{\varepsilon}_{AB}.
\end{eqnarray}
These represent the submatrices of a covariance matrix $\tilde{\tilde{\Sigma}}$ if
\begin{equation}
\left|\sqrt{\det\tilde{\Sigma}} - g(2+\tilde{\alpha}+\tilde{\beta}) -1\right| \geq (1+g)\left| \tilde{\alpha} - \tilde{\beta}\right|
\end{equation}
is satisfied. This condition both ensures the positivity of the covariance matrix and that the uncertainty relation is satisfied. For this, we have assumed that $\tilde{\Sigma}_{A}=\tilde{\alpha}\Id_{2}$, $\tilde{\Sigma}_{B}=\tilde{\beta}\Id_{2}$, and $\tilde{\varepsilon}_{AB} = \tilde{\gamma}\sigma_{z}$. The problem is that this convention only works for the symmetric state, and not for the asymmetric one. For the latter, we write
\begin{eqnarray}
\nonumber \tilde{\tilde{\Sigma}}_{A} &=& \frac{1}{1+g}\left(\tilde{\Sigma}_{A} - kg\Id_{2}\right), \\
\nonumber \tilde{\tilde{\Sigma}}_{B} &=& \frac{1}{1+g}\left(\tilde{\Sigma}_{B} - (2-k)g\Id_{2}\right), \\
\tilde{\tilde{\varepsilon}}_{AB} &=& \frac{1}{1+g}\tilde{\varepsilon}_{AB}.
\end{eqnarray}
Since we have seen that a symmetric re-Gaussified state is viable, we impose the same balanced partition on the re-Gaussification of the asymmetric state. From $\tilde{\tilde{\Sigma}}_{A}=\tilde{\tilde{\Sigma}}_{B}$, we obtain $k = 1+(\tilde{\alpha}-\tilde{\beta})/2g$, which leads to the submatrices
\begin{eqnarray}\label{2PS_asym_eff_submatrices}
\nonumber \tilde{\tilde{\Sigma}}_{A} &=& \frac{1}{1+g}\left(\frac{\tilde{\Sigma}_{A}+\tilde{\Sigma}_{B}}{2} - g\Id_{2}\right), \\
\nonumber \tilde{\tilde{\Sigma}}_{B} &=& \frac{1}{1+g}\left(\frac{\tilde{\Sigma}_{A}+\tilde{\Sigma}_{B}}{2} - g\Id_{2}\right), \\
\tilde{\tilde{\varepsilon}}_{AB} &=& \frac{1}{1+g}\tilde{\varepsilon}_{AB}.
\end{eqnarray}
The condition these terms need to satisfy is
\begin{equation}
\left|\sqrt{\det\tilde{\Sigma}} + \frac{1}{4}(\tilde{\alpha}-\tilde{\beta})^{2} -g(\tilde{\alpha}+\tilde{\beta}) + g^{2} - 1 \right| \geq 0,
\end{equation}
which is naturally met. In Appendix~\ref{app_D}, a graphical proof that these conditions are met is provided.

As a result of these redefinitions, we effectively mask the non-Gaussian corrections in the expression of the fidelity as further corrections to the submatrices of the covariance matrix of an entangled resource, which is now Gaussian, while maintaining the same fidelity we obtained with the photon-subtracted states. This treatment has shown that we are using a resource that, in the regions in which photon subtraction is beneficial, shows higher entanglement than the bare resource, as expected. In Fig.~\ref{fig10}~(e) and Fig.~\ref{fig10}~(f), we subtract the logarithmic negativity $E_{\mathcal{N}}=\log_{2}(2\mathcal{N}+1)$ of the bare resource (TMST) from those of the heuristic (solid) and the probabilistic (dashed) two-photon-subtracted states. In Fig.~\ref{fig10}~(e), we display the symmetric states, and in Fig.~\ref{fig10}~(f), the asymmetric ones. Note that the gain in negativity is lost around the same points as the gain in fidelity. As discussed before, the fidelities corresponding to the symmetric and asymmetric states are equal at first order in $\mu L\ll 1$, and the same behavior can be observed initially in the negativities of both states (see Fig.~\ref{fig10}~(b)). However, while the points at which the fidelities of the symmetric and asymmetric states reach the classical limit differs by centimeters, the points at which entanglement is lost for these states differ by tens of meters. This region where negativity is lost is highlighted with a pale red background. Although the entanglement in the asymmetric state reaches further, the symmetric photon-subtraction protocol we envision works better when applied on the symmetric state. The logarithmic negativity of heuristic photon-subtracted states presents a 46\% increase with respect to the value for the bare state at $x=0$, while probabilistic photon-subtracted states only present an initial gain of 28\%.

\subsection{Fidelity with entanglement swapping}
We consider the case in which both Alice and Bob produce two-mode squeezed states, and each sends one mode to Charlie, who is equidistantly located from the two parties. Then, he performs entanglement swapping using the two modes he has received, which have been degraded by thermal noise and photon losses. If Alice and Bob use the remaining entangled resource they share for teleporting an unknown coherent state, the fidelity of the protocol will be given by
\begin{equation}
\overline{F}_{\text{es}} = \frac{1}{1+\alpha-\frac{\gamma^{2}}{\beta}}, 
\end{equation}
where now we have
\begin{eqnarray}
\nonumber \alpha &=& (1+2n)\cosh 2r, \\
\nonumber \beta &=& (1+2N_{\text{th}})\eta_{\text{eff}}  + (1+2n)(1-\eta_{\text{eff}})\cosh 2r, \\
\gamma &=& (1+2n)\sqrt{1-\eta_{\text{eff}}}\sinh 2r,
\end{eqnarray}
and $\eta_{\text{eff}} = 1-e^{-\mu L/2}(1-\eta_{\text{ant}})$, since the total distance has been reduced by half due to the presence of a third, equidistant party. 

This fidelity is represented as the orange curve in Fig.~\ref{fig9}, where it shows a gain in fidelity for large distances, right before the classical limit of $\overline{F}=0.5$ is reached. The extended distance represents 14\% of the maximum distance for the bare TMST state. This will be advantageous when the distance at which the classical limit occurs can be extended, for example in the case of quantum communication between satellites.  

\section{Inter-Satellite Quantum Communication Model}\label{section:satellites}
An important application of the protocols and the technology addressed in this manuscript is quantum communication between satellites \cite{Sidhu2021, Pirandola2021, Pirandola2021_2}. Given the security inherent to quantum-based communication protocols, many of the motivations for the use of submillimiter microwaves, i.e., frequencies in the range 30-300 GHz, which is a trend in classical communication between satellites orbiting low earth orbits (LEOs), fade away, and it seems reasonable to aim at maximizing the distances between linked satellites~\cite{Sanz2018}. We consider a greatly simplified model for free-space microwave communication, assuming unpolarized signals and hence ignoring the effects of scintillation and polarization rotation, among others. This means that whenever we discuss entanglement, it will be understood that we are talking about particle number entanglement. Polarization entanglement, even if perhaps more natural when considering the physics of antennae, is lost whenever the signal enters a coplanar waveguide, hence making it not a good candidate for quantum communication between one-dimensional superconducting chips. Moreover, we assume that the communication is done within the same altitude, i.e., that the two satellites are in similar orbits, which is typically the case when building satellite constellations. This means that the atmospheric absorption, if any, will remain constant during the time of flight of the signals. Additionally, we ignore Doppler effects caused by relative speeds between the orbits.

There are four main families of satellite orbits: GEO, HEO, MEO, and LEO, corresponding to geosynchronous, high, medium, and low earth orbits, respectively. It is customary to define LEOs as orbits with altitudes in the range 700-2000 km; MEOs would then range between 2000-35786 km; and HEOs in 35 786-$d_M/2$, where $d_M$ is the distance from the Earth to the Moon. The seemingly arbitrary altitude separating MEOs and HEOs is actually the average altitude for which the period equals one sidereal day (23 h 56 m 4 s), and this is precisely where GEOs sit. This altitude is more than 3 times the point at which the exosphere, the last layer of the atmosphere, is observed to fade. GEOs and HEOs are hence ``true'' free-space orbits, in the sense that there is hardly any gas, and temperature is dominated by the cosmic microwave background --which peaks at 2.7 K. The MEO region is the least populated one, since it is home to the Van Allen belts, which contain charged particles moving at relativistic speeds due to the magnetic field of the Earth, and that can destroy unshielded objects. LEOs, on the other hand, are `cheap' orbits, where most of the satellites orbiting our planet live. Their low altitudes simplify the problems arising from delays between earth-based stations and the satellites.

In this section we will be concerned only with two satellites orbiting either the same GEO or the same LEO, as a simple case study of expected losses and entanglement degradation. There are essentially two kinds of loss one must take into account: atmospheric loss and free-space path loss (FSPL). Total loss will then be simply given by
\begin{equation}
    L = L_\text{A} L_{\text{FSPL}}.
\end{equation}

Atmospheric absorption loss is caused by light-matter interactions. These strongly depend on the altitude of the orbits considered, among other parameters such as polarization, frequency, or weather conditions. Atmospheric loss can range from almost negligible (up in the exosphere and beyond) to very significant in the lower layers of the atmosphere, especially when water droplets and dust are present. Atmospheric loss has to be taken into account when considering the case of up- and downlinks, i.e., when linking a satellite with an earth-based station. However, for relatively high altitudes--that is, any altitude where there are satellites--absorption loss is so low in microwaves that it can be taken to vanish as a first approximation, so we set $L_\text{A} =1$.

FSPL is due to the inevitable spreading of a signal in three dimensions; they are often referred to as geometric losses. FSPL is maximal when there is no beam-constraining mechanism, such as a wave guide, or a set of focalizing lenses, i.e., when the signal spreads isotropically: $L_\text{FSPL}=\left({\lambda}/{4\pi d } \right)^{-2}$.

\begin{figure}
\includegraphics[width=0.95 \textwidth]{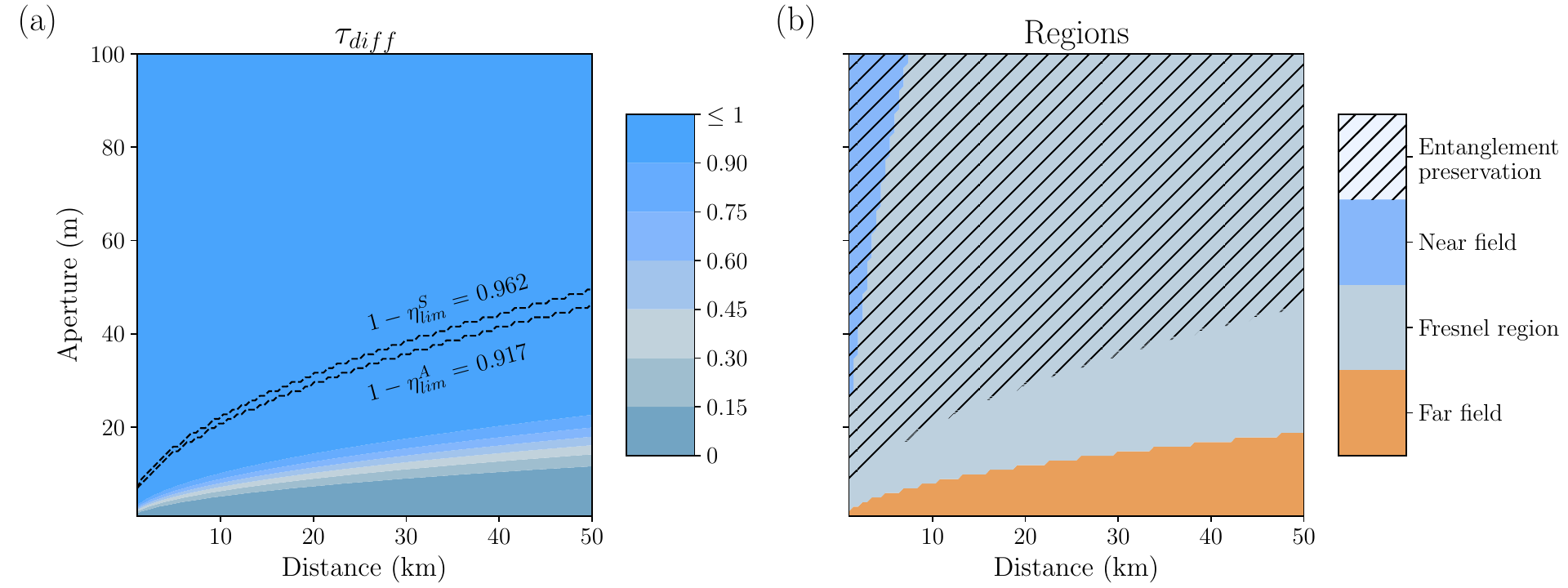}
\caption[Contour plot of the transmissivity associated with diffraction, $\breve{\tau}_\text{diff}$, against the aperture radius of the antenna and the traveled distance; and contour plot of the regions of free space as delimited by the relation between the aperture radius of the antenna and the distance at which the signal is observed]{(a) Contour plot of the transmissivity associated with diffraction, $\breve{\tau}_\text{diff}$, against the aperture radius of the antenna and the traveled distance. We can observe that losses are greatly reduced with the aperture of the antenna. (b) Contour plot of the regions of free space as delimited by the relation between the aperture radius of the antenna and the distance at which the signal is observed: near field (blue) $\varpi_{0} > (\lambda d/0.62)^{2/3}$, Fresnel (gray) $\sqrt{\lambda d/2} < \varpi_{0} < (\lambda d/0.62)^{2/3}$, and far field (orange) $\varpi_{0} < \sqrt{\lambda d/2}$. With dashed lines, we represent the region where entanglement can be preserved. Parameters are $\lambda = 6 \text{ cm}$, $a_{R}=2\varpi_{0}$. \textit{Figure credit: T. G.-R. \& \textbf{M. C.}}}
\label{fig11}
\end{figure}

Suppose that two comoving satellites are separated by a linear distance $d$, and that the emitter sends a quasimonochromatic signal with power $P_e$ centered at frequency $\nu=\omega/2\pi = c/\lambda$. The receiver gets a power $P_r$ such that their ratio defines a transmission coefficient that is the product of the loss and gains (or directivities) of the antennae. The resulting equation for long, `far-field' distances is sometimes referred to as Friis' equation \cite{Friis1946, Hogg1993, Shaw2013}, which is the compromise between gain (or directivity) and loss:
\begin{equation}
\frac{P_e}{P_r} =\frac{D_e D_d}{L_\text{A} L_{\text{FSPL}}}= D_e D_r  \left(\frac{\lambda}{4\pi d } \right)^2 \equiv\breve{\tau}_\text{path}.
\end{equation}
Here $D_e$ and $D_r$ are the directivities of the emitter and receiver antennas, and we set $L_\text{A}=1$ as discussed before. The directivity of an antenna is the maximized gain in power in some preferred direction with respect to a hypothetical isotropic antenna, at a fixed distance from the source, and assuming that the total radiation power is the same for both antennas: $D = \max_{\theta, \phi} D(\theta, \phi)$. It is a quantity that strongly depends on the geometry design, but that can be enhanced in a discrete fashion by means of antenna arrays. Indeed, given $N$ identical antennas with directivity gain $D(\theta, \phi)$, a phased array consists of an array of such antennas, each preceded by a controlled phase shifter. This diffraction problem essentially gives $D_\text{array}(\theta, \phi) = A^2_N(\bm{\varepsilon}) D(\theta, \phi)$, where $A_N$ is the so-called $N$-array factor that symbolically depends on the phases via some vector $\bm{\varepsilon}$~\cite{Balanis2005}. In three dimensions, phase arrays are two-dimensional grids of antennas, so that the main lobe of the resulting signal becomes as sharp as possible. We assume that we have an array of small coplanar antennas as discussed in Section~\ref{subsection:antenna_model}, adding up to a radiation pattern mimicking that of a parabolic antenna. We also assume that both emitter and receiver have the same design, $D_e = D_r \equiv D$ with
\begin{equation}
D=\left(\frac{\pi a}{\lambda}\right)^2 e_a
\end{equation}
where $0 \leq e_a \leq 1$ is the aperture efficiency, defined as the ratio between the effective aperture $A_e$, and the area of the antenna's actual aperture, $A_\text{phys}$, and $a$ is the diameter of the parabola, such that $A_{\text{phys}}=\pi a^{2}/4$. With this, the parabolic path transmissivity becomes
\begin{equation}
\breve{\tau}_\text{path} = \left( \frac{\pi a^2 e_a }{4 d \lambda}\right)^{2}.
\end{equation}
The effect of path losses can alternatively be described by a diffraction mechanism, affecting the spot size of the signal beam,
\begin{equation}
\varpi = \frac{\varpi_{0}}{\sqrt{2}}\sqrt{\left(1-\frac{d}{R_{0}}\right)^{2} + \left(\frac{d}{d_{R}} \right)^{2}},
\end{equation}
given an initial spot size $\varpi_{0}$, curvature of the beam $R_{0}$, and Rayleigh range $d_{R}=\pi\varpi_{0}^{2}\lambda^{-1}/2$. Given the aperture radius $a_{R}$ of the receiver antenna, the diffraction-induced transmissivity can be computed as~\cite{Pirandola2021,Pirandola2021_2}
\begin{equation}
\breve{\tau}_{\text{diff}} = 1 - e^{-2a_{R}^{2}/\varpi^{2}}.
\end{equation}
Note that, in the far-field approximation, we can recover the result for $\breve{\tau}_\text{path}$, 
\begin{equation}
\breve{\tau}_{\text{diff}} \approx \left( \frac{\pi\varpi_{0}a_{R}}{\lambda d}\right)^{2},
\end{equation}
by setting $a_{R}=\varpi_{0}=a/2$, $R_{0}=d$, and assuming that $e_{a}=1$. Setting $\lambda = 6 \text{ cm}$ and $a_{R}=2\varpi_{0}$, we plot the transmissivity associated with diffraction versus the distance $d$ for different values of the aperture $\varpi_{0}$ in Fig.~\ref{fig11}~(a), observing that losses are reduced as a result of an increase in the aperture.

We address entanglement preservation in TMST states distributed through open air by considering that the dominant source of error will be diffraction, as opposed to attenuation, which we describe by means of a beam splitter with a thermal input. We introduce $N_{\text{th}}\sim 11$ as the number of thermal photons in the environment at $2.7$ K. Considering this loss mechanism, entanglement preservation is achieved for reflectivities that satisfy $\eta<(1+N_{\text{th}})^{-1}\sim0.083$ for lossy TMST asymmetric states, and $\eta<[1+N_{\text{th}}(1+\coth r)]^{-1}\sim0.038$ for symmetric states, assuming that $n\approx 0$ and $\breve{\tau}=1-\eta$. Given this diffraction channel, entanglement is preserved in the regime $a_{R}\varpi_{0}/d > (\lambda/\pi)\sqrt{-\log\eta_{\text{lim}}} \sim 0.035$, for $\lambda=6$ cm and $\eta_{\text{lim}}=0.038$. This implies that, for two satellites that are separated by $d=1$ km, the product of the apertures of emitter and receiver antennae must be $a_{R}\varpi_{0} > 35 \text{ m}^{2}$ in order to have entanglement preservation. In Fig.~\ref{fig11}~(b), we represent the regions of free space as delimited by the relation between the distance at which the signal is detected and the aperture of the emitting antenna, taking $a_{R}=2\varpi_{0}$, and depicting the region in which entanglement is preserved with a dashed line. This shows that the radius of the antennae of emitter and receiver satellites will be large, as is usually the case for microwave communications. In order to correct the effects of diffraction with microwaves, it would also be useful to study focalizing techniques and the incorporation of beam collimators.

\section{Conclusions}
In this manuscript, we study the feasibility of microwave entanglement distribution in open air with two-mode squeezed states. We study these as a resource for the Braunstein-Kimble quantum-teleportation protocol adapted to microwave technology, reviewing the steps involved in this process and the possible experimental realization. First, we review the generation of two-mode squeezed states using JPAs that was experimentally demonstrated in Ref.~\cite{Fedorov2018}. Then, we look at an antenna model~\cite{GonzalezRaya2020} for optimal transmission of these states into open air, while also discussing the maximum possible reach of entanglement and its degradation due to interactions with the environment. The reach is found to extend up to 550 m for asymmetric states with ideal weather conditions. We adapt entanglement distillation and entanglement swapping to microwave technology to counteract degradation at different stages. In particular, we study photon subtraction, an entanglement distillation protocol that works for short distances and low squeezing, allowing up to a 46\% increase in the logarithmic negativity. Entanglement distillation, on the other hand, contributes to extending the reach of entanglement by up to 14\%. Since these operations require homodyne detection, as well as photocounting, we discuss recent advances in microwave technologies that permit these operations. We then test the efficiency of open-air entanglement distribution, including the different enhancement techniques, with the Braunstein-Kimble protocol adapted to microwaves: we compute the average fidelities of open-air microwave quantum teleportation of coherent states using the open-air-distributed states as the entangled resource. We conclude with a study of the applicability and efficiency of these techniques for quantum communication between satellites, a field where, with the proper directivity, the reach of entanglement can be greatly increased, given the low absorption rates. 

Efficient information retrieval from an open-air distribution of microwave quantum states is a key component of open-air quantum communications, which requires the design of a receiver antenna. The device achieving this target may resemble that in Section~\ref{subsection:antenna_model}, but it calls for a different type of termination into open air in order to, for instance, reduce diffraction losses. Since the lack of an amplification protocol considerably limits the entanglement transmission distance through open air, it seems necessary to develop a theory of quantum repeaters for microwave signals, following the ideas shown in Ref.~\cite{DiCandia2015}. To this end, entanglement distillation and entanglement swapping techniques discussed in this manuscript are useful. 

Since superconducting circuits naturally work in the microwave regime, it is desirable to explore realizations of photon subtraction that use devices specific to this technology. In particular, a possible deterministic photon-subtraction scheme can be studied, making use of circuit QED for nondemolition detection of itinerant microwave photons~\cite{Lescanne2020}. In this paper, the detection of a previously unknown microwave photon is triggered by a transmon qubit jumping to its excited state, indicating a successful photon-subtraction event.

\section{Appendix A: Amplification}\label{app_A}
In this manuscript, we have not discussed an amplification protocol for entangled signals because it cannot increase quantum correlations. Amplification of signals is an essential feature in classical microwave communication in open air, as increasing the number of photons that Alice sends will improve the chances Bob has of detecting that signal. In the quantum regime, cryogenic high electronic mobility transistor (HEMT) amplifiers are suited for experiments with microwaves. These are able to greatly enhance signals in a large frequency spectrum, while introducing a significant amount of thermal photons. This noise is reflected in the input-output relation 
\begin{equation}
a_{\text{out}}=\sqrt{g_H}a_{\text{in}}+\sqrt{g_H-1}h_H,
\end{equation}
with $a_{\text{in}}$, $a_{\text{out}}$, and $h_H$ the annihilation operators of the input field, output field and noise added by the amplifier, respectively. From this formula, we can see that amplification is a procedure that acts individually on the modes of a quantum state, which means that we can increment the number of photons of that mode, but they will not be entangled with the other ones. That is why we say that amplification cannot increase quantum correlations. If anything, the introduction of thermal noise can lead to entanglement degradation. 

HEMTs normally work at 4 K temperatures, which implies that the number of thermal photons they introduce is around $n_{H} \sim 10 - 20$, for 5 GHz frequencies. It is the number of thermal photons that determines the thermally-radiated power~\cite{Clerk2010}, $P = \hbar \omega N B$, meaning that the excess output noise produces a flux of $N$ photons per second in a bandwidth of $B$ Hz. The gain is given by the ratio between output and input powers, and for a constant bandwidth, it is just $g_{H}=n_{H}/n$, the ratio between the number of thermal photons introduced by the HEMT and the number of photons in the input state. Considering the antenna described above, a HEMT of these characteristics produces a gain of $g_{H}=2\times 10^{3}$ when acting on a TMST state with $n\sim 10^{-2}$, which completely destroys entanglement. In order for entanglement to survive such an amplification process, the HEMT must be placed at temperatures below 100 mK. In any case, they do not present an advantage. 

\section{Appendix B: Heuristic photon subtraction}\label{app_B}
In this section, we describe the heuristic photon-subtraction operation, in which annihilation operators are applied to a quantum state in order to reduce the number of photons in each mode. Consider that we initially have a TMSV state,
\begin{equation}
|\psi\rangle_{AB} = \sqrt{1-\lambda^{2}}\sum_{n=0}^{\infty}\lambda^{n}|n,n\rangle_{AB},
\end{equation}
with $\lambda=\tanh r$ and squeezing parameter $r$. Assume that we apply the operator $a_{A}^{k}a_{B}^{l}$, which implies subtracting $k$ photons in mode $A$ and $l$ photons in mode $B$. The resulting state is 
\begin{eqnarray}
&& |\psi^{(k+l)}\rangle_{AB} = \sqrt{1-\lambda^{2}}\times \\
\nonumber &&\sum_{n=\max\{k,l\}}^{\infty}\frac{\lambda^{n}n!}{\sqrt{(n-k)!(n-l)!}}|n-k,n-l\rangle_{AB}.
\end{eqnarray}
Considering symmetric photon subtraction ($k=l$), we can write this state as
\begin{equation}
|\psi^{(2k)}\rangle_{AB} = \sqrt{1-\lambda^{2}}\sum_{n=0}^{\infty}\frac{\lambda^{n+k}(n+k)!}{n!}|n,n\rangle_{AB},
\end{equation}
where we have shifted $n\rightarrow n+k$. This state needs to be normalized, and the normalization is given by
\begin{eqnarray}
N_{2k} &=& (1-\lambda^{2})\lambda^{2k}(k!)^{2}\sum_{n=0}^{\infty}\lambda^{2n}\begin{pmatrix}n+k \\ k\end{pmatrix}^{2} \\
\nonumber &=& (1-\lambda^{2})\lambda^{2k}(k!)^{2} \, _2F_1\left(k+1,k+1;1; \lambda^2\right).
\end{eqnarray}
The negativity of this state can be computed as
\begin{equation}
\mathcal{N}(\rho^{(2k)}) = \frac{A_{k}-1}{2},
\end{equation}
with $\rho^{(2k)}=|\psi^{(2k)}\rangle\langle\psi^{(2k)}|$ and
\begin{equation}
A_{k} \equiv \left( \sqrt{\frac{1-\lambda^{2}}{N_{2k}}}\sum_{n=0}^{\infty} \lambda^{n+k}\frac{(n+k)!}{n!}\right)^{2}.
\end{equation}
This leads to
\begin{equation}
\mathcal{N}(\rho^{(2k)}) = \frac{1}{2} \left(\frac{(1-\lambda)^{-2 (k+1)}}{\, _2F_1\left(k+1,k+1;1; \lambda^2\right)}-1\right)
\end{equation}
from which we can recover the negativity of the TMSV state by setting $k=0$,
\begin{equation}
\mathcal{N}(\rho^{(0)}) = \frac{\lambda}{1-\lambda} = \frac{e^{2r}-1}{2}.
\end{equation}

Now, we go beyond the TMSV state case, and explore the heuristic photon-subtraction protocol applied on a general Gaussian state. The application of the single-photon annihilation operators on both modes of a bipartite quantum state $\rho$ is equivalent to applying
\begin{equation}
\Theta_{1}\Theta_{2}\chi(\alpha,\beta)
\end{equation}
to its the characteristic function~\cite{Serafini2017}, with
\begin{equation}
\Theta_{i} = \partial_{x_{i}^{2}} + \partial_{p_{i}^{2}} + \frac{x_{i}^{2}}{4} + \frac{p_{i}^{2}}{4} + x_{i}\partial_{x_{i}} + p_{i}\partial_{p_{i}} + 1,
\end{equation}
for $i=\{1,2\}$. Given that $\rho = \frac{1}{\pi^{2}}\int \diff^{2}\alpha\int \diff^{2}\beta \chi(\alpha,\beta)D_{1}(-\alpha)D_{2}(-\beta)$, and assuming that $\rho$ is a Gaussian state with covariance matrix 
\begin{equation}
\Sigma = \begin{pmatrix} \Sigma_{A} & \varepsilon_{AB} \\ \varepsilon^{\intercal}_{AB} & \Sigma_{B}\end{pmatrix},
\end{equation}
then we can write

{\small
\begin{eqnarray}
&& \Theta_{1}\Theta_{2}\chi(\alpha,\beta) = N \Big[ \left( m_{B} + \vec{\beta} M_{B} \vec{\beta}^{\intercal} + \vec{\alpha}M_{BC}\vec{\beta}^{\intercal} + \vec{\alpha} M_{C} \vec{\alpha}^{\intercal} \right) \left( m_{A} + \vec{\alpha} M_{A} \vec{\alpha}^{\intercal} + \vec{\alpha}M_{AC}\vec{\beta}^{\intercal} + \vec{\beta} M_{C} \vec{\beta}^{\intercal} \right) \\
\nonumber && + m_{C} - \vec{\alpha}M_{AC}\Omega \varepsilon_{AB}\Omega^{\intercal}\vec{\alpha}^{\intercal} + 2\vec{\beta}M_{C}\left(\Id_{2} - \Omega \Sigma_{B}\Omega^{\intercal}\right)\vec{\beta}^{\intercal} + \vec{\alpha}\left[M_{AC}\left(\Id_{2}-\Omega \Sigma_{B}\Omega^{\intercal}\right) - 2\Omega\varepsilon_{AB}\Omega^{\intercal}M_{C}\right]\vec{\beta}^{\intercal} \Big] \chi(\alpha,\beta),
\end{eqnarray}
}
where we have defined 
\begin{eqnarray}
\nonumber m_{A} &=& 1 - \frac{1}{2}\tr \Sigma_{A}, \\
\nonumber m_{B} &=& 1 - \frac{1}{2}\tr \Sigma_{B}, \\
\nonumber m_{C} &=& \frac{1}{2}\tr \left(\varepsilon_{AB}^{\intercal}\varepsilon_{AB}\right), \\
\nonumber M_{A} &=& \frac{1}{4}\left( \Id_{2} -2\Omega\Sigma_{A}\Omega^{\intercal} + \Omega \Sigma_{A}^{2}\Omega^{\intercal}\right), \\
\nonumber M_{B} &=& \frac{1}{4}\left( \Id_{2} -2\Omega \Sigma_{B}\Omega^{\intercal} + \Omega \Sigma_{B}^{2}\Omega^{\intercal}\right), \\
\nonumber M_{C} &=& \frac{1}{4}\Omega \varepsilon_{AB}^{\intercal}\varepsilon_{AB}\Omega^{\intercal}, \\
\nonumber M_{AC} &=& \frac{1}{2}\left( \Omega \Sigma_{A}\varepsilon_{AB}\Omega^{\intercal} - \Omega\varepsilon_{AB}\Omega^{\intercal} \right), \\
M_{BC} &=& \frac{1}{2}\left( \Omega\varepsilon_{AB}\Sigma_{B}\Omega^{\intercal} - \Omega\varepsilon_{AB}\Omega^{\intercal}\right).
\end{eqnarray}
Here, $N^{-1}=m_{A}m_{B}+m_{C}$ is a normalization constant, which enforces $\Theta_{1}\Theta_{2}\chi(\alpha,\beta)|_{\alpha=0,\beta=0}=1$. Keep in mind that we have assumed that the submatrices $\Sigma_{A}$, $\Sigma_{B}$, and $\varepsilon_{AB}$ of the covariance matrix are symmetric.

If we use these states as the entangled resource on a CV quantum teleportation protocol of an unknown coherent state, we get that the average fidelity can be written as
\begin{equation}
\overline{F} = \frac{1+ h}{\sqrt{\det\left[\Id_{2}+\frac{1}{2}\Gamma\right]}},
\end{equation}
with $\Gamma \equiv \sigma_Z \Sigma_{A}\sigma_Z + \Sigma_{B} -\sigma_Z \varepsilon_{AB}-\varepsilon_{AB}^\intercal\sigma_Z$. Here, we have defined $h$ as

\begin{eqnarray}
\nonumber h &=& \frac{1}{E_{0}}\bigg\{\tr\left[\Omega\left(\Id_{2}+\frac{1}{2}\Gamma\right)^{-1}\Omega^{\intercal}E_{1}\right] -\frac{2}{\det\left(\Id_{2}+\frac{1}{2}\Gamma\right)}\tr\left(\Omega E_{2}^{A}\Omega^{\intercal}E_{2}^{B}\right) \\
&+& 3\tr\left[ \Omega\left(\Id_{2}+\frac{1}{2}\Gamma\right)^{-1}\Omega^{\intercal}E_{2}^{A}\right]\tr\left[ \Omega\left(\Id_{2}+\frac{1}{2}\Gamma\right)^{-1}\Omega^{\intercal}E_{2}^{B}\right]\bigg\},
\end{eqnarray}

together with
\begin{eqnarray}
\nonumber E_{0} &=& m_{A}m_{B} + m_{C}, \\
\nonumber E_{1} &=& m_{A}\left( M_{B} + \sigma_{z}M_{C}\sigma_{z}+ \sigma_{z}M_{BC}\right) \\
\nonumber &+& m_{B}\left( \sigma_{z}M_{A}\sigma_{z} + M_{C} + \sigma_{z}M_{AC}\right) \\
\nonumber &+& \left(2M_{C}+\sigma_{z}M_{AC}\right)\Omega\left( \Id_{2} + \sigma_{z}\varepsilon_{AB} - \Sigma_{B} \right)\Omega^{\intercal} \\
\nonumber E_{2}^{A} &=& M_{C} + \sigma_{z}M_{AC} + \sigma_{z}M_{A}\sigma_{z}, \\
E_{2}^{B} &=& M_{B} + \sigma_{z}M_{BC} + \sigma_{z}M_{C}\sigma_{z}.
\end{eqnarray}
We can identify $h$ as the non-Gaussian corrections to the fidelity, and mask them as corrections to the covariance matrix of a Gaussian state with the same fidelity. This is done by defining 
\begin{equation}
\tilde{\Gamma} = \frac{\Gamma-2h\Id_{2}}{1+h} \equiv \sigma_Z \tilde{\Sigma}_{A}\sigma_Z + \tilde{\Sigma}_{B} -\sigma_Z \tilde{\varepsilon}_{AB}-\tilde{\varepsilon}_{AB}^\intercal\sigma_Z.
\end{equation}
For the symmetric resource we define
\begin{eqnarray}\label{H2PS_sym_eff_submatrices}
\nonumber \tilde{\Sigma}_{A} &=& \frac{1}{1+h}\left(\Sigma_{A} - h\Id_{2}\right), \\
\nonumber \tilde{\Sigma}_{B} &=& \frac{1}{1+h}\left(\Sigma_{B} - h\Id_{2}\right), \\
\tilde{\varepsilon}_{AB} &=& \frac{1}{1+h}\varepsilon_{AB},
\end{eqnarray}
whereas for the asymmetric one we require $\tilde{\Sigma}_{A}=\tilde{\Sigma}_{B}$, such that
\begin{eqnarray}\label{H2PS_asym_eff_submatrices}
\nonumber \tilde{\Sigma}_{A} &=& \frac{1}{1+h}\left(\frac{\Sigma_{A}+\Sigma_{B}}{2} - h\Id_{2}\right), \\
\nonumber \tilde{\Sigma}_{B} &=& \frac{1}{1+h}\left(\frac{\Sigma_{A}+\Sigma_{B}}{2} - h\Id_{2}\right), \\
\tilde{\varepsilon}_{AB} &=& \frac{1}{1+h}\varepsilon_{AB}.
\end{eqnarray}
These ``re-Gaussified'' covariance matrices need to satisfy positivity and the uncertainty principle, meaning that $|\sqrt{\det\tilde{\Sigma}}-1|\geq |\tilde{\alpha}-\tilde{\beta}|$, assuming that we can write $\tilde{\Sigma}_{A}=\tilde{\alpha}\Id_{2}$, $\tilde{\Sigma}_{B}=\tilde{\beta}\Id_{2}$, and $\tilde{\varepsilon}_{AB} = \tilde{\gamma}\sigma_{z}$. Furthermore, if $\Sigma_{A}=\alpha\Id_{2}$, $\Sigma_{B}=\beta\Id_{2}$, and $\varepsilon_{AB} = \gamma\sigma_{z}$, this condition can be expressed as
\begin{equation}
\left|\sqrt{\det\Sigma}-h(2+\alpha+\beta)-1 \right| \geq (1+h)|\alpha-\beta|
\end{equation}
for the symmetric state, and as
\begin{equation}
\left|\sqrt{\det\Sigma}+\frac{1}{4}(\alpha-\beta)^{2} - h(\alpha+\beta) + h^{2} - 1 \right| \geq 0
\end{equation}
for the asymmetric one. A graphical proof that these conditions are met can be found in Appendix~\ref{app_D}.

\section{Appendix C: Probabilistic photon subtraction}\label{app_C}
In this section, we present the definitions we have used to shorten the notation of the modified submatrices of the covariance matrix of a Gaussian state that has undergone a symmetric two-photon-subtraction protocol (with beamsplitters and photodetectors). Assuming that $\Sigma_{A}$, $\Sigma_{B}$, and $\varepsilon_{AB}$ are symmetric, the general expression for the submatrices of the covariance matrix of the resulting state is
\begin{eqnarray}
\nonumber \tilde{\Sigma}_{A} &=& \tau \Sigma_{A} + (1-\tau)\Id_{2} - 2\left( J_{1}X_{A}^{-1}J_{1}^{\intercal} + K_{1}Y^{-1}K_{1}^{\intercal}\right), \\
\nonumber \tilde{\Sigma}_{B} &=& \tau \Sigma_{B} + (1-\tau)\Id_{2} - 2\left( J_{2}X_{A}^{-1}J_{2}^{\intercal} + K_{2}Y^{-1}K_{2}^{\intercal}\right), \\
\tilde{\varepsilon}_{AB} &=& \tau \varepsilon_{AB} - 2\left( J_{1}X_{A}^{-1}J_{2}^{\intercal} + K_{1}Y^{-1}K_{2}^{\intercal} \right),
\end{eqnarray}
whose success probability is given by
\begin{equation}
P = \frac{m_{1}m_{2}+m_{3}}{\sqrt{\det X_{A}\det Y}}. 
\end{equation}
Here, we have used
\begin{eqnarray}
\nonumber X_{A} &=& \frac{1}{2}\Omega \left[ (1-\tau)\Sigma_{A} + (1+\tau)\Id_{2} \right]\Omega^{\intercal}, \\
\nonumber X_{B} &=& \frac{1}{2}\Omega \left[ (1-\tau)\Sigma_{B} + (1+\tau)\Id_{2} \right]\Omega^{\intercal}, \\
\nonumber H &=& -\frac{1}{2}(1-\tau)\Omega \varepsilon_{AB}\Omega^{\intercal}, \\
\nonumber Y &=& X_{B} - HX_{A}^{-1}H, \\
\nonumber W_{X,M} &=& X^{-1}\tr(X^{-1}M) - \frac{\Omega M \Omega^{\intercal}}{\det X}, \\
\nonumber m_{1} &=& 1-\frac{1}{2}\tr Y^{-1}, \\
\nonumber m_{2} &=& 1-\frac{1}{2}\tr X_{A}^{-1} -\frac{1}{2}\tr\left(Y^{-1}HW_{X_{A},\Id_{2}}H\right), \\
\nonumber m_{3} &=& \frac{1}{2}\tr\left(W_{Y,\Id_{2}}HW_{X_{A},\Id_{2}}H\right), \\
\nonumber K_{1} &=& \frac{1}{2}\sqrt{\tau(1-\tau)}\left[ \varepsilon_{AB}\Omega^{\intercal} + (\Sigma_{A}-\Id_{2})\Omega^{\intercal}X_{A}^{-1}H \right], \\
\nonumber K_{2} &=& \frac{1}{2}\sqrt{\tau(1-\tau)}\left[ (\Sigma_{B}-\Id_{2})\Omega^{\intercal} + \varepsilon_{AB}\Omega^{\intercal}X_{A}^{-1}H \right], \\
\nonumber J_{1} &=& \frac{1}{2}\sqrt{\tau(1-\tau)}(\Sigma_{A}-\Id_{2})\Omega^{\intercal}, \\
J_{2} &=& \frac{1}{2}\sqrt{\tau(1-\tau)}\varepsilon_{AB}\Omega^{\intercal}.
\end{eqnarray}

\begin{landscape}

The characteristic function of the resulting non-Gaussian state is

\begin{eqnarray}
&& \chi^{(1,1)}(\alpha,\beta) = \frac{e^{-\frac{1}{4}\left[\vec{\alpha}\Omega \tilde{\Sigma}_{A}\Omega^{\intercal}\vec{\alpha}^{\intercal} + \vec{\beta}\Omega \tilde{\Sigma}_{B}\Omega^{\intercal}\vec{\beta}^{\intercal} + 2\vec{\alpha}\Omega \tilde{\varepsilon}_{AB}\Omega^{\intercal}\vec{\beta}^{\intercal} \right]}}{m_{1}m_{2}+m_{3}} \times \\
\nonumber && \left[ \left( m_{1} + \vec{\alpha}P_{1}\vec{\alpha}^{\intercal} + \vec{\beta}P_{2}\vec{\beta}^{\intercal} + \vec{\alpha}P_{12}\vec{\beta}^{\intercal} \right) \left(m_{2} + \vec{\alpha}Q_{1}\vec{\alpha}^{\intercal} + \vec{\beta}Q_{2}\vec{\beta}^{\intercal} + \vec{\alpha}Q_{12}\vec{\beta}^{\intercal} \right) + m_{3} + \vec{\alpha}R_{1}\vec{\alpha}^{\intercal} + \vec{\beta}R_{2}\vec{\beta}^{\intercal} + \vec{\alpha}R_{12}\vec{\beta}^{\intercal} \right].
\end{eqnarray}

Furthermore, when computing the average fidelity (Eq.~\eqref{eq:fidelity_2PS}), we obtain the non-Gaussian corrections defined by

\begin{eqnarray}
\nonumber g &=& \frac{1}{m_{1}m_{2}+m_{3}}\bigg[ m_{1}\tr\left[\Omega\left(\Id_{2}+\frac{1}{2}\tilde{\Gamma}\right)^{-1}\Omega^{\intercal}\left(\sigma_{z} Q_{1}\sigma_{z} + Q_{2} + \sigma_{z}Q_{12}\right)\right] + m_{2}\tr\left[\Omega\left(\Id_{2}+\frac{1}{2}\tilde{\Gamma}\right)^{-1}\Omega^{\intercal}\left(\sigma_{z} P_{1}\sigma_{z} + P_{2} + \sigma_{z}P_{12}\right)\right] \\
\nonumber && +\tr\left[\Omega\left(\Id_{2}+\frac{1}{2}\tilde{\Gamma}\right)^{-1}\Omega^{\intercal}\left(\sigma_{z} P_{1}\sigma_{z} + P_{2} + \sigma_{z}P_{12}\right)\right]\tr\left[\Omega\left(\Id_{2}+\frac{1}{2}\tilde{\Gamma}\right)^{-1}\Omega^{\intercal}\left(\sigma_{z} Q_{1}\sigma_{z} + Q_{2} + \sigma_{z}Q_{12}\right)\right] \\
&& + \tr\left[\Omega\left(\Id_{2}+\frac{1}{2}\tilde{\Gamma}\right)^{-1}\Omega^{\intercal}\left(\sigma_{z} R_{1}\sigma_{z} + R_{2} + \sigma_{z}R_{12}\right)\right] + 2\tr\left[ W_{\Omega\left(\Id_{2}+\frac{1}{2}\tilde{\Gamma}\right)\Omega^{\intercal},\sigma_{z} P_{1}\sigma_{z} + P_{2} + \sigma_{z}P_{12}}\left(\sigma_{z} Q_{1}\sigma_{z} + Q_{2} + \sigma_{z}Q_{12}\right)\right] \bigg]
\end{eqnarray}
which can be computed using
\begin{eqnarray}
\nonumber P_{1} &=& -\frac{1}{2}\Omega K_{1}W_{Y,\Id_{2}}K_{1}^{\intercal}\Omega^{\intercal}, \\
\nonumber P_{2} &=& -\frac{1}{2}\Omega K_{2}W_{Y,\Id_{2}}K_{2}^{\intercal}\Omega^{\intercal}, \\
\nonumber P_{12} &=& -\Omega K_{1}W_{Y,\Id_{2}}K_{2}^{\intercal}\Omega^{\intercal}, \\
\nonumber Q_{1} &=& -\frac{1}{2}\Omega\left( J_{1}W_{X_{A},\Id_{2}}J_{1}^{\intercal} +2J_{1}W_{X_{A},\Id_{2}}HY^{-1}K_{1}^{\intercal} + K_{1}W_{Y,HW_{X_{A},\Id_{2}}H}K_{1}^{\intercal}\right)\Omega^{\intercal}, \\
\nonumber Q_{2} &=& -\frac{1}{2}\Omega \left(J_{2}W_{X_{A},\Id_{2}}J_{2}^{\intercal} +2J_{2}W_{X_{A},\Id_{2}}HY^{-1}K_{2}^{\intercal} + K_{2}W_{Y,HW_{X_{A},\Id_{2}}H}K_{2}^{\intercal}\right)\Omega^{\intercal}, \\
\nonumber Q_{12} &=& -\Omega \left(J_{1}W_{X_{A},\Id_{2}}J_{2}^{\intercal} + J_{1}W_{X_{A},\Id_{2}}HY^{-1}K_{2}^{\intercal} + K_{1}Y^{-1}HW_{X_{A},\Id_{2}}J_{2}^{\intercal}  + K_{1}W_{Y,HW_{X_{A},\Id_{2}}H}K_{2}^{\intercal}\right)\Omega^{\intercal}, \\
\nonumber R_{1} &=& \frac{1}{2}\Omega\bigg[ J_{1} W_{X_{A},\Id_{2}}HW_{Y,\Id_{2}}K_{1}^{\intercal} \\
\nonumber &+& K_{1}\left(W_{Y,\Id_{2}}\tr\left(Y^{-1}HW_{X_{A},\Id_{2}}H\right) + Y^{-1}\tr\left(W_{Y,\Id_{2}}HW_{X_{A},\Id_{2}}H\right) - \frac{\Omega HW_{X_{A},\Id_{2}}H\Omega^{\intercal}}{\det Y}\tr Y^{-1} \right) K_{1}^{\intercal} \bigg]\Omega^{\intercal}, \\
\nonumber R_{2} &=& \frac{1}{2}\Omega\bigg[ J_{2} W_{X_{A},\Id_{2}}HW_{Y,\Id_{2}}K_{2}^{\intercal} \\
\nonumber &+& K_{2}\left(W_{Y,\Id_{2}}\tr\left(Y^{-1}HW_{X_{A},\Id_{2}}H\right) + Y^{-1}\tr\left(W_{Y,\Id_{2}}HW_{X_{A},\Id_{2}}H\right) - \frac{\Omega HW_{X_{A},\Id_{2}}H\Omega^{\intercal}}{\det Y}\tr Y^{-1} \right) K_{2}^{\intercal} \bigg]\Omega^{\intercal}, \\
\nonumber R_{12} &=& \frac{1}{2}\Omega\bigg[ J_{1} W_{X_{A},\Id_{2}}HW_{Y,\Id_{2}}K_{2}^{\intercal} + K_{1} W_{Y,\Id_{2}}HW_{X_{A},\Id_{2}} J_{2}^{\intercal} \\
\nonumber &+& 2 K_{1}\left(W_{Y,\Id_{2}}\tr\left(Y^{-1}HW_{X_{A},\Id_{2}}H\right) + Y^{-1}\tr\left(W_{Y,\Id_{2}}HW_{X_{A},\Id_{2}}H\right) - \frac{\Omega HW_{X_{A},\Id_{2}}H\Omega^{\intercal}}{\det Y}\tr Y^{-1} \right) K_{2}^{\intercal} \bigg]\Omega^{\intercal}.
\end{eqnarray}

\end{landscape}

\section{Appendix D: Positivity and uncertainty principle for covariance matrices}\label{app_D}
\begin{figure}[H]
\centering
\includegraphics[width=0.7 \textwidth]{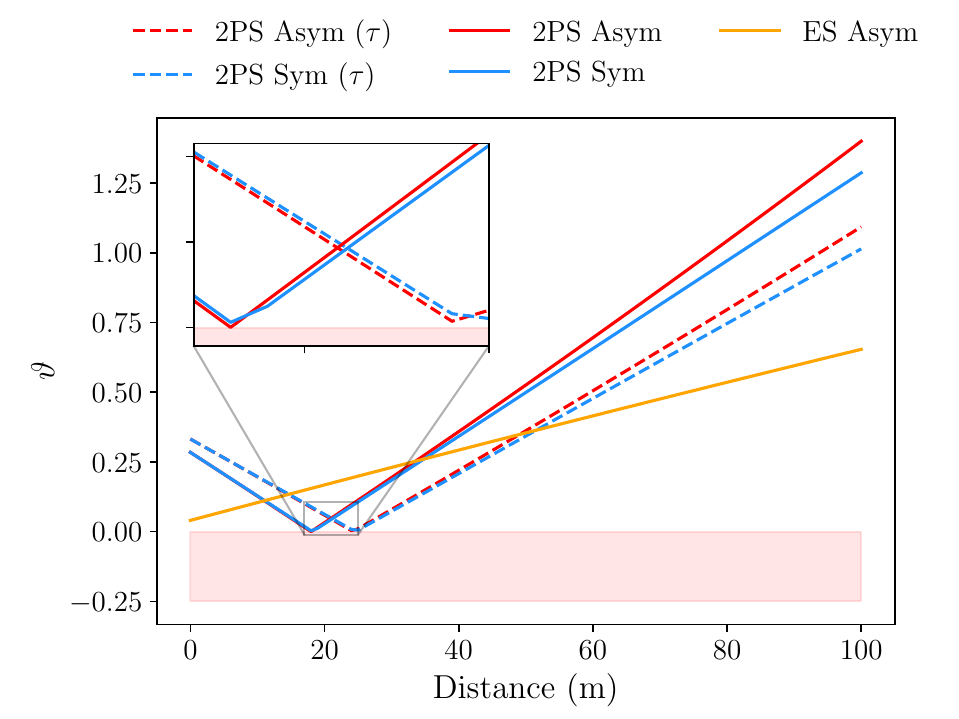}
\caption[Graphical representation of the curves constructed from Eq.~\eqref{check} against the traveled distance that, if positive, prove the submatrices used to compute this quantity characterize a covariance matrix, which satisfies a positivity condition, as well as the uncertainty principle.]{Graphical representation of the curves constructed from Eq.~\eqref{check} against the traveled distance that, if positive, prove the submatrices used to compute this quantity characterize a covariance matrix, which satisfies a positivity condition, as well as the uncertainty principle. In orange, we represent the curve associated with the submatrices in Eqs.~\eqref{ES_submatrices} that result from entanglement swapping. The blue and red solid curves correspond to the heuristic two-photon-subtracted ``re-Gaussified'' symmetric and asymmetric states, respectively, described in Eqs.~\eqref{H2PS_sym_eff_submatrices} and~\eqref{H2PS_asym_eff_submatrices}. The blue and red dashed curves correspond to the probabilistic two-photon-subtracted ``re-Gaussified'' symmetric and asymmetric states, respectively, described in Eqs.~\eqref{2PS_sym_eff_submatrices} and~\eqref{2PS_asym_eff_submatrices}. Inset: enlarged view of the region of short distances, in which we observe that the condition $\vartheta > 0$ is still met. \textit{Figure credit: T. G.-R.}}
\label{fig12}
\end{figure}
In this section, we discuss the conditions that a covariance matrix must satisfy in order for it to describe a quantum state. Then, we apply this criterion to the covariance matrices obtained after photon subtraction and entanglement swapping. The first condition is the positivity of the covariance matrix
\begin{equation}
\Sigma = \begin{pmatrix} \Sigma_{A} & \varepsilon_{AB} \\ \varepsilon_{AB}^{\intercal} & \Sigma_{B} \end{pmatrix} > 0,
\end{equation}
and the second one is preservation of the uncertainty principle,
\begin{equation}
\begin{pmatrix} \Sigma_{A} & \varepsilon_{AB} \\ \varepsilon_{AB}^{\intercal} & \Sigma_{B} \end{pmatrix} + i\begin{pmatrix} \Omega & 0 \\ 0 & \Omega \end{pmatrix} \geq 0.
\end{equation}
If we consider $\Sigma_{A}=\alpha\Id_{2}$, $\Sigma_{B}=\beta\Id_{2}$, and $\varepsilon_{AB}=\gamma\sigma_{z}$, the positivity condition reduces to
\begin{eqnarray}
\alpha &>& 0, \\
\nonumber \det\Sigma &>& 0,
\end{eqnarray}
whereas the uncertainty principle can be written as
\begin{eqnarray}
\alpha &\geq& 1, \\
\nonumber \det\Sigma &\geq& \alpha^{2} + \beta^{2} - 2\gamma^{2} - 1.
\end{eqnarray}
Note that the latter imposes a more restrictive condition. Given that any covariance matrix requires $\alpha\geq 1$ and $\beta\geq 1$, we can summarize all conditions as
\begin{equation}\label{check}
\vartheta \equiv \left|\sqrt{\det\Sigma} - 1\right| - \left|\alpha-\beta\right| \geq 0.
\end{equation}
In Fig.~\ref{fig12}, we investigate whether this condition is satisfied for different modified covariance matrices by representing $\vartheta$ against the traveled distance: submatrices in Eqs.~\eqref{ES_submatrices} due to entanglement swapping (orange); in Eqs.~\eqref{H2PS_sym_eff_submatrices} and~\eqref{H2PS_asym_eff_submatrices} due to re-Gaussified heuristic photon subtraction (symmetric shown with a blue line, asymmetric shown with a red line); in Eqs.~\eqref{2PS_sym_eff_submatrices} and~\eqref{2PS_asym_eff_submatrices}, due to re-Gaussified probabilistic photon subtraction (symmetric shown with a blue dashed line, asymmetric shown with a red dashed line). Note that all five cases satisfy both positivity and uncertainty principle conditions, confirming that they are indeed covariance matrices. In the inset, we present an enlarged view of the short distance behavior, where the curves approach the region in which $\vartheta < 0$ (highlighted in a pale red background). As we can see, even in that area $\vartheta > 0$ is satisfied.

\section{Appendix E: Imperfect photodetection and homodyne detection}\label{app_E}
In this section, we investigate different sources of error that affect the protocols described in this paper. By using parameters taken from recent experimental benchmarks in microwave quantum technologies, we are able to estimate how inefficiencies and imperfections surrounding microwave photocounting and homodyne detection affect photon subtraction, quantum teleportation and entanglement swapping. We believe that this provides a closer relation to state-of-the-art experiments with quantum microwaves. 


First, let us introduce a parameter for the efficiency of the microwave photodetectors. In Ref.~\cite{Dassonneville2020}, a circuit-QED-based microwave photon counter was presented, which could detect between zero and three photons, with fidelities ranging from 99 \% to 54 \%. Such a device is particularly useful for the photon-subtraction scheme investigated in the manuscript, in which we only consider single-photon subtraction in each mode of a bipartite entangled state. Then, we need to look at the success probability of detecting a single photon, which in this experiment is 76 \%. An imperfect detector can be modeled as a pure-loss channel, which is represented by a beam splitter that mixes the signal traveling towards the detector with a vacuum state, and whose transmissivity determines the efficiency. In this case, we have $\tau_{\text{detector}} = 0.76$. This characterizes the detection probability in the heuristic photon-subtraction case, but in the probabilistic description, this will be given by $\tau_{\text{total}} = \tau_{\text{detector}}\tau = 0.72$, since we have considered $\tau=0.95$.

Taking into account the detector efficiency, we observe that the maximum negativity associated with the re-Gaussified photon-subtracted states is, at least, 47 \% of the maximum negativity of the photon-subtracted states with ideal detector efficiency. This means that, taking into account this source of error, the negativity of the states obtained through this entanglement distillation technique is almost cut in half. Furthermore, these values are below the negativity of the bare state in the ideal case, which means that performing photon subtraction leads to entanglement degradation. In order for it to be advantageous, with the parameters we have considered throughout this manuscript, we would need to have detection efficiencies above 85 \% for the heuristic protocol, and above 90 \% for the probabilistic one. 

Now, let us discuss homodyne detection, a measurement technique that is widely used in many quantum communication protocols, and in this article is required for quantum teleportation and for entanglement swapping. Homodyne detection aims at extracting information about the quadratures of a signal by mixing it with a coherent light source with a large number of photons in a balanced beam splitter, and then measuring the photocurrents at both ends. From the difference between the currents, we can get information about either quadrature of the signal. This measurement works ideally if the source has an infinite number of photons (infinite gain), which is not generally the case. Therefore, we can consider the case of finite gain in homodyne detection. 

The theoretical description of these measurements corresponds to a projection onto a state that is infinitely-squeezed in $x$ (or in $p$) in phase space. That is, an eigenstate of the position operator (or the momentum operator) whose eigenvalue corresponds to the signal's $x$ (or $p$) quadrature value. In the symplectic formalism, this measurement operator has a covariance matrix 
\begin{equation}
\Gamma = \begin{pmatrix} e^{-2\xi} & 0 \\ 0 & e^{2\xi} \end{pmatrix} \equiv \begin{pmatrix} \frac{1}{\sqrt{G}} & 0 \\ 0 & \sqrt{G} \end{pmatrix}.
\end{equation}
In the limit $G\rightarrow\infty$, we will recover the usual homodyne detection scheme. We have obtained that the fidelity of teleporting an unknown coherent state with $|\theta|^{2}$ photons using a bipartite entangled state with covariance matrix
\begin{equation}
\Sigma = \begin{pmatrix} \alpha\Id_{2} & \gamma \sigma_{z} \\ \gamma \sigma_{z} & \beta\Id_{2} \end{pmatrix}
\end{equation}
is given by
\begin{landscape}
\begin{eqnarray}
\overline{F} &=& \frac{2\left[2+\frac{1}{\sqrt{G}}(1+\alpha)\right]}{4\left(1+\frac{\alpha+\beta-2\gamma}{2}\right)+\frac{1}{\sqrt{G}}\left[\alpha(5+\beta) + \beta -(\gamma-1)(\gamma+5)\right] + \frac{2}{G}(1+\alpha)} \times \\
\nonumber && \text{exp}\left[-\frac{\frac{2}{G}(1-\alpha+\gamma)^{2}|\theta|^{2}}{\left[2+\frac{1}{\sqrt{G}}(1+\alpha)\right]\left[ 4\left(1+\frac{\alpha+\beta-2\gamma}{2}\right)+\frac{1}{\sqrt{G}}\left[\alpha(5+\beta) + \beta -(\gamma-1)(\gamma+5)\right] + \frac{2}{G}(1+\alpha)\right]}\right],
\end{eqnarray}
\end{landscape}

In the limit $G\rightarrow\infty$ we recover $\overline{F} = \left(1+\frac{\alpha+\beta-2\gamma}{2}\right)^{-1}$, which is the usual result. 
Notice that, while the average teleportation fidelity for an unknown coherent state with ideal homodyne detection does not depend on the value of the displacement for said state, we find that the first order corrections do include this dependence in the value of $\theta$. The average fidelity associated with a resource with increasing entanglement asymptotically tends to 1 when considering ideal homodyne detection. In this case, it tends to the value $\frac{1}{1 + \frac{1}{\sqrt{G}}}$, which becomes closer to 1 as $G$ increases. 

In a recent paper, continuous-variable quantum teleportation in the microwave regime was performed~\cite{Fedorov2021}, where the optimal gain considered was 21 dB, which implies that $1/G\approx0.008$. Using this value, and considering we want to teleport a vacuum state ($\theta=0$), we observe that the fidelity reaches the maximum classical fidelity at 434 m for the asymmetric state, and at 429 m for the symmetric one, while this distance is 479 m with ideal homodyne detection for both kinds of states. 

Here, we also consider the effect of finite-gain homodyne detection on the states that result from entanglement swapping. As a generalization of Eq.~\ref{ES_submatrices}, these states can be characterized by a covariance matrix with submatrices
\begin{eqnarray}
\nonumber \tilde{\Sigma}_{A} &=& \left( \alpha - \frac{\gamma^{2}\left( 1 + \frac{1}{\sqrt{G}}2\beta + \frac{1}{G}\right)}{2\left[\beta + \frac{1}{\sqrt{G}}\left(1+\beta^{2}\right) + \frac{\beta}{G}\right]}\right)\Id_{2}, \\
\nonumber \tilde{\Sigma}_{D} &=& \left( \alpha - \frac{\gamma^{2}\left( 1 + \frac{1}{\sqrt{G}}2\beta + \frac{1}{G}\right)}{2\left[\beta + \frac{1}{\sqrt{G}}\left(1+\beta^{2}\right) + \frac{\beta}{G}\right]}\right)\Id_{2}, \\
\tilde{\varepsilon}_{AD} &=& \frac{\gamma^{2}\left( 1 - \frac{1}{G}\right)}{2\left[\beta + \frac{1}{\sqrt{G}}\left(1+\beta^{2}\right) + \frac{\beta}{G}\right]}\sigma_{z}.
\end{eqnarray}
With entanglement swapping, the maximum classical fidelity is reached at 416 m, which is smaller than the reach of the bare states taking into account finite-gain homodyne detection. This is natural, since the effects of the finite gain come both from entanglement swapping and from quantum teleportation. Nevertheless, we have seen that finite-gain effects are not significantly detrimental to the measures we have computed in this manuscript, and this means that, if larger optimal gains can be engineered, errors can then be reduced. At the end of the day, we have observed that entanglement distillation and entanglement swapping suffer from errors associated with photon counting and homodyne detection. However, we believe that these errors can be easily overcome by technological improvements. Furthermore, by the time all the pieces necessary for experiments in open-air microwave quantum communication arrive, we expect these errors to be further reduced. Meanwhile, entanglement distribution and quantum teleportation with microwaves, in the realistic open-air scenario, are still viable despite the errors we considered in this section.

\chapter{Propagating Quantum Microwaves: \\Towards  Applications in Communication and Sensing}
\label{chapter:Roadmap}
This chapter is based on the paper Mateo Casariego et al.  Quantum Sci. Technol. \textbf{8} 023001 (2023), \cite{Casariego2022}. I lead the project of the manuscript production, and coordinated all the contributions. Figures which have not been my own production are credited. I take the opportunity again to thank all of my co-authors.

The abstract reads as follows:

The field of propagating quantum microwaves is a relatively new area of research that is receiving increased attention due to its promising technological applications, both in communication and sensing. While formally similar to quantum optics, some key elements required by the aim of having a controllable quantum microwave interface are still on an early stage of development. Here, we argue where and why a fully operative toolbox for propagating quantum microwaves will be needed, pointing to novel directions of research along the way: from microwave quantum key distribution to quantum radar, bath-system learning, or direct dark matter detection. The article therefore functions both as a review of the state-of-the-art, and as an illustration of the wide reach of applications the future of quantum microwaves will open.

\section{Introduction}
Despite the fact that photons are described by the same formalism regardless of their frequency, propagating quantum microwave technology is some 20 years behind the quantum optics at visible and infrared wavelengths. This is because up until recently we did not need quantum properties in a microwave field, although it is also true that the five orders of magnitude difference in energy have not made things easier: the way light interacts with matter naturally depends on its wavelength, and counting, or even detecting photons out of a field that does not trigger a photoelectric current is hard.

In classical domains such as radar, imaging, or mobile communication, microwave open-air technology is well established and omnipresent in everyday life. Hence, the question of how to extend such technology to the quantum regime is quite natural. In this context, one unexpected but nevertheless very important reason to study quantum microwaves comes from the field of quantum computing. There, one of the most promising platforms is based on superconducting circuits, which operate at microwave frequencies between \SIrange{1}{10}{\giga\hertz}. Superconducting quantum computing devices have reached a state of maturity where more than 100 coupled qubits can be controlled with high fidelity gates~\cite{IBM-Eagle}. Over the years, the requirement of cryogenic temperatures on the order of \SI{10}{\milli\kelvin} to preserve the quantum coherence of the qubits has been mitigated by robust commercial cryogenic technology. Just like in classical high-performance computing systems, the power of a quantum computer can be enhanced by further integration, and distributed via networked architectures. Regarding the latter, a microwave-to-optical transduction platform with high quantum efficiency would be desirable, but it is still quite out of reach of present-day technology. Hence, it is natural to consider propagating quantum microwaves for this task, because they intrinsically have zero frequency conversion loss, promising high remote-gate fidelities. This reasoning motivates the study of microwave quantum communications, a field that offers security protocols such as quantum key distribution (QKD) of paramount importance for the long-dreamed quantum internet. Superconducting quantum chips communicating efficiently according to the laws of quantum mechanics represent the first major step in this direction: fully-microwave, operative quantum local area networks (QLANs).

Another motivation for using propagating microwaves in open-air without converting to optical comes from the fact that some of the current telecom infrastructure relies on these frequencies. The atmosphere has a transparency window for them, and their absorption is less impacted by unfavourable weather conditions than it is for so-called telecom frequencies. Combined, microwave and telecom links could further enable the reach of quantum communications.
For the same reason, the remote quantum sensing community has started to think about advantages of quantum microwaves in different metrology tasks. A prominent example here is the ongoing work towards a first demonstration of a quantum radar~\cite{macconeRadar, lanzagorta}, where the realisation of microwave quantum illumination in an open-air setting could represent a proof of principle experiment, although as we will argue, other approaches can be more practical in real-life scenarios.

Here, we give an overview of recent efforts to understand and tame propagating quantum microwaves, hinting at some promising directions along the way, and with an eye put on real-life applications for both communications and sensing. 

The paper is organized as follows: in Section \ref{buildingBlocks} we review the state-of-the-art in quantum microwave technology: state generation, guided and non-guided propagation, Gaussian transformations, detection and state characterization techniques, signal amplifiers, and circulators. Section \ref{QCommunication} is devoted to the challenges in the field of quantum communications. We review some of the most recent results, such as quantum teleportation of an unknown microwave coherent state, and then move on to comment on the difficulties and benefits of proof-of-principle experiments such as inter-fridge QKD with microwave states.
Section \ref{QSensing} discusses quantum sensing, a field that we expect to benefit from recent developments in microwave photon-counting techniques. We discuss quantum illumination, quantum radar and imaging, and then mention two novel directions in quantum sensing: quantum thermometry, and direct detection of axionic dark matter. Finally, Section \ref{sec:Conclusions} closes with a quantum sensing and communication roadmap for the upcoming years, where we condense the technological and theoretical requirements for each of the key elements that will shape the future of propagating quantum microwaves.

\section{Building blocks of propagating quantum microwaves}\label{buildingBlocks}
The theory underlying propagating quantum microwaves is no other than quantum optics:
the quantized electromagnetic field is, in short, a continuous collection of frequency-dependent harmonic oscillators $\lbrace \hat{a}_\omega, \hat{a}^\dagger_\omega\rbrace_{\omega}$ with $\omega \in \mathbbm{R}^{+}$, where creation and annihilation operators satisfy the commutation relation $[\hat{a}_\omega, \hat{a}^\dagger_{\omega^\prime}] = \delta(\omega-\omega^\prime) \hat{\Id}$. Equivalently, and setting $\hbar = 1$, one can use two (dimensionless) orthogonal field quadratures for each mode $\omega$: $\hat{x}\equiv (\hat{a}+\hat{a}^\dagger)/\sqrt{2}$ and $\hat{p}\equiv (\hat{a}-\hat{a}^\dagger)/\sqrt{2}$. These quadratures are quantum continuous variables (CVs), in the sense that their expected values span the reals: $(x,p)\in \mathbbm{R}^2$ while satisfying the canonical commutation relation $[\hat{x}, \hat{p}] =i\hat{\Id}$. Additionally, photons carry a polarization degree of freedom, which can as well be used as a quantum information carrier. Often, the Wigner function --one of the quasi-probability distributions associated to a quantum state $\rho$, is used as an alternative description to the density operator. This function's operational interpretation lies closer to the CVs spirit, since its integral over some field quadrature is proportional to the probability of measuring the orthogonal one: $\int_\mathbbm{R} \diff p W(x,p) \propto \tr[\hat{x} \rho]$. This fact is used when performing Wigner-tomography, \textit{i.e.} reconstruction of a quantum state from the measured values of two orthogonal quadratures. Although in theory both quantum optics and quantum microwaves are described by the same formalism, \textit{e.g.} observables are obtained as some power series of $\hat{a}_\omega$ and $\hat{a}^\dagger_\omega$, the five orders of magnitude difference in energy makes the corresponding technology substantially different. In this section we review the most important steps and techniques related to the generation, transformation, and detection of propagating quantum microwaves, mentioning along the way some of the directions we expect to provide improvements in the short-term future of real-life applications in communications and sensing.

\subsection{State generation}\label{subsec:stategen}
Entanglement plays a central role in quantum technology. Typically bipartite, it arises when more than one mode of the electromagnetic field is considered. The word `mode' can mean different things, and essentially refers to the labels one uses to distinguish between Hilbert spaces: spatial modes give path-entanglement, frequency (or time of emission/arrival) modes give frequency-entanglement (or time-bins), polarization modes give discrete, polarization entanglement, and so on. All these types of entanglement are not necessarily mutually exclusive. Bipartite, mixed state entanglement, is commonly measured through the negativity, an entanglement monotone defined as $2\mathcal{N}(\rho):=\norm{\tilde{\rho}}_1-1$, where $\norm{\tilde{\rho}}_{1}:=\Tr\sqrt{\tilde{\rho}^\dagger \tilde{\rho}}$ is the trace norm of the partially transposed density operator.
Two-mode squeezed (TMS) states are natural quantum CVs candidates for communication and sensing protocols that require entanglement. They are routinely produced in labs. The most used devices to perform the squeezing operation in microwaves are Josephson parametric amplifiers (JPA) \cite{Eichler2011, Flurin2014}, which require a cryogenic environment to operate.  Symmetric two-mode squeezed states can be obtained either by directly pumping a parametric down-conversion term or by single-mode-squeezing two vacuum states -- which in practice are in thermal states -- in orthogonal directions, and then using a beam splitter to combine them. In either case, the resulting state is described by a two-mode squeezed thermal (TMST) state with mean photon number $N_\text{TMST} = 2n_\mathrm{th} \cosh(2r) + 2 \sinh^2(r)$, where $n_\mathrm{th}$ is the thermal photon number, and $r\in \mathbbm{R}$ is the squeezing parameter. For frequency-degenerate TMST states, a negativity $\mathcal{N}=3.9$ has been experimentally observed \cite{Fedorov:2018a}, with corresponding entangled-bit rate of $4.3\times 10^6 \text{ ebit}\cdot \text{s}^{-1}$, while in the frequency non-degenerate case, a rate of $6\times 10^6 \text{ ebit}\cdot \text{s}^{-1}$ has been reported \cite{Flurin2012}. 
Other approaches use traveling wave parametric amplifiers \cite{Esposito2021}, which is attractive for broad-band applications. Another way to generate TMS states consists in using a dc-biased Josephson junction in presence of two resonators. The entanglement production rate was more than $100\times 10^6 \text{ ebit}\cdot \text{s}^{-1}$ in Ref.~\cite{Peugeot2021}.

Cat states represent another source for CVs quantum entanglement with potential applications in microwave quantum technologies. In Ref.~\cite{Ma2019} these states are used to generate an entangled coherent state of two superconducting microwave resonators. 
Time-bin encoding in propagating microwaves has also been experimentally demonstrated \cite{Kurpiers2019}. These approaches are particularly interesting in scenarios where decoherence plays a role, since time bins are well known for their resilience against loss. Resilience of entanglement when these states propagate in open-air needs to be further studied in order to experimentally assess their utility for real-life applications.

Finally, polarization is a degree of freedom of  practical interest for propagating quantum microwaves. Current classical antennae designs already contemplate linearly or circularly polarized signals. However, microwaves in superconducting circuits have their polarization suppressed, as coplanar waveguides naturally imply a projection. The microwave-equivalent to an optical fibre, where the polarization vector could rotate freely, seems highly impractical due to the physical dimensions these fibres would need to have. Advances in 3D superconducting technology will be required to solve the issue.

\subsection{Propagation of quantum microwaves}\label{subsec:propagation}
\subsubsection{Guided}

Superconducting Niobium-Titanium coaxial cables are commonly used nowadays for low-loss guiding of microwave signals at cryogenic temperatures below $10$\,K. Such cables typically have a $50$\,$\Omega$ characteristic impedance and exhibit absorption losses on the order of $10^{-3}$\,dB/m for $\nu \simeq $\SI{5}{\giga \hertz}. These losses are mainly limited by the loss tangent of respective dielectrics, such as PTFE, and surface quality of superconducting material itself. Their diameter varies typically between $\sim$ 1\,mm and 3\,mm. In combination with crimped SMA connectors connectors, these cables offer a flexible, robust, and commercially available way to interface various devices in cryogenic environments. 

A somewhat alternative way for low-loss guiding of quantum microwave signals lies via using various rigid waveguides made of Niobium or Aluminum. Due to larger sizes of such waveguides, and therefore larger inner volumes, electromagnetic fields are strongly diluted in these systems, which reduces coupling to various dissipative channels. Additionally, these waveguides do not require the use of inner dielectrics. These factors lead to lower absorption losses below $5\cdot 10^{-4}$\,dB/m in the microwave waveguides. However, the price for this reduction comes in the form of their inflexible designs, connectors, and large sizes.

\subsubsection{Non-guided}

Non-guided propagation of microwave signals represents the typical scenario in open-air. This will be the result of transmitting a quantum state out of the cryostat by means of a quantum antenna. This device can be thought of as an inhomogeneous medium that smoothly connects two very different environments; a finite cavity that achieves impedance matching between the cryostat and the open-air. Let us briefly point out a very important difference between optical and microwave frequencies, noise-wise. Photons follow Bose-Einstein statistics: the walls of a cavity in thermal equilibrium at temperature $T$ are expected to produce an average photon number given by
\begin{equation}\label{eq:BEStats}
    n(\nu, T)= \frac{1}{e^{h\nu/k_B T}-1}
\end{equation}
 per unit volume at frequency $\nu$. At room temperature ($T=300$\,K), this gives an insignificant $n\sim  10^{-28}$--$10^{-55}$ in the optical domain (400-790\,THz), while a very noisy $n\sim 6250$--$625$ thermal microwave photons with $\nu$ in \SIrange{1}{10}{\giga \hertz}.

In a simple case study of a quantum antenna~\cite{GonzalezRaya2020}, the impedance function was optimized for the transmission of Gaussian microwave quantum states (two-mode squeezed states, in this case), from a cryostat at 50\,mK and associated impedance of 50\,$\Omega$, to open-air at 300\,K and associated impedance of 377\,$\Omega$. An exponential shape of the impedance, a well known result in the classical case, was found to reduce the reflectivity below $10^{-9}$. This showed that reducing losses in the antenna could lead both to the maximization of energy transmission, as well as to the preservation of quantum correlations.

Entanglement preservation was addressed by considering the antenna as a beamsplitter with a thermal noise input (using Eq.~\eqref{eq:BEStats} we find $ n(5\,\text{GHz}, 300\,\text{K}) \sim 1250$ photons), the latter being the main source of entanglement degradation. In a similar fashion, the reach of entanglement was studied in Ref.~\cite{GonzalezRaya2022}, considering an attenuation channel to describe absorption losses in a thermal environment. It was computed that entanglement could be transferred up to \SI{550}{\meter} in a realistic open-air setting through a two-mode squeezed thermal state generated at \SI{50}{\milli\kelvin} and with squeezing parameter $r=1$, considering oxygen molecules in the environment as the largest source of attenuation. Yet, a rigorous theoretical study of the atmospheric channel capacity \cite{Shapiro2005} in the microwave regime of interest for applications in communications and sensing is still missing to the best of our knowledge. 

\begin{figure}
\begin{center}
\includegraphics[width=0.7 \textwidth]{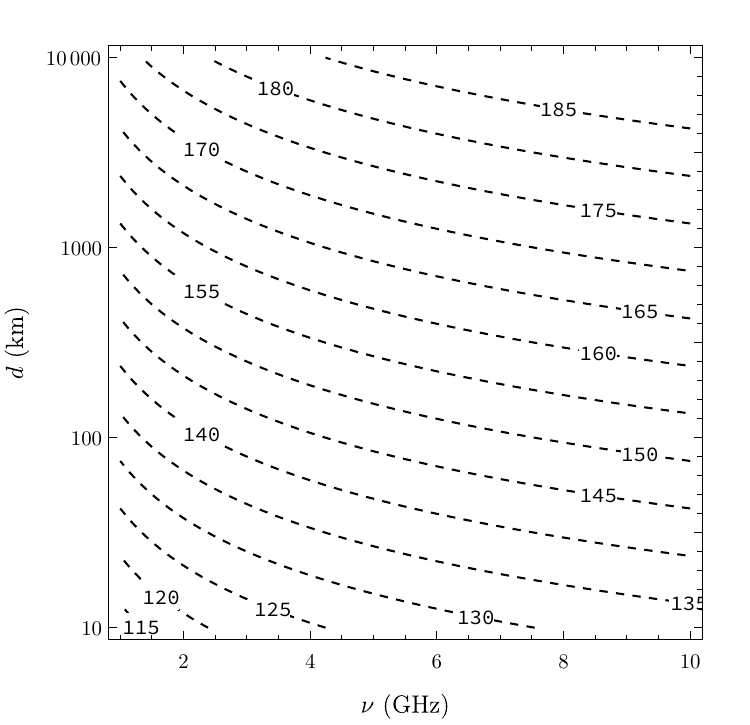}
\caption[Contour lines of the free-space path loss (FSPL) expressed in decibels as a function of the signal frequency $\nu$ (in the \SIrange{1}{10}{\giga\hertz} interval) and the distance $d$ between the emitter and the receiver in the range \SIrange{10}{10000}{\kilo\meter}: $L_\text{FSPL} (\text{dB})=20 \log_{10}\left(4\pi d\nu/c\right)$.]{Contour lines of the free-space path loss (FSPL) expressed in decibels as a function of the signal frequency $\nu$ (in the \SIrange{1}{10}{\giga\hertz} interval) and the distance $d$ between the emitter and the receiver in the range \SIrange{10}{10000}{\kilo\meter}: $L_\text{FSPL} (\text{dB})=20 \log_{10}\left(4\pi d\nu/c\right)$. FSPL quantifies a purely geometric  phenomenon arising from the natural three-dimensional spread of a signal in the far-field limit. The quadratic dependence of the FSPL on the frequency $\nu$, could make microwaves suitable for long-range broadcasting communications.}\label{fig:FSPLmw}
\end{center}
\end{figure}

Open-air and free-space propagation of quantum microwaves \cite{Pirandola2021, Pirandola2021-2, Kaltenbaek2021} requires addressing additional loss mechanisms: signal diffraction due to the natural spreading of an electromagnetic pulse in three dimensions; atmospheric attenuation; and wheather conditions. First, diffraction: We shall refer to this strictly geometrical loss as free-space path loss (FSPL). This is particularly relevant when large distances are considered, for example in inter-satellite or Earth-to-satellite links, or open-air communications on Earth. High Earth orbits (HEOs) are defined in the altitude range of 35786-$d_M/2$\, km, where $d_M$ is the distance from the Earth to the Moon. Space temperature in HEOs is roughly \SI{2.7}{\kelvin}, corresponding to the peak of the cosmic microwave background (CMB). These low temperatures could motivate the use of mechanically obtained cryogenics in satellites, as well as focalization lenses \cite{azad2017}. 
If an emitter and a receiver are separated a distance $d$ in a vacuum, with $d$ large enough so that the far-field approximation holds, the power ratio is given by Frii's transmission formula \cite{friis, hogg, shaw}:
\begin{equation}\label{eq:friis}
\frac{P_e}{P_r} =\frac{D_e D_d}{L_\text{A} L_{\text{FSPL}}},
\end{equation}
where $L_A$ is the absorption loss (that depends on wheather conditions), $L_{\text{FSPL}}=(4\pi d\nu/c )^2$ is  the geometric FSPL,  $\nu$ and $c$ are the frequency and the (vacuum) speed of the signal, respectively, and $D_{e,r }$ are the directivities of the emission and reception antennas. Directivity is defined as the maximized gain with respect to some preferred direction in space. It heavily depends on the design, ranging from no directivity at all (isotropic antenna with $D=1$) to high directional gains like the ones obtained with parabolic designs $D_\text{parabolic}= e_a (\pi a \nu /c)^2 $, where $a$ is the aperture, and $e_a$ is the aperture efficiency, a dimensionless parameter typically lying in \SIrange{0.6}{0.8}{} for commercial devices. Directivity gains are commonly given in $\text{dBi}=10\log_{10} D$. Importantly, the expression for $L_\text{FSPL}$ assumes a polarization-matching between emitter and receiver antennas. If this is not applicable, additional loss due to a polarization mismatch should be accounted for.  In Figure \ref{fig:FSPLmw} we plot the FSPL in dBs (i.e. the function $10 \log_{10}\left(L_{\text{FSPL}}\right)$) as a function of $\nu$ and $d$. 
Using Frii's formula alone as a comparison between optics and microwave signals is not completely fair, since it seems to indicate that microwaves are simply better for open-air communications. This is not necessarily the case. Traditionally, so-called `telecom' wavelengths are grouped in the following sets: near infra-red (NIR) with \SIrange{400}{207}{\tera\hertz}, short infra-red (SIR) with \SIrange{214}{100}{\tera\hertz}, mid infra-red (MIR) with \SIrange{100}{37}{\tera\hertz}, long infra-red (LIR) with \SIrange{37}{20}{\tera\hertz}, and far infra-red (FIR) with \SIrange{20}{0.3}{\tera\hertz} \cite{Kaushal2018}. In particular, the wavelengths \SIrange{780}{850}{\nano\meter} and \SIrange{1520}{1600}{\nano\meter} that belong to NIR and SIR ranges, respectively, enjoy of a very low atmospheric absorption loss of the order of \SI{0.1}{\dB \per \kilo\meter} in optimal wheather conditions. However, in the presence of dust, haze, or rain, Rayleight and/or Mie scattering needs to be accounted for. In Ref.~\cite{Fesquet2022} a full comparison between telecom and microwave regimes was made, concluding that the latter are more robust against unfavourable wheather conditions. In addition to this, the strong interaction of microwaves with non-linear elements give a higher entanglement rate production than telecom frequencies, which could justify the larger loss factor associated with the lack of highly directive emitters that is possible with telecom lasers.

\subsection{Gaussian transformations}\ref{subsec:Gaussian transformations}
The manipulation of propagating electromagnetic signals via Gaussian transformations (50:50 beam splitting, displacement, phase rotation, squeezing) is well established at optical frequencies since many decades. There, many components used to guide and manipulate classical light are also known to work for quantum signals. However, despite the same theory description, the physical construction of components such as phase shifts, 50:50 beam splitters, displacers, or squeezers is different in the microwave regime of \SIrange{1}{10}{\giga\hertz}, because, there, the wavelength is roughly five orders of magnitude larger, i.e., on the order of a few centimeters.

When the research community started to think about the quantum properties of propagating microwaves in the late 2000s, first linear operations had been addressed. Due to the convenient wavelength, phase shifts can be achieved rather trivially via short pieces of delay line. The construction of microwave beam splitters and displacers is more involved; they are interference devices with typical dimensions on the order of the operation wavelength. When respecting the relevant boundary conditions of unitarity (absence of loss), impedance matching, and isolation between ports, a 50:50 beam splitter is always a four-port device. In terms of quantum mechanics, these conditions ensure energy conservation and commutation relations. Experimentally, vacuum fluctuations as fundamental  as the minimal noise added to the input signal of a 50:50 beam splitter were first discussed in Ref.~\cite{Mariantoni:2010a}. Interestingly, even devices with only three signal connectors were shown to have a fourth internal port there. 50:50 beam splitters are used to create superpositions or path entanglement between two microwave beams~\cite{Menzel:2012a}, in microwave interferometers~\cite{Eder:2018a}, for dual-path tomography~\cite{Menzel:2010a,dicandia2014}, and in the feedforward mechanism of quantum teleportation~\cite{Fedorov:2021a}. In these experiments, often commercial conducting devices (\textit{i.e.} non-superconducting) with little dissipation on the order of \SI{0.3}{\deci\bel} are enough. For more delicate situations where even little dissipation is harmful, superconducting beam splitters were developed~\cite{Hoffmann:2010a,Ku:2011a,Eder:2018a}.

The second important linear Gaussian transformation is the displacement operation $\hat{D}=\exp(\alpha\hat{a}^\dag-\alpha^*\hat{a})$. The name of this transformation is inspired by the fact that it actually displaces the Wigner function of the state of a bosonic mode $\hat{a}$ in phase space by the vector $\alpha$. Experimentally, the displacement is implemented by weakly coupling a coherent drive of complex amplitude $\alpha$ to the propagating mode $\hat{a}$ via a strongly asymmetric (typically 99:1) beam splitter. The corresponding microwave device is commonly called a directional coupler and, again, exists in low-loss normal-conducting commercial~\cite{Fedorov:2016a}, and laboratory-made superconducting~\cite{Ku:2011a} variants. Since its demonstration for propagating microwaves~\cite{Fedorov:2016a}, displacement is being routinely used in the feedforward mechanism of microwave continuous-variable quantum communication protocols~\cite{Pogorzalek:2019a,Fedorov:2021a}.

\begin{figure}
\begin{center}
\includegraphics[width=0.4 \textwidth]{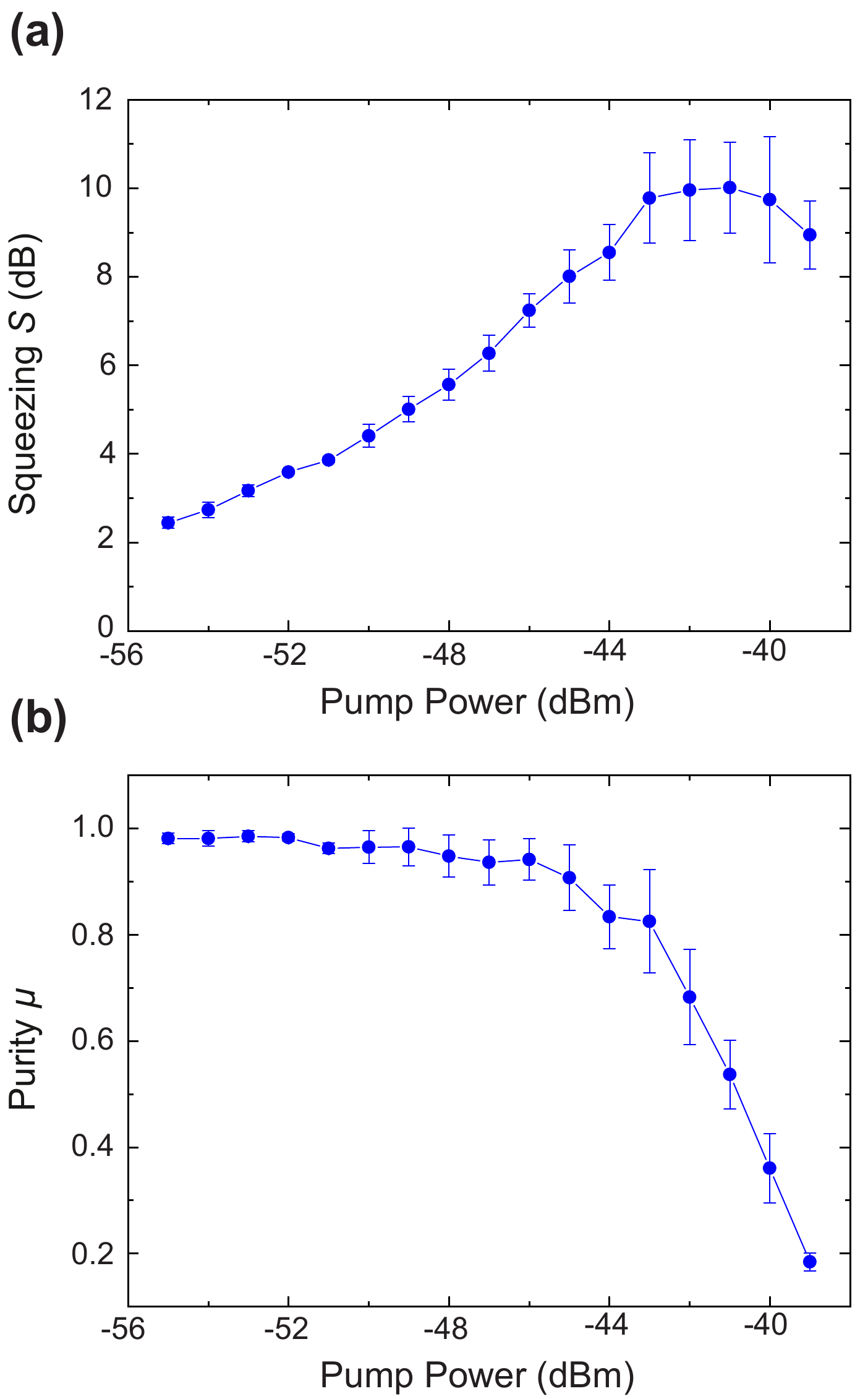}
\caption{(\textbf{a}) Squeezing as a function of pump power for a single-SQUID niobium JPA. The pump power is referred to the JPA input. (\textbf{b}) Corresponding purity of the reconstructed squeezed states as a function of the pump power. \textit{Figure credit: M. R., M. P. \& D. D.}}
\label{fig:SqueezingPurity}
\end{center}
\end{figure}

The generic nonlinear Gaussian transformation for a single mode is called squeezing, because one quadrature variance of a Gaussian state is reduced below the corresponding variance of the vacuum state. In turn, the orthogonal quadrature must be enlarged to respect the Heisenberg principle. In a Wigner function picture, the name can be intuitively understood because the circular blob of the vacuum state is squeezed into a narrow elliptic shape. The single-mode squeeze operator $\hat{S}=\exp\left(\frac{1}{2}\xi^*\hat{a}^2-\frac{1}{2}\xi\left(\hat{a}^\dag\right)^2\right)$ describes the squeezing operation on the mode $\hat{a}$. Here, the complex parameter $\xi$ is related to the squeezing parameter $r$ via $\xi = r e^{i\theta}$, and controls the suppression (``squeezing'') and enlargement (``anti-squeezing'') of two orthogonal field quadratures, as well as the orientation of the squeezed quadrature in phase space. In general, squeezing is created via a parametric process, where a system parameter is modulated with a suitable high-frequency signal. Squeezed microwave signals are commonly generated from a superconducting LC circuit, where part of the inductance is a current- or flux-tunable Josephson inductance. The parametrically induced nonlinearity can be conveniently amplified in narrowband resonant devices such as Josephson parametric amplifiers (JPA)~\cite{Zhong:2013a} or Josephson parametric converters (JPC)~\cite{Bergeal:2010a}. When broadband operation is desired, a successful solution has been to use an open transmission line with many Josephson devices and a suitable phase matching. These devices, which are known as Josephson travelling-wave parametric amplifiers (JTWPAs)~\cite{Perelshtein:2021a,Macklin:2015a,White:2015a}, are also important for the high-fidelity single-shot qubit readout in superconducting quantum computing architectures. Most experiments based on flux-driven JPAs~\cite{Zhong:2013a} employ propagating microwave modes with up to \SI{9}{\deci\bel} of squeezing~\cite{Fedorov:2018a}. However, for squeezing beyond \SI{3}{\deci\bel}, induced noise results in a significant thermal contribution and, hence, a significantly reduced purity of the squeezed state~\cite{Pogorzalek:2019a}. Recent experiments on the amplification properties of JPAs suggest pump-induced noise as a limiting factor for quantum efficiencies~\cite{Renger:2021a}. Figure\,\ref{fig:SqueezingPurity} shows an exemplary plot for the pump power dependence of reconstructed single-mode squeezing (\textbf{a}) and purity (\textbf{b}) for a single-SQUID niobium JPA, fabricated by VTT. Here, we define the squeezing level as $S = -10 \log(\sigma_\mathrm{s}^2/0.25)$, where $\sigma_\mathrm{s}^2$ denotes the squeezed variance. The purity has been calculated from the corresponding reconstructed covariance matrix $V$ by $\mu = 1/(4\sqrt{\det V})$.

By means of a symmetric beam splitting operation, the quantum correlations inherent to squeezed states can be transferred to entanglement correlations -- also called two-mode squeezing -- between the split beams. For frequency-degenerate two-mode squeezing, the two operations are usually executed in series~\cite{Menzel:2012a,Fedorov:2018a}. Here, the entanglement must be distributed among two distinct physical paths (``path entanglement''). State-of-the-art devices exhibit an entanglement of formation on the order of a few Mebits/s over the full JPA bandwidth, which would be usable for quantum communication purposes~\cite{Fedorov:2018a}. Note that an ``ebit'' is a logical unit of bipartite entanglement. A maximally-entangled pair of qbits, for example in a Bell state, carry one ebit. For frequency-nondegenerate two-mode squeezing, both operations can be combined into a single device such as the above-mentioned ~JTWPA or a JPC. In the former case, the two-mode squeezing coexists in the same beam at two different frequencies. For a JPC, the entanglement is between two different beams at two different frequencies.

Recently, single-mode and two-mode squeezing have been obtained not only with narrowband superconducting parametric devices, but also with a JTWPA. Operating in dual-pumped, non-degenerate four-wave mixing, the device provides \SI{-11.35}{\deci\bel} of single-mode squeezing and an average of \SI{-6.71}{\deci\bel} of two-mode squeezing over a bandwidth of \SI{1.75}{\giga\hertz}.
The temperature regime of these experiments is on the order of \SI{10}{\milli\kelvin}, to avoid thermal fluctuations at the input of the device.

Another important device class for experiments with propagating microwaves are circulators and isolators, where the latter can be described by a circulator with one \SI{50}{\ohm}-terminated port. A circulator is an $n$-port device with a time-reversal symmetry breaking mechanism. As a consequence, from each port, an incoming signal can only propagate towards one of the two neighboring ports, thereby forming a directive element. In other words, if inputs are called $I_k$ and outputs $O_k$, with $k\in [1,\ldots, n]$, then a circulator satisfies $I_k = O_{k+1}$. Circulators are of eminent practical importance for the currently known communication and sensing protocols with propagating microwaves. In a cryogenic environment, passive circulators based on ferrite materials biased by permanent magnets are widely used because of their favorable combination of sufficient bandwidth, insertion loss, isolation, input power tolerance, and commercial availability. Their main drawbacks are the requirement of significant magnetic bias fields, incompatibility with on-chip integration into superconducting circuits, and bulkiness due to the interference concept and magnetic shielding requirements. Furthermore, even the small insertion loss of \SI{0.3}{\deci\bel} from their normal conducting microwave circuitry adds up when using multiple circulators in series in more complex experimental settings. As a way out, circulators based on Josephson or nanomechanical elements have been proposed and implemented~\cite{Chapman:2017a,Ranzani:2019a}. Although these devices can be on-chip integrated with other superconducting quantum circuits and have potential to operate at the quantum limit, they still suffer from complex fabrication, demanding flux tuning, or low input power tolerance.
Finally, and closely related to circulators, the quantum microwaves toolbox should also include microwave switches, for which we have witnessed important advances in the last years \cite{Naaman2016, Chapman2016, Pechal2016, Chang2020, Rosenthal2021}.

\subsection{Detection of quantum microwaves}\label{subsec:detection}
State characterization and quantum tomography techniques typically rely on extracting information from a large sample of states generated under the same conditions. The way this information is extracted depends on the detection techniques one has available, and this naturally varies depending on the frequency of the field under observation. 
Quantum sensing and communication protocols requiring feedforward control are more challenging, as a high signal-to-noise ratio and fast signal processing are required for making a decision in a single shot, i.e., without ensemble averaging. 
In this section, we review the state-of-the art in detection of quantum microwaves. This also involves the step of amplifying, and, or downconverting the signals. The results discussed here are all related to cryogenic settings. The problem of open-air detection at higher temperatures still remains a technological challenge. 
\subsubsection{Homodyne, and heterodyne detection}
Photodetectors, not to be confused with photon-counters, are devices capable of transforming a photocurrent $\hat{i} \propto \hat{a}^\dagger \hat{a}$ into an electric current. In microwaves, photodetection has been experimentally demonstrated in a variety of settings  \cite{milford2006, romero2009, chen2011, peropadre2011, sathyamoorthy2016, Govenius2016}.
Homodyne detection relies on this concept in order to extract information about a single quadrature of the electromagnetic field. Heterodyne detection, on the other hand, measures two orthogonal quadratures at the same time, allowing a complete quantum state reconstruction. Most quantum microwave tomography techniques use some form of heterodyne measurements \cite{Eichler2011, menzel2012}.
In Ref. \cite{eichler2012}  a single-shot heterodyne detection scheme is proposed for microwaves, which makes use of a linear phase-insensitive cryogenic amplifier, a mixer, and analog-to-digital converters (ADC). State tomography for single Fock itinerant microwave states has been experimentally demonstrated with linear amplifiers and ADCs \cite{Eichler2011-2}. Using an NV-center as mixer, in Ref.~\cite{Meinel2021} a heterodyne measurement scheme with spectral resolution below \SI{1}{\mega \hertz} for a \SI{4}{\giga \hertz} signal is proposed.
 Closely related to heterodyne detection are CVs Bell measurements. In Ref.~\cite{Fedorov:2021a} these are realized using two phase-sensitive amplifiers together with two hybrid rings, and a directional coupler. Bell measurements are required in most implementations of CVs quantum teleportation, as well as in situations where one needs to guarantee a loophole-free Bell test scenario.
 
It is expected that advances both in microwave photon-counting and amplifiers will improve the quality of single-shot homodyne measurements, and consequently of Bell measurements.

\subsubsection{Qubit-based photon counting}
Many scenarios in quantum communication, sensing, or computation,  require the use of non-Gaussian operations and measurements. The most notable one is photon counting. As an example, as described below, quantum illumination requires photon counting to obtain a quantum advantage. Photodetectors can also be used to implement the full state tomography of a propagating quantum state~\cite{besse2020}. And to distribute entanglement between remote qubits using a joint measurement~\cite{Narla2016}. Furthermore, photodetectors allow one to build heralded noiseless amplifiers, which can restore lost entanglement in dissipative transmission lines or owing to imperfect generation~\cite{Xiang2010}.

First, it is instructive to distinguishing counters of stationary versus propagating modes. The first kind has been demonstrated since 2007 with Rydberg atoms and with superconducting circuits~\cite{Gleyzes2007,Guerlin2007,Johnson2010,Leek2010,Sun2014} and is now a standard component of the circuit-QED toolbox. In contrast, the detection of propagating microwave photons is more challenging and many implementations have been proposed~\cite{Helmer2009,Koshino2013,Sathyamoorthy2014,Fan2014,Sathyamoorthy2016,Gu2017,Wong2017,Leppakangas2018,Royer2018}. First experimental realizations used direct photon absorption by a Josephson junction~\cite{Chen2011,Inomata2016}, which leads to a destructive photodetector. More recently, quantum non-demolition detectors have been demonstrated either by encoding the parity in the phase of a qubit~\cite{Besse2018,Kono2018} or by encoding the presence of a single photon in a qubit excitation~\cite{Narla2016}, also combined with reservoir engineering~\cite{Lescanne2019}. 

Until recently, and despite several proposals of a photocounter -- a microwave photodetector able to resolve the photon number -- for a propagating mode~\cite{Royer2018,Kono2018,Sokolov2020,grimsmo2020quantum}, an experimental realization was missing. In Ref.~\cite{Dassonneville2020}, a photon-counter working within a \SI{20}{\mega\hertz} band centered around \SI{10.220}{\giga\hertz} is able to distinguish between 0, 1, 2 and~3 photons, with detection efficiencies of $99\%$ for no photons, $(76\pm3)\%$ for a single photon, $(71\pm3)\%$ for two photons, and $(54\pm2)\%$ for three, with a dark count probability of $(3\pm0.2)\%$, and an average dead time of \SI{4.5}{\micro\second}. Microwave photon detectors and counters enable a variety of applications. In Ref.~\cite{Balembois2021}, a state-of-the-art photodetector has been used to improve the sensitivity of electron spin resonance detection. 

\subsubsection{Bolometers for photon counting}\label{subsubsec:bolometers}
A bolometer is a device that detects heat deposited by absorbed radiation and converts it into an electric signal. In contrast to qubit detectors, a bolometer typically has a resistive input, and hence can operate on a broad input frequency band. A bolometer that has a fast readout and low enough noise, can be used as a calorimeter, i.e., to measure the energy of an incoming wave packet. If the frequency of the input photons is known, such a calorimeter also works as a photocounter. In addition, a bolometer is typically read out continuously with no dead time. However, the challenge in using bolometers in the single-photon microwave experiments has been their too high noise and slow thermal time constant, which has been greatly relieved thanks to recent developments~\cite{Kokkoniemi2020}. This motivates us to discuss the recent advancements in bolometers.

The most promising bolometers for cQED consist of a two-dimensional graphene flake~\cite{Kokkoniemi2020, Lee2020} or a metallic gold-palladium nanowire~\cite{lowNEP} connected to a temperature-dependent effective inductance implemented using the superconductor proximity effect~\cite{govenius2014microwave, proximity2017}. Changes in the inductance change the resonance frequency of an $LC$ tank circuit and hence the reflection coefficient of the roughly \SI{600}{\mega\hertz} probe signal. Thus, by continuously digitizing the probe signal, one can continuously monitor the temperature or the absorbed power at the bolometer input.

Owing to the electrothermal feedback, i.e., the probe tone also heating the bolometer depending on the resonance frequency, the $LC$ tank circuit exhibits a bistable regime~\cite{zeptoj} at a certain range of input powers and probe frequencies. Thus, by fine-tuning the probe parameters we can optimize for high signal-to-noise ratio or speed of thermal relaxation. The optimal operation point typically lies close but outside the regime of bistability. 

The noise equivalent power (NEP) is a typical figure of merit for bolometers and equals to the noise spectral density in the bolometer readout signal in the units of the input power to the bolometer. The recently discovered metallic bolometers have showed an NEP as low as $20$~zW/$\sqrt{\text{Hz}}$ \cite{lowNEP} with a thermal time constant of $30$~ms, leading to an extracted bolometer energy resolution of $h\times400$~GHz. On one hand, this energy resolution and speed are not satisfactory for single-photon microwave counting, but on the other hand, they are not many orders of magnitude off. Fortunately, using graphene as the proximity superconductor, in order to lower the bolometer heat capacity, and hence to maximize the induced temperature change due to an absorbed photon, the obtained results are further improved: with a similar NEP of $30$~zW/$\sqrt{\text{Hz}}$, and a much improved time constant in the hundred-nanosecond scale, and an extracted energy resolution as low as $h\times30$~GHz~\cite{Kokkoniemi2020}. Although the energy resolution is still not meeting the requirements of photon counting for typical frequencies in cQED, $\lesssim10$~GHz, these numbers seem satisfactory to start using the bolometer as a readout device for superconducting qubits, where the information of the qubit state is encoded into a microwave pulse of several photons.

The above-described bolometer is a robust and easy to operate device. Its footprint on the chip is not larger than that of typical transmon qubits and it is capable of detecting absorbed photons in a broad bandwidth in real time. Neither photon shaping nor knowledge of its arrival time is required. Thus, some further optimization of the graphene bolometer may lead to a convenient device for photon counting for the cQED. In addition to its wide range of applications in photon sensing in general, this device can be modified to detect the heat deposited by a dc current, also allowing one to calibrate for absolute microwave power or possibly low currents at millikelvin temperatures~\cite{girard2021cryogenic}.

\subsubsection{Low-noise cryogenic amplifiers}
In most applications, readout signals from the quantum circuits need to be eventually read out and processed at room temperature. Thus, a low-noise and high-gain amplification chain is required to amplify the weak signals originating from quantum circuits, typically located at millikelvin temperatures, before they pass through room-temperature components.
Typical output powers originating from quantum circuits range between roughly \SI{-150}{\dBm} for single-photon signals and  \SI{-120}{\dBm} used in dispersive qubit readout.
The first stage of amplification is usually provided by superconducting parametric amplifiers, as they have been demonstrated to provide near-quantum-limited noise performance as well as sufficient bandwidth and power handling for most applications. The next-stage amplification is typically provided by low-noise cryogenic high-electron-mobility transistor (HEMT) amplifiers at around \SIrange{3}{4}{\kelvin}, followed by a final post-amplification stage at room temperature.
The gain and noise temperature of each amplification stage is usually chosen such that system noise, calculated using Friis formula, remains close to the noise added by the first amplifier.

Regarding the cryogenic low-noise solid-state amplifier technologies, the lowest noise temperatures are achieved based on Indium-Phosphide (InP) HEMTs. In Ref.~\cite{Schleeh2012} a \SIrange{4}{8}{\giga\hertz}, three-stage, hybrid low-noise amplifier operating at \SI{10}{\kelvin} is proposed, providing an average noise temperature of \SI{1.6}{\kelvin}. The gain of the amplifier across the entire band is \SI{44}{\deci\bel}, consuming \SI{4.2}{\milli\watt} of DC power. Future quantum computing applications demand a number of readout channels, integrated within a single cryogenic system, of the order of $10^3$. Improvements in power consumption will be key to efficiently enable future quantum technologies.  Recently, an ultra-low power \SIrange{4}{8}{\giga\hertz} InP HEMT cryogenic low-noise amplifier (LNA) has been demonstrated \cite{Cha2020}, achieving an average noise of \SI{3.2}{\kelvin} with \SI{23}{\deci\bel} gain, and an ultra-low power consumption of just \SI{300}{\micro\watt}. Apart from InP HEMTs, Silicon-Germanium (SiGe) heterojunction bipolar transistors (HBTs) are appearing as promising cryogenic LNAs candidates because they are compatible with complementary metal–oxide–semiconductor (CMOS) technology, making them more appropriate for uses in large-scale systems. A \SIrange{4}{8}{\giga\hertz} SiGe cryogenic LNA has been implemented using the BiCMOS8HP process \cite{Montazeri2017}, providing \SI{26}{\deci\bel} of gain while dissipating \SI{580}{\micro\watt} of DC power. The noise temperature was \SI{8}{\kelvin} across the frequency band, so significant research is still required to optimize the cryogenic SiGe HBTs performance.

At cryogenic temperatures superconducting parametric amplifiers, such as JPAs, play an important role for microwave quantum technology. These devices exhibit ultra-low power dissipation in the relevant temperature range due to their superconducting properties. Moreover, these amplifiers can be operated close the fundamental quantum amplification limit of $1/2$ added noise photons, also known as the standard quantum limit (SQL) of phase-insensitive amplification. Fundamentally, the SQL originates from the commutation relation of bosonic operators describing electromagnetic fields in quantum mechanics. It should be noted that in the framework of QMiCS, sub-GHz JPAs have also been used to improve the performance of the microwave calorimeters described in Sect. \ref{subsubsec:bolometers}  (see also Ref. \cite{lowNEP}). 

Resonator-based JPAs typically suffer from a limited gain-bandwidth product, often on the order of \SIrange{0.1}{5}{\giga\hertz}. This constraint limits their applications in many practically relevant applications, such as broadband frequency-multiplexed qubit readout or generation of cluster states, among others. Furthermore, conventional superconducting JPAs used to suffer from limited \SI{1}{\deci\bel} compression point, on the  order of \SI{-120}{\dBm}. In recent years, it has been shown that the latter limitation can be circumvented by using multiple superconducting nonlinear elements, such as SQUIDs or SNAILs \cite{Frattini:2017}, in combination with the CPW resonators. Alternatively, one can exploit superconducting materials with high kinetic inductance, which can push the \SI{1}{\deci\bel} compression point as high as \SI{-50}{\dBm} for typical microwave frequencies around \SI{5}{\giga\hertz}.

In applications requiring higher bandwidth, superconducting traveling-wave amplifiers are used. Although parametric amplifiers were originally proposed even before the transistor amplifier, the recent rise of popularity has happened in last decade. 
JTWPAs introduced in \label{subsec:Gaussian transformations} consists of array of Josephson junction-based circuit elements\cite{Macklin:2015a,Miano:2019,Perelshtein:2021a,Frattini:2017} fabricated on a superconducting transmission line which acts as a 'non-linear media' facilitating wave mixing and amplification. Similarly, other efforts have fabricated the non-linear media using high-kinetic inductance material \cite{Malnou:2021}, which facilitates non-linear inductance. The later devices are often named as Kinetic Inductance TWPA (KI-TWPA). In JTWPAs, in addition to the Josephson elements various dispersion engineering or frequency domain band engineering are implemented in order to achieve phase matching \cite{Esposito:2021a}. Whereas in KI-TWPA such dispersion engineering are achieved by periodically loading the transmission line with capacitive elements. 
Early realization of JTWPAs operating in degenerate four-wave mixing mode (4WM) have produced gain in excess of \SI{15}{\deci\bel} and noise performance close to standard quantum limit \cite{Macklin:2015a}, with a disadvantage that the pump signal resides in the same frequency band as the amplified quantum signals, making filtering cumbersome and risking saturation of subsequent components in the readout chain. In contrast, three-wave mixing (3WM) TWPAs by device design have the advantage of the pump signal being situated outside the signal band, facilitating easy filtering \cite{Ranadive:2021,Perelshtein:2021a}.
Recently, a non-degenerate 4WM TWPA with two pump tones has been demonstrated \cite{JackY.Qiu:2022}, providing multiple GHz of signal bandwidth between the pump tones without sacrificing gain and noise performance. 
Similarly, 3WM implementations of both JTWPAs \cite{Perelshtein:2021a} and KITWPAs \cite{Malnou:2021} have produced gain in excess of \SI{15}{\deci\bel} in the \SIrange{4}{8}{\giga \hertz} frequency range with near-quantum-limited noise performance in the same frequency band.
Although the gain performance as a first stage amplifier in readout chain provided by both technology platform are comparable, KITWPAs provide significantly larger \SI{1}{\deci\bel}-compression power of roughly \SI{-60}{\dBm} \cite{Malnou:2021}, which is almost \SI{20}{\deci\bel} higher compared to that of state-of-the-art JTWPAs. The higher compression power provides advantages in multiplexed read out of detectors, sensors and other devices.  
Satisfactory gain and noise performance has been achieved with broadband superconducting TWPAs, in view of Friis formula for system noise. However, compared to conventional semiconductor amplifiers, TWPAs still suffer from relatively high ripple in the signal band, generation of spurious frequency tones, less standardized fabrication, and a delicate tune up procedure. Furthermore, superconducting TWPA performance is strongly dependent on the quality of broadband matching provided by RF components at the input and output of the JTWPA.Hence, future improvement of the superconducting parametric amplifier as a first-stage amplifier platform would require improvement in associated microwave components.    

\section{Quantum Communication}\label{QCommunication}

Quantum communication promises unconditional security in private communications, as long as nature behaves in accordance with the laws of quantum physics. This has been recently demonstrated in proof-of-principle experiments~\cite{Nadlinger2021,Zhang2021,Liu2021}. Such remarkable proof-of-principle experiments became feasible thanks to many developments such as the experiments that closed the locality and detection loopholes in 2015~\cite{Hensen2015,Shalm2015,Giustina2015}, theoretical advances in understanding the certification of quantum correlations~\cite{Arnon2018,Nadlinger2021}, and major experimental achievements in linear optics~\cite{Liu2021} and light-matter interaction for example with trapped ions~\cite{Stephenson2020,Nadlinger2021}. Quantum communication has clearly come a long way since its inception in 1984 with the celebrated paper by C. Bennett and G. Brassard~\cite{Bennett84}, describing the first quantum key distribution protocol: BB84. 

Since then, a panoply of quantum communication protocols have been developed, employing different types of degrees of freedom of particles, such as position in space, time, frequency, polarization, and spin. For example, with free-space linear optics one often uses polarization for its simplicity and availability of polarization optics. Nevertheless in fiber networks, where polarization losses can be important, a degree of freedom such as time is much more convenient. More generally, one may distinguish discrete degrees of freedom from continuous ones, giving rise to DV-QKD and CV-QKD, respectively. 

A long list of protocols have been proposed for DV-QKD, some of which are: E91~\cite{Ekert1991}, B92~\cite{Bennett1992}, COW~\cite{Stucki2005}, variations of BB84 such as the ones using decoy states~\cite{Lo2005,Grunenfelder2018}. 
Although CV-QKD is a more recent topic, there exist many protocols, making use of various quantum resources, i.e., states and measurements. On the one hand, security proofs are harder to obtain with respect to the DV case because of the need to work with infinite dimensional states and unbounded measurement operators. On the other hand, the technology for CV QKD, i.e., coherent detection, is more readily available. Both approaches are therefore very promising.

Finally, note that now quantum communication has been taken to space, with spectacular experiments such as 1200 km quantum teleportation~\cite{Ren2017,Liao2017}, and entanglement-based QKD demonstration over the same distance~\cite{Yin2020}.

Although quantum communication experiments have seen an impressive flourishing over the past four decades thanks to a better understanding of light and light-matter interaction, the idea of microwave quantum communication has not seen such a dramatic development. In fact, many questions remain: What is the feasibility of CV and DV quantum communication protocols with microwaves? How does one efficiently transmit a quantum state to open-air, given the dramatic differences in temperature and impedance between the two environments? Can microwave quantum communication be easier implemented in some situations? For example, can it be advantageous for satellite communication? 
Is it better to communicate with microwave or optical photons in a turbulent/polluted atmosphere? 
Some of these questions have been partly answered already in Section \ref{subsec:propagation}. In this section, we will review the state of the art, and explore what is being done currently to accelerate the development of microwave quantum communication. 

The section is organized as follows. First, in Section~\ref{subsec:swap} we show experiments that generate and manipulate quantum resources, such as entanglement swapping and teleportation experiments. Then, in Section~\ref{subsec:qkd} we present some QKD experiments, while in Section~\ref{subsec:qLAN} quantum local area networks and the scaling to a quantum internet are finally discussed.

\subsection{Distribution of quantum resources}
\label{subsec:swap}

As discussed in Section \ref{subsec:stategen} and illustrated with the example of linear polarization, the generation of quantum resources for CV QKD is more straightforward than for DV QKD. For this reason, it would be of great value to develop techniques to generate DV quantum states, not only in polarization, but also in other degrees of freedom which could be easier to implement, as for example it has been already demonstrated for time-bins \cite{Kurpiers2019}.


The first microwave quantum teleportation demonstrations were performed in an intra-fridge setting, usually over distances shorter than $1$ meter. Quantum teleportation in the microwave regime has been performed both with DV \cite{Wallraff2014} and CV \cite{DiCandia2015,Fedorov:2021a}. Intra-fridge quantum state transfer has been also demonstrated \cite{Kurpiers2018}. 

In the inter-fridge scenario, quantum states can be transferred from a cryostat to another one using a microwave quantum link both with DV and CV. In the DV case (transmon qubits), one can both prepare quantum states and measure them on the other end of the channel, and share entanglement (Bell states) between both ends, with fidelities around $85.8\%$ and $79.5\%$ respectively \cite{Magnard2020}. In the CV case, the experiment has not been performed yet.

Recently, a study about the feasibility of open-air microwave entanglement distribution for quantum teleportation of CVs was presented  \cite{GonzalezRaya2022}. There, absorption losses in a thermal environment are taken into account, to obtain an upper bound of \SI{550}{\meter} for the maximum distance that a TMST state generated at \SI{50}{\milli\kelvin} with squeezing parameter $r=1$ can propagate before completely losing its entanglement. To overcome this, non-Gaussian subroutines like photon-subtraction are proposed, arguing that these that could enhance significantly the degree of two-mode CV entanglement and therefore the overall quality of the protocol.

\subsection{Quantum key distribution}
\label{subsec:qkd}

Recently, theoretical studies taking into account the recent developments in propagating microwaves have indicated the significant potential brought by quantum microwaves for both short distance quantum communication \cite{Diamanti2016}, and long distance in the context of satellite communications \cite{Sanz2018,Fesquet2022}. Furthermore, QKD comes in two different flavors, DV- and CV-QKD \cite{Pirandola2020}. Due to the premature stage in which MW photon counters are, CV-QKD should be easier to implement with current MW technology \cite{Laudenbach}, although DV-QKD is expected to enable longer distance secure quantum key distribution.

In general, the motivations for CV-QKD are manifold. In traditional DV-QKD protocols such as BB84, one deals with single photons. Since there are no perfect single photons and single photon counters in the laboratories, and the security proof for BB84 relies on those, new protocols with relaxed security assumptions has been invented, such as measurement-device independent (MDI), device-independent (DI) and even more recently twin-field (TF) QKD. One may then argue whether it would make sense to invest on CV QKD, because we only rely on standard resources of the telecommunication industry: coherent states and (heterodyne) detection. In fact, there is much more interest from the telecom industry to develop coherent control than single photon states/detectors, as it would be useful also for classical telecommunications. 



Quantum key distribution with microwaves has not been realized yet. Seeing the performances obtained for time-bin generation \cite{Kurpiers2019} and photon counting (see Section \ref{subsec:detection}), it would be in principle possible to implement already the COW protocol with microwaves. As previously described, for microwave DV-QKD it is still early to use the polarization degree of freedom. Nevertheless, developments in this front would be extremely useful, as they would enable the use of more recent QKD protocols such as the three-state one-decoy BB84 \cite{Grunenfelder2018}. CV-QKD should be possible to realize today, as demonstrated by the recent publication \cite{Fesquet2022}. In this article, the authors show that using the protocol from Ref.~\cite{Cerf2001} based on Gaussian encoding of squeezed states, microwave CV-QKD can perform better under imperfect weather conditions.

\subsection{Scaling and integration}
\label{subsec:qLAN}

The realization of a scalable QLAN for quantum communication is an important milestone for the implementation of distributed quantum computing and for quantum internet applications \cite{Awschalom2021}. Such a network should provide an all-to-all connectivity between distant superconducting quantum processors and simultaneously maintain quantum coherence of interactions, including a possibility for entanglement distribution. There has been already significant efforts towards QLAN implementation in the optical domain at telecom frequencies over fiber networks and in the free-space environment. However, after years of research efforts, microwave-to-optics transduction experiments still have not reached sufficiently good efficiencies. The latter remain on the order of $10^{-5}$ in the single-photon regime \cite{Mirhosseini2020} after accounting for post-selection probabilities or added noise photons. Furthermore, the materials compatibility of such converters with that of high-performance superconducting qubits also remains to be an outstanding problem. In order to avoid these problems, we consider a direct realization of microwave quantum networks by connecting spatially separated dilution refrigerators via a cryogenic microwave link \cite{Deppe2020}. This approach has two important advantages. First, microwaves are the natural frequency scale of superconducting quantum circuits and, therefore, frequency conversion losses trivially vanish in this scenario. Second, since superconducting quantum processors require cooling to millikelvin temperatures, the cryogenic requirements are reduced to the development of a suitable millikelvin interface between dilution fridges containing distant superconducting circuits. We have implemented this goal within the EU quantum Flagship project QMiCS together with  Oxford Instruments Nanotechnology Ltd. (OINT). Our cryogenic link with a total length of 6.6\,m reaches temperatures below $\SI{35}{\milli \kelvin}$ and connects a home-built dry dilution refrigerator with a commercial Triton500 dilution refrigerator from OINT \cite{Batey2009}, both dilution refrigerators reaching below $\SI{20}{\milli \kelvin}$. During the design phase of the cryogenic link, several unique features were incorporated. One of these is a cold ‘junction box’ (a cold network node (CNN)) at the mid-point of the link. In addition, due to the large radiative heat load from such a system -- \textit{i.e.}, two dilution refrigerators connected by a long cryogenic link -- the cooling power at the higher temperature stages is important. This radiative heat load can be reduced by adding multi-layer insulation, but in order to add significant cooling power, an extra Pulse Tube Refrigerator (PTR) or 1 K pot can be added to the CNN. Other unique design features will be discussed in further publications \cite{Renger2022}. An inter-fridge quantum communication channel within the cryolink is realized with a superconducting Niobium-Titanium coaxial cable with characteristic losses around $0.001$\,dB/m. The CNN furthermore allows to connect additional link arms and, therefore, enables scaling to a two-dimensional quantum area network. All integral parts for such a system were supplied commercially by OINT and could be supplied for future projects. This ensures straightforward scalability to longer distances and allows near-future implementation of distributed superconducting quantum computing platforms operating in the microwave regime \cite{Krinner2019,Magnard2020b}.

\begin{figure*}[t!]
\centering
\includegraphics[width=\linewidth]{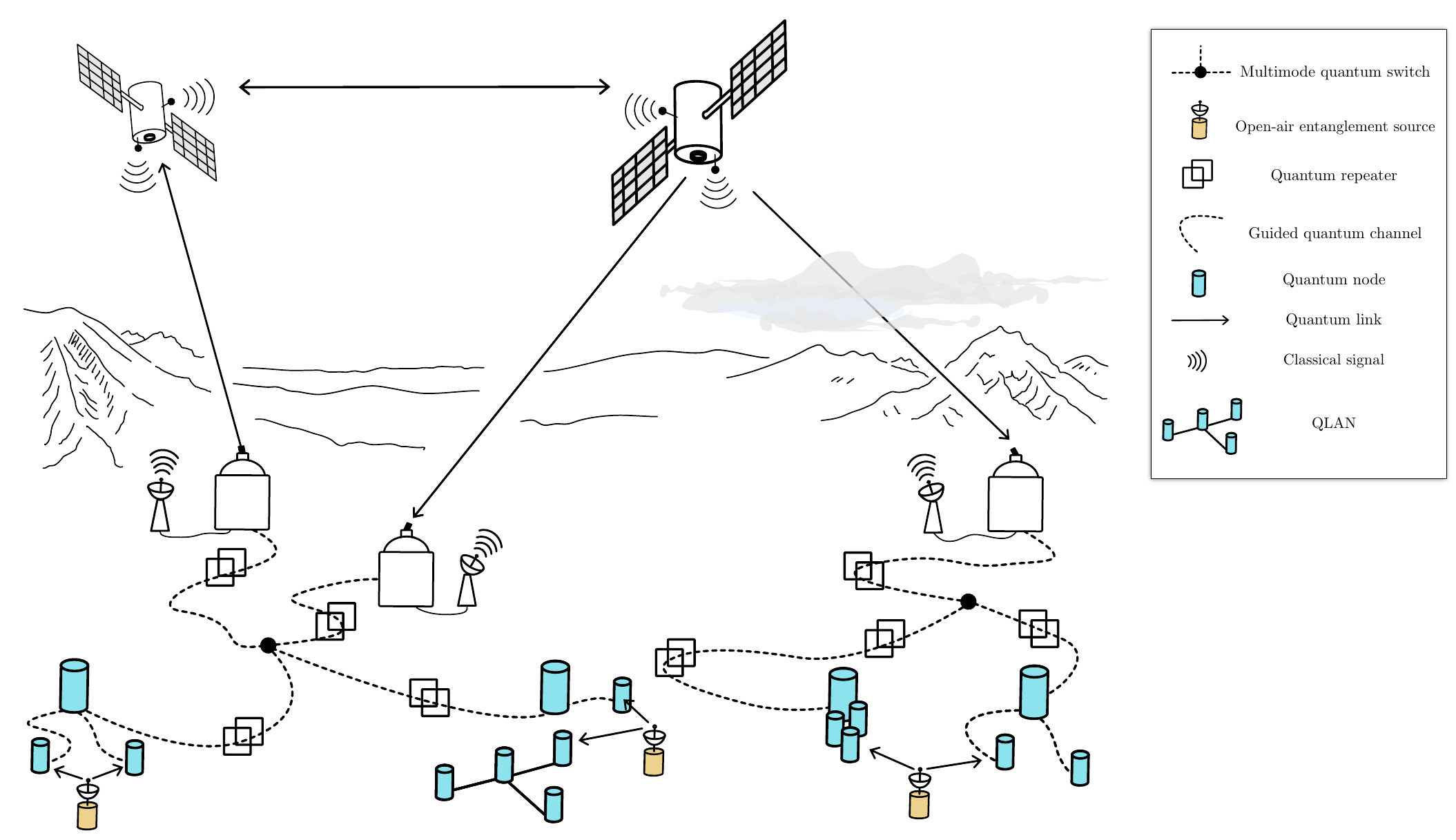}
\caption[Idealized view of a section of an interconnected, large-scale quantum communications network that combines optical and microwave links.]{Idealized view of a section of an interconnected, large-scale quantum communications network that combines optical and microwave links. Two low-orbit satellites capable of performing QKD  protocols are optically connected to distant Earth-based stations. Quantum repeaters are then used to distribute the quantum resources throughout an entire network, making use of trusted quantum switches to choose which node(s) receive the quantum information. Locally, open-air microwave entanglement sources are used to entangle remote but relatively close stations (up to \SI{200}{\meter}), making use of the benefits of microwaves in open-air transmissions with unfavorable weather conditions (see Refs.~\cite{GonzalezRaya2022, Fesquet2022}), and a simplified version of a QLAN, where cryolinks are used to connect different cryogenic stations to perform local quantum communication or sensing experiments.}
\label{fig:network}
\end{figure*}

A key, elusive element for the task of scaling-up a quantum communication network to cover large distances is the quantum repeater (QR). In Figure \ref{fig:network} we show an idealized quantum network that makes use of QRs for long-range quantum communications. Classically, a repeater, in its simplest form, is a device sitting between two nodes of a network that receives an incoming noisy signal, measures it, reproduces a noiseless copy of the signal, and sends it to the other node. Given the fact that quantum information can't be cloned, this notion of repeater has to be revised. In order to be compatible with all the communication advantages the quantum realm offers, the most notable is its intrinsic security, quantum tomography is not a solution. This is because, in short, a receive-measure-send scheme would destroy any possible initial entanglement between the two active nodes. To overcome this, three generations of QRs have been identified, depending on the techniques they use to improve the communication rate between nodes, in a sophistication-increasing order. The first generation uses heralded entanglement generation (\textit{e.g.}, via entanglement swapping), heralded entanglement purification/distillation, or a mix of both. Essentially, these repeaters rely on the ability to perform Bell measurements and two-way classical communication between the nodes, which already pose a challenge in microwaves, as argued in Section \ref{subsec:detection}. Additionally, Gaussian entanglement distillation requires many copies of the state, which may not be possible in some scenarios. The following proposals for QRs all apply to the CV case. The approach of noiseless linear amplification, weak measurements, and postselection of Ref.~\cite{DiCandia2015} have been conceived specifically for microwaves. A QR based on quantum scissors and non-deterministic, non-Gaussian entanglement swapping was proposed in Ref.~\cite{Seshadreesan2020}. The entanglement distillation, and swapping techniques discussed in Ref.~\cite{GonzalezRaya2022} will also prove relevant for this task. Single-shot QRs will be required at some point, but this requirement most likely will imply the need for microwave quantum memories~\cite{Wu2010,Kubo2012,Saito2013,Yin2013,Wenner2013,Flurin2014,Bao2021,Palomaki2013,Palomaki2013a}, as explained in Ref.~\cite{Dias2020}. Still, models for quantum communications without quantum memories have been proposed \cite{Munro2012}. The second and third generations of quantum repeaters use classical and quantum error correction, respectively, to solve additional sources of imperfection in the incoming signal, and are still beyond current reach, to the best of our knowledge. Further, in the specific case of propagating quantum microwaves, much work needs to be done, both theoretically and experimentally to produce operative quantum repeaters, even of first generation.


\section{Quantum Sensing}\label{QSensing}
Quantum sensing and metrology address a fundamental aspect of quantum mechanics
from a practical viewpoint: what is the maximum precision available at any given situation? Typically, quantum-enhanced sensing seeks an advantage provided by the resorting to quantum properties: entanglement, squeezing, or simply the use of single-photon states turn into \textit{resources} in this context. In this section we review some of the most important sensing protocols with potential applications in
quantum microwave technology.
\subsection{Quantum illumination}
Quantum illumination (QI) \cite{Lloyd1463, PhysRevLett.101.253601, guha2009,borre, shapiroStory, Pirandola2018, alsing2019} is a protocol that beats the classical signal-to-noise ratio (SNR) in the problem of detecting a low-reflectivity object embedded in a bright thermal bath by using signal-idler path entanglement as a resource. Importantly, QI preserves its quantum advantage even when the noisy, lossy channel is entanglement-breaking: The remaining quantum correlations (quantum discord) between the probe signal and the reference idler mode are enough to beat any classical scheme. Actually, the quantum discord directly quantifies the quantum advantage~\cite{Pirandola2016, Bradshaw2017}. 

In simple terms, the ideal case of Gaussian QI works as follows: $M \gg 1$ copies of a signal-idler symmetric two-mode squeezed vacuum state $\ket{\psi}_{SI}$ are prepared. The signal pulses, each containing $N_S$ photons on average, are sent to probe a given region at a given distance (both known) where there may be a low-reflectivity object, while the idlers are coherently kept in the lab. After the time associated with the assumed distance to the target, the receiver gets either just noise (the object is not there), or the reflected signals mixed with noise, in case the object is there. Assuming no bias in the presence/absence of the object, the problem reduces to a discrimination between two equally likely
states: one containing signals, one that doesn't. Theoretically, it is possible to obtain an enhancement of \SI{6}{\deci\bel} in the SNR, equivalent to an error probability four times smaller than any classical strategy. The full advantage of QI over the optimal classical approach using Gaussian states and detectors, namely the use of a coherent state as probe followed by any detection scheme using Gaussian operations and field quadratures measurements.
However, this \SI{6}{\deci\bel} enhancement requires a joint measurement between all signal and idler pairs, a very challenging task that essentially requires a quantum computer~\cite{zhuang2017,guha2009}. More modest approaches, which require only local operations and classical communication (LOCC), can obtain up to \SI{3}{\deci\bel} \cite{Sanz2017}, and could in principle be realised with efficient photon-counters.  

QI has been experimentally demonstrated in the optical regime \cite{genovese}. The main problem there is idler storage loss, which we address in this manuscript (Section \ref{subsec:qradar+imaging}). Experimental schemes for QI with microwaves have been proposed, but a fully-microwave implementation is  missing to the best of our knowledge: in Ref.~\cite{pirandolaMicrowave} the authors up-convert the signal to optical frequencies, with the corresponding conversion losses. In Ref.~\cite{shabir} a digital receiver is used, without any joint measurement and thus no absolute quantum advantage: in fact the experiment showed the advantage over coherent states generated via the same cryogenic setup, but not with respect to the possibility to generate microwave coherent states directly at room temperature, which is a non-trivial technological challenge at the low-photon number regime \cite{messaoudi2020, Karsa2021}. { An interesting alternative is the one proposed in Ref.~\cite{chang2019}, where a modified QI protocol that does not require idler storage is proposed, and demonstrated in the microwave domain, hence showing potential applications in radar physics.}

Still, recent developments in photon-counting techniques, such as qubit-based detectors proposed in Ref.~\cite{Dassonneville2020} or ultrasensitive bolometers developed in Ref.~\cite{Kokkoniemi2020}, could represent an important step towards the experimental realization of QI in microwaves. Indeed, in Ref.~\cite{Sanz2017} it was shown that using a local approach based on linear operations and photon-counting the TMSV state obtains the \SI{3}{\deci\bel} gain mentioned above. A microwave-oriented study based on TMST states, and including decoherence in the idler mode due to imperfections in the delay line is needed in order to experimentally test these ideas. Another challenge for the realization of microwave QI is the fact that the number of pulses $M = T W$, where $T$ is the duration of each signal-idler pulse and $W$ its phase-matching bandwidth, has to be very large ($M \sim 10^6$), because the theoretical advantage of QI is only guaranteed when the Chernoff bound is achieved, and this happens asymptotically with $M$. While this requirement is easily met by optical systems, it may be more challenging with microwaves.
However, the main caveat for a end-to-end microwave demonstration of QI remains: finding a way to perform a joint measurement on the signal and idler, which requires a measurement apparatus at low temperature. 

\subsection{Quantum radar and imaging}\label{subsec:qradar+imaging}
The \SIrange{1}{10}{\giga\hertz} transparency window of the atmosphere
has motivated for many years now the use of microwave frequencies for radar applications. The term `quantum radar' \cite{macconeRadar, lanzagorta, Pirandola2018, KarsaThesis} refers to a theoretical device that would outperform any classical radar by resorting to quantum effects, typically entanglement. For some time, the QI protocol has been the main theoretical candidate for enabling such device, for two reasons: QI works best in the very low average signal photon number $N_S \ll 1$ and high noise regime $N_\text{th} \gg 1$. The first condition makes one think immediately of applying QI to a radar scheme, since radars typically want to detect without being detected; the second, points towards a  \textit{microwave} QI-based quantum radar because the atmosphere is naturally bright enough to meet $N_\text{th} \gg 1$ while being almost transparent: the expected number of photons at $T=300$ K and \SI{5}{\giga\hertz} is roughly 1250. So QI seems to be the perfect fit for enabling a quantum-enhanced radar, capable of obtaining an unprecedented detection precision, or even to `unveil' electromagnetically-cloaked objects, invisible to a classically-conceived apparatus \cite{Lasheras2017}. However, one must be careful before claiming that QI \textit{is} the way to go to construct a quantum radar solving the detection problem. First, a microwave QI-based quantum radar would need to solve the issue of idler storage loss:  just \SI{6}{\deci\bel} of loss in the idler delay line would automatically destroy the quantum advantage. The bound is twice as strict if the local approach providing \SI{3}{\deci\bel} of enhancement is used. Indeed, in Ref.~\cite{shabir} a bound of \SI{11.25}{\kilo\meter} on the maximum target distance was obtained, using an optical fiber delay line and up-conversion schemes. Second, as argued before, the pulse-time/bandwidth product requirement, $TW \gg 1$ appears very challenging with current microwave technology: as argued in Ref.~\cite{Sorelli2020}, the phase matching bandwidth for a microwave signal at \SI{10}{\giga \hertz} is $W \sim 100$ MHz, which gives a time per pulse of $T\sim10$ ms. Third, QI assumes that the location and velocity of the object is known. This translates in each pulse interrogating a single polarization-azimuth-elevation-range-Doppler region at a time. If the strategy is to be split into different bins, the quantum advantage decays logarithmically \cite{shapiroStory}. Still, $T\sim$ 10 ms is far too long to safely assume that the target has not moved. Fourth, even if all the above difficulties are circumvented,  microwave QI will still require cryogenics, which significantly increases the payload of the radar, something that may not be justified by the modest advantage of QI over classical target detection. A synthetic aperture quantum radar (SAqR) could find microwave imaging applications in high-orbit satellite-based stations, where the cooling power needed to achieve cryogenics is greatly reduced by the already cold temperature of space (2.7 K). We thus conclude that a commercial microwave quantum radar based solely on quantum illumination for target detection will not appear in the near future. Still, an inter-fridge experimental proposal for QI, where a TMST state is generated inside one of the cryostats, sending one mode through a cryolink equipped with controlled thermal noise, and interacting with a low-reflectivity target at the other cryostat, may allow to better test the limitations of QI in the microwave regime.
Other proposals for quantum radar, like the one in Ref.~\cite{macconeRadar} seem rather impractical in the microwave regime, due to the high sensitivity of the protocol to loss and noise. This is not the case of the Doppler quantum radar presented in Ref.~\cite{Reichert2022}. 
There, a signal beam is sent towards and reflected by a moving target, which shifts the signal frequency due to the Doppler effect. By measuring the frequency of the reflected signal, the radial velocity can be inferred. The protocol uses  signal-idler beams, with both frequency entanglement and squeezing as enhancing quantum resources. Contrary to QI, the protocol's quantum advantage grows with increasing signal power for most parameter regimes, that is, it beats the SQL and even reaches the Heisenberg limit for the majority of parameters. Another advantage over QI is that the idler beam does not need to be stored, which as we have argued can make QI problematic for real-life target detection in certain frequency intervals. The main results are for the lossless and noiseless case. However, a study of loss for the high-squeezing regime has shown that the protocol is loss resilient: a constant factor quantum advantage is achieved, higher than 3 dB in the variance of the estimator, given a roundtrip lidar-to-target-to-lidar transmissivity larger than 50\%, which makes the protocol promising for short range applications. The main challenge for a realization of the protocol in the microwave regime is the implementation of the optimal measurement: a frequency-resolved photon counter for the signal and idler beams. Broad-band microwave photon-counters are thus required. The protocol can be adapted to range estimation, for which the optimal measurement is the arrival time of the individual signal and idler photons, which may be easier to implement for microwaves.
In this line, some recent work has focused on using QI for the ranging problem, with encouraging theoretical results of an improvement of $\mathcal{O}(10)$\,dB over the classical strategy \cite{Zhuang2021}. Despite all these important efforts, the gap between theory proposals and actual commercial devices is currently too big to expect operative microwave quantum radars in the near future.

`Quantum imaging' encompasses any technique benefiting from quantum effects -- such as entanglement, superposition or squeezing -- in order to enhance the contrast or resolution of an image. An { important} example is ghost imaging~\cite{simon2016}.
Biological tissue is mostly comprised of water, limiting microwave penetration depths to few cm. Shorter microwave wavelengths provide lower resolution widths, require smaller antennas, and face lower thermal background, at the cost of lower penetration depths.
To date, radar-like techniques have been successfully applied in diagnose imaging of breast cancer~\cite{fear2002, felicio2019}, benefiting from the low water content of breast fat and distinct dielectric properties between healthy and malignant tissue. 
Quantum advantages in the sensing of dielectric properties could enhance these protocols. A first proof of advantage in the computation of reflectivity gradients, using bi-frequency entangled probes, has been theoretically proposed~\cite{casariego2020}.
A QI-based medical imaging device avoids the first and third objections to the quantum radar in the previous paragraph, with idler storage times 5 orders of magnitude shorter (from km to cm), and a static target. Nevertheless, the high in-tissue attenuation and corresponding low penetration depth is a new obstacle to overcome. 
A transmission image of an object with transparent-opaque contrasts has been obtained with QI, achieving an advantage in the rejection of background noise, in optical frequencies~\cite{gregory}. In principle, this protocol could be adapted to microwave frequencies.

In general, we expect that taking advantage of the quantumness of microwaves in order to perform a quantum imaging protocol will only be justified in systems that can sustain cryogenic temperatures. As an example, recent developments in microwave photon-counting techniques could have an impact in magnetic resonance experimental sensitivities, like the effort of resorting to squeezed states in order to perform electron spin resonance spectroscopy~\cite{Bienfait2017, albertinale2021}.

\subsection{Inference of quantum system-environment interactions}


Advances in propagating quantum microwaves could be further boosted by the sensing of the way the external environment dynamically affects nominal working conditions. In other words, one would ask for sensing protocols able to infer quantum system-environment interactions. In this context, one of the most promising applications is quantum thermometry, that may play a crucial role to improve accuracy in carrying out communication and computing in the quantum regime~\cite{KurpiersPRApp2019,SultanovAPL2021,DanilinArXiv2021, Gasparinetti2021}. It is generally known, indeed, that the unavoidable influence of the environment can lead both to a temperature increase and energy fluctuations in terms of heat losses~\cite{ManzanoPRX2018,BatalhaoChapter}, which is expected to prevent the correct functionality of the physical mechanism one is investigating. In this regard, it is worth mentioning recent studies about irreversible losses in quantum logic gates~\cite{CimininpjQI2020} and quantum annealers~\cite{GardasSciRep2018,BuffoniQST2020}. Of course, a similar evidence also holds if one deals with quantum microwaves. In fact, one may consider an optical quantum memory~\cite{Morton2008,Simon2010} in solid state that is a perfect example of a quantum system, based on microwave transitions, subject to environmental degrees of freedom. A quantum memory usually consists of a $\Lambda$-system with a microwave transition that is used for quantum storage. A generalized  $n$-level $\Lambda$-system is  described in the rotating frame by a Hamiltonian of the form $H=\sum_{k=1}^n \Omega_k \ket{e}\bra{k} + \Omega_k^* \ket{k}\bra{e}$, where $\lbrace \Omega_k \rbrace_k$ are the Rabi frequencies associated with the transition from the $k$-th level $\ket{k}$ to the excited state $\ket{e}$. The name $\Lambda$-system comes from the way the levels are arranged in the simplest case of $n=2$. For its functionality, one needs to reduce the effect of decoherence on the spin systems that compose the memory. Such decoherence mechanisms can have different origins; among the most relevant we recall the interaction between magnetic dipoles, and the coupling of the spins to phonons. In general, for solid state media, dynamical decoupling enhanced magnetometry~\cite{Pham2012,Baumgart2016} already provides information about locally varying magnetic fields causing decoherence. In such a framework, we are confident that decoherence and relaxation processes can be generally inferred by means of quantum sensing protocols. From this point of view, quantum thermometry itself can be seen effectively as a branch of quantum sensing in the sense that one aims at deciding whether the quantum system of interest is in contact, or not, with the external environment that leads to thermal fluctuations.

With quantum propagating microwaves in mind, we propose a quantum estimation strategy with a direct application to thermometry, by taking inspiration from recent results in Refs.~\cite{RossiPRL2020,MontenegroPRR2020,SongPRA2021}. Our proposal is based on using a cavity-system that is resonantly driven by a coherent laser pulse. The quantum system that one aims to investigate is inserted in the cavity. The system is assumed to be in contact with an external environment exhibiting macroscopic features, as for example a thermal bath. The interaction between the quantum system and the environment leads to fluctuations of physical quantities, such as energy and/or temperature. Here, it is worth observing that the assumption of considering the presence of the environment (within the open quantum system framework) stems from our impossibility to isolate the dynamics of the quantum process of interest. One of the main reasons for that is the inaccessibility of the latter, which may be physically embedded within a body with larger dimension, \textit{e.g.}, a mesoscopic system. Concrete examples are actually provided by quantum systems with microwave transitions, as rare-earth ions in solids~\cite{Tittel2010,ZambriniCruzeiro2017}, nitrogen vacancy (NV) centers~\cite{Shim2013,Heshami2014}, silicon vacancy centers~\cite{Sukachev2017}, or even molecules~\cite{Rabl2006}. 

For all these quantum systems, the scope of the cavity-based setup is to scan -- and possibly reconstruct -- the way the system is in contact with the environment, and how it leads the system to relax towards a steady-state. The thermalization towards a steady-state with a well-defined temperature is a special case of the latter. We recall that in the case of systems with microwave transitions in solid media, this kind of sensing strategy is expected to be useful in studying the spin lattice relaxation process~\cite{Orbach1961}, which is important to improve quantum information processing devices like quantum memories~\cite{ZambriniCruzeiro2017}. Specifically, the scanning using a cavity-system is enabled by performing a continuous sequence of weak measurements, without completely destroying coherence in the measurement basis~\cite{WallraffNature2004,MurchNature2013,TanPRL2015}. The output field from the cavity is then physically monitored by means of a detector that depends on the specific quantum measurement observable. For example, in~\cite{RossiPRL2020} a homodyne receiver is employed, while in~\cite{SongPRA2021} energy projective measurements are performed. In the context of propagating quantum microwaves, it is already known that homodyne detection can be employed. However, as pointed out in Ref.~\cite{TycJPAMG2004}, also a projective measurement of energy might be implemented, thanks to the evidence that a balanced homodyne detection, under specific conditions, can reproduce the effect of performing projective measurements of the quadrature phase of the output signal field (\textit{i.e.}, CVs Bell measurements). Furthermore, one could be also interested in addressing quantum sensing in non-Gaussian regimes, for which non-Gaussian measurements are required. For such a task, homodyne detection has to be replaced or supplemented by photon-counting~\cite{QiOE2020}; in this context, the photon-counting device in~\cite{Wang2019}, for the sensing of microwave radiation at the sub-unit-photon level, could be a strong candidate.

Overall, the proposed method based on continuous quantum measurements is expected to efficiently acquire information about the environment to which the analyzed system is in contact (or in which it is embedded) by making use of a sequence of outcomes that are continuously recorded by the monitoring process. Such an information is granted, at the price of taking into account both the quantum measurement back-action (below denoted as ``mba'') and the stochastic contribution (``stoc'') to the dynamics induced from conditioning upon the measurement records~\cite{TanPRL2015,RossiPRL2020,SongPRA2021}, described respectively by the super-operators $\mathcal{L}_{\rm mba}$ and $\mathcal{L}_{\rm stoc}$. Refer to Refs.~\cite{TanPRL2015,RossiPRL2020} for a detailed description of the latter that holds independently on the system one monitors (in our case, a mechanism related to propagating quantum microwaves as for example the one detailed below when discussing the system operator $\hat{A}$). Hence, formally one has to consider the conditional dynamics of the quantum state $\rho$, governed by a \emph{stochastic master equation} of the form 
\begin{equation}\label{eq:stoc_master_eq}
    \diff \rho = \left(\mathcal{L}_{\rm bath} + \mathcal{L}_{\rm mba} + \mathcal{L}_{\rm stoc}\right)\rho \, \diff t \,. 
\end{equation}
In Eq.~(\ref{eq:stoc_master_eq}), $\mathcal{L}_{\rm bath}$ denotes the super-operator modelling the interaction between the quantum system and the bath (with macroscopic features) in its surrounding, since for the sake of simplicity we are here considering the particular case that the environment is a thermal bath. A common expression for $\mathcal{L}_{\rm bath}\rho$ is the following:
\begin{equation}\label{eq:L_bath}
\mathcal{L}_{\rm bath}\rho = \Gamma \left(n_{\rm bath} + 1\right)\mathcal{D}[\hat{A}]\rho + \Gamma\, n_{\rm bath} \mathcal{D}[\hat{A}^{\dag}]\rho 
\end{equation}
where $\Gamma$ denotes the energy damping rate and $n_{\rm bath}$ is the bath occupancy. Moreover, in Eq.~(\ref{eq:L_bath}), $\mathcal{D}[\hat{A}]\rho \equiv \hat{A}\rho\hat{A}^{\dag} - \{\hat{A}^{\dag}\hat{A},\rho\}/2$ (usual super-operator in Lindblad form with $\{\cdot,\cdot\}$ denoting the anti-commutator), and $\hat{A}$ is the quantum system operator on which the bath acts. A concrete example for the operator $\hat{A}$, involving quantum microwaves, can be found in Ref.~\cite{NorambuenaPRB2018} where the spin-lattice relaxation of individual solid-state spins in diamond NV centers is studied. In fact, it is there shown that dissipative spin-lattice dynamics -- induced both by phonon interactions and noise due to magnetic impurities -- act isotropically on the angular momentum operators that rule the spin transitions along each space coordinates. Thus, once known the system-bath interactions (even on average), the goal of the sensing strategy we are here proposing is to extract information about $\Gamma$ and $n_{\rm bath}$ from the weak measurement record $r(t)$ (time-varying signal) that is measured as output field of the cavity-system. A possible expression for $r(t)$, valid also if we deal with quantum microwaves, is $r(t)={\rm Tr}\left[\rho(t)\sigma_z\right] + \diff\mathcal{W}$ with $\diff\mathcal{W}$ denoting the zero-mean Gaussian distributed Wiener increment and $\sigma_z$ the Pauli matrix $Z$. This procedure, which well fulfills with the present roadmap, will be the subject of a forthcoming research activity. 

To conclude, we stress that this formalism can take into account also fluctuation terms, since it is able to directly scan an open quantum dynamics at the single-trajectory level. Hence, by means of such a high degree of resolution, it is expected to provide the sufficient amount of information to: \textit{(i)} infer how a quantum system physically interacts with the external environment, and \textit{(ii)} reconstruct some key bath parameters, as for example $\Gamma$ and $n_{\rm th}$ in the model of Eq.\,(\ref{eq:stoc_master_eq}). Finally, notice also that the proposed method works for both discrete and continuous variable quantum systems. 

\subsection{Direct Dark Matter detection}\label{subsec:DarkMatter}
To illustrate the reach of microwave quantum sensing, we close with a discussion of novel ways to look for 
dark matter candidates.
Dark matter is predicted to make up to 85\% of the matter in the Universe, based on gravitational observations compatible with General Relativity. If interpreted within the context of particle physics theories, these imply a wide range of possible masses for dark matter particles. For example, currently many new technologies are pursued in order to search for Weakly Interacting Massive Particles (WIMPs) with masses in the meV-GeV range; for a recent review, see Ref.~\cite{Essig2022}.

Within the context of this paper, two especially interesting dark matter candidates are axion like particles and dark photons. The former is originally motivated as a solution to the strong CP problem in the Standard Model of elementary particle interactions, while the latter generally arises when extending the Standard Model with a new U(1) gauge symmetry. 

The axion is predicted to have a very weak coupling with electromagnetism, and this allows to search for a weak narrow band signal at a frequency corresponding to the unknown mass of the axion in cavity based detectors known as haloscopes. The dark photon, on the other hand, is constrained by observation to have only extremely weak coupling with electromagnetism and axion haloscopes can therefore be used to search for dark photons, as well. Haloscopes have been used to search for axionic dark matter in the \SIrange{1.8}{24}{\micro e\volt} mass range~\cite{Semertzidis2022,Chadha-Day2022}, and the results have also been applied to constrain dark photons in this mass range~\cite{Ghosh:2021ard}.

This method, however, is mostly hindered by the fact that these searches can be slow and that it is difficult to overcome the standard quantum limit, where cavity measurements introduce noise due to quantum uncertainty.

Several improvements for this technique are now being considered. These may bring relevant results not only for axion searches,
but also for general metrology beyond the standard quantum limit.
Backes et al.~\cite{Backes2021} introduce quantum enhancements to the HAYSTAC, which allows for a small speed improvement in the search
and exploration beyond the quantum limit with the microwave-frequency field prepared in a squeezed state. Wurtz et al.~\cite{Wurtz2021} suggest and implement a way of amplifying the axion signal through mode squeezing and state swapping interactions before any noise from the quantum limit actually contaminates the signal. 
Dixit et al.~\cite{Dixit2021} introduce a new method for cavity searches which circumvents the quantum uncertainty generated by performing measurements in
the cavity by instead performing a photon counting technique which does not destroy the photon. This is possible with the use of a superconducting qubit, as discussed above, and they demonstrate the technique for a dark photon search.
Recent developments in ultrasensitive bolometers \cite{lowNEP,Kokkoniemi2020, Lee2020} and calorimeters \cite{Karimi2020} show promise in overcoming the quantum noise in the cavity measurements especially in the higher mass end \cite{Lamoreaux2012}, as they have demonstrated sensitivities with potential for single photon detection down to tens of GHz frequencies. These detectors are also particularly appealing candidates for experiments that rely on photon-counting due to their ability to absorb radiation from a wide band. A plan for a new generation of haloscope experiments employing thermal detectors for axion search in the mass range of $10^{-3}$--1~\SI{}{e\volt} has been proposed~\cite{LiuAxion2021}. 

The profound observable for precision cosmology is the cosmic microwave background (CMB). The main focus of current observational efforts is in the so called polarization B-modes, as their observation would provide evidence for the exponential expansion (inflation) of the very early Universe. Bolometric interferometry to measure these modes has been proposed in several projects currently combined in the QUBIC experiment~\cite{QUBIC:2020kvy}. Precision measurements of the CMB may also be relevant for axion searches: oscillations of the axion field have recently been predicted to lead to two different effects in the polarization of the CMB \cite{Fedderke2019}. A uniform reduction of the polarization is predicted to result from multiple oscillations of the axion field occuring during the CMB decoupling epoch in the early universe. On the other hand, present day oscillations of the axion field are predicted to lead to a real-time AC oscillation of the polarization of the CMB. The authors of \cite{Fedderke2019} state that observing the latter effect would be a particularly convincing evidence for the existence of axions in the lowest mass range, but these experiments require dedicated time-series analysis of the CMB signal. Compared with the detectors used in the previous CMB experiments \cite{Pirro2017}, bolometers with improved sensitivities may help in faster mapping of the CMB background across the sky.

\section{Conclusions}\label{sec:Conclusions}
 \begin{figure*}[t!]
\includegraphics[width=\linewidth]{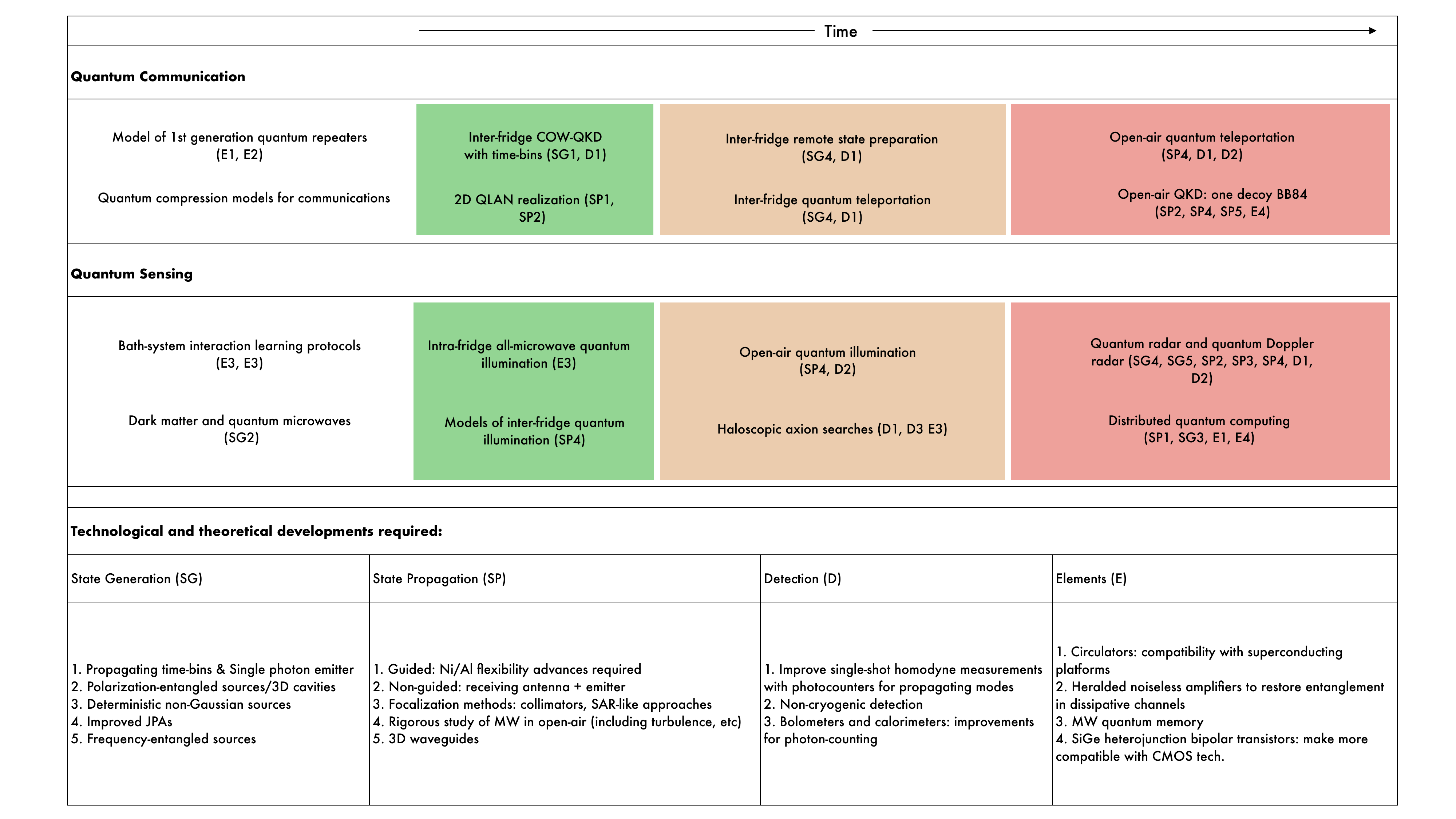}
\caption[Quantum microwaves communication and sensing roadmap for the upcoming years.]{Quantum communication and sensing roadmap for the upcoming years. The first part of the table consists in the quantum communication and sensing protocols and results we think should be addressed in the mid-term future. The first box on the left contains proposals for experiments that are typically on an early stage, for example the quantum microwaves-dark matter detection interplay we discussed in Section \ref{subsec:DarkMatter}. Then, three colored boxes contain experimental proposals with an increasing difficulty factor to the right. In green, we have experiments that can already be performed, with very minor adjustments to current technology; in orange, we include experiments that should become proof-of-principle results for future real-life technologies and applications of quantum microwaves. In red, finally, we have included some of the necessary results discussed along this manuscript, but that we think will require years. Importantly, the second part of the table consists on technical results, roughly divided by quantum-mechanical categories, that will enable the technologies described before. To link the two tables, we have included which are the requirements of each of the quantum communication and sensing experiments, hoping to compress the ideas and to trigger motivation for doing more research in the field. { To this aim, we have used a compressed notation: for example, Inter-fridge remote state preparation requires SG4 and D1, which stand for item 4 of the `State generation'  box (\textit{Improved JPAs}), and item 1 of the `Detection' box (\textit{Improve single-shot homodyne measurements with photocounters for propagating modes}).} }
\label{fig:roadmap}
\end{figure*}
In this manuscript we have addressed the challenges that the field of propagating quantum microwaves faces in order to find real-life applications in quantum communications and sensing. To this end, we started with a state-of-the-art description of the essential stages that are a common denominator for both communications and sensing protocols, such as state generation (including entanglement rates), guided and non-guided propagation, amplification, beam splitting, detection schemes and photon-counting devices, among others. Along the way, we pointed the aspects that require further research, both from a theoretical and the experimental point of view. Then, we discussed quantum communications in length, with a focus on the need for both continuous variables and discrete variables implementations of microwave quantum key distribution, for which the existence of an operative quantum local area network (QLAN) making use of cryolinks to connect remote cryogenic refrigerators will represent a major step towards proof-of-principle experiments. A discussion on the scalability of such QLANs and their integration within the quantum internet followed, concluding that both an efficient microwave-to-optical, and a fully-microwave platforms are needed and expected to coexist. Then we moved on to discuss quantum sensing, where quantum illumination and its relation to quantum radar were treated. Here, the conclusions were, first, that a fully-microwave demonstration of quantum illumination is still missing but the recent advances in photon-counters for propagating microwaves will surely solve this, and second, that a quantum radar based solely on this protocol is still far of reach. A brief discussion of some areas where quantum imaging protocols may benefit from the new results in photon-counting followed, though only in scenarios where the system can sustain a cryogenic temperature, ruling out medical imaging. Finally, we proposed two novel ideas in two fields that are wildly apart: thermometry or, more generally, the inference and characterization of system-environment interactions, and axionic dark matter search. In the first, we argued that a continuous sequence of weak measurements making use of the recent advances in both single-shot homodyne detection and photon-counting could lead to an efficient method for the quantum estimation of parameters such as temperature, or the expected number of photons present in a thermal bath. As for axionic dark matter, we investigated two novel ways to infer the existence of the elusive particle: one where the axion may generate  detectable microwave photons, and other where its interaction with the cosmic microwave background leads to two measurable effects in the polarization. 

As a closing remark, we stress what are in our view the most urgent needs for the future of applicable quantum microwave technologies: to solve the impedance-matching problem that the open-air transmission poses in order to design quantum-capable emission and reception antennae; to rigorously investigate the quantum-channel capacity of the atmosphere from a quantum-theoretic perspective, in order to identify the scenarios where quantum microwaves are expected to beat the telecom frequencies (visible and near IR), and taking into account the high entanglement production rates that come from the strong interactions of microwaves with non-linear devices; to make a proof-of-principle demonstration of a QLAN with operative superconducting chips for distributed quantum computing; and to realistically assess the validity and usefulness of microwave quantum illumination for quantum radar by making an open-air experimental proposal. To summarize, in Figure~\ref{fig:roadmap} we have condensed the results of the manuscript in a roadmap fashion, hoping to give a bigger picture and to trigger new ideas and research in this fascinating field of propagating quantum microwaves.

\bibliographystyle{ieeetr}
\addcontentsline{toc}{chapter}{Bibliography}
\UseRawInputEncoding
\bibliography{bibliography/biblio_UTF-8.bib}

\appendix



\chapter{Coefficients for the optimal quantum observable}\label{appendix:OptimalObs}
Here we give the general expressions of the coefficients of the optimal observable for the TMS state:
\begin{equation}
\hat{O}_Q = L_{11}\hat{a}_1^\dagger \hat{a}_1 + L_{22}\hat{a}_2^\dagger \hat{a}_2 + L_{12}\left( \hat{a}_1^\dagger \hat{a}_2^\dagger + \hat{a}_1 \hat{a}_2\right) + L_0 \Id_{12},
\end{equation}
where the coefficients can be seen in the next page.
\begin{landscape}
{\tiny
\begin{align*}
\begin{split}
L11 &=-\frac{2 \eta _1 N_\text{S} \left(2 N_\text{S}+1\right) \left(2 N_{\text{th}}+1\right)}{-A+B-C+D}\\
L22 &= \frac{4 \eta _1 \left(2 \eta _1-1\right) N_\text{S}^2 \left(2 N_{\text{th}}+1\right)+2 N_\text{S} \left(\eta _1-2 N_{\text{th}} \left(\left(\eta _1-3\right) \eta _1+\left(\eta _1-1\right) \left(3 \eta _1-1\right) N_{\text{th}}+1\right)-1\right)+N_{\text{th}} \left(2 \left(\eta _1-1\right) N_{\text{th}} \left(\left(\eta _1-1\right) N_{\text{th}}-1\right)+1\right)}{A-B+C-D}\\
L12 &= -\frac{\sqrt{2} \sqrt{N_\text{S} \left(2 N_\text{S}+1\right)} \left(\eta _1^2 \left(N_\text{S} \left(4 N_{\text{th}}+2\right)-N_{\text{th}}^2\right)+N_{\text{th}} \left(N_{\text{th}}+1\right)\right)}{A-B+C-D}\\
L_0 &=-\frac{4 \eta _1^3 \left(N_{\text{th}}^2-2 N_\text{S} \left(2 N_{\text{th}}+1\right)\right){}^2-2 \eta _1^2 \left(6 N_{\text{th}}+5\right) F \left(N_\text{S} \left(4 N_{\text{th}}+2\right)-N_{\text{th}}^2\right)+4 \eta _1 \left(N_{\text{th}}+1\right) \left(N_\text{S}^2 \left(8 N_{\text{th}}+4\right)-2 N_\text{S} \left(N_{\text{th}} \left(6 N_{\text{th}}+5\right)+1\right)+N_{\text{th}}^2 \left(3 N_{\text{th}}+2\right)\right)+\left(2 N_{\text{th}}+3\right) G F}{E-4 \eta _1 \left(2 N_{\text{th}}+1\right) F \left(-4 N_\text{S} N_{\text{th}}+N_\text{S} \left(2 N_\text{S}-1\right)+N_{\text{th}}^2\right)-8 N_\text{S}^2 \left(3 N_{\text{th}} \left(N_{\text{th}}+1\right)+1\right)+4 N_\text{S} N_{\text{th}} \left(N_{\text{th}} \left(4 N_{\text{th}}+3\right)+1\right)-2 N_{\text{th}}^2 G},
\end{split}
\end{align*}
}%
\end{landscape}
where 
\begin{align*}
\begin{split}
A &\equiv 8 \left(\eta _1-1\right) \eta _1 N_\text{S}^3 \left(2 N_{\text{th}}+1\right)\\
B &\equiv 4 N_\text{S}^2 \left(-\eta _1+\left(\eta _1+3 \eta _1 N_{\text{th}}\right){}^2-\eta _1 N_{\text{th}} \left(10 N_{\text{th}}+7\right)+3 N_{\text{th}} \left(N_{\text{th}}+1\right)+1\right)\\
C &\equiv 2 N_\text{S} N_{\text{th}} \left(-\eta _1+N_{\text{th}} \left(\eta _1 \left(3 \eta _1-8\right)+4 \left(\eta _1-1\right) \left(2 \eta _1-1\right) N_{\text{th}}+3\right)+1\right)\\
D &\equiv N_{\text{th}}^2 \left(2 \left(\eta _1-1\right) N_{\text{th}} \left(\left(\eta _1-1\right) N_{\text{th}}-1\right)+1\right)\\
E &\equiv 4 \eta _1^2 \left(-4 N_\text{S} N_{\text{th}}+N_\text{S} \left(2 N_\text{S}-1\right)+N_{\text{th}}^2\right) \left(N_\text{S} \left(4 N_{\text{th}}+2\right)-N_{\text{th}}^2\right)\\
F &\equiv 2 N_\text{S}-N_{\text{th}}\\
G &\equiv  2 N_{\text{th}} \left(N_{\text{th}}+1\right)+1
\end{split}
\end{align*}
In the high reflectivity case $\eta_1 \rightarrow 1$ we find:
\begin{eqnarray*}
\lim_{\eta_1\rightarrow 1}L_{11} &=-\frac{2 N_\text{S} (2 N_\text{S}+1) (2 N_\text{th}+1)}{N_\text{S}^2 (8 N_\text{th} (N_\text{th}+1)+4)+4 N_\text{S} N_\text{th}^2+N_\text{th}^2}\\
\lim_{\eta_1 \rightarrow 1}L_{22} &=-\frac{4 N_\text{S} (2 N_\text{S} N_\text{th}+N_\text{S}+N_\text{th})+N_\text{th}}{N_\text{S}^2 (8 N_\text{th} (N_\text{th}+1)+4)+4 N_\text{S} N_\text{th}^2+N_\text{th}^2}\\
\lim_{\eta_1 \rightarrow 1}L_{12} &=\frac{2 \sqrt{2} \sqrt{N_\text{S} (2 N_\text{S}+1)} (N_\text{S} (4 N_\text{th}+2)+N_\text{th})}{N_\text{S}^2 (8 N_\text{th} (N_\text{th}+1)+4)+4 N_\text{S} N_\text{th}^2+N_\text{th}^2}\\
\lim_{\eta_1 \rightarrow 1}L_0 &=\frac{-2 N_\text{S} (N_\text{S} (8 N_\text{th}+4)+6 N_\text{th}+1)-3 N_\text{th}}{8 N_\text{S}^2 (2 N_\text{th} (N_\text{th}+1)+1)+8 N_\text{S} N_\text{th}^2+2 N_\text{th}^2}
\end{eqnarray*}
Additionally, as shown in the main text, in the noiseless case we get
\begin{align*}
 \hat{O}_Q ^{\text{Lim}}:=\lim_{\substack{N_\text{th}\rightarrow 0 \\ \eta_1 \rightarrow 1}} \hat{O}_Q 
 &=
-\mu^2\hat{a}_1^\dagger \hat{a}_1 - \hat{a}_2^\dagger \hat{a}_2 
+ \mu\left( \hat{a}_1^\dagger \hat{a}_2^\dagger + \hat{a}_1 \hat{a}_2\right)  -\nu \Id_{12}\\
&\equiv \hat{b}_1^\dagger \hat{b}_1 - 1,
\end{align*}
where $\mu^2 \equiv \left(1+1/2 N_\text{S}\right)$ and $\nu \equiv \left(1+1/4 N_\text{S}\right)$ and  $
\hat{b}_1\equiv -i \left(\hat{a}_2^\dagger - \mu\hat{a}_1\right)$.

\if\includeGlossary 1
\printglossary
\fi

\end{document}